\documentclass[a4paper,12pt]{article}
 
\PassOptionsToPackage{original}{pict2e}
\usepackage{amsmath}
\usepackage{amssymb}
\usepackage{amsfonts}
\usepackage{mathrsfs}
\usepackage{bm}
\usepackage{mathtools}
\usepackage{geometry}
\usepackage{dsfont}
\usepackage{lmodern}
\usepackage{calc}
\usepackage[english]{babel}
\usepackage{color}
\usepackage{xcolor}
\usepackage{upgreek}
\usepackage{graphicx,pspicture}
\usepackage{subfig}
\usepackage{psfrag}
\usepackage{afterpage}
\usepackage{float}
\usepackage{array}
\usepackage{tabularx}
\usepackage{arydshln}
\usepackage{dashrule}
\usepackage[margin=10pt,font=small,labelfont=bf,labelsep=period]{caption}
\usepackage{upref} 
\usepackage{cite} 
\usepackage[bottom]{footmisc} 
\usepackage{indentfirst} 
\usepackage{textcomp}
\usepackage{hyperref} 
 
\definecolor{darkred}{rgb}{0.5,0,0}
\definecolor{lightblue}{rgb}{0.5,0.5,1}
\definecolor{darkblue}{rgb}{0,0,0.5} 

\numberwithin{equation}{section}

\renewcommand{\baselinestretch}{1.258}
\date{\vspace{-5ex}}
\include{Definitions} 

\begin{document}

\begin{titlepage}
 \enlargethispage{1cm}
\renewcommand{\baselinestretch}{1.1} 
\title{\begin{flushright}
\vspace{-2cm}
\hspace{2cm}
\normalsize{MITP/14-027}
\end{flushright}
En route to Background Independence:\\
Broken split-symmetry, and how to restore it\\ with bi-metric average actions}
\author{Daniel Becker and Martin Reuter\\
{\small Institute of Physics, University of Mainz}\\[-0.2cm]
{\small Staudingerweg 7, D-55099 Mainz, Germany}}
\maketitle\thispagestyle{empty}
\begin{abstract} 
 The most momentous requirement a quantum  theory of gravity must satisfy is Background Independence, necessitating in particular  an ab initio derivation of the arena all non-gravitational physics takes place in, namely spacetime.
Using the background field technique, this requirement translates into the condition of an unbroken split-symmetry  connecting the (quantized) metric fluctuations to the (classical) background metric. 
If the regularization scheme used violates split-symmetry during  the quantization process it is mandatory to restore it in the end at the level of observable physics.
In this paper we present a detailed investigation of split-symmetry breaking and restoration within the Effective Average Action (EAA) approach to Quantum Einstein Gravity (QEG) with a special emphasis on the Asymptotic Safety conjecture.
In particular we demonstrate for the first time in a non-trivial setting that the two key requirements of Background Independence and Asymptotic Safety can be satisfied simultaneously.
Carefully disentangling fluctuation and background fields, we employ a `bi-metric' ansatz for the EAA and project the flow generated by its functional renormalization group equation on a truncated theory space spanned by two separate Einstein-Hilbert actions for the dynamical and the background metric, respectively.
A new  powerful method is used to derive the corresponding renormalization group (RG) equations for the Newton- and cosmological constant, both in the dynamical and the background sector.
We classify and analyze their solutions in detail, determine their fixed point structure, and identify an attractor mechanism which turns out instrumental in the split-symmetry restoration.
We show that there exists a subset of RG trajectories which are both asymptotically safe and split-symmetry restoring:
In the ultraviolet they emanate from a non-Gaussian fixed point, and in the infrared they loose all symmetry violating contributions inflicted on them by the non-invariant functional RG equation.
As an application, we compute the scale dependent spectral dimension which governs the fractal properties of the effective QEG spacetimes at the bi-metric level.
Earlier tests of the Asymptotic Safety conjecture almost exclusively employed `single-metric truncations' which are blind towards the difference between quantum and background fields.
We explore in detail under which conditions they  can be reliable, and we discuss how the single-metric  based picture of Asymptotic Safety needs to be revised in the light of the new results.
We shall conclude that the next generation of truncations for quantitatively precise predictions (of critical exponents, for instance) is bound to be of the bi-metric type. 
\end{abstract}
\end{titlepage}

\setcounter{page}{1}
\pagenumbering{roman}
\tableofcontents
\clearpage
\setcounter{page}{1}
\pagenumbering{arabic}
\section{Introduction}
One of the key requirements every candidate for a quantum theory of the gravitational interaction and spacetime geometry should satisfy is {\it Background Independence}. 
The theory's basic kinematical rules and dynamical laws should be formulated without reference to any distinguished spacetime such as Minkowski space, for instance. 
Rather, the possible states of a `quantum spacetime' should be a prediction of the theory. 
In addition, it must provide us with a set of special observables which, by means of  their expectation values in a given state, `interpret'  this state in terms of classical geometry, or a generalized notion thereof. Among them the expectation value of the metric would play a significant role. 
If non-degenerate, smooth and approximately flat, on large length scales at least, the underlying state might appear like a classical spacetime macroscopically, possibly similar to the real Universe we live in. 
We can then try to match the predictions against concrete measurements and observations.\cite{Kiefer,A,R,T,hamber,pad-book,pad-review}

However, in general one would also expect states without any interpretation in terms of concepts from classical General Relativity, Riemannian geometry in particular. 
A simple example are situations in which the metric has an expectation value which is degenerate, identically vanishing, for instance.
While the gravitational physics implied by such states is certainly very different from the one we know, they might realize a `symmetric phase' of gravity which arguably is easier to understand than the broken phase we live in. 
In the latter, diffeomorphism symmetry is broken down to the stability group of the metric expectation value, the Poincar\'{e} group in the flat case.

There exist two fundamentally different approaches to deal with the requirement of Background Independence.
They differ in particular in the way they deal with the rather severe conceptual and technical difficulties which are caused by this requirement and are of a kind never encountered in conventional matter field theories on Minkowski space:

\noindent {\bf(i)} The most obvious strategy is to literally employ no background geometry at all in setting up the foundational structures of the theory, then work out its quantum dynamical consequences, and try to find states on which appropriate geometric operators (metric etc.) signal the existence of almost classical spacetimes at the expectation value level.  
Examples of such literally Background Independent settings include statistical field theory models like the Causal Dynamical Triangulations (CDT) \cite{CDT}, and Loop Quantum Gravity (LQG) \cite{A,R,T}. 
Following this route, there are (at least) two characteristic difficulties one has to cope with. 
First, since most of the traditional tools of quantum (field) theory presuppose a rigid background metric, considerable conceptual problems must be overcome. 
The proposed solutions are usually outside the realm of quantum field theory. 
Often they encode the fundamental degrees of freedom in new types of variables which replace the quantum fields, triangulated spaces or spin-foams being typical examples.
Second, at the computational level the difficulty arises that an {\it ab initio} explanation of a (non-degenerate, smooth, approximately flat) macroscopic metric has to bridge an enormous gap of scales if one starts from a microscopic theory governing the `atoms of geometry'. For Planck sized building blocks of spacetime, say, even the typical scales of particle physics are about 20 orders of magnitude away. This calls for an application of Wilson's renormalization group but, again, most of the existing tools are inapplicable in the background-free context \cite{bianca}.

\noindent{\bf (ii)} Rather than using literally no background at all,  `Background Independence' can also be achieved in the diametrically opposite way, namely by actually taking advantage of background structures in formulating and evaluating the theory, but making sure that {\it all possible backgrounds are treated on a completely equal footing}. This latter requirement implies in particular that the background geometry may appear only in intermediate steps of the calculations while predictions for observable quantities may never depend on it.\cite{dewittbook}

This second approach is realized for instance by the Effective Average Action for gravity \cite{mr}, a scale dependent analogue of the standard effective action. In the simplest case of metric gravity it involves a background-quantum field split of the dynamical metric $\hat{g}_{\mu\nu}$, the integration variable in the underlying functional integral, in the form $\hat{g}_{\mu\nu}=\bar{g}_{\mu\nu}+\hat{h}_{\mu\nu}$. 
Here $\bar{g}_{\mu\nu}$ is the arbitrary classical background metric, and $\hat{h}_{\mu\nu}$ a nonlinear fluctuation which plays the role of the quantum field, i.e. it is functionally integrated over (or promoted to an operator, in the canonical formulation).
Thanks to the presence of the arbitrary but, by assumption, non-degenerate background metric $\bar{g}_{\mu\nu}$ the original problem appears in a new, somewhat different guise now: it consists in the conceptually easier task of quantizing the, now matter-like field $\hat{h}_{\mu\nu}$ `living' on the classical background geometry given by $\bar{g}_{\mu\nu}$. 

At this point the technical challenge resides in the fact that this background is completely generic and enjoys no special symmetries in particular. 
However, the availability of $\bg_{\mu\nu}$ opens the door for the application of a considerable arsenal of quantum field theory techniques, namely basically all those that have been developed for matter systems on Minkowski space or on curved classical backgrounds. 
The price we pay for this enormous advantage is that {\it in the $\hat{h}_{\mu\nu}$-theory we must have complete control over the $\bg_{\mu\nu}$-dependence of all expectation values}, the $n$-point functions of $\hat{h}_{\mu\nu}$ in particular. 
Their 1PI version, with an infrared (IR) cutoff at the scale $k$, is generated by the Effective Average Action $\EAA_k[\flcb_{\mu\nu};\bg_{\mu\nu}]$. 
As usual this functional depends on the expectation value field $\flcb_{\mu\nu}\equiv\langle \hat{h}_{\mu\nu}\rangle$, but also the background metric, $\bg_{\mu\nu}$, which here acquires the status of an indispensable second argument of $\EAA_k$. The expectation value of the full metric is
\begin{align*}
 g_{\mu\nu}\equiv \langle \hat{g}_{\mu\nu} \rangle= \bg_{\mu\nu}+\langle \hat{h}_{\mu\nu}\rangle \equiv \bg_{\mu\nu}+ \flcb_{\mu\nu}
\end{align*}
Sometimes it is more natural to consider $g_{\mu\nu}$ and $\bg_{\mu\nu}$, rather than the pair $(\flcb_{\mu\nu},\bg_{\mu\nu})$ as the independent variables on which the Effective Average Action (EAA) depends, and to set 
\begin{align*}
 \EAA_k[g_{\mu\nu},\bg_{\mu\nu}]\equiv \EAA_k[\flcb_{\mu\nu};\bg_{\mu\nu}]\big|_{\flcb=g-\bg}
\end{align*}
In this formulation the intrinsic {\it bi-metric character of the EAA} becomes manifest. We refer to the second argument of $\EAA_k[g_{\mu\nu},\bg_{\mu\nu}]$ as its {\it extra $\bg_{\mu\nu}$-dependence} since, contrary to the $\bg$-dependence within $g_{\mu\nu}\equiv\bg_{\mu\nu}+\flcb_{\mu\nu}$, this `extra' dependence does not combine with $\flcb_{\mu\nu}$ to form a full dynamical metric. 

As is well known, the EAA satisfies a functional renormalization group equation (FRGE) which governs its $k$-dependence \cite{wett-mr}, and in principle the functional integral over $\hat{g}_{\mu\nu}$ can be evaluated indirectly by solving the FRGE instead \cite{elisa1}. 
For this to be possible, and the FRGE to exist in the first place, it is unavoidable to employ a running action functional which admits an arbitrary extra $\bg_{\mu\nu}$-dependence and, in fact, depends also on Faddeev-Popov ghost fields $\Ghx^{\mu}$ and $\GhAx_{\mu}$, respectively: $\EAA_k[g_{\mu\nu},\bg_{\mu\nu},\Ghx^{\mu},\GhAx_{\mu}$]. (See ref. \cite{mr} for further details.)

The approaches {\bf (i)} and {\bf (ii)} have complementary advantages and disadvantages. In {\bf(i)} the strategy of literally avoiding any background, and starting from a `vacuum' state with no spacetime interpretation at all, it is comparatively easy to describe a possible phase of unbroken diffeomorphism invariance. But the corresponding broken phase is a very hard problem since it is due to  the cooperative effect of a huge number of `atoms of spacetime'. 
Conversely, in {\bf (ii)}, the broken phase is the easier one since, at least when the background is chosen self-consistently so that $g_{\mu\nu}=\bg_{\mu\nu}$, the quantum fluctuations can be relatively weak, with vanishing expectation value $\flcb_{\mu\nu}\equiv\langle\hat{h}_{\mu\nu}\rangle=0$.
The symmetric phase, on the other hand, requires huge quantum effects as  $g_{\mu\nu}\equiv \bg_{\mu\nu}+\langle\hat{h}_{\mu\nu}\rangle$ is supposed to vanish identically even though $\bg_{\mu\nu}$ stays always non-degenerate.

\vspace{3mm}
In this paper we shall realize Background Independence in the second way, i.e., loosely speaking, by quantizing the fluctuations of the metric in all possible background spacetimes at a time.
We shall employ the framework of the gravitational EAA in order to explore its renormalization group (RG) evolution with a particular emphasis on how the amount of `extra' $\bg$-dependence the functional $\EAA_k$ suffers from depends on the RG scale $k$.
The extra $\bg$-dependence is an entirely unphysical artifact present only at intermediate computational steps. 
To establish Background Independence it must be turned to zero at the end. In this paper we shall analyze in detail whether this is always possible and how it can be done.

In the present paper, the main application will be to the Asymptotic Safety conjecture according to which Quantum Einstein Gravity (QEG) is non-perturbatively renormalizable at a non-Gaussian RG fixed point \cite{wein,  mr, oliver1,oliver2,oliver3,oliver4,frank1, frank2,frank+friends,frank-fR,frank-ghost,astrid-ghost,frank-sig,vacca,prop,max-pert,max,creh1,creh2,h1-2,h3,jan1,jan2,elisa1,elisa2,elisa-Reconst,MRS1,MRS2,NJP,percacci,percadou,percacci-pagani,JE1,JEUM,QEG+QED,JEe-omega,stefan-frankfrac,litimPRL,andi1,andi-MG,codello,daniel1,daniel-MG}.
The crucial issue we shall be concerned with is whether Asymptotic Safety and Background Independence can be achieved {\it simultaneously}.

The extra $\bg$-dependence is most conveniently discussed in terms of the {\it background quantum split-symmetry,} its breaking, and its restoration. This symmetry transformation changes $\flcb_{\mu\nu}$ and $\bg_{\mu\nu}$ according to 
\begin{align}
 \dEps \flcb_{\mu\nu}=\epsilon_{\mu\nu}(x)\, , \qquad \dEps \bg_{\mu\nu}=-\epsilon_{\mu\nu}(x) 
\label{eqn:int_001}
\end{align}
and it has an obvious action on the functionals%
\footnote{We suppress the ghosts when they are inessential for the discussion. On them, $\dEps \Ghx^{\mu}=0$ and $\dEps \GhAx_{\mu}=0$.}
$\EAA_k[\flcb;\bg]\equiv \EAA_k[g,\bg]$.
Here $\epsilon_{\mu\nu}(x)$ is an arbitrary symmetric tensor field. Clearly, the full dynamical metric $g=\bg+\flcb$ is invariant under $\dEps$, while $\bg$ is not. 
As a consequence, $\EAA_k$ is invariant under split transformations, $\dEps\EAA_k=0$, precisely when the two-metrics functional $\EAA_k[g,\bg]$ is actually independent of its second argument $\frac{\delta}{\delta \bg}\EAA_k[g,\bg]=0$, or equivalently, when $\EAA_k[\flcb;\bg]$ happens to be a functional of the sum of its arguments only, viz. $\bg+\flcb\equiv g$.

In order to understand were the unavoidable extra $\bg$-dependence of the EAA comes from recall that the EAA derives from a functional integral which, after the split $\hat{g}_{\mu\nu}\rightarrow \bg_{\mu\nu}+\flcb_{\mu\nu}$, has the form \cite{mr}
\begin{align}
 \int \mathcal{D}\hat{h}_{\mu\nu} \int \mathcal{D}\xi^{\mu}\mathcal{D}\bar{\xi}_{\mu} \, \exp\Big( -\SW[\bg+\hat{h}]-\SW_{\text{gf}}[\hat{h};\bg] -\SW_{\text{gh}}[\hat{h},\xi,\bar{\xi};\bg] -\Delta\SW_k [\hat{h},\xi,\bar{\xi};\bg]	\Big)
\label{eqn:int_002}
\end{align}
While the (arbitrary) bare action $\SW[\bg+\hat{h}]$ is indeed $\dEps$-invariant, the same is {\it not} true for the gauge fixing term $\SW_{\text{gf}}$, the ghost action $\SW_{\text{gh}}$, and in particular the cutoff action $\Delta\SW_k$; it implements the IR cutoff in the familiar way by mode suppression terms which are quadratic in the quantum fields \cite{wett-mr}. 
This shows that the very concept of an EAA, namely the idea to `coarse grain' (that is, smoothly regularize in the infrared) a gauge-fixed functional integral hinges in a crucial way on the availability of $\bg_{\mu\nu}$: neither the gauge fixing, nor the `cutting out' of the IR modes  necessary to derive a functional RG equation could be implemented without promoting the background metric to an independent entity, different from $g_{\mu\nu}-\flcb_{\mu\nu}$ in general. That this is indeed necessary is easily traced back to the gravitational field's special status among the carriers of the fundamental interactions, namely its very relation to the geometry of spacetime.%
\footnote{Note that there exists no analogous `Background Independence' issue in Yang-Mills gauge theories on Minkowski space. There, an `extra $\bar{A}_{\mu}$-dependence'  can always be avoided in a trivial way, namely by simply not using a gauge fixing condition involving a background gauge field, whereupon no such field will appear anywhere. Not so in gravity: even if we were to give up background gauge invariance and use a $\bg_{\mu\nu}$-free gauge fixing condition, a mode suppression term $\Delta\SW_k$ with the necessary properties  could not even be written down without having a second metric at our disposal. \cite{mr}}

We also recall that the EAA employs a gauge fixing condition which belongs to the special class of {\it background gauges}, as a result of which $\SW_{\text{gf}}+\SW_{\text{gh}}$ is invariant under the so-called background gauge transformations, $\dBgt$, not be confused with the `quantum gauge transformations' to be fixed, of course.
Furthermore, $\Delta\SW_k$ is constructed so as to enjoy the same invariance property.
For a diffeomorphism generated by any vector field $v$, the background gauge transformation $\dBgt$ acts as the Lie derivative $\mathcal{L}_v$ on both the dynamical fields $(\hat{g},\xi,\bar{\xi})$ or $(\hat{h},\Ghx,\GhAx)$, {\it and on $\bg_{\mu\nu}$}.
Ultimately this leads to an EAA which is an invariant functional of its arguments, $\dBgt \EAA_k[\flcb,\Ghx,\GhAx;\bg]=0$, and its RG flow preserves this property.

Indeed, starting from \eqref{eqn:int_002} and defining the EAA functional as in \cite{mr}, one finally arrives at the following FRGE which controls its dependence on $k$ or, equivalently, on the RG-time $t\equiv \ln k$:
\begin{align}
\partial_t \EAA_k[\flcb,\Ghx,\GhAx;\bg] = \frac{1}{2}\Tr \Big[ \big(\EAA_k^{(2)}+ \mathcal{R}_k\big)^{-1} \partial_t \mathcal{R}_k\Big] \label{eqn:int_01}
\end{align}
Here $\EAA_k^{(2)}$ is the Hessian of $\EAA_k$ with respect to $(\flcb,\Ghx,\GhAx)$ at fixed $\bg$, and $\mathcal{R}_k$ is the mode suppression operator defining the quadratic form $\Delta\SW_k$. Symbolically, $\Delta\SW_k\propto \int (\flcb,\Ghx,\GhAx) \mathcal{R}_k  (\flcb,\Ghx,\GhAx)^{\text{T}}$.

In the limit $k\rightarrow0$, where the IR regulator is removed, $\mathcal{R}_k$ and $\Delta\SW_k$ vanish by construction, and as a result $\EAA\equiv \lim_{k\rightarrow0}\EAA_k$ coincides with the ordinary effective action, for the specific (background-type) gauge chosen.

While $\EAA$, like $\EAA_k$ at any $k>0$, is perfectly $\dBgt$-invariant, it is still not $\dEps$-invariant: While one source of split-symmetry violation, the one due to $\Delta\SW_k$, disappears at $k=0$, the other, the gauge fixing and ghost sector, still precludes complete $\dEps$-invariance. However, in a sense, this is a very weak violation since it concerns the gauge modes only, and should disappear, too, upon going on-shell \cite{joos,back}.

Nevertheless, at intermediate scale $k>0$ the RG flow generates in principle all possible, generically $\dEps$-violating functionals $\EAA_k[g,\bg,\Ghx,\GhAx]$ of four independent arguments. They are constrained only by their built-in $\dBgt$-invariance, and proper approximation schemes for solving the functional flow equation such as truncations of theory space must take account of this fact.

To date, almost all available RG studies of the Asymptotic Safety scenario still involve the same type of approximation to the exact EAA which was used very early on as the first testing ground for the gravitational FRGE \cite{mr}, namely a so-called {\it single-metric truncation}. Truncations of this general class project the RG flow implied by the (exact!) equation \eqref{eqn:int_01} on the subspace spanned by functionals of the form
\begin{align}
 \EAA_k[g,\bg,\Ghx,\GhAx]=\EAA_k^{\sm}[g]+ (\SW_{\text{gf}}+\SW_{\text{gh}})[g,\bg,\Ghx,\GhAx]
\label{eqn:int_003}
\end{align}
While $\EAA_k^{\sm}$ is a generic (diffeomorphism invariant) functional of the dynamical metric only, $\SW_{\text{gf}}$ and $\SW_{\text{gh}}$ are the classical gauge fixing and ghost actions%
\footnote{With the running field renormalization of $\flcb_{\mu\nu}$ as implied by $G_k$ included usually.},
depending on the expectation value fields now. An example of \eqref{eqn:int_003} is the Einstein-Hilbert truncation \cite{mr, oliver1, frank1} in which $\EAA_k^{\sm}[g]$ is specialized further to contain the two invariants $\int \sqrt{g}\SR$ and $\int \sqrt{g}$, only, with $k$-dependent prefactors involving a running Newton constant $G_k$ and cosmological constant $\Kkbar_k$, respectively.

Because of their quite substantial technical complexity, the work on more general truncations that would go beyond \eqref{eqn:int_003} started only recently. In \cite{elisa2} a first `bi-metric' truncation with separate $g_{\mu\nu}$ and $\bg_{\mu\nu}$-dependence was analyzed in conformally reduced gravity and, in \cite{MRS1}, for matter induced gravity (in the large $N$-limit).
The first bi-metric investigation of fully fledged Quantum Einstein Gravity employed a truncation ansatz with two separate Einstein-Hilbert terms for the dynamical (`$\dyn$') and the background ($`\background$') metric \cite{MRS2}. It consists of the following `graviton' (`grav') part, added to  the classical gauge fixing and ghost terms whose RG evolution is still neglected :
\begin{align}
\EAA_k^{\text{grav}}[g,\bg]
&= - \frac{1}{16\pi G_k^{\dyn}} \int \md^d x\sqrt{g}\, \big(\SR(g)-2\Kkbar_k^{\dyn}\big) \nonumber \\
&\quad - \frac{1}{16\pi G_k^{\background}} \int \md^d x\sqrt{\bg}\, \big(\SR(\bg)-2\Kkbar_k^{\background}\big) 
\label{eqn:int_004}
\end{align}
(Generalizing the truncation in a different direction, single-metric ans\"atze with a running ghost sector were studied in \cite{frank-ghost, astrid-ghost}.)

The present paper has two main purposes:

\noindent {\bf (1)}
On the technical side we develop a new computational strategy for deriving the explicit beta-functions related to bi-metric truncations such as the `bi-Einstein-Hilbert truncation' of eq. \eqref{eqn:int_004}.
This is a very hard problem in general since  it requires evaluating (`projecting') the functional supertrace on the RHS of the FRGE in dependence on {\it two} independent metrics. 
Unfortunately there exist basically no standard tools (such as general heat kernel expansions, etc.) available for this task.
This is one of the reasons why, despite their obvious significance, there are still almost no results on bi-metric truncations available today. 
The new strategy which we present here involves  a new type of gauge fixing which when applied to an action like \eqref{eqn:int_004} allows for a much simpler evaluation of the corresponding traces.
We shall see that in the case of  the `double Einstein-Hilbert truncation' \eqref{eqn:int_004} the derivation of the beta-functions of its running coupling constants simplifies considerably in comparison to the approach followed in \cite{MRS2}. 
There, a transverse-traceless (TT) decomposition of the fluctuation field $\flcb_{\mu\nu}$ had been necessary, something that can be completely avoided in the new approach, reducing the computational effort quite considerably.

\noindent {\bf (2)}
The second purpose of this paper is to perform a detailed investigation of the RG flow related to the bi-metric Einstein-Hilbert truncation using the RG equations derived with the new method.
The analysis will be far more comprehensive than the preliminary one in ref. \cite{MRS2}.
We shall be specifically interested in {\it global} properties of the flow, in particular the question as to whether Asymptotic Safety can coexist with Backround Independence, that is, the restoration of split-symmetry at the physical level.
This question is not easy to answer as it requires control over the fully extended RG trajectories, their limits $k\rightarrow\infty$ and $k\rightarrow0$ in particular.
Coexistence of Asymptotic Safety with Background Independence implies that there exists at least one  RG trajectory $k\mapsto \EAA_k[g,\bg]$ which is non-singular for all $k\in[0,\infty)$, and which approaches a non-Gaussian fixed point at its upper end, i.e.  for $k\rightarrow\infty$, while at the lower end, in the `physical' limit $k\rightarrow0$ where the EAA equals the ordinary effective action, it is split-symmetry-restoring, i.e. there $\EAA_k$ looses its extra $\bg$-dependence.%
\footnote{Except the one due to the gauge modes, to be precise.} 
 In particular on a truncated theory space it is by no means obvious that such trajectories  do indeed exist and both requirements can be met simultaneously.

One of our main results will be that within the bi-Einstein-Hilbert truncation Asymptotic Safety and Background Independence can indeed coexist: {\it Some}, but not all RG trajectories which emanate from a non-Gaussain fixed point in the UV also restore split-symmetry in the IR.
It will be instructive to uncover the very elegant mechanism of how the concrete RG differential equations bring about this symmetry restoration;
we shall see it involves a {\it moving attractor} in the background part of theory space.
We shall see that from a 4-parameter family of asymptotically safe trajectories a 2-parameter subset is symmetry restoring.

The existence of this subset of symmetry restoring trajectories is good news for the Asymptotic Safety program for two independent reasons:
{\bf (i)} It shows that at least in this truncation Asymptotic Safety and Background Independence are not mutually exclusive.
{\bf (ii)} Since a physically meaningful theory can only be based on a RG trajectory from this subset it follows that the predictive power of the Asymptotic Safety scenario is actually {\it higher} than what one would expect by just counting relevant perturbations at the fixed point;
the indispensable subsidiary condition of Background Independence narrows down the possibilities of constructing physically inequivalent theories at a given fixed point.

Recently the crucial importance of split-symmetry has also been demonstrated in a very impressive way within a 3-dimensional scalar toy model \cite{morris-dietz}: being careless about its restoration, one can even destroy the Wilson-Fisher fixed point!

The plan for the following parts of this paper is as follows.
In {\bf Section \ref{sec:trunc}} we present and justify in detail our new computational strategy, and apply it to obtain the new beta-functions for the bi-Einstein-Hilbert truncation.
The idea is that in future investigations similar strategies can be applied to more general truncations which are too complex to be dealt with by standard methods.

In {\bf Section \ref{sec:03}} we present and discuss the general properties of the concrete RG equations based on the new beta-functions. 
While these equations should describe the same physics as those obtained in the approach using the TT-decomposition, their mathematical appearance is quite different.
Here, and throughout the rest of the main body of this paper (except Section \ref{sec:dOther}), we focus on $d=4$ spacetime dimensions; the main results for arbitrary dimensionalities can be found in the {\bf Appendix}, in tabulated form.
The main physics result we shall find in this section is that, according to the bi-metric analysis, and in sharp contradiction to all single-metric results, there is no gravitational anti-screening in the semiclassical regime.

In {\bf Section {\ref{sec:fps}}} we start the detailed investigation of the RG flow by finding and classifying all its fixed points, exploring their properties, and assessing their eligibility for the Asymptotic Safety construction.

Then, in {\bf Section \ref{sec:splitandAS}}, we present a detailed analysis of the global properties of the RG flow.
In particular we demonstrate that Asymptotic Safety and Background Independence are indeed compatible by explicitly constructing RG trajectories which restore split-symmetry in the IR.
We explain how the special properties of these trajectories come about and we uncover the role played by a moving attractor point in theory space.

{\bf Section \ref{sec:sm_bm}} contains a very detailed comparison of the bi-metric Einstein-Hilbert truncation with its single-metric approximation. 
The goal of this section is to find out under what conditions the latter is sufficiently reliable, and under what circumstances it is mandatory to employ the former.
We shall find the somewhat `miraculous' result, unexplained by any general principles, that the single-metric approximation seems to perform best near a non-Gaussian fixed point.

Since Section \ref{sec:sm_bm} is rather technical, and mostly intended to establish the lessons one can learn from the present calculation and which can help to design future truncations other approximations optimally, this section can be skipped by the reader who is mostly interested in the new results.

{\bf Section \ref{sec:spctDim}} contains a first application of the bi-metric RG equation: we compute the scale dependent spectral dimension of the effectively fractal spacetimes QEG gives rise to.

In {\bf Section \ref{sec:dOther}} we briefly comment on the bi-metric RG flow in $d=2+\epsilon$ and $d=3$ spacetime dimension.
Near $2$ dimensions, where all Newton constants become dimensionless, their beta-functions contain a universal leading term which we compute. The resulting {\it universal} discrepancies between the single- and the bi-metric results are rather striking. 
In this setting it becomes particularly obvious that single-metric approximations can be very misleading even at the qualitative level already.

Finally, {\bf Section \ref{sec:conclusion}} contains a short summary and our conclusions.
\section{The `bi-Einstein-Hilbert truncation'} \label{sec:trunc}
This section is devoted to the steps leading from the truncation ansatz for the effective average action to the beta-functions for the running couplings it contains. 
An important role will be played by the new calculations scheme based upon the `deformed $\alpha=1$ harmonic gauge' we are going to employ here.

\subsection[The ansatz for \texorpdfstring{$\EAA_k$}{the EAA} ]{The ansatz for ${\bm{\EAA_k}}$}
In the following we shall employ a truncation ansatz for the effective average action (EAA) which is a sum of the following three parts:
\begin{align}
 \EAA_k[g,\bg,\Ghx,\GhAx]&= \EAA^{\text{grav}}_k[g,\bg]+\EAA_k^{\text{gf}}[g,\bg]+ \EAA^{\text{gh}}_k[g,\bg,\Ghx,\GhAx] \label{eqn:trA02}
\end{align}
The first term of the EAA, $\EAA^{\text{grav}}_k[g,\bg]$, is an arbitrary functional of both metrics, $g_{\mu\nu}$ and $\bg_{\mu\nu}$, whose form will be fixed later on.
To define the second term, the gauge fixing contribution, we introduce the differential operator 
\begin{align}
 \mathcal{F}_{\mu}^{\alpha\beta}[\bg]&\equiv \delta^{\beta}_{\mu}\bg^{\alpha\gamma} \bZ_{\gamma}-\varpi \bg^{\alpha\beta}\bZ_{\mu} 	\label{eqn:trA03}
\end{align}
which defines the gauge fixing condition: $\mathcal{F}_{\mu}^{\alpha\beta}[\bg]\flcb_{\alpha\beta}\stackrel{!}{=}0$. This operator, containing a free parameter $\varpi$, covers both the familiar harmonic gauge ($\varpi=1\slash 2$) as well as the `anharmonic gauge' used in \cite{MRS2} which has $\varpi=1\slash d$. The `square' of $ \mathcal{F}_{\mu}^{\alpha\beta}[\bg]\flcb_{\alpha\beta}$ yields our ansatz for the running gauge fixing action which is of the classical form essentially:
\begin{align}
\EAA_k^{\text{gf}}[g,\bg]&= \frac{1}{32\pi\agf_k} \int\md^d x \sqrt{\bg} \,\,\bg^{\mu\nu} \Big[\mathcal{F}_{\mu}^{\alpha\beta}[\bg] \left(g_{\alpha\beta}-\bg_{\alpha\beta}\right) \Big] \Big[\mathcal{F}_{\nu}^{\rho\sigma}[\bg] \left(g_{\rho\sigma}-\bg_{\rho\sigma}\right) \Big]		\label{eqn:trA04}
\end{align}
Here $\hat{\alpha}_k$ is a running $\alpha$-parameter; our assumptions about its $k$-dependence will be spelled out in a moment.
The third term in eq. \eqref{eqn:trA02} is taken to be the classical ghost action associated to the above gauge fixing by the Faddeev-Popov trick. 
Hence, it is bilinear in the ghost fields,
\begin{align}
\EAA_k^{\text{gh}}[g,\bg,\Ghx,\GhAx]&= -\sqrt{2} \rhoGk \int\md^d x \sqrt{\bg} \,\, \GhAx_{\mu}\, \mathcal{M}[g;\bg]^{\mu}_{\phantom{\mu}\nu}\,\Ghx^{\nu}	\label{eqn:trA05}
\end{align}
It involves the Faddeev-Popov operator $\mathcal{M}[g;\bg]$ which is defined to act on contravariant vectors $\Ghx$ according to
\begin{align}
 \mathcal{M}[g;\bg]^{\mu}_{\phantom{\mu}\nu}\,\Ghx^{\nu}&= \bg^{\mu\lambda} \mathcal{F}_{\lambda}^{\rho \sigma}[\bg] \Lie_{\Ghx} g_{\rho\sigma} \label{eqn:trA06}
\end{align}
where $\Lie_{\Ghx}$ denotes the Lie derivative with respect to $\Ghx$. In the ghost action \eqref{eqn:trA05} the overall prefactor $\rhoGk$ is an a priori $k$-dependent generalized coupling.

In the sequel, we further narrow down our truncation by making a specific ansatz for $\EAA_k^{\text{grav}}$. Since the effective average action contains all possible field monomials, a suitable ansatz may contain any number of curvature invariants built from both $g_{\mu\nu}$ and $\bg_{\mu\nu}$, as well as their associated covariant derivatives. 
In the purely dynamical, i.e. the $g_{\mu\nu}$-sector the Einstein-Hilbert truncation, which we are going to use in the following, is the first choice for at least two reasons: 
First, in a systematic derivative expansion it constitutes all invariants up to second order on manifolds without boundary. 
Second, General Relativity is known to be a very good description  of gravity in the infrared (IR). 
Thus, its action should be included in any truncation that wants to reproduce GR as an effective field theory at low scales $k$. 

As for the $\bg_{\mu\nu}$-dependence, there are various obvious field monomials with 2 derivatives that could accompany the Einstein-Hilbert term in the $g_{\mu\nu}$-sector: besides $\sqrt{\bg}\SRb$ there exists an infinite number of mixed terms, such as $\left(\frac{\sqrt{g}}{\sqrt{\bg}}\right)^n \sqrt{g}\SRb$ and $\left(\frac{\sqrt{\bg}}{\sqrt{g}}\right)^n \sqrt{ \bg}\SR$, for instance. However, following ref. \cite{MRS2} we will focus on the $\sqrt{\bg}\SRb$ term here and discard the other possibilities in the sequel.%
\footnote{Using different methods, a calculation distinguishing $g_{\mu\nu}$ and $\bg_{\mu\nu}$ has also been reported in ref. \cite{codello-closure}. 
However, we disagree with respect to its identification of `physical' ($=$dynamical) and `unphysical' ($=$background) couplings. (Indeed, the fixed point found in \cite{codello-closure} is numerically close to our `unphysical' \fpnB\fpC\fpgD-\fpL{}, see below.)
In ref. \cite{paw-rodigast} a very general momentum dependence for $\flcb_{\mu\nu}$ was retained (vertex expansion), but to become technical feasible the background had to be restricted to $\bg_{\mu\nu}=\delta_{\mu\nu}$, and the status of split-symmetry (and general covariance) is unclear.} 
 Thus we consider only actions which are of Einstein-Hilbert type with respect to both $g_{\mu\nu}$ and $\bg_{\mu\nu}$. 

More precisely, we shall compute RG flows on the 4-dimensional  subspace of theory space which is described by actions of the form:
\begin{align}
  \EAA_k^{\text{grav}}[g,\bg]&= - \frac{1}{16\pi \nkD } \int\md^d x \sqrt{g} \left(\SR(g) - 2 \KkbarD\right)\nonumber  \\
&\quad   - \frac{1}{16\pi \nkbB }  \int\md^d x \sqrt{\bg} \left(\SR(\bg)- 2 \KkbarB\right) \label{eqn:trA07}
\end{align}
The 4 running coupling constants $\nkD$ ($\KkbarD$) and $\nkbB$ ($\KkbarB$) correspond to the Newton's (cosmological) constants for the dynamical and the background sector, respectively.
Throughout the present paper we shall consider the truncation with \eqref{eqn:trA04}, \eqref{eqn:trA05}, and \eqref{eqn:trA07}, respectively.

\subsection{The level expansion}
It will be very stimulating to reformulate the ansatz $\EAA_k= \EAA^{\text{grav}}_k+\EAA_k^{\text{gf}}+ \EAA^{\text{gh}}_k$ in terms of the fluctuation field $\flcb_{\mu\nu}=g_{\mu\nu}-\bg_{\mu\nu}$ along with $\bg_{\mu\nu}$, instead of $g_{\mu\nu}$ with $\bg_{\mu\nu}$. Hence we have, up to terms of second order in $\flcb_{\mu\nu}$:
\begin{align}
  \EAA_k^{\text{grav}}[\flcb;\bg]&= - \frac{1}{16\pi G_k^{(0)} }  \int\md^d x \sqrt{\bg} \left(\SR(\bg) - 2\Kkbar_k^{(0)}\right)  \nonumber  \\
&\quad - \frac{1}{16\pi G_k^{(1)} }  \int\md^d x \sqrt{\bg}\, \Big[-\bar{G}^{\mu\nu}-\Kkbar_k^{(1)} \bg^{\mu\nu}\Big] \flcb_{\mu\nu} \nonumber \\
&\quad - \frac{1}{2}  \int\md^d x \sqrt{\bg}\ \,\flcb^{\mu\nu} \,\,\EAA^{\text{grav}\,(2)}_k[\bg,\bg]\,\, \flcb_{\rho\sigma}
+ \Order{\flcb^3} \label{eqn:trA08}
\end{align}
Here $\EAA^{\text{grav}\,(2)}_k$ is the Hessian of the functional \eqref{eqn:trA07}, the matrix of its second functional derivatives with respect to the metric $g_{\mu\nu}$, evaluated at $g=\bg$; it contains $G_k^{(2)}$ and $\Kkbar_k^{(2)}$. 

Eq. \eqref{eqn:trA08} displays the first three orders of what we call the {\it level expansion} of the EAA. 
By definition, the `level' of a field monomial equals the number of $\flcb_{\mu\nu}$-factors it contains. 
The running couplings at level $p$ are denoted by a superscript $(p)$ throughout. 

Comparing \eqref{eqn:trA07} to \eqref{eqn:trA08} we can read off  the coefficients at level zero for instance:
\begin{align}
 \frac{1}{G_k^{(0)}}=\frac{1}{G_k^{\background}}+\frac{1}{G_k^{\dyn}}\,,\qquad \quad 
\frac{\Kkbar_k^{(0)}}{G_k^{(0)}}=\frac{\Kkbar_k^{\background}}{G_k^{\background}}+\frac{\Kkbar_k^{\dyn}}{G_k^{\dyn}}\, .
\label{eqn:trA08B}
\end{align}
The ansatz \eqref{eqn:trA07} with its 4 independent couplings $\{G_k^{\background},\,\Kkbar_k^{\background},\,G_k^{\dyn},\,\Kkbar_k^{\dyn}\}$ is equivalent to an ansatz of the type \eqref{eqn:trA08}; generically it contains infinitely many different couplings $G_k^{(p)}$, $\Kkbar_k^{(p)}$, $\cdots$. In the special case of the `bi-Einstein-Hilbert truncation' they are not independent however but satisfy
\begin{subequations}
\begin{align}
& G_k^{(0)}=\frac{\nkD \nkbB}{ \nkD+\nkbB}\,, \qquad \quad 
 G_k^{(1)}=G_k^{(2)}=\cdots=\nkD \, ,  \label{eqn:trA09}\\
& \Kkbar_k^{(0)}= \frac{\KkbarD \,\nkbB + \KkbarB\,\nkD}{\nkD+\nkbB} \,,\qquad \quad \Kkbar^{(1)}=\Kkbar^{(2)}=\cdots =\KkbarD \,. \label{eqn:trA10}
\end{align}
\end{subequations}
It will often be instructive to switch back and forth between the $(g,\bg)$-language employing the `$\background$' and `$\dyn$' parameters, and the $(\flcb;\bg)$-language with $G_k^{(p)}$, $\Kkbar_k^{(p)}$, $p=0,1,2,\cdots$.

In case of fully intact split-symmetry the terms of the different levels stem from a Taylor expansion of $\EAA^{\text{grav}}_k[\bg+\flcb,0]$, i.e. the EAA can be rewritten as a functional of one metric only, $g=\bg+\flcb$, i.e. it has no `extra $\bg$-dependence' \cite{mr}. This situation corresponds for the EAA of eq. \eqref{eqn:trA08} to the special case where
\begin{align}
 G_k^{(0)}=G_k^{(1)}=G_k^{(2)}=\cdots \, , \quad \text{and}\quad \, \Kkbar_k^{(0)}=\Kkbar_k^{(1)}=\Kkbar_k^{(2)}=\cdots \quad \text{(split-sym.)}  \label{eqn:trA11}
\end{align}
In the $(g,\bg)$-language, these relations are satisfied if the `$\background$' part of the EAA in \eqref{eqn:trA07} is completely vanishing, 
\begin{align}
 \frac{1}{\nkbB} =0 \,, \quad \text{and}\quad \frac{\KkbarB}{\nkbB}=0 \qquad \quad \text{(split-sym.)}\label{eqn:trA12}
\end{align}
or, stated differently, when $\EAA_k^{\text{grav}}[g,\bg]$ is actually independent of its second argument, $\bg$.

In general split-symmetry is broken by the gauge fixing and the coarse graining terms in the action, and  there is no guarantee for the existence of trajectories satisfying the conditions \eqref{eqn:trA11} or equivalently \eqref{eqn:trA12} for some interval of scales $k$. 
However, there might be regions in theory space and scales where these conditions are fulfilled, or at least fulfilled approximately.
In these cases a single-metric description is well applicable.  From eq. \eqref{eqn:trA12} we see that if $1\slash G_k^{\background}$ and $\KkbarB\slash G_k^{\background}$ are very small, in particular in comparison to their `$\dyn$' counterparts ($G_k^{\background}\gg\nkD$ and $\KkbarB\lessapprox \KkbarD$), split-symmetry is realized at an approximate level.

\subsection{The conformal projection technique}
A truncated theory space is in general not invariant under the RG flow, due to the fact that the FRGE generates terms not present in the original ansatz for $\EAA_k$. In the present case the RHS of the FRGE contains invariants with higher powers of $\SR$, $\SRb$, $(\Ghx \GhAx)$, $\cdots$, for instance, while its LHS consists of only a few of them, namely those invariants retained in the truncation ansatz \ref{eqn:trA07}. 
Thus, in order to complete the specification of the truncation it is necessary to define a projection of the RHS onto exactly those field monomials. 
When applying it to the supertrace we need a technique  to distinguish between contributions to $\SRb$ and $\SR$. In this paper we define the projection in the same way as in ref. \cite{MRS2}. The essential step is to conformally relate the two metrics involved as follows:  $g_{\mu\nu}(x)=e^{2\cf}\bg_{\mu\nu}(x)$. Then, after having eliminated $g_{\mu\nu}$ in favor of $\cf$ and $\bg_{\mu\nu}$, we expand the supertrace in the flow equation in powers of the $x$-independent conformal factor $\cf$. For $\cf=0$ we obtain the invariants built from $\bg$, whereas the level-(1)  invariants $\propto\flcb$ are identified as those linear in $\cf$. 

For the $t$-derivative of $\EAA_k^{\text{grav}}$ on the LHS of the FRGE this `conformal projection' yields, to the required order in $\cf$,
\begin{align}
\partial_t\EAA^{\text{grav}}_k[g=e^{2\cf}\bg,\bg]&= \left(\tfrac{1}{8\pi}\partial_t\,[\Kkbar_k^{(0)}\slash G_k^{(0)}] +\tfrac{ d\,\cf }{8\pi}\, \partial_t\,[\KkbarD \slash \nkD\right)  \int\md^d x \sqrt{\bg}\,  \nonumber \\
&\quad
- \left(\tfrac{1 }{16\pi}\partial_t\,[1\slash{G_k^{(0)}}]+\tfrac{(d-2)\,\cf }{16\pi}\,\partial_t\,[1\slash {\nkD}] \right) \int\md^d x \sqrt{\bg}\,\SRb \,+ \Order{\cf^2}  \label{eqn:trA13}
\end{align}
Under the conformal projection the gauge fixing action is identical zero, $\EAA_k^{\text{gf}}[e^{2\cf}\bg,\bg]= 0$. Thus, the running of the gauge parameter $\agf_k$ can not be resolved with this method. 
Finally, in the ghost sector the LHS of the flow equation reduces under the conformal projection to
\begin{align}
&\partial_t\EAA_k^{\text{gh}}[g=e^{2\cf}\bg,\Ghx,\GhAx,\bg]\nonumber \\
&\qquad= -\sqrt{2}\, (1+2\cf)\, [\partial_t \rhoGk]\, \int\md^d x \sqrt{\bg} \,\, \GhAx_{\mu}\,  	\big[  \id^{\mu}_{\nu}\,\bZ^2+(1-2\varpi)  \bZ^{\mu} \bZ_{\nu}+\Ricb\ud{\mu}{\nu}    \big]\,	\Ghx^{\nu}	 \label{eqn:trA14}
\end{align}

Incidentally, note that for the harmonic gauge, $\varpi=1\slash 2$, the ghost action, and therefore the entire truncation ansatz, does not contain any covariant derivative $\bZ_{\mu}$ that would not be contracted to a Laplacian $\bZ^2\equiv \bg^{\mu\nu}\bZ_{\mu}\bZ_{\nu}$.

In the following we consider only the more special truncations where the ghost's wave function renormalization factor is assumed constant: $\rhoGk\equiv \rhoG$. Then, $\partial_t \EAA_k^{\text{gh}}=0$, yielding a LHS of the FRGE that does not depend on the ghost fields any longer. 
This simplifies matters considerably since it is now sufficient to evaluate the RHS of the flow equation for $\Ghx=0=\GhAx$, after having performed the functional derivatives. (A more detailed investigation of ghost sector can be found in \cite{frank-ghost, astrid-ghost}.)

\subsection{Structure of the Hessian operator}
Next, let us turn our attention to the functional traces on the  RHS of the FRGE \eqref{eqn:int_01}. The first step towards the beta-functions is the evaluation of the Hessian of the EAA with respect to the dynamical fields. In the truncated theory space considered, we achieve a significant simplification of $\EAA_k^{(2)}$ by exploiting that we may set $\Ghx=0=\GhAx$ after the corresponding differentiations, yielding
\begin{align}
 &\EAA^{(2)}_k[g=e^{2\cf}\bg,\bg,\Ghx=0,\GhAx=0] \label{eqn:trA15} \\
&=\left. \begin{pmatrix} \big(\EAA^{(2)}_k\big)_{gg}  &  \big(\EAA^{(2)}_k\big)_{g\Ghx} &   \big(\EAA^{(2)}_k\big)_{g \GhAx}  \\   \big(\EAA^{(2)}_k\big)_{\Ghx g} & 0  &  \big(\EAA^{(2)}_k\big)_{\GhAx\Ghx}  \\   \big(\EAA^{(2)}_k\big)_{\GhAx g}  &   \big(\EAA^{(2)}_k\big)_{\Ghx\GhAx} & 0  
\end{pmatrix} \right|_{ \substack{{\scriptstyle g=e^{2\cf}\bg,}\\{\scriptstyle \Ghx=0=\GhAx}}} 
=\left. \begin{pmatrix} \big(\EAA^{(2)}_k\big)_{gg}  &  0 &   0  \\  0 & 0  &  \big(\EAA^{(2)}_k\big)_{\GhAx\Ghx}  \\   0 &   \big(\EAA^{(2)}_k\big)_{\Ghx\GhAx} & 0  
\end{pmatrix} \right|_{g=e^{2\cf}\bg} \nonumber 
\end{align}
The mixed variations with respect to $g$ and $\Ghx$ or $\GhAx$, i.e. $\big(\EAA^{(2)}_k\big)_{g \Ghx},\,\big(\EAA^{(2)}_k\big)_{g \GhAx},\cdots$, are linear in either the ghost or anti-ghost field. After inverting $\EAA_k^{(2)}+\mathcal{R}_k$  they yield contributions of at least linear order in $(\Ghx\GhAx)$ and thus are beyond the scope of our truncation ansatz. So, it is sufficient to set  $\Ghx=0=\GhAx$ on the RHS, yielding a block diagonal structure of $\EAA_k^{(2)}$ and, provided $\mathcal{R}_k$ is chosen appropriately, for the entire operator under the supertrace which no longer couples the graviton to the ghost sector.

In the graviton (`$gg$') block, the Hessian of the gauge fixing and dynamical action combine to
\begin{align}
 \EAA^{(2)}_k[e^{2\cf}\bg,\bg]^{\mu\nu\rho\sigma}_{gg}&=    
- \frac{1}{32\pi\agf_k} \Big[\bg^{\rho\mu}\bg^{\sigma\nu}  + (2\varpi^2-1)\bg^{\mu\nu}\bg^{\rho\sigma}  \Big]\bZ^2
\nonumber \\
&\quad
- \frac{1}{32\pi\agf_k}(1-2\varpi) \Big[\bg^{\rho\sigma} \bZ^{\mu}\bZ^{\nu} + \bg^{\mu\nu}\bZ^{\sigma} \bZ^{\rho}\Big]
\nonumber \\
&\quad
+\frac{1}{32\pi}\Big[ \frac{1}{\agf}_k- \frac{e^{(d-6)\cf}}{\nkD} \Big]\mathcal{K}^{\mu\nu\rho\sigma}[\bg]  
\nonumber \\
&\quad
+\frac{1}{16\pi\agf_k}\Big[\bg^{\nu\sigma} \Ricb^{\mu\rho}+\Rmb^{\rho\mu\nu\sigma}\Big]+\, \frac{e^{(d-6)\cf}}{32\pi\,\nkD}\,\mathcal{U}^{\mu\nu\rho\sigma}[\bg]   \label{eqn:trA16}
\end{align}
Here we already inserted the conformal relation $g=e^{2\cf}\bg$. The first three terms in equation \eqref{eqn:trA16} contain differential operators of various types, some contracted to Laplacians $\bZ^2$, others with uncontracted open indices, $\bZ^{\mu}\bZ^{\nu}$. The operator $\mathcal{K}$ in the third term is a mixture of both types:
\begin{align}
 \mathcal{K}^{\mu\nu\rho\sigma}[\bg]&\equiv  - 2\bg^{\nu\sigma}\bZ^{\rho} \bZ^{\mu}  + \left(\bg^{\mu\nu}\bZ^{\rho} \bZ^{\sigma} + \bg^{\rho\sigma}\bZ^{\mu}\bZ^{\nu}\right) - \left(\bg^{\mu\nu}\bg^{\rho\sigma}-\bg^{\mu\rho}\bg^{\nu\sigma}\right)\bZ^2 \label{eqn:trA17}
\end{align}
The remaining terms in \eqref{eqn:trA16}  are ultra-local, potential-type operators containing the Riemann tensor and its contractions. The first one $\propto 1\slash \agf_k$ stems from the gauge fixing Hessian and the second, proportional to
\begin{align}
 \mathcal{U}^{\mu\nu\rho\sigma}[\bg]&\equiv 
\big(\SRb -2e^{2\,\cf}\KkbarD \big)\left(\bg^{\mu\rho}\bg^{\nu\sigma}-\frac{1}{2}\bg^{\rho\sigma}\bg^{\mu\nu} \right) 
 \nonumber \\
&\quad \quad 
 -2 \big(\bg^{\mu\rho} \Ricb^{\nu\sigma} + \bg^{\nu\rho}\Ricb^{\mu\sigma} \big)  + \left(\bg^{\mu\nu} \Ricb^{\rho\sigma}+\bg^{\rho\sigma} \Ricb^{\mu\nu}\right) \label{eqn:trA18}
\end{align}
stems from the gravitational Hessian in the `$\dyn$' sector.

In the ghost-ghost block of the matrix \eqref{eqn:trA15} we encounter the matrix elements
\begin{subequations}
\begin{align}
   {\EAA^{(2)}_k[e^{2\cf}\bg,\bg]_{\Ghx\GhAx}}^{\mu}_{\phantom{\mu}\nu}&= \sqrt{2} \, e^{2\cf}\left\{\Ricb\ud{\mu}{\nu} + (1-2\varpi)\bZ^{\mu}\bZ_{\nu} 
  +\id^{\mu}_{\nu}\bZ^2 \right\} -  \sqrt{2} \, e^{2\cf}\,(1-2\varpi)\Ricb\ud{\mu}{\nu} \label{eqn:trA19} \\
 {\EAA^{(2)}_k[e^{2\cf}\bg,\bg]_{\GhAx\Ghx}}^{\mu}_{\phantom{\mu}\nu}
&= -\sqrt{2}\, e^{2\cf}\left\{ \Ricb\ud{\mu}{\nu} +(1-2\varpi)  \bZ^{\mu} \bZ_{\nu}   + \id^{\mu}_{\nu}\,\bZ^2 \right\} \label{eqn:trA20}
\end{align}
\end{subequations}
Notice that the operators \eqref{eqn:trA19} and \eqref{eqn:trA20} are indeed Hermitian adjoints of each other w.r.t. the usual $L^2$ inner product.
The source of the term  $\sqrt{2} \, e^{2\cf}\,(1-2\varpi)\Ricb\ud{\mu}{\nu} $ that spoils the simple relation $ \left(\EAA_k^{\text{gh}}\right)^{(2)}_{\GhAx\Ghx}= -\left(\EAA_k^{\text{gh}}\right)^{(2)}_{\Ghx\GhAx}$ is the uncontracted  derivative contribution proportional to $(1-2\varpi)$ that appears when reordering $\bZ^{\mu}\bZ^{\nu}$ after the integration by parts. 

In the case of  $\varpi=1\slash 2$ these contributions vanish. 
As a result, the RHS of the FRGE, in the  ghost sector, contains covariant derivatives only as fully contracted Laplacian $\bZ^2$. 
Next we shall try to fix the gauge in such a way that an analogous simplification occurs in the `gg'-sector as well.

\subsection[The `\texorpdfstring{$\cf$}{conformally} deformed' \texorpdfstring{$\alpha=1$}{alpha = 1} harmonic gauge]{The `${{\cf}}$ deformed' ${{\alpha=1}}$ harmonic gauge}
The remaining task is to actually compute the RHS of the truncated FRGE, thereby retaining only the field monomials which are contained in the ansatz for the EAA. 
We would like to take advantage of known asymptotic expansions for the heat kernels of the operators involved.
This would be particularly easy if those operators were functions of the Laplacian $\bZ^2$.
A priori this is not the case, however.
There are also operators such as $\bZ_{\mu} \bZ_{\nu}$ with uncontracted indices.
One strategy to deal with them is to  decompose the fields $\flcb_{\mu\nu}$, $\Ghx^{\mu}$,  and $\GhAx_{\mu}$ into a sum of differentially constrained fields along the lines of the transverse-traceless (TT)- or York-decomposition. 
This was the strategy used in ref. \cite{MRS2}.

In the present paper, we instead employ a different method that uses the gauge freedom to cancel all non-Laplacian differential operator terms. We will call it the `{\it $\cf$ deformed $\alpha=1$ harmonic gauge}', for reasons we shall explain in the following.

\subsubsection{The motivation}
We have already noticed that the harmonic gauge $\varpi=1\slash2$ reduces the operators occurring in  the flow equation, in the ghost sector, to a function of the fully contracted Laplacian operator only. 
All operators of the type $\bZ^{\mu}\bZ^{\nu}$ are proportional to $(1-2\varpi)$ and drop out in the harmonic gauge.
 
In the graviton sector, the action $\EAA_k^{\text{grav}}$ itself is free of such operators, but on the RHS of the flow equation they appear in its Hessian. In fact, employing the harmonic gauge $\varpi=1\slash2$ the second term of equation \eqref{eqn:trA16}, that contains uncontracted differential operators, vanishes.  However, there is another source of troublesome covariant derivatives, namely those contained in $\mathcal{K}^{\mu\nu\rho\sigma}[\bg]$. They are proportional to $\big( \agf^{-1}_k- e^{(d-6)\cf} \slash \nkD \big)$. As a result, they disappear if we pick the following gauge parameter:
\begin{align}
\agf_k=  \nkD  e^{-(d-6)\cf} =\nkD \big[1-(d-6)\cf +\Order{\cf^2}\big]  \label{eqn:trA21}
\end{align}

Before continuing note that $\agf_k$ has the same mass dimension $(2-d)$ as the Newton constants, $G_k^{\dyn}$ in particular, it is natural to define the {\it dimensionless} gauge fixing parameter
\begin{align}
 \alpha_k \equiv \frac{\agf_k}{G_k^{\dyn}}  \label{eqn:trA21B}
\end{align}
Then the preferred choice \eqref{eqn:trA21} corresponds to the value, up to $\Order{\cf^2}$ terms,
\begin{align}
 \alpha_k = 1- (d-6)\cf \equiv \alpha \label{eqn:trA21C}
\end{align}
Note that  this $\alpha$ happens to be scale independent.

Indeed, in all single-metric (`$\sm$') calculations following ref. \cite{mr} the coefficient of the gauge fixing action has been parametrized as 
\begin{align}
 \agf_k&=\alpha^{\sm} G^{\sm}_k  \label{eqn:trA22} \, ,
\end{align}
and it was then the choice $\alpha_k^{\sm}\equiv\alpha=1$ that removed the uncontracted derivatives. 

In order to achieve the same effect in the dynamical sector of the bi-metric setting we would have to generalize this choice in a $\cf$-dependent way, namely $\alpha= e^{(6-d)\cf}$. Since only terms linear in $\cf$ are relevant, we can think of 
\begin{align}
 \alpha=1 - (d-6)\cf + \Order{\cf^2} \label{eqn:trA22B}
\end{align}
as being infinitesimally close to unity. It is this choice, eq. \eqref{eqn:trA22B}, which we refer to as the `$\cf$ deformed $\alpha=1$ gauge'.

\subsubsection{The justification}
It should be clear that in general disposing of $\alpha$ as in \eqref{eqn:trA22B}, i.e. making it $\cf$ dependent, is by no means legitimate from the conceptual point of view:
The factor $\cf$ represents the dynamical metric, $g=e^{2\cf}\bg$, it cannot appear in the final answer for the beta functions, but is rather a book keeping parameter we should expand in to disentangle terms with different powers of $\flcb_{\mu\nu}$. The beta functions at levels (0) and (1), respectively, are obtained from the Taylor expansion of the traces:
\begin{align}
 \Tr[\cdots]=\Tr[\cdots]\big|_{\cf=0}+ \cf\, \frac{\md}{\md\cf}\,\Tr[\cdots]\big|_{\cf=0}+ \Order{\cf^2} \label{eqn:trA22C}
\end{align}
If we decide to eliminate the uncontracted covariant derivatives by giving the formally $\cf$-dependent value \eqref{eqn:trA22B} to $\alpha$ we pay a price, namely we neglect a certain contribution to the linear term in \eqref{eqn:trA22C}, $\cf \frac{\md}{\md \cf} \Tr[\cdots]$, which arises through the $\cf$-dependence of $\alpha\equiv e^{(6-d)\cf}$. By the chain rule, it has the form 
\begin{align}
( 6-d)\cf \frac{\md}{\md \alpha} \Big\{ \Tr[\cdots]\big|_{\cf=0}\Big\}\Big|_{\alpha=1} \label{eqn:trA22D}
\end{align}
Now, the crucial observation is that the trace at $\cf=0$ whose $\alpha$-derivative is taken here is to justify our choice, equals precisely the trace which  appears in the single-metric computation. Its $\alpha$-dependence has been investigated in detail in ref. \cite{oliver1} and found to be rather small. In particular varying $\alpha$ over the interval from $\alpha=0$ to $\alpha=1$ causes changes in the flow properties which are smaller than those due to the cutoff scheme dependence and truncation error which, in fact, supplied the justification for setting $\alpha_k$ equal to a $k$-independent constant. This implies that, at the level of accuracy we may expect on the basis of the truncations already made, the term \eqref{eqn:trA22D} is indeed negligible because $\Tr[\cdots]\big|_{\cf=0}$ has no significant $\alpha$-dependence near $\alpha=1$. Thus, {\it in a truncation which neglects the running of $\alpha$ also the piece missed by the deformed gauge, \eqref{eqn:trA22D}, may be omitted.}

\subsubsection{A second, independent justification}
Formally using a $\cf$-dependent $\alpha$-parameter calls for careful consideration of the resulting mixing of level-(0) with higher level contributions, which has to be kept as small as possible. Indeed, there is  a way to justify the choice \eqref{eqn:trA22B} from a different perspective. Assume we had started with a slightly different truncation ansatz in which the gauge fixing action, $\EAA^{\text{gf}}_k$, is substituted with 
\begin{align}
\EAA_k^{(\rhoA)}[g,\bg] &= \, \left(1-\tfrac{1}{\rhoA}\right)\,\tfrac{\agf_k}{\ags_k}\EAA_k^{\text{gf}}[g,\bg] +  \tfrac{1}{\rhoA}\, \EAA_k^{\text{subs}}[g,\bg]	\label{eqn:trA22E}
\end{align}
Here $\rhoA \neq 0$ is a constant introduced for later convenience. 
Furthermore, $\EAA_k^{\text{subs}}[g,\bg]$ is a new functional which under the conformal projection ($g=e^{2\cf}\bg$), is required to obey  $\EAA_k^{\text{subs}}[g=e^{2\cf}\bg,\bg]=0$, and $(\delta^2_g\, \EAA_k^{\text{subs}})[g=e^{2\cf}\bg,\bg] = e^{\rhoA(d-6)\cf}\,(\agf_k \slash \ags_k)(\delta^2_g\, \EAA_k^{\text{gf}})[g=e^{2\cf}\bg,\bg]$. The first condition guarantees that $\partial_t \EAA_k$  remains unchanged, and the second requirement assures that the old and the new Hessians $\EAA_k^{(2)}$  agree at linear order in $\cf$ for a certain choice of $\ags_k$. 
Expanding the Hessian of \eqref{eqn:trA22E} in powers of $\cf$, 
\begin{align}
 (\delta^2_g\, \EAA_k^{(\rhoA)})[g=e^{2\cf}\bg,\bg] &= \,  \big(1+ (d-6)\cf \big)\,(\agf_k\slash \ags_k)(\delta^2_g\, \EAA_k^{\text{gf}})[g=e^{2\cf},\bg]+ \Order{\cf^2}	\label{eqn:trA22F}
\end{align}
we find it proportional to  the `old' gauge fixing contribution. The additional prefactor can be absorbed by a redefinition of the gauge parameter as follows:
\begin{align}
\agf_k\equiv \big(1- (d-6)\cf + \Order{\cf^2} \big)\,\ags_k\, .	\label{eqn:trA22G}
\end{align}
The Hessians are equal then: $ (\delta^2_g\, \EAA_k^{(\rhoA)})[g=e^{2\cf}\bg,\bg] =(\delta^2_g\, \EAA_k^{\text{gf}})[g=e^{2\cf},\bg]+\Order{\cf^2}$. This in turn implies that the above steps for removing the uncontracted covariant derivatives work with all choices of $\EAA_k^{(\rhoA)}$, however now  with the parameter
\begin{align}
 \ags_k = \nkD \quad \Leftrightarrow \quad \alpha^{\text{new}}_k=1	\label{eqn:trA22H}
\end{align}
Within the generalized truncations \eqref{eqn:trA22E}, a {\it $\cf$-independent} choice for $\ags_k$ is sufficient to account for the same simplification on the LHS of the flow equation that we used in the original ansatz with only $\EAA_k^{\text{gf}}$ containing the $\cf$-dependent coupling $\agf_k$. This $\cf$-dependence of a coupling in the truncation ansatz is thus merely an artifact of the conformal projection  on the set of all possible invariants whereby different level-$(p)$ monomials lead to the same Hessian. In this sense the $\cf$-dependence of $\agf_k$ can be thought of as  a method for implicitly including additional field monomials in the truncation ansatz that lead to a removal  of uncontracted covariant derivatives without introducing any new effects, otherwise. 

There are several suitable candidates for $\EAA_k^{\text{subs}}[g,\bg]$ of which we consider here only those of the form
\begin{align}
 \EAA_k^{\text{subs}}[g,\bg]&= \frac{1}{32\pi \ags_k}\int \sqrt{\bg}\, \bg^{\mu\nu} \, \widetilde{F}^{(\rhoA)}_{\mu}[g,\bg]\, \widetilde{F}^{(\rhoA)}_{\nu}[g,\bg]
\label{eqn:trA22I}
\end{align}
For the choice $\rhoA=2$ the prefactor of the gauge fixing action in eq. \eqref{eqn:trA22E} is $1\slash 2$. The `missing' half in the Hessian of the RHS of the flow equation now stems from the level-(2) contribution of $ \EAA_k^{\text{subs}}[g,\bg]$. 
All higher level terms are new, additional field monomials with identical prefactors fixed by $\ags_k$. Possible choices of the structure given in \eqref{eqn:trA22I} with $\rhoA=2$ include
\begin{subequations}
\begin{align}
 \widetilde{F}^{2;\,\text{\Rmnum{1}}}_{\mu}[g,\bg]&= \frac{\sqrt{g}}{\sqrt{\bg}}\, g^{\alpha\lambda}g^{\beta\rho}g^{\kappa\sigma}\Big[\bg_{\mu\rho}\bg_{\lambda\sigma}-\varpi\, \bg_{\mu\sigma}\bg_{\lambda\rho}\Big]\bZ_{\kappa} \big(g_{\alpha\beta}-\bg_{\alpha\beta}\big)	\label{eqn:trA22JA}\\
 \widetilde{F}^{2;\,\text{\Rmnum{2}}}_{\mu}[g,\bg]&= \frac{\sqrt{g}}{\sqrt{\bg}}\, g^{\beta\lambda}\bg_{\lambda\rho}\bg_{\mu\sigma}\Big[g^{\alpha\kappa}g^{\sigma\rho}-\varpi\, g^{\alpha\rho}g^{\kappa\sigma}\Big]\bZ_{\kappa} \big(g_{\alpha\beta}-\bg_{\alpha\beta}\big) \label{eqn:trA22JB}
\end{align}\label{eqn:trA22J}
\end{subequations}
It is straightforward to verify that  all requirements on $\EAA_k^{\text{subs}}[g,\bg]$ are indeed fulfilled by these examples. (Notice that additional terms appearing in the Hessian associated to the action \eqref{eqn:trA22I} are zero when evaluated for the conformally related metrics, $g=e^{2\cf}\bg$.)

Another interesting option is $\rhoA=1$. In this case one actually replaces the gauge fixing action of the original truncation ansatz \eqref{eqn:trA04} with a new action containing the gauge fixing condition $\widetilde{F}^{1}_{\mu}[g,\bg]=0$. 
This choice requires however a new Faddeev-Popov operator $\widetilde{\mathcal{M}}^{\mu}_{\phantom{\mu}\nu}[g,\bg]$ that corresponds to this gauge fixing. 
If we want to leave the RHS of the flow equation unchanged, we have to adopt an additional modification  in the ghost sector. 

First of all notice that with the choice \eqref{eqn:trA22I} the requirements put on $ \EAA_k^{\text{subs}}[g,\bg] $ can be transferred to requirements on $\widetilde{F}^{\rhoA}_{\mu}[g,\bg]$. 
We obtain $ \widetilde{F}^{\rhoA}_{\mu}[e^{2\cf}\bg,\bg]\stackrel{!}{=}0$ from the vanishing of $ \EAA_k^{\text{subs}}[g,\bg]$ under the conformal projection, and $ (\delta_g^1 \widetilde{F}^{\rhoA}_{\mu} )[e^{2\cf}\bg,\bg]=e^{\rhoA(d-6)\cf \slash 2}(\delta_g^1 F^{\rhoA}_{\mu} )[e^{2\cf}\bg,\bg]$ follows from the second criterion and by using $\widetilde{F}^{\rhoA}_{\mu}[e^{2\cf}\bg,\bg]\stackrel{!}{=}0$. Neglecting ghost contributions on the RHS of the flow equation allows us to rewrite the new Faddeev-Popov operator as being proportional to the original one, namely
\begin{align}
 \widetilde{\mathcal{M}}^{\mu}_{\phantom{\mu}\nu}[g=e^{2\cf}\bg,\bg]&= e^{\rhoA(d-6)\cf \slash 2}\,\mathcal{M}^{\mu}_{\phantom{\mu}\nu}[g=e^{2\cf}\bg,\bg]	\label{eqn:trA22K}
\end{align}
This change of $\mathcal{M}$ could be compensated by a $\cf$-dependent ghost coupling $\rhoG$ in the original truncation. 
Thus, if there is a  gauge fixing condition of the type
\begin{align}
  \widetilde{F}^{1;\Rmnum{1}}_{\mu}[g,\bg]&=\left(\frac{\sqrt{g}}{\sqrt{\bg}}\, g^{\lambda\kappa}\bg_{\lambda\kappa}\right)^{1\slash2}\, \big(\delta_{\mu}^{\beta}g^{\alpha\rho}\bZ_{\rho} - \varpi g^{\alpha\beta}\bZ_{\mu}\big)(g_{\alpha\beta}-\bg_{\alpha\beta}) \label{eqn:trA22L}
\end{align}
for instance, it would correspond to the choice   $\agf_k=\nkD \big[1+(6-d)\cf +\Order{\cf^2}\big] $ and $\rhoG=\big[1+\tfrac{\rhoA}{2}(d-6) \cf\big]\rho_{\text{gh}}^{\text{new}}$ for the gauge parameter and the ghost coupling, respectively, in a truncation with the gauge fixing action of eq. \eqref{eqn:trA04}. 
We shall not continue the analysis of these possibilities in the present paper.

\subsubsection{The Hessian simplified}
From now on we shall adopt the deformed $\alpha=1$ harmonic gauge ($\varpi=1\slash2$) which then eliminates the technically difficult uncontracted covariant derivatives. 
Later on in this paper we shall provide another  independent justification of this procedure by comparing our results to those of the calculation in ref. \cite{MRS2} where the uncontracted derivatives had been dealt with by a transverse-traceless decomposition of $\flcb_{\mu\nu}$.

When we employ the deformed $\alpha=1$ gauge in our truncation the resulting Hessian simplifies considerably. 
In the $gg$-sector it is given by
\begin{align}
 \EAA^{(2)}_k[e^{2\cf}\bg,\bg]^{\mu\nu\rho\sigma}_{gg}&=    
 \frac{e^{(d-6)\cf}}{32\pi\nkD} \Big[\big(\bg^{\rho\mu}\bg^{\sigma\nu}  -\frac{1}{2}\bg^{\mu\nu}\bg^{\rho\sigma}  \big) \big(-\bZ^2+\SRb -2e^{2\,\cf}\KkbarD \big) \nonumber \\
&\quad  \phantom{=  \frac{e^{(d-6)\cf}}{32\pi\nkD}==\,} 
+2\Rmb^{\rho\mu\nu\sigma} -2 g^{\mu\rho} \Ric^{\nu\sigma}  + g^{\mu\nu} \Ric^{\rho\sigma}+g^{\rho\sigma} \Ric^{\mu\nu} \Big] \label{eqn:trA23}
\end{align}
The cancellation of all uncontracted differential operators in the ghost-sector leads to 
\begin{align}
   {\EAA^{(2)}_k[e^{2\cf}\bg,\bg]_{\Ghx\GhAx}}^{\mu}_{\phantom{\mu}\nu}&= \sqrt{2} \, e^{2\cf}\left\{\Ricb\ud{\mu}{\nu} + \id^{\mu}_{\nu}\bZ^2 \right\}   \label{eqn:trA24}
\end{align}
and its Hermitian adjoint operator reads
\begin{align}
 {\EAA^{(2)}_k[e^{2\cf}\bg,\bg]_{\GhAx\Ghx}}^{\mu}_{\phantom{\mu}\nu}
&= -\sqrt{2}\, e^{2\cf}\left\{ \Ricb\ud{\mu}{\nu}  + \id^{\mu}_{\nu}\,\bZ^2 \right\} \label{eqn:trA25}
\end{align}
Now all operators occurring under the traces on the RHS of the FRGE are functions of the Laplacian $\bZ^2$ alone, and we can easily apply the standard heat kernel techniques to compute the beta-functions for the four couplings of interest, $G_k^{\dyn}$, $\KkbarD$, $G_k^{\background}$, and $\KkbarB$.

\subsection{Projecting onto the relevant invariants}
To obtain the beta-functions we can now follow the lines of ref. \cite{mr} where the corresponding single-metric calculation had been performed. 
The only change  in the FRGE are the ($x$-independent!) factors of $\cf$ which indicate the level a certain term belongs to. 

To evaluate the functional traces we first  decompose the metric fluctuations into its traceless and its trace part: $\flcb_{\mu\nu}=\flcb_{\mu\nu}^{T}+\frac{1}{d}\,\bg_{\mu\nu}(\bg^{\rho\sigma}\flcb_{\rho\sigma})$. In the next step we then express the  Ricci and Riemann tensor everywhere in terms of the scalar curvature $\SRb$ by inserting the maximally symmetric metric of a round $S^d$, exactly as in ref. \cite{mr}. 
Since the scalar curvature occurs in the resulting $\EAA_k^{(2)}$, we Taylor-expand the traces in terms of $\SRb$ explicitly, and then apply the heat kernel expansion
\begin{align}
\Tr\left[e^{s\bZ^2}\right]&=\frac{ \tr(\Id)}{\left(4\pi s\right)^{d\slash 2}} \left\{ \int\md^d x\, \sqrt{\bg} 
+\frac{1}{6}s \int\md^d x\, \sqrt{\bg}\, \SRb +\Order{s^{2}}\right\}   \label{eqn:trA26}
\end{align}
in order to take account of the $\bg$-dependence residing in the Laplacian $\bZ^2$.  In the end, we compare  the LHS of the flow equation to its  evaluated and projected RHS and read off the $k$-derivatives $\partial_t G_k^{\dyn}$, $\cdots$ of all 4 running couplings, $G_k^{\dyn}$, $\KkbarD$, $G_k^{\background}$, and $\KkbarB$.

To disentangle the dynamical and the $\background$-couplings, we hereby expand both sides in terms of $\cf$; the zeroth order in $\cf$ identifies the level-(0) beta-functions, and the linear order those at all higher levels $p=1,2,\cdots$. 

Finally we introduce  dimensionless running couplings:
\begin{align}
 \tg_k^{\cix} = k^{d-2}G^{\cix}_k\,, && \lambda_k^{\cix}=k^{-2}\Kkbar^{\cix}_k\,, \qquad \cix\in\{\dyn,\,\background,\, (0),\,(1),\,(2),\,\cdots\} \label{eqn:trA27}
\end{align}
The superscript $\cix$ stands for either `$\dyn$' or `$\background$', or for the level $(p)$, depending on the language chosen. 
Furthermore, for each one of the Newton constants we define an associated anomalous dimension by
\begin{align}
 \eta^{\cix}\equiv \partial_t \ln G_k^{\cix}\,, \qquad \cix \in \{\dyn,\background,(0),(1),(2),\cdots\} 
\label{eqn:trA27B}
\end{align}
When using \eqref{eqn:trA27} and \eqref{eqn:trA27B} to rewrite  the dimensionful beta-functions in dimensionless terms we have reached our goal and derived the  system of four coupled, autonomous differential equations which describes the RG flow of the EAA in the bi-Einstein-Hilbert truncation. 

\subsection{The truncated system of  RG differential equations}
In the $\dyn$-$\background$ formulation, the four independent differential equations describe the $k$-dependence of the dimensionless couplings $\{\tg_k^{\dyn},\KkD_k,\tg_k^{\background},\KkB_k\}$. 
Following the steps described above, one finds that their general structure is as follows:
\begin{subequations}
\begin{align}
 &\partial_t \tg_k^{\dyn}= \beta_{\tg}^{\dyn}(\tg_k^{\dyn},\KkD_k)\equiv \big[d-2+\aDz(\tg_k^{\dyn},\KkD_k)\big]\tg_k^{\dyn}\label{eqn:trA27CA}\\
&\partial_t \KkD_k= \beta_{\Kk}^{\dyn}(\tg_k^{\dyn},\KkD_k)\label{eqn:trA27CB}\\
&\partial_t \tg_k^{\background}= \beta_{\tg}^{\background}(\tg_k^{\dyn},\KkD_k,\tg_k^{\background})\equiv \big[d-2+\eta^{\background}(\tg_k^{\dyn},\KkD_k,\tg_k^{\background})\big]\tg_k^{\background}\label{eqn:trA27CC}\\
&\partial_t \KkB_k= \beta_{\Kk}^{\background}(\tg_k^{\dyn},\KkD_k,\tg_k^{\background},\KkB_k)\label{eqn:trA27CD}
\end{align}\label{eqn:trA27C}
\end{subequations}
We observe an important hierarchy among the four RG equations:
The two  equations for the `$\dyn$' sector close among themselves and do not contain the `$\background$' couplings, while conversely the `$\dyn$' couplings do  appear in the beta-functions of the `$\background$' sector.

The equivalent system of equations in level language can be written  in terms of the 4 independent couplings $\{\tg_k^{(0)},\Kk^{(0)}_k,\tg_k^{(1)},\Kk_k^{(1)}\}$ since, within the present truncation, all remaining ones are equal to those at level one: $\tg_k^{(p)}=\tg_k^{(1)}$ and $\lambda_k^{(p)}=\lambda_k^{(1)}$ for all $p\geq 1$.
Clearly, for the dimensionless couplings too we can switch easily from  the $\dyn$-$\background$-description to the level-language using 
eqs. \eqref{eqn:trA09}, \eqref{eqn:trA10} and \eqref{eqn:trA27}: 
\begin{align}
 &\tg_k^{(0)}= \frac{\tg_k^{\dyn}\tg_k^{\background}}{\tg_k^{\dyn}+\tg_k^{\background}}\,\quad && \text{ and }  \quad
&& \tg_k^{(p)}= \tg_k^{\dyn}\quad\text{ for all }p\geq1 \,, \nonumber \\
&\Kk_k^{(0)}= \frac{\KkD_k \,\tg_k^{\background} + \KkB_k\,\tg_k^{\dyn}}{\tg_k^{\dyn}+\tg_k^{\background}} \,\quad &&\text{ and } \quad
&& \Kk_k^{(p)}=\KkD_k\quad \text{ for all }p\geq1 \,.
\label{eqn:res4D_016}
\end{align}
The ensuing relation among the anomalous dimensions reads
\begin{align}
 \frac{\eta^{(0)}}{\tg_k^{(0)}}=\frac{\eta^{\dyn}}{\tg_k^{\dyn}}+\frac{\eta^{\background}}{\tg_k^{\background}} \quad \text{ and }\quad \frac{\eta^{(p)}}{\tg_k^{(p)}}=\frac{\eta^{\dyn}}{\tg_k^{\dyn}} \, \quad \text{ for all } p\geq 1\,. \label{eqn:trA27D}
\end{align}

The RG equations of  all Newton type couplings can be cast into the form 
\begin{align*}
\partial_t \tg_k^{\cix}=(d-2+\eta^{\cix})\tg_k^{\cix} \quad \text{ for all }\, \cix\in \{\dyn,\,\background,\, (0),\,(1),\,(2),\,\cdots\}
\end{align*}
where all non-canonical $k$-dependence is encoded in the anomalous dimension $\eta^{\cix}$. 

The differential equations for all cosmological constants have the general form 
\begin{align}
\partial_t \lambda_k^{\cix}= (\eta^{\cix}-2)\lambda_k^{\cix} + \tg_k^{\cix}\, \kAx(\{\lambda^{\mathcal{J}},\tg^{\mathcal{J}}\})
\label{eqn:trA100D}
\end{align}
In the present truncation, the functions $\kAx$ depend on the $\dyn$-couplings only.

For an arbitrary spacetime of dimension $d$, we tabulate the rather complicated explicit formulae for the four beta-functions in Appendix \ref{sec:BetaD} to which the reader could turn at this point. 
In the main body of this paper, we shall analyze the RG equations in  $d=4$ dimensions in the next sections. 
Only in Section \ref{sec:dOther} we shall have a brief look at $d=2+\epsilon$ and $d=3$.

 \section{The new RG equations} \label{sec:03}
The rest of this paper is mainly devoted to a detailed analysis of the flow equations  of the bi-metric EH-truncation, eqs. \eqref{eqn:trA27C}, mostly in $d=4$ dimensions. 
The focus will be on the crucial question as to whether Asymptotic Safety in the UV can coexist with Background Independence, that is, with unbroken split-symmetry at the physical point $k=0$. In particular, we shall uncover a novel attractor mechanism leading to a complete restoration of split-symmetry for $k\rightarrow0$. Technically, the question of coexisting Asymptotic Safety and Background Independence is a very hard one since it concerns the {\it global} features of the RG trajectories: Asymptotic Safety concerns their UV behavior ($k\rightarrow\infty$), while Background Independence, interpreted as split-symmetry restoration at $k=0$, is an IR property.

In this Section we start the investigation by presenting and discussing the explicit form of the new RG equations for $d=4$. Before displaying them in detail, various remarks concerning their general structure are in order. 

\noindent {\bf (A)} The size of all physical RG-effects is controlled by the $\dyn$-couplings only, more precisely by couplings of at least level-(2) which, in the present case, are equal to $G_k^{\dyn}$ or $\KkbarD$, respectively. The level-(0) and level-(1) couplings, and correspondingly the $\background$-couplings, cannot enter the RHS of the FRGE, for they are at most linear in the dynamical fields and hence drop out in the calculation of the Hessian operator. 

\noindent {\bf (B)} When considering the various terms contributing to the beta-functions for general spacetime dimensions $d$ one observes that, quite remarkably, {\it all terms proportional to $\KkD$ are also proportional to $(d-4)$}. Hence in $d=4$ these terms drop out, and the remaining  $\KkD$-dependence is via the threshold functions only.
In fact, all beta-functions can be expressed in terms of the same well-known threshold functions $\ThrfA{p}{n}{{\textstyle w}}$ and $\ThrfB{p}{n}{{\textstyle w}}$ already introduced for the single-metric case \cite{mr}.\footnote{See eqs. (4.32) of \cite{mr} for their definition.}
Furthermore, the threshold functions with non-zero arguments will always appear in the combination 
\begin{align}
\qA{p}{n}{{\textstyle w}}\equiv \ThrfA{p}{n}{{\textstyle w}} -\frac{1}{2}\aDz\, \ThrfB{p}{n}{{\textstyle w}}\label{eqn:res4D_000B}
\end{align}
Note that it is always $\aDz$ that enters $\qA{p}{n}{{\textstyle w}}$. 
Its background counterpart $\eta^{\background}$ cannot make its appearance here since, as we said already, all contributions $\propto G_k^{\background}$ drop out from the trace upon calculating the Hessian.

\noindent {\bf (C)} In the following subsections we list the  beta-functions of the dynamical couplings, the background couplings, and the  beta-functions of the level description in turn. 
Similar to the prefactor of the ghost action, $\rhoG$, all formulae contain also a parameter  $\rhoP$ $\in\{0,1\}$ which we merely introduced as a book keeping device. Contributions proportional (not proportional) to  $\rhoP$ originate from the {\bf paramagnetic} ({\bf diamagnetic}) interaction terms\footnote{By definition \cite{andi1,andi-MG}, and in accordance with the identical nomenclature in Yang-Mills theory, derivative (non-derivative) interaction terms coupling metric fluctuations $\flcb_{\mu\nu}$ to the background $\bg_{\mu\nu}$ are referred to as diamagnetic (paramagnetic), typical structures being $\flcb \bZ^2 \flcb$ ($\flcb \SRb \flcb$).}.\cite{andi1} We refer to all terms that are {\it not} proportional to $\rhoG$ as `graviton' contributions.
In the nomenclature of ref. \cite{andi1} those are either of `paramagnetic' or `diamagnetic' origin and correspondingly  carry a factor of $\rhoP$ or are missing a factor of $\rhoP$, respectively. In view of the physical picture of Asymptotic Safety developed in ref. \cite{andi1} it will be instructive to keep the two types of contributions separately.
Later on, when we are not interested in this distinction we `turn on' all terms, putting $\rhoP=\rhoG=1$, unless stated otherwise.

\subsection[The \texorpdfstring{$\dyn$}{Dyn} sector: beta-functions for \texorpdfstring{$\tg^{\dyn}$, $\KkD$}{the Dyn-couplings}]{The $\bm{\dyn}$ sector: beta-functions for $\bm{\tg^{\dyn}}$, $\bm{\KkD}$}
The RG equation for the dynamical Newton constant $\tg_k^{\dyn}$ is governed by the anomalous dimension $\aDz$.
The FRGE provides the following relation for it:
\begin{align}
\eta^{\dyn}(\tg^{\dyn},\,\KkD)&=
 +  \frac{1}{\pi}\,\Big\{\tfrac{5}{3}  \, \qA{2}{2}{-2\KkD}
-6   \rhoP\,
   \big[  4 \, \qA{3}{3}{-2\KkD} 
-  \qA{2}{2}{-2\KkD} \big] 
 \, \Big\}\,\tg^{\dyn} \nonumber \\ &\quad
+\frac{2 }{\pi} \rhoG \,\Big\{ \big(\tfrac{2}{3}-\rhoP\big) \ThrfA{2}{2}{0} + 4 \rhoP\ThrfA{3}{3}{0} \Big\}\,\tg^{\dyn}
\label{eqn:res4D_004}
\end{align}
It has a global factor of $\tg_k^{\dyn}$ and depends  on $\KkD_k$ through the threshold functions. In addition, $\qA{p}{n}{w}$ contains another term proportional to $\aDz$. As a consequence, eq. \eqref{eqn:res4D_004} is an implicit equation for the anomalous dimension. It can be solved to yield the final result:
\begin{align}
\eta^{\dyn}(\tg^{\dyn},\,\KkD)&= \frac{\kAD(\KkD;4)\,\tg^{\dyn}}{1-\kBD(\KkD;4)\,\tg^{\dyn}}
\label{eqn:res4D_005}
\end{align}
We obtain a non-polynomial dependence of $\aDz$ on $\tg^{\dyn}$, and  $\kAD(\KkD;4)$ and $\kBD(\KkD;4)$ are functions of the cosmological constant, here specialized for $d=4$. The first one contains graviton as well as ghost contributions, 
\begin{align}
\kAD(\KkD;4)&=
  \frac{1}{\pi}\,\Big\{\tfrac{5}{3}  \, \ThrfA{2}{2}{-2\KkD}
-6   \rhoP\,
   \big[  4 \, \ThrfA{3}{3}{-2\KkD} 
-  \ThrfA{2}{2}{-2\KkD} \big] 
 \, \Big\} \nonumber \\ &\quad
+\frac{2 }{\pi} \rhoG \,\Big\{ \big(\tfrac{2}{3}-\rhoP\big) \ThrfA{2}{2}{0} + 4 \rhoP\ThrfA{3}{3}{0} \Big\}
\label{eqn:res4D_006}
\end{align}
whereas the second one stems entirely from the  graviton sector:
\begin{align}
\kBD(\KkD;4)&=
  -\frac{1}{2\pi}\,\Big\{\tfrac{5}{3}  \, \ThrfB{2}{2}{-2\KkD}
-6   \rhoP\,
   \big[  4 \, \ThrfB{3}{3}{-2\KkD} 
-  \ThrfB{2}{2}{-2\KkD} \big] 
 \, \Big\}
\label{eqn:res4D_007}
\end{align}
Furthermore the running of the dynamical cosmological constant $\KkD_k$ is described by
\begin{align}
\partial_t \KkD_k
&= (\aDz-2)\,\KkD_k + \tg^{\dyn} \frac{1}{\pi}\left[ 5\,
 \qA{2}{3}{-2\KkD_k}
  +4  \, \rhoG  \,\ThrfA {2}{3}{0}  \right]
\label{eqn:res4D_011}
\end{align}

\paragraph{Remarks:}
The beta-functions for the `$\dyn$' couplings exhibit certain properties that are reminiscent of the single-metric truncation \cite{mr, frank1}: First, notice that the threshold functions $\ThrfA{p}{n}{-2\KkD}$ and $\ThrfB{p}{n}{-2\KkD}$ have singularities when the argument approaches $-1$. 
This leads to a boundary of theory space at $\KkD=1\slash2$ in both types of truncations. 
Second, in the single- and the bi-metric truncation we find a further divergence in $\aDz$ that can restrict the physically relevant part of theory space even stronger, namely a  boundary caused by the singular curve $1-B_2^{\sm} \tg^{\sm}=0$ or $1-\kBD \tg^{\dyn}=0$, respectively. 
A certain class of RG trajectories, later referred to  as type (IIIa)$^{\dyn}$ trajectories, terminate on these lines. 
We find that even though quantitatively the singularity properties change when moving from the single- to the bi-metric beta-functions, the qualitative picture remains the same.

\subsection[The  \texorpdfstring{$\background$}{B}-sector: beta-functions for \texorpdfstring{$\tg^{\background}$, $\KkB$}{the B-couplings}]{The  $\bm{\background}$-sector: beta-functions for $\bm{\tg^{\background}}$, $\bm{\KkB}$}
In the $\background$-sector, the essential RG running is inherited from $\tg_k^{\dyn}$ and $\KkD_k$. The $\background$-couplings themselves appear on the RHS of their differential equations only in the  trivial canonical term since $\Tr[\cdots]$ did not depend on them. In particular, Newton's constant $\tg^{\background}$ enters its own anomalous dimension $\eta^{\background}$ only as a global factor. Besides that it depends on the dynamical couplings only:
\begin{subequations}
\begin{align}
 \eta^{\background}(\tg^{\dyn},\,\KkD,\,\tg^{\background})&= \kB(\tg^{\dyn},\KkD)\, \tg^{\background} \label{eqn:res4D_002A} \\
\kB(\tg^{\dyn},\KkD)&=
 \frac{1}{\pi}\, \Big\{
\tfrac{5}{3} \qA{1}{1}{-2\KkD}
- \big(\tfrac{5}{3} +12\rhoP\big)\, \qA{2}{2}{-2\KkD}
+24\,  \rhoP\,
   \qA{3}{3}{-2\KkD} \, \Big\} \nonumber \\ &\quad
-\frac{4}{\pi} \rhoG \,\Big\{ \tfrac{1}{3}  \ThrfA{1}{1}{0}    + \tfrac{1}{3} \ThrfA{2}{2}{0} + 2 \rhoP\ThrfA{3}{3}{0} \Big\}
\label{eqn:res4D_002B}
\end{align}\label{eqn:res4D_002}
\end{subequations}
Notice that eq. \eqref{eqn:res4D_002A} has no denominator analogous to its $\dyn$ counterpart \eqref{eqn:res4D_005}. 
Note also that $\eta^{\background}$, via the $q^p_n$-functions \eqref{eqn:res4D_000B}, indirectly depends on $\aDz=\aDz(\tg^{\dyn},\KkD)$.

The running of the cosmological constants in the $\background$-sector is described by the differential equation
\begin{subequations}
\begin{align}
 \partial_t \KkB_k &= (\eta^{\background}-2)\KkB_k 
+ \kA(\KkD_k,\tg_k^{\dyn}) \,\tg_k^{\background} 
\label{eqn:res4D_008}
\end{align}
Here  the RG-effects that are not already covered by $\eta^{\background}$ are encoded in the function
\begin{align}
 \kA(\KkD_k,\tg_k^{\dyn})&\equiv \frac{1}{\pi}
\Big\{		
 5\big(\qA{1}{2}{-2\KkD_k}
-   \qA{2}{3}{-2\KkD_k}\big)
- 4 \rhoG \big(\ThrfA {1}{2}{0} +  \,\ThrfA {2}{3}{0}  \big)
\Big\} \label{eqn:res4D_008B}
\end{align}
\end{subequations}

\paragraph{Remarks:}
The beta-functions of the `$\background$' (or level-$(0)$) couplings are free from any additional singularities that would further reduce the physical part of theory space. 
In particular, as it is apparent from  \eqref{eqn:res4D_002}, the anomalous dimension $\eta^{\background}$ is well-defined for any value of $\tg^{\background}$ on which it depends linearly. 
Notice also that the threshold functions are evaluated at $-2\KkD_k$, and never at $-2\KkB_k$, so that there is {\it no restriction in the $\KkB$ direction of theory space} that would be analogous to the $\KkD=1\slash2$ boundary. 
The same holds true for the level-$(0)$ plane.

We thus have all four beta-functions at hand and can, at least in principle, solve the full system of differential equations. 
However, before embarking on that let us discuss a number of further general aspects.

\subsection[The level-description: beta-functions for \texorpdfstring{$\tg^{(p)}$, $\Kk^{(p)}$}{the level-couplings}]{The level-description: beta-functions for $\bm{\tg^{(p)}}$, $\bm{\Kk^{(p)}}$}
The $\{\background,\,\dyn\}$-language is equivalent to the one based upon the level number where the a priori different higher levels $p=1,2,\cdots$ happens to be identical within the present truncation. The higher level Newton constants $\tg^{(p)}_k$ have the same running for all $p=1,2,3,\cdots$, for instance, and their beta-functions coincide in turn with that of $\tg_k^{\dyn}$. Hence 
\begin{align}
 \eta^{(1)}=\eta^{(2)}=\cdots = \eta^{\dyn} \label{eqn:res4D_003B}
\end{align}
with $\eta^{\dyn}$ given in equation \eqref{eqn:res4D_005}.
For the level-(0) anomalous dimension, i.e. the anomalous dimension corresponding to the running prefactor of the $\sqrt{g}\SR$-term $1\slash G_k^{(0)}=1\slash G_k^{\background}+1\slash G_k^{\dyn}$, we obtain instead:
\begin{align}
 \eta^{(0)}(\tg^{\dyn},\KkD,\tg^{(0)})&=
 \frac{2}{\pi}\, \left[\frac{5}{6} \qA{1}{1}{-2\KkD}
- 3\rhoP\, \qA{2}{2}{-2\KkD}- \rhoG \big(  \tfrac{2}{3}  \ThrfA{1}{1}{0}    +\rhoP \,\ThrfA{2}{2}{0}\big)\right]\tg^{(0)}
\label{eqn:res4D_003}
\end{align}
Note that $\eta^{(0)}\slash \tg^{(0)}$ is a function of $\KkD_k$ and $\tg_k^{\dyn}$ only. 

For the cosmological constant at level-(0) we find 
\begin{align}
\partial_t \Kk_k^{(0)}
&= (\eta^{(0)}-2)\Kk^{(0)}+  \tg^{(0)}\,  \frac{1}{\pi}\left[ 5 
 \, \qA{1}{2}{-2\KkD}  -4  \rhoG \ThrfA {1}{2}{0} \right]\,,
\label{eqn:res4D_009}
\end{align}
while for the higher cosmological constants
\begin{align}
 \lambda_k^{(1)}=\lambda_k^{(2)}=\cdots=\KkD_k \, .
\label{eqn:res4D_010}
\end{align}
At all  levels $p=1,2,3,\cdots$ their  running is locked to that of $\KkD_k$, which in turn is governed by equation \eqref{eqn:res4D_011}.

\subsection{The hierarchical structure of the truncated RG equations}\label{sec:03_04}
No matter whether we employ the $\{\background, \dyn\}$ or the $\{(0),(1),(2),\cdots\}$ language, the truncated FRGE always comprises a coupled system of ordinary differential equation in four variables,  $\{\tg_k^{\background},\tg_k^{\dyn},\KkD_k,\KkB_k\}$ or $\{\tg_k^{(0)},\tg_k^{(p)}\equiv\tg_k^{\dyn},\lambda_k^{(0)},\lambda_k^{(p)}\equiv\KkD_k\,,\,p\geq1\}$, respectively. 
In either case {\it the dynamical couplings $\tg_k^{\dyn}$ and $\KkD_k$ influence the beta-functions of the $\background$ or level-(0) couplings, but there is no back-reaction of $\{\tg_k^{\background},\,\lambda_k^{\background}\}$ or $\{\tg_k^{(0)},\,\lambda_k^{(0)}\}$  on the `$\dyn$' variables; furthermore $\Kk_k^{\background\slash (0)}$ does not backreact on $\tg_k^{\background\slash(0)}$.} The system of equations decomposes in the following hierarchical way:
\begin{align}
 \{\tg_k^{\dyn},\KkD_k\} \longrightarrow \tg_k^{\background \slash (0)} \longrightarrow \lambda_k^{\background\slash (0)}\,.
\label{eqn:res4D_012}
\end{align}

This observation fixes the strategy for the (mostly numerical) solution of the RG equations on which we embark in the following sections:
First we solve the non-trivially coupled $\tg_k^{\dyn}$-$\KkD_k$ system, then we insert its solution into the single differential equation of the background (or alternatively, level-(0) ) Newton coupling, solve for its $k$-dependence, and finally use all three solution functions obtained already to get the running of the cosmological constant in the background sector (or at level-(0)).

The hierarchy \eqref{eqn:res4D_012} raises the following question: 
Is it really necessary for the couplings in the background sector (a) to assume fixed point values for $k\rightarrow\infty$, and (b) to restore split-symmetry at $k=0$?
After all, the $\background$-couplings could even diverge in the UV without causing any problems for the $\dyn$-couplings and the $\flcb_{\mu\nu}$-correlation functions given by $(\delta\slash \delta \flcb)^n\, \EAA_k[\flcb;\bg]\big|_{\flcb=0}$.
Or equivalently, should we insist that the level-($0$) part of $\EAA_k$, that is, $\EAA_k[\flcb;\bg]\big|_{\flcb=0}$, is well defined, too?

The answer to these questions is `yes'.
In fact, the EAA at level zero is related to the partition function $Z_k[\bg]$ whose $\bg$-dependence is indeed of physical interest, for instance, when it is used as a tool for `counting' the field modes which are already integrated out at a given scale \cite{daniel1, daniel-MG}. This quantity is similar to a state sum and makes its appearance when one studies quantum gravity effects in Black Hole Thermodynamics, for example. (See ref. \cite{daniel1} for further details.)
Therefore we want also the level-($0$) part of the EAA to be asymptotically safe, and to be linked to the higher levels in a split-symmetric way.

Moreover, the present paper is supposed to pave the way for future work on more complicated truncations which also allow lifting the degeneracy
 among all higher levels $p=1,2,\cdots$ which is still assumed here.
In this sense our analysis of the level-($0$) \slash level-($1$) interplay is supposed to have the character of a model for the general case.

\subsection{No anti-screening in the semiclassical regime}
In the following sections we shall present a comprehensive analysis of the RG equations using the strategy of the previous subsection. Here, as a warm up, we present a simple analytic solution to these equation which is valid in the semiclassical regime, i.e. at scales in, and slightly above those of the classical regime where all dimensionful Newton and cosmological constants are strictly $k$-independent. 

The technical details related to this approximation, over and above the approximation due to the truncation, are relegated to Appendix \ref{sec:AppSemiclassical}. There we find that if $\tg^{\cix}_{k}\ll1$ and $\KkD_k\ll 1$ all of the dimensionful quantities $G_k^{\cix}$ and $\Kkbar_k^{\cix}$, $\cix\in \{\dyn, \background, (p)\}$, behave as 
\begin{subequations}
\begin{align}
 G_k^{\cix}&\approx G_0^{\cix}\,\big[ 1- \omega_d^{\cix}\, G_0^{\cix}\, k^{d-2}\big] \\
\Kkbar_k^{\cix}&\approx \Kkbar_0^{\cix}+ \nu_d^{\cix} \, G_0^{\cix}\, k^d 
\end{align}\label{eqn_res4D_0SC}
\end{subequations}
The dimension-dependent constants $\omega_d^{\cix}$ and $\nu_d^{\cix}$ are tabulated in the Appendix. 
The general structure of the solution \eqref{eqn_res4D_0SC} is quite familiar; it also obtains in the single-metric (`$\sm$') Einstein-Hilbert truncation. 
There it was found that the crucial $\omega$-coefficient which governs the running of $G_k\equiv G_k^{\sm}$ is {\it positive} in the most interesting case $d=4$.
With $\omega_4^{\sm}>0$ Newton's constant decreases for increasing $k$, and this was interpreted as a kind of gravitational anti-screening \cite{mr}.

Within the approximation \eqref{eqn_res4D_0SC}, the anomalous dimension is given by
\begin{align}
 \eta^{\cix}&= - (d-2) \omega_d^{\cix}\, \tg^{\cix} \label{eqn_res4D_0SCC}
\end{align}
so that $\omega_d^{\cix}>0$, i.e. anti-screening, corresponds to a negative anomalous dimension $\eta^{\cix}<0$. Therefore, the positive (negative) sign of $\omega_d^{\sm}$ ($\eta^{\sm}$) was highly welcome from the Asymptotic Safety point of view since at a non-trivial fixed point of the equation $\partial_t \tg = \big[d-2+\eta\big]\,\tg$ the anomalous dimension is negative, too:%
\footnote{Here and in the following we always assume $d>2$.} 
 $\eta_* = -(d-2)<0$.

It is therefore somewhat discomforting to discover that in the semi-classical regime the anti-screening sign is {\it not} re-obtained within the bi-metric truncation. 
In fact, in  the bi-metric setting it is the dynamical Newton constant $G_k^{\dyn}$ that should be compared to $G_k^{\sm}$. 
Therefore the hallmark of anti-screening is now $\aDz<0$, that is, $\omega_d^{\dyn}>0$ in the semiclassical approximation. 
However, using the explicit equations in Appendix \ref{sec:AppSemiclassical} for $d=4$, we find that, with any cutoff, $\omega_4^{\dyn}<0$.
Hence {\it the dynamical Newton constant shows a screening rather than anti-screening behavior in the semiclassical regime: $\eta^{\text{\rm \dyn}}\big|_{\text{\rm semiclass}}>0$.}
Instead, the background Newton constant $G_k^{\background}$ runs in the opposite direction, $\eta^{\background}\big|_{\text{semiclass}}<0$, but this has no direct physical meaning.

Fortunately later on we shall also discover that in other parts of theory space $\eta^{\dyn}$ does become negative actually, and that even non-trivial fixed points form. 
Nevertheless, this simple example is a severe warning showing the limitations of the single-metric truncations.

\section{The fixed points of the RG flow} \label{sec:fps}
Next we turn to the task of solving the truncated flow equations \eqref{eqn:trA27C} without further approximations. 
We follow the strategy outlined in Subsection \ref{sec:03_04} based on their hierarchical structure. In this Section we begin by searching for fixed points of the $4$ dimensional flow.

In all numerical calculations of the following sections we shall employ the `optimized' shape function \cite{litimPRL} $R^{(0)}(z)=(1-z) \Theta(1-z)$ for which the structure functions $\ThrfA{p}{n}{w}$ and $\ThrfB{p}{n}{w}$ \cite{mr} assume a simple rational form, see eqs. \eqref{eqn:appx_BetaD_020} in the Appendix.

\subsection{From the dynamical to the background sector}

\paragraph{The $\dyn$-sector.}
So let us start with the dynamical couplings $\tg^{\dyn}$, $\KkD$ and try to find common zeros of their beta-functions: $\beta^{\dyn}_{\tg}=0=\beta^{\dyn}_{\lambda}$. Setting $\rhoP=1$, $\rhoG=1$, i.e. including all (dia- and paramagnetic, as well as ghost) contributions, we find indeed three fixed points $(\tg^{\dyn}_*,\KkD_*)$: Besides a Gaussian fixed point at $\KkD_*=\tg_*^{\dyn}=0$, henceforth denoted \fpgD-\fpL{}, there are two non-Gaussian fixed points at which both $\tg^{\dyn}$  and $\KkD$ are positive and negative, respectively; they will be denoted \fpnD-\fpL{} and \fpnDn-\fpL{} in the following. 
The fixed point coordinates are given by
\begin{align}
\begin{matrix}[lc|c|c]
& \text{\fpnD-\fpL{}} & \text{\fpnDn-\fpL{}} & \text{\fpgD-\fpL{}}
\\\cline {2-4}
\tg_*^{\dyn} =	&	0.703 &	 -3.54  	& 0
\\ 
 \KkD_* =		& 	0.207 &	-0.302	& 0
\end{matrix} 
\label{eqn:res4D_013A}
\end{align}
We recall that for the same shape function the well-known fixed point of the single-metric truncation \cite{oliver1, frank1} is located at $(\tg_*^{\sm}=0.707 ,\lambda_*^{\sm}=0.193)$. These coordinates are remarkably close to those of the \fpnD-\fpL{}.

An RG trajectory which `sits' at the Gaussian fixed point \fpgD-\fpL{} for all $k\in[0,\infty)$ constitutes a simple solution of both $\dyn$-equations:
\begin{align}
 \tg_k^{\dyn}=0\,, && \KkD_k = \KkD_{k_0}\, (k_0\slash k)^2 && (0\leq k <\infty)
\label{eqn:res4D_013}
\end{align}
Here $\KkD_{k_0}$ denotes the value of the cosmological constant at $k=k_0$.
This trajectory separates the positive $\tg^{\dyn}>0$ from the negative $\tg^{\dyn}<0$ regime. 
No trajectory ever crosses the $\tg^{\dyn}=0$ plane.
The  picture  for the background and  the level-(0) couplings is similar. 
The trajectory that separates the positive from the negative $\tg^{\background\slash (0)}$ regime is
\begin{align}
 \tg_k^{\background\slash (0)}=0\,, && \Kk^{\background\slash (0)}_k = \Kk_{k_0}^{\background\slash (0)}\, (k_0\slash k)^2\,,
\label{eqn:res4D_014}
\end{align}
and no trajectory ever passes through $\tg^{\background}=0$ or $\tg^{(0)}=0$. 
This implies that  {\it there exists no crossover trajectory connecting \fpnD-\fpL{} to \fpnDn-\fpL{}}. 
More generally, there is no trajectory along which any of the various Newton constants $\tg^{\cix}$ would  cross zero.

\paragraph{The $\background$-sector.}
So far, we have seen that there are two non-Gaussian and one Gaussian fixed point solutions in the {\it dynamical} sector. 
Next, we insert their coordinates into the beta-functions of the {\it background} quantities $(\tg^{\background},\KkB)$ and look for zeros which would generalize the `$\dyn$' fixed points to the full 4 dimensional theory space. 
Such associated zeros $(\tg_*^{\background},\, \KkB_*)$ do indeed exist, and we find precisely one with $(\tg^{\background}_*,\KkB_*)\neq0$  for each of the fixed points in the $\dyn$-sector. 
Those related to the three non-Gaussian $\dyn$-FPs are located at
\begin{align}
\begin{matrix}[lc|c|c]
& \text{\fpnD-\fpL{}} & \text{\fpnDn-\fpL{}} & \text{\fpgD-\fpL{}}
\\\cline {2-4}
\tg_*^{\background} =	&	8.18 &	 1.531  	& 1.396
\\ 
 \KkB_* =		& 	-0.008 &	-0.12	& -0.111
\end{matrix}
\label{eqn:res4D_015}
\end{align}
Furthermore, for $\tg^{\background} =0= \KkB$ the beta-functions for the $\background$-sector vanish as well, for any value of the $\dyn$-couplings.
Thus, also a Gaussian fixed point $(\tg_*^{\background},\KkB_*)=(0,0)$ exists in the $\background$-sector, and it can be combined with each one of the $\dyn$ fixed point in \eqref{eqn:res4D_013A}. 

In total we have six fixed points therefore: one which is purely Gaussian, having both $(\tg^{\dyn}_*,\KkD_*)=0$ and $(\tg_*^{\background},\KkB_*)=(0,0)$, three mixed ones , and two are purely non-Gaussian ones. 
The table \eqref{eqn:res4D_017B} gives a summary of all six combined fixed points and introduces the notation we shall use for them.
\begin{align}
\begin{tabular}[!ht]{l||c|c|c}
	d=4			  & $\substack{\text{\fpnD-\fpL{}}\\ (\tg_*^{\dyn}= 0.7,\,\KkD_*= 0.2)}$ & $\substack{\text{\fpnDn-\fpL{}}\\ (\tg_*^{\dyn}= -3.5,\,\KkD_*=-0.3)}$	&	$\substack{\text{\fpgD-\fpL{}}\\ (\tg_*^{\dyn}= 0,\,\KkD_*= 0)}$ \\[2.2ex] \hline\hline
&&& \\[0pt] 
\text{\fpnB-\fpL{}}	 &  	$\substack{\text{\fpnB\fpC\fpnD-\fpL{}}\\ (\tg_*^{\background}= 8.2,\,\KkB_*= -0.01)}$		&$\substack{\text{\fpnB\fpC\fpnDn-\fpL{}}\\ (\tg_*^{\background}= 1.5,\,\KkB_*= -0.1)}$ &$\substack{\text{\fpnB\fpC\fpgD-\fpL{}}\\ (\tg_*^{\background}= 1.4,\,\KkB_*= -0.1)}$\\[2.2ex] \hlinewd{0.2pt}
&&& \\[1pt] 
\text{\fpgB-\fpL{}} 	 &  $\substack{\text{\fpgB\fpC\fpnD-\fpL{}}\\ (\tg_*^{\background}= 0,\,\KkB_*=0)}$			& $\substack{\text{\fpgB\fpC\fpnDn-\fpL{}}\\ (\tg_*^{\background}=0,\,\KkB_*=0)}$ & $\substack{\text{\fpgB\fpC\fpgD-\fpL{}}\\ (\tg_*^{\background}= 0,\,\KkB_*= 0)}$\\[2.2ex]
\end{tabular}	\label{eqn:res4D_017B}
\end{align}
An analogous table in the level language will be given below.

\paragraph{The fixed points for level-(0) couplings.}
In the level-language the dynamical couplings, i.e. those at the higher levels $p=1,2,3,\cdots$, influence the beta-functions at level-(0), but not vice versa. The fixed points found in \eqref{eqn:res4D_013A} for the `$\dyn$' couplings entail  $(\tg_*^{(p)},\Kk^{(p)}_*)=(\tg_*^{\dyn},\KkD_*)$ for the levels $p\geq1$. 

Regarding $p=0$, the  fixed points listed in \eqref{eqn:res4D_017B} give rise to three non-trivial ones for the level-(0) couplings. 
Two of them, namely \fpnB\fpC\fpnD-\fpL{} and \fpnB\fpC\fpnDn-\fpL{} are easily computed using the conversion rules for the dimensionless couplings of eq. \eqref{eqn:res4D_016}.
To obtain the third one, \fpnB\fpC\fpgD-\fpL{}, related to the Gaussian fixed points of the `$\dyn$' sector, one must be careful:
When directly using relation \eqref{eqn:res4D_016} any of the remaining 4 fixed points would seem to be located at  $\tg_*^{(0)}=0=\Kk_*^{(0)}$. 
Because of a division by zero this is incorrect, however. 
Going back to the coupled system \eqref{eqn:res4D_003}-\eqref{eqn:res4D_010} it turns out that the correct coordinates of \fpnB\fpC\fpgD-\fpL{} are actually non-zero:
\begin{align}
 \text{\fpnO\fpC\fpgD-\fpL{}}:\qquad \tg_*^{(0)}=1.713 \,,\quad \Kk^{(0)}_*=0.068
\label{eqn:res4D_018}
\end{align}
They are obtained  if we directly solve for zeros in  the full $\{\tg_k^{(0)},\tg_k^{\dyn},\lambda_k^{(0)},\KkD\}$ system. 

In total we thus have found the following three non-trivial fixed point values for the level-(0) couplings:
\begin{align}
\begin{matrix}[lc|c|c]
& \text{\fpnD-\fpL{}} & \text{\fpnDn-\fpL{}} & \text{\fpgD-\fpL{}} 
\\\cline {2-4}
\tg_*^{(0)} =	&	0.647 &	 2.697 	& 1.713
\\ 
 \Kk^{(0)}_* =		& 	0.190 &	0.0168 & 0.068	
\end{matrix}
\label{eqn:res4D_017}
\end{align}
As a result, the list of all six combined fixed points looks as follows:
\begin{align}
\begin{tabular}[!ht]{l||c|c|c}
	d=4			  & $\substack{\text{\fpnD-\fpL{}}\\ (\tg_*^{\dyn}= 0.7,\,\KkD_*= 0.2)}$ & $\substack{\text{\fpnDn-\fpL{}}\\ (\tg_*^{\dyn}= -3.5,\,\KkD_*=-0.3)}$	&	$\substack{\text{\fpgD-\fpL{}}\\ (\tg_*^{\dyn}= 0,\,\KkD_*= 0)}$ \\[2.2ex] \hline \hline
&&& \\[0pt] 
\text{\fpnO-\fpL{}}	 &  	$\substack{\text{\fpnO\fpC\fpnD-\fpL{}}\\ (\tg_*^{(0)}= 0.65,\,\Kk^{(0)}_*=0.2)}$		&$\substack{\text{\fpnO\fpC\fpnDn-\fpL{}}\\ (\tg_*^{(0)}=2.7,\,\Kk^{(0)}_*=0.02)}$ &$\substack{\text{\fpnO\fpC\fpgD-\fpL{}}\\ (\tg_*^{(0)}= 1.7,\,\Kk^{(0)}_*=0.1)}$\\[2.2ex] \hlinewd{0.2pt}
& & &
\\[0pt]
\text{\fpgO-\fpL{}} 	 &  $\substack{\text{\fpgO\fpC\fpnD-\fpL{}}\\ (\tg_*^{(0)}= 0,\,\Kk^{(0)}_*=0)}$			& $\substack{\text{\fpgO\fpC\fpnDn-\fpL{}}\\  (\tg_*^{(0)}= 0,\,\Kk^{(0)}_*=0)}$ & $\substack{\text{\fpgO\fpC\fpgD-\fpL{}}\\  (\tg_*^{(0)}= 0,\,\Kk^{(0)}_*=0)}$
\end{tabular}	\label{eqn:res4D_017C}
\end{align}
\noindent {\bf Summary:} Each fixed point in the level description has an analog in the $\background$ - $\dyn$ system.
Regardless of whether we consider the $\{\background, \dyn\}$ or the $\{(0),(1),(2),\cdots\}$ language, we find a total of six fixed points. Two of them lie in the negative $\tg^{\dyn}$ half-plane and will be not relevant for our further investigation. The remaining four are a doubly non-Gaussian fixed point \fpnB\fpC\fpnD-\fpL{}, a purely Gaussian one, and two mixed fixed points at which  either the dynamical or the background  couplings vanish.
Their potential relevance to the Asymptotic Safety construction will be explored in the following sections.

\subsection{Critical exponents and scaling fields}
In the following we list for all 6 fixed points their critical exponents $\theta_j$ and the corresponding eigenvectors $V^{(j)}$ of the stability matrix, the `scaling fields'.
In our conventions, $\text{Re}\, \theta_j>0$ corresponds to an IR-relevant scaling field. Hence, the dimensionality $s_{\text{UV}}$ of the UV-critical manifold $\cUV$ of a certain fixed point equals the number of $\theta$'s  with a positive real part.
This dimensionality in turn is equal to the number of undetermined parameters in an asymptotically safe theory based upon the fixed point in question \cite{NJP}.

The Tables \ref{tab:fp1A}, \ref{tab:fp2A}, and \ref{tab:fp3A}, in turn refer to the fixed points which descend from the $\dyn$-fixed points \fpgD-\fpL{}, \fpnDn-\fpL{}, and \fpnD-\fpL{}, respectively. 
Besides the critical exponents $\theta_j$, $j=1,\cdots,4$, the negative eigenvalues of the stability matrix, each Table contains the related 4-component eigenvectors $V^{(j)}$ for both the $\background$-$\dyn$ and the level-presentation. 
Here the $\hat{e}$'s are Cartesian unit vectors; $\hat{e}_{\tg}^{\dyn}\equiv \hat{e}^ {(1)}_{\tg}= \hat{e}^ {(2)}_{\tg}=\cdots$ points in the direction of the dynamical (or level-($p$), $p\geq1$) Newton constant.

As can be seen in Tables \ref{tab:fp1A} and \ref{tab:fp2A}, the fixed points related to \fpgD-\fpL{} and \fpnDn-\fpL{} have real critical exponents throughout. For the dimensionality of their UV critical hypersurface we read off $s_{\text{UV}}=2$ for the `doubly Gaussian' \fpgB\fpC\fpgD-\fpL{},  $s_{\text{UV}}=3$ for the mixed ones, 
\fpnB\fpC\fpgD-\fpL{}, \fpgB\fpC\fpnDn-\fpL{}, and $s_{\text{UV}}=4$ for the `doubly non-Gaussian' fixed points, \fpnB\fpC\fpnDn-\fpL{} and  \fpnB\fpC\fpnD-\fpL{} .

According to Table \ref{tab:fp3A}, the fixed points which stem from \fpnD-\fpL{}, located in the physically relevant $\tg^{\dyn}>0$ half-space, are special in that some of their critical exponents are complex. 
\begin{table}[!ht]
\begin{center}
  \centering
  \caption{Critical exponents and scaling fields  of the 4 fixed points related to \fpgD-\fpL{}}
 \begin{tabular}{ |r|l| }
\hline
          \multicolumn{2}{ |c| }{\fpgB\fpC\fpgD-\fpL{} ${\scriptstyle (\tg^{\background}_*,\Kk^{\background}_*)=(0,0)}$} \\ 
    \hline \hline
  $\theta_j$ &eigenvectors $V^{(j)}$\\
   \hline
    $ -2$ &$ \frac{8\pi}{3}\hat{e}_{\tg}^{\dyn}+\hat{e}_{\Kk}^{\dyn}  $ 
\\
$ 2 $& $  +\hat{e}_{\Kk}^{\dyn}  $    
\\
 $ -2$ & $ -4\pi\hat{e}_{\tg}^{\background} +\hat{e}_{\Kk}^{\background} $     
\\
  $2$ & $ \hat{e}_{\Kk}^{\background}  $      
\\
    \hline
    \end{tabular}%
$\phantom{H}$
 \begin{tabular}{ |r|l| }
\hline
          \multicolumn{2}{ |c| }{\fpnB\fpC\fpgD-\fpL{} ${\scriptstyle (\tg^{\background}_*,\Kk^{\background}_*)=(1.40,-0.11)}$} \\
    \hline \hline
 $\theta_j$ &eigenvectors $V^{(j)}$\\
   \hline
 $-2 $& $ 0.99\hat{e}_{\tg}^{\dyn}+0.12\hat{e}_{\Kk}^{\dyn}+0.01\hat{e}_{\tg}^{\background}+0.01\hat{e}_{\Kk}^{\background}  $
\\
$2$ &   $0.94\hat{e}_{\Kk}^{\dyn}+0.35\hat{e}_{\Kk}^{\background}$
\\
$2 $&   $1.00\hat{e}_{\tg}^{\background}-0.08\hat{e}_{\Kk}^{\background}$
\\
$4$ &   $\hat{e}_{\Kk}^{\background}$
\\
    \hline
    \end{tabular}%
  \label{tab:fp1A}%
\end{center}
\end{table}%
\begin{table}[!ht]
\begin{center}
 \begin{tabular}{ |r|l| }
\hline
          \multicolumn{2}{ |c| }{\fpgO\fpC\fpgD-\fpL{} ${\scriptstyle (\tg^{(0)}_*,\Kk^{(0)}_*)=(0,0)}$} \\
    \hline \hline
 $\theta_j$ &eigenvectors $V^{(j)}$\\
   \hline
 $-2$& $ \frac{8\pi}{3}\hat{e}_{\tg}^{\dyn}+\hat{e}_{\Kk}^{\dyn}  $    
\\
$ 2$ &   $+\hat{e}_{\Kk}^{\dyn}$
\\
$-2$ &     $8\pi\hat{e}_{\tg}^{(0)}+ \hat{e}_{\Kk}^{(0)}$
\\
$2 $&  $\hat{e}_{\Kk}^{(0)}$
\\
    \hline
    \end{tabular}%
$\phantom{H}$
 \begin{tabular}{ |r|l| }
\hline
          \multicolumn{2}{ |c| }{\fpnO\fpC\fpgD-\fpL{} ${\scriptstyle (\tg^{(0)}_*,\Kk^{(0)}_*)=(1.7,0.07)}$} \\
    \hline \hline
 $\theta_j$ &eigenvectors $V^{(j)}$\\
   \hline
$-2$&  $ 0.97\hat{e}_{\tg}^{\dyn}+0.12\hat{e}_{\Kk}^{\dyn}-0.23\hat{e}_{\tg}^{(0)}+0.03\hat{e}_{\Kk}^{(0)}  $  
\\
$2$ &  $5\times 10^{-17}\hat{e}_{\Kk}^{\dyn}+1.00\hat{e}_{\tg}^{(0)}+0.38\hat{e}_{\Kk}^{(0)}$
\\
$2$ & $1.00\hat{e}_{\tg}^{(0)}+0.04\hat{e}_{\Kk}^{(0)}$
\\
$ 4 $ &   $\hat{e}_{\Kk}^{(0)}$
\\
    \hline
    \end{tabular}%
  \label{tab:fp1B}%
\end{center}
\end{table}%
\begin{table}[!ht]
\begin{center}
  \centering
  \caption{Critical exponents and scaling fields  of the 4 fixed points related to \fpnDn-\fpL{}}
 \begin{tabular}{ |r|l| }
\hline
          \multicolumn{2}{ |c| }{\fpgB\fpC\fpnDn-\fpL{} ${\scriptstyle (\tg^{\background}_*,\Kk^{\background}_*)=(0,0)}$} \\
    \hline \hline
 $\theta_j$ &eigenvectors $V^{(j)}$\\
   \hline
$2.21$ &$ 1.0\,\hat{e}_{\tg}^{\dyn}+0.1\,\hat{e}_{\Kk}^{\dyn} $
\\
$5.12$ & $  -0.6\,\hat{e}_{\tg}^{\dyn}+0.8\,\hat{e}_{\Kk}^{\dyn}$ 
\\
 $-2$ & $ 1.0\,\hat{e}_{\tg}^{\background} -0.1\,\hat{e}_{\Kk}^{\background} $   
\\
$2$ & $ \hat{e}_{\Kk}^{\background}  $  
\\
    \hline
    \end{tabular}%
$\phantom{H}$
\begin{tabular}{ |r|l| }
\hline
          \multicolumn{2}{ |c| }{\fpnB\fpC\fpnDn-\fpL{} ${\scriptstyle (\tg^{\background}_*,\Kk^{\background}_*)=(1.53,-0.12)}$} \\
    \hline \hline
 $\theta_j$ &eigenvectors $V^{(j)}$\\
   \hline
$2.21$ &$ -0.93\hat{e}_{\tg}^{\dyn}-0.06\hat{e}_{\Kk}^{\dyn}-0.35\hat{e}_{\tg}^{\background}+0.04\hat{e}_{\Kk}^{\background}  $   
\\
$5.12$ & $ 0.4\hat{e}_{\tg}^{\dyn}-0.6\hat{e}_{\Kk}^{\dyn}-0.4\hat{e}_{\tg}^{\background}+0.5\hat{e}_{\Kk}^{\background}  $
\\
 $2$ & $1.00\hat{e}_{\tg}^{\background}-0.08\hat{e}_{\Kk}^{\background}$
\\
$4$ & $\hat{e}_{\Kk}^{\background}$
\\
    \hline
    \end{tabular}%
  \label{tab:fp2A}%
\end{center}
\end{table}%
\begin{table}[!h]
\begin{center}
 \begin{tabular}{ |r|l| }
\hline
          \multicolumn{2}{ |c| }{\fpgO\fpC\fpnDn-\fpL{} ${\scriptstyle (\tg^{(0)}_*,\Kk^{(0)}_*)=(0,0)}$} \\
    \hline \hline
 $\theta_j$ &eigenvectors $V^{(j)}$\\
   \hline
$2.21$ &$ 1.0\,\hat{e}_{\tg}^{\dyn}+0.1\,\hat{e}_{\Kk}^{\dyn} $
\\
$5.12$ & $  -0.6\,\hat{e}_{\tg}^{\dyn}+0.8\,\hat{e}_{\Kk}^{\dyn}$ 
\\
 $2$ & $1.00\hat{e}_{\tg}^{(0)}+ 0.01\hat{e}_{\Kk}^{(0)}$
\\
$4$ & $\hat{e}_{\Kk}^{(0)}$  
\\
    \hline
    \end{tabular}%
$\phantom{H}$
\begin{tabular}{ |r|l| }
\hline
          \multicolumn{2}{ |c| }{\fpnO\fpC\fpnDn-\fpL{} ${\scriptstyle (\tg^{\background}_*,\Kk^{\background}_*)=(1.53,-0.12)}$} \\
    \hline \hline
 $\theta_j$ &eigenvectors $V^{(j)}$\\
   \hline
$2.21$ &$ -0.50\hat{e}_{\tg}^{\dyn}-0.03\hat{e}_{\Kk}^{\dyn}-0.88\hat{e}_{\tg}^{(0)}-0.01\hat{e}_{\Kk}^{(0)}  $ 
\\
$5.12$ & $ 0.2\hat{e}_{\tg}^{\dyn}-0.3\hat{e}_{\Kk}^{\dyn}-0.5\hat{e}_{\tg}^{(0)}+0.8\hat{e}_{\Kk}^{(0)}  $  
\\
 $-2$ & $1.00\hat{e}_{\tg}^{(0)}+0.01\hat{e}_{\Kk}^{(0)}$
\\
$2$ &  $\hat{e}_{\Kk}^{(0)}$
\\
    \hline
    \end{tabular}%
  \label{tab:fp2B}%
\end{center}
\end{table}%
\begin{table}[!ht]
\begin{center}
  \centering
  \caption{Critical exponents and scaling fields  of the 4 fixed points related to  \fpnD-\fpL{}}
\begin{tabular}{ |r|l| }
\hline
          \multicolumn{2}{ |c| }{\fpgB\fpC\fpnD-\fpL{} ${\scriptstyle (\tg^{\background}_*,\Kk^{\background}_*)=(0,0)}$} \\
    \hline \hline
 $\theta_j$ &eigenvectors $V^{(j)}$\\
   \hline
$3.6+4.3\Ii$ &$ -0.99\hat{e}_{\tg}^{\dyn}-(0.04+0.15\Ii)\hat{e}_{\Kk}^{\dyn}  $
\\
$ 3.6-4.3\Ii$ & $  -0.99\hat{e}_{\tg}^{\dyn}-(0.04-0.15\Ii)\hat{e}_{\Kk}^{\dyn} $ 
\\
 $2$ &  $ \hat{e}_{\tg}^{\background} -9\times10^{-4}\hat{e}_{\Kk}^{\background} $  
\\
$2$ & $ \hat{e}_{\Kk}^{\background}  $   
\\
    \hline
    \end{tabular}%
\begin{tabular}{ |r|l| }
\hline
          \multicolumn{2}{ |c| }{\fpgO\fpC\fpnD-\fpL{} ${\scriptstyle (\tg^{(0)}_*,\Kk^{(0)}_*)=(0,0)}$} \\
    \hline \hline
 $\theta_j$ &eigenvectors $V^{(j)}$\\
   \hline
$3.6+4.3\Ii$ &$ -0.99\hat{e}_{\tg}^{\dyn}-(0.04+0.15\Ii)\hat{e}_{\Kk}^{\dyn}  $
\\
 $ 3.6-4.3\Ii$ & $  -0.99\hat{e}_{\tg}^{\dyn}-(0.04-0.15\Ii)\hat{e}_{\Kk}^{\dyn} $
\\
 $2$ &   $0.96\hat{e}_{\tg}^{(0)}+ 0.28\hat{e}_{\Kk}^{(0)}$
\\
$2$ & $\hat{e}_{\Kk}^{(0)}$   
\\
    \hline
    \end{tabular}%
  \label{tab:fp3A}%
\end{center}
\end{table}%
\begin{table}[!ht]
\begin{center}
\begin{tabular}{ |r|l| }
\hline
          \multicolumn{2}{ |c| }{\fpnB\fpC\fpnD-\fpL{} ${\scriptstyle (\tg^{\background}_*,\Kk^{\background}_*)=(8.18,-0.01)}$} \\
    \hline \hline
 $\theta_j$ &eigenvectors $V^{(j)}$\\
   \hline
$3.6+4.3\Ii$ & $ -(1.6-0.2\Ii)10^{-2}\hat{e}_{\tg}^{\dyn} -(0.9+2.4\Ii)10^{-2} \hat{e}_{\Kk}^{\dyn}+1.0\hat{e}_{\tg}^{\background}+(1.7-2.3\Ii)10^{-2}\hat{e}_{\Kk}^{\background}  $
\\
$ 3.6-4.3\Ii$ &  $ -(1.6+0.2\Ii)10^{-2}\hat{e}_{\tg}^{\dyn} -(0.9-2.4\Ii)10^{-2} \hat{e}_{\Kk}^{\dyn}+1.0\hat{e}_{\tg}^{\background}+(1.7+2.3\Ii)10^{-2}\hat{e}_{\Kk}^{\background}  $  
\\
 $2$ &   $ \hat{e}_{\tg}^{\background} -9\times10^{-4}\hat{e}_{\Kk}^{\background} $
\\
$4$ &  $\hat{e}_{\Kk}^{\background}$   
\\
    \hline
    \end{tabular}%
\\ $\phantom{H}$\\
\begin{tabular}{ |r|l| }
\hline
          \multicolumn{2}{ |c| }{\fpnO\fpC\fpnD-\fpL{} ${\scriptstyle (\tg^{(0)}_*,\Kk^{(0)}_*)=(0.65,0.19)}$} \\
    \hline \hline
 $\theta_j$ &eigenvectors $V^{(j)}$\\
   \hline
$3.6+4.3\Ii$ &  $ 0.86\hat{e}_{\tg}^{\dyn}+(0.04+0.13\Ii)\hat{e}_{\Kk}^{\dyn}+(0.40-0.03\Ii)\hat{e}_{\tg}^{(0)}-(0.2-0.2\Ii)\hat{e}_{\Kk}^{(0)}  $
\\
 $ 3.6-4.3\Ii$ & $ 0.86\hat{e}_{\tg}^{\dyn}+(0.04-0.13\Ii)\hat{e}_{\Kk}^{\dyn}+(0.40+0.03\Ii)\hat{e}_{\tg}^{(0)}-(0.2+0.2\Ii)\hat{e}_{\Kk}^{(0)}  $  
\\
 $2$ &   $0.96\hat{e}_{\tg}^{(0)}+ 0.28\hat{e}_{\Kk}^{(0)}$
\\
$4$ & $\hat{e}_{\Kk}^{(0)}$
\\
    \hline
    \end{tabular}%
  \label{tab:fp3B}%
\end{center}
\end{table}%
\noindent 
They lead to the characteristic spirals of the RG trajectories approaching the FP for $k\rightarrow \infty$. This behavior is well known to occur in the single-metric Einstein-Hilbert truncation \cite{frank1, oliver1}. In fact, the complex conjugate pair $\theta_{1\slash2}$ of Table \ref{tab:fp3A} is very similar to what one finds in the single-metric computation. This close numerical similarity to the single-metric result lends further credit to the conjecture that {\it the single-metric fixed point should correspond to one of the two fixed points of the full system which are related to the  \fpnD-\fpL{}}. We will see later on that this is in fact true.

\subsection{Recovering the mechanism of paramagnetic dominance} \label{subsec:paramagnetic}
In the single-metric analysis of \cite{andi1} it was found that in $d=4$ the NGFP owes its existence entirely to the paramagnetic interactions of the $\flcb_{\mu\nu}$ fluctuations with the background metric; the diamagnetic interactions disfavor the formation of a fixed point instead. (Below $d=3$ the situation changes, showing that the respective physical mechanism behind the NGFPs in $d=2+\epsilon$ and $d=4$ are quite different \cite{andi1}.) What is the picture in the bi-metric case?

In the $\tg^{\dyn}$-$\KkD$-regime of interest, all qualitative properties of $\aDz$ in eq. \eqref{eqn:res4D_005} are well described by its linear approximation $\eta^{\dyn}\approx\kAD(\KkD)\,\tg^{\dyn} $. 
The contribution of the denominator in \eqref{eqn:res4D_005}, involving $\kBD(\KkD)$, influences the resulting anomalous dimension only weakly. So let us focus on $\kAD(\KkD)$. 
Assuming, as always, a positive dynamical Newton constant, the negative anomalous dimension which is indicative of anti-screening and is necessary for a NGFP, requires a negative $\kAD$.

Now, what we find is that the paramagnetic interactions indeed drive $\kAD$ negative, but the diamagnetic ones have an antagonistic effect trying to make $\kAD$ and $\aDz$ positive.
In fact, switching off the paramagnetic contributions yields, in the bi-metric setting,
\begin{align}
 &\kAD(\KkD)\big|_{\rhoP=0} >0 \quad \text{ for all values of }\KkD\,.
\end{align}
Instead including the paramagnetic parts we find that there is a crossover from negative to positive values of the anomalous dimension at a certain critical value $\KkD_{\text{crit}}$, that is
\begin{align}\kAD(\KkD)\big|_{\rhoP=1} 
\begin{cases}
 > 0 & \quad \forall \, \KkD < \KkD_{\text{crit}} \\
< 0 & \quad \forall \, \KkD > \KkD_{\text{crit}}
\end{cases} \label{eqn:res4D_ParaFP01}
\end{align}
The precise value of the critical cosmological constant, $\KkD_{\text{crit}}$, is cutoff scheme dependent, but in any scheme we find $0<\KkD_{\text{crit}}<\KkD_*$.

The behavior \eqref{eqn:res4D_ParaFP01} makes it explicit that, first, the absence of anti-screening in the semi-classical regime ($\kAD(0)>0$) can be reconciled with a non-trivial fixed point existing simultaneously ($\kAD(\KkD_*)<0$) and, second, {\it the NGFP can form only because in some part of theory space the paramagnetic interactions are stronger than the diamagnetic ones.} 

Thus we essentially recover the single-metric picture according to which `paramagnetic dominance' is the physical mechanism responsible for Asymptotic Safety. 
However, the bi-metric analysis suggests that this mechanism can occur only for a non-zero, positive cosmological constant.%
\footnote{The same conclusion can be drawn from the beta-functions found in \cite{MRS2} with a bi-metric truncation too.} 
The implications of this result will be further discussed elsewhere \cite{daniel-prep}.

\subsection{Impact of the ghosts' wave function normalization}
It is instructive to look at the dependence of the fixed point data on the ghost contributions to the beta-functions. Since we kept $\rhoG$, the prefactor of the ghost kinetic term in $\EAA_k$ as a free parameter it is easy to assess the importance of ghosts relative to the gravitons by varying this parameter away from $\rhoG=1$, the value used in the previous subsections. This will give us an idea of the numerical accuracy one may expect within the present approximation which neglects the running of the ghosts' wave function renormalization.

\begin{figure}[!ht]
\centering
\psfrag{r}{$\rhoG$}
\psfrag{g}{$\tg^{\dyn}_*$}
\psfrag{l}{$\KkD_*$}      
\psfrag{f}{$\scriptscriptstyle {\text{\fpgD-\fpL{}}}$}
\psfrag{m}{$\scriptscriptstyle {\text{\fpnDn-\fpL{}}}$}
\psfrag{n}{$\scriptscriptstyle {\text{\fpnD-\fpL{}}}$}
 \subfloat[The  dependence of $\tg_*^{\dyn}$ on $\rhoG$.]{\label{fig:rhoGA}\includegraphics[width=0.4\textwidth]{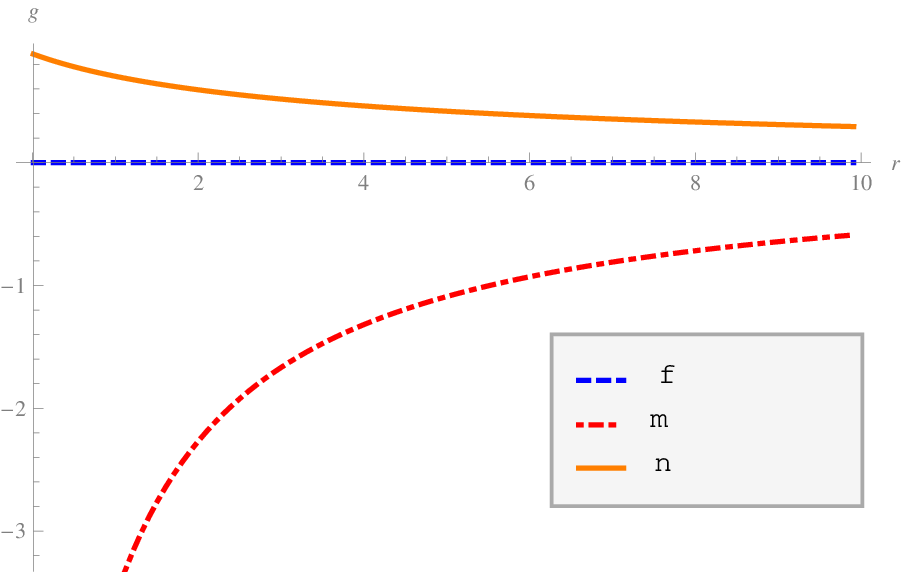}}\hspace{0.1\textwidth}
 \subfloat[The dependence of $\KkD_*$ on $\rhoG$.]{\label{fig:rhoGB}\includegraphics[width=0.4\textwidth]{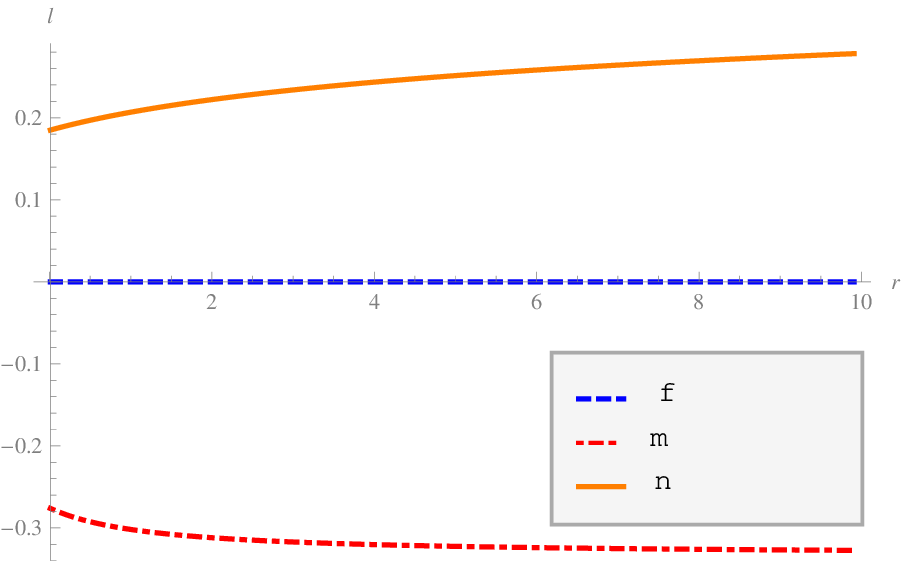}}
\caption{The  dependence of $\tg_*^{\dyn}$ and $\KkD_*$ on $\rhoG$ for the three types of fixed points, \fpgD-\fpL{}, \fpnDn-\fpL{}, and \fpnD-\fpL{}. 
While the Gaussian one is insensitive to independent on $\rhoG$ the $\KkD_*$ values of the others only slightly change when increasing $\rhoG$. 
As for the Newton constant $\tg*^{\dyn}$, the fixed point in the lower half-plane $(\tg^{\dyn}<0)$ is by far more sensitive to $\rhoG$ than the one in the upper,  \fpnD-\fpL{}.} \label{fig:rhoG}
\end{figure}

The fixed point coordinates  in the dynamical sector depend on the $\rhoG$ non-trivially as shown  in Fig. \ref{fig:rhoG}.
While the trivial fixed point \fpgD-\fpL{} is independent of $\rhoG$, the non-Gaussian ones \fpnD-\fpL{} and \fpnDn-\fpL{} are not. 
However, it turns out that $\tg_*^{\dyn}$ of \fpnDn-\fpL{} is very sensitive to a change in $\rhoP$, especially going from $\rhoG=0$ to $\rhoG=1$. 
This suggests that \fpnDn-\fpL{} is likely to be a truncation artifact.
On the other hand, the fixed point values of \fpnD-\fpL{} are quite stable when changing the strength of the ghost-contributions.

From the point of view of the Asymptotic Safety scenario both of these facts are encouraging: 
First, the fixed point at positive dynamical Newton constant, \fpnD-\fpL{}, seems well described within the present approximation and hardly could be a truncation artifact. 
If we define a $k\rightarrow\infty$ limit there, since the running $\tg_k^{\dyn}$ never changes its sign, the resulting theory has a positive $G_0^{\dyn}$, as it should be. 
Second, the fixed point at negative $\tg^{\dyn}$, \fpnDn-\fpL{}, is much less stable and more likely to be a truncation artifact. If so, this would actually be welcome since then no UV limit could be taken there, and no asymptotically safe theory with $G^{\dyn}_0<0$ could be constructed, and the result of positive $G_0^{\dyn}$ implied by \fpnD-\fpL{} is a true prediction.

For the background couplings the picture is as follows. 
The Gaussian solution $\tg_*^{\background}=0=\KkB_*$, leading  to the three fixed points \fpgB\fpC\fpgD-\fpL{}, \fpgB\fpC\fpnDn-\fpL{}, and \fpgB\fpC\fpnD-\fpL{}, is independent of $\rhoG$, and so the same is true for their $(\tg^{\background},\KkB)$-coordinates.
\begin{figure}[!ht]
\centering
\psfrag{r}{$\rhoG$}
\psfrag{g}{$\tg^{\background}_*$}
\psfrag{l}{$\KkB_*$}      
\psfrag{f}{$\scriptscriptstyle {\text{\fpgB\fpC($\cdots$)$^{\dyn}$-\fpL{}}}$}
\psfrag{m}{$\scriptscriptstyle {\text{\fpnB\fpC\fpnDn-\fpL{}}}$}
\psfrag{n}{$\scriptscriptstyle {\text{\fpnB\fpC\fpnD-\fpL{}}}$}
\psfrag{o}{$\scriptscriptstyle {\text{\fpnB\fpC\fpgD-\fpL{}}}$}
 \subfloat[The  dependence of $\tg_*^{\background}$ on $\rhoG$.]{\label{fig:rhoGC}\includegraphics[width=0.4\textwidth]{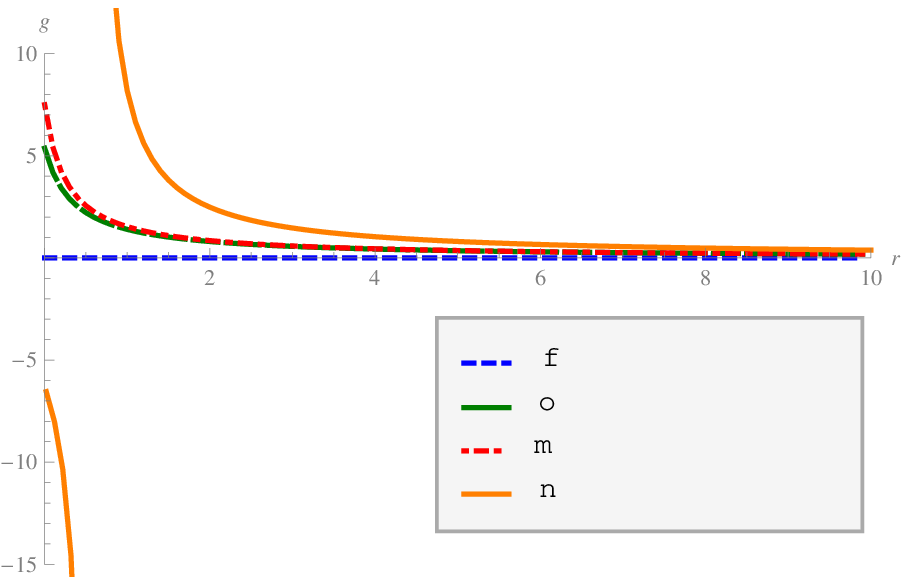}}\hspace{0.1\textwidth}
 \subfloat[The dependence of $\KkB_*$ on $\rhoG$.]{\label{fig:rhoGD}\includegraphics[width=0.4\textwidth]{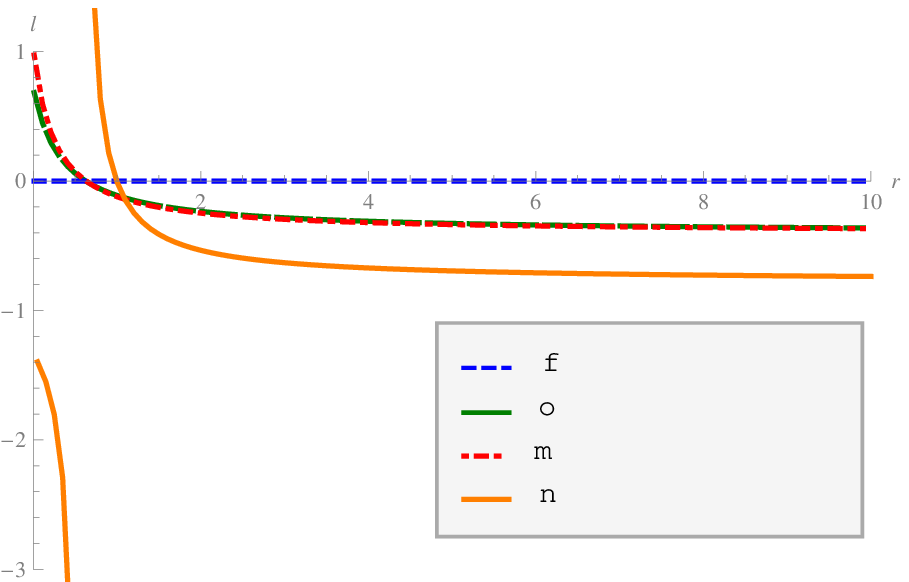}}
\caption{The dependence of $\KkB_*$ and $\tg_*^{\background}$  on $\rhoG$  for the six  fixed points. 
There are three Gaussian and three non-Gaussian fixed points in the $\background$-sector, one for each $\dyn$-fixed point.  The Gaussian ones are independent on $\rhoG$. In contrast to the $\dyn$-part of these fixed points, the $\background$ fixed point values vary  most strongly for \fpnB\fpC\fpnD-\fpL{}. Especially for small $\rhoG$ it starts out negative, diverges and then turns positive for $\rhoG<1$.}\label{fig:rhoGn}
\end{figure}
The fixed point that is situated in the negative $\tg^{\dyn}$ half-plane, \fpnB\fpC\fpnDn-\fpL{}, is again very sensitive to the influence of the ghost sector. The background fixed point values exhibit a strong dependence on $\rhoG$ even for \fpnB\fpC\fpnD-\fpL{} which has the positive $\tg_*^{\dyn}$. The $\background$-fixed point values actually change the sign and turn negative below $\rhoG <0.6$, see Fig. \ref{fig:rhoGn}.

We do not think that this apparent instability of $(\tg_*^{\background},\KkB_*)$ sheds any doubt on the viability of the Asymptotic Safety construction at the doubly non-Gaussian \fpnB\fpC\fpnD-\fpL{}. In fact, the $\background$-couplings are {\it entirely unphysical} and owe their existence only to the extra $\bg$-dependence of $\EAA_k$, hence to the split-symmetry violation in unobservable quantities.
 
\section[Can split-symmetry coexists with Asymptotic Safety?]{Can split-symmetry coexists with\\ Asymptotic Safety?}\label{sec:splitandAS}
Next, we are going to study the fully fledged RG flow on the four dimensional theory space  by analyzing the system of coupled  differential equations \eqref{eqn:trA27C} with both analytical and numerical methods. 
In this section we are mostly interested in the { global properties of the RG flow}. In particular we investigate to what extent split-symmetry at $k=0$ can be realized by a judicious choice of the trajectory's `initial' conditions.
In practice they are imposed at an intermediate scale $k_0$, and the differential equations are solved then both in the upward and downward direction. 
The all decisive question we try to answer is: 
{\bf Do there exist RG trajectories for which split-symmetry, i.e. Background Independence for $k\rightarrow0$  coexists with Asymptotic Safety?} What makes this question hard is its global nature: Background Independence and Asymptotic Safety  concern exactly the opposite ends of the RG trajectories, the limits of very small and very large scales, respectively.

Since the $\dyn$-couplings are not affected by the $\background$-sector but, conversely, enter the $\background$-beta-functions, we first  solve the eqs. \eqref{eqn:trA27CA} and \eqref{eqn:trA27CB} for $\tg_k^{\dyn}$ and $\KkD_k$,  and then substitute the solutions into the beta-functions of $\tg_k^{\background}$ and $\KkB_k$ in eqs. \eqref{eqn:trA27CC} and \eqref{eqn:trA27CD} to obtain the `$\background$'-components of the trajectory. We describe the results of the two steps in turn.

\subsection[Trajectories of the `\texorpdfstring{$\dyn$}{Dyn}' sub-system]{Trajectories of the `$\bm{\dyn}$' sub-system}
The set of solutions for the $\background$-independent $(\tg_k^{\dyn},\KkD_k)$-system \eqref{eqn:trA27CA}, \eqref{eqn:trA27CB} decomposes into a subset with positive $\tg_k^{\dyn}$ for all $k$,  one with  $\tg_k^{\dyn}<0$ always, and  a single trajectory with $\tg^{\dyn}_k=0$ $\forall k$ that separates the two regions. The sign of the Newton coupling never changes along any trajectory. 
We know already that  the $\dyn$-system allows for three fixed points: the Gaussian one, \fpgD-\fpL{}, which is located on the separating trajectory, and  two non-Gaussian ones, \fpnDn-\fpL{} and \fpnD-\fpL{}, that lie below $(\tg^{\dyn}<0)$ or above it $(\tg^{\dyn}>0)$, respectively.  We will focus in the sequel on the positive $\tg^{\dyn}$ domain.
\begin{figure}[!ht]
\centering
\psfrag{g}{$\tg^{\dyn}$}
\psfrag{l}{$\KkD$}     
\includegraphics[width=0.6\textwidth]{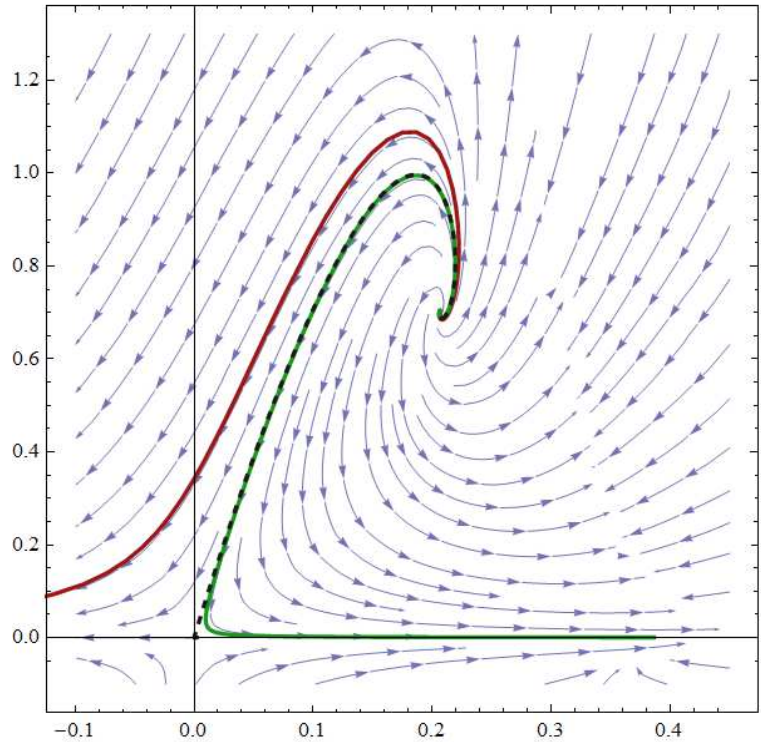}
\caption{The phase portrait of the bi-metric `$\dyn$'-sector. The vertical (horizontal) axis corresponds to $\tg^{\dyn}$ ($\KkD$). Note the remarkable similarity with the phase portrait of the single-metric Einstein-Hilbert truncation \cite{frank1}.}
\label{fig:flowD4Traj}
\end{figure}

Fig.  \ref{fig:flowD4Traj} shows the phase portrait on the $\tg^{\dyn}$-$\KkD$ plane which we obtained numerically.%
\footnote{Here and in all similar diagrams the  arrows always point from the UV towards the IR.}
We see that it is impressively similar to the well-known phase portrait of the single-metric Einstein-Hilbert truncation \cite{frank1}.

 Following ref. \cite{frank1} the trajectories in the upper half-plane in Fig. \ref{fig:flowD4Traj} can be classified in the same way as their single-metric relatives, namely as of type (Ia)$^{\dyn}$, type (IIa)$^{\dyn}$, and type (IIIa)$^{\dyn}$ respectively, depending on whether  the cosmological constant $\KkD_k$ approaches $-\infty$, $0$, or  $+\infty$ in the far IR, i.e. for $k\rightarrow 0$. The type (IIa)$^{\dyn}$ solution separates the two regimes and is, henceforth, also called separatrix.
(In Fig. \ref{fig:flowD4Traj} it is represented by  a dashed line, and we have also highlighted a representative of the other types,  a (Ia)$^{\dyn}$ and a (IIIa)$^{\dyn}$ trajectory.)
The separatrix `crosses over' from the \fpnD-\fpL{} in the UV to the  \fpgD-\fpL{} in the IR. Notice that  \fpnD-\fpL{} is UV-attractive in both directions, and thus all trajectories are pulled into this fixed point when $k\rightarrow \infty$. 
Due to the imaginary part of the  critical exponents they form spirals.

\subsection[Solving the non-autonomous `\texorpdfstring{$\background$}{B}' system]{Solving the non-autonomous `$\bm{\background}$' system}
 Each one of the $\dyn$ trajectories obtained above can now be substituted into the two RG equations of the $\background$- or level-(0) couplings. 
After fixing initial conditions they can be solved to give the $k$-dependence of the two remaining coordinates of the 4-dimensional trajectories, namely $\tg_k^{\background\slash(0)}$ and $\lambda_k^{\background\slash(0)}$. 

Let us start by investigating their qualitative behavior for arbitrary initial conditions in the $\background$-sector. Once a solution $(\tg_k^{\dyn},\KkD_k)$ is picked, the beta-functions in eqs. \eqref{eqn:trA27CC}, \eqref{eqn:trA27CD} for the $\background$-couplings are polynomials with known, but {\it scale dependent} coefficients $\kA(k)\equiv \kA (\tg_k^{\dyn},\KkD_k)$ and $\kB(k)\equiv \kB(\tg_k^{\dyn},\KkD_k)$:
\begin{subequations}
\begin{align}
\partial_t \tg_k^{\background}&=\beta_{\tg}^{\background}(\tg_k^{\background},\KkB_k;k)=2\tg_k^{\background} + \kB(k) (\tg_k^{\background})^2 \label{eqn:res4D_021A} \\
 \partial_t \KkB_k &=\beta_{\Kk}^{\background}(\tg_k^{\background},\KkB_k;k)= -2 \KkB_k +\kA(k)\,\tg^{\background}_k + \kB(k)\,\tg_k^{\background } \KkB_k \label{eqn:res4D_021B}
\end{align}\label{eqn:res4D_021}
\end{subequations}%
It is important to appreciate that when the functions $\kA(k)$ and $\kB(k)$ are given, the eqs. \eqref{eqn:res4D_021} form a closed coupled system for the two remaining $(\tg_k^{\background},\, \KkB_k)$ which, however, contrary to that for $(\tg_k^{\dyn},\,\KkD_k)$, is {\it not autonomous}.  The beta-functions on the RHS of eqs. \eqref{eqn:res4D_021A} and \eqref{eqn:res4D_021B} possess an {\it explicit} dependence on $k$. Hence the vector field on the $\tg^{\background}$-$\KkB$-plane they give rise to and, as a consequence, the entire phase portrait on this plane are RG-time dependent. This will complicate the analysis considerably.

The other fact to be appreciated is that the beta-functions of \eqref{eqn:res4D_021} are {\it polynomial} in the unknowns $\tg^{\background}$ and $\KkB$. This is in sharp contradistinction to the dynamical sector which involves complicated threshold functions $\ThrfA{p}{n}{-2\KkD}$.
For later use we also mention that, in terms of the dimensionful quantities $1\slash G^{\background}$ and $\Kkbar^{\background}\slash G^{\background}$, the system \eqref{eqn:res4D_021} has the following general solution:
\begin{subequations}
\begin{align}
\frac{1}{G_k^{\background}}=\frac{1}{G_{k_0}^{\background}}-\int_{k_0}^k\md k^{\prime}\, k^{\prime} \kB(k^{\prime})  \label{eqn:res4D_021dimA} \\
 \frac{\Kkbar_k^{\background}}{G_k^{\background}}=\frac{\Kkbar_{k_0}^{\background}}{G_{k_0}^{\background}}+\int_{k_0}^k\md k^{\prime}\, k^{\prime\,3} \kA(k^{\prime})  \label{eqn:res4D_021dimB}
\end{align}\label{eqn:res4D_021dim}
\end{subequations}%
Here, as always, $k_0$ denotes an arbitrary (initial, or intermediate) scale somewhere along the trajectory.

\paragraph{The `$\background$' Newton constant.}
Returning to dimensionless quantities the hierarchy among the two equations \eqref{eqn:res4D_021} allows to first solve \eqref{eqn:res4D_021A} for $\tg_k^{\background}$, and then  determine the $k$-dependence of $\KkB_k$ from \eqref{eqn:res4D_021B}. 
The differential equation \eqref{eqn:res4D_021A} is of Bernoulli type. It determines $\tg_k^{\background}$ and is independent of $\KkB_k$. Besides the trivial solution $\tg_k^{\background}=0$ there exists a non-trivial one, namely
\begin{align}
 \tg_k^{\background}&=  \frac{k^2\, \tg_{k_0}^{\background}}{k^2_0 -  \tg_{k_0}^{\background} \int_{k_0}^k \,k^{\prime} \kB(k^{\prime})\md k^{\prime}} \label{eqn:res4D_022}
\end{align}

\begin{figure}[!ht]
\centering
\psfrag{k}[t]{${\scriptstyle k}$}
\psfrag{j}[br]{${\scriptstyle \text{UV}}$}
\psfrag{i}[bl]{${\scriptstyle \text{IR}}$}   
\psfrag{z}[cl]{$\int {\scriptstyle k \kB(k)}$}  
 \subfloat[Type (Ia)$^{\dyn}$]{   \label{fig:res4DkBT1}\includegraphics[width=0.40\textwidth]{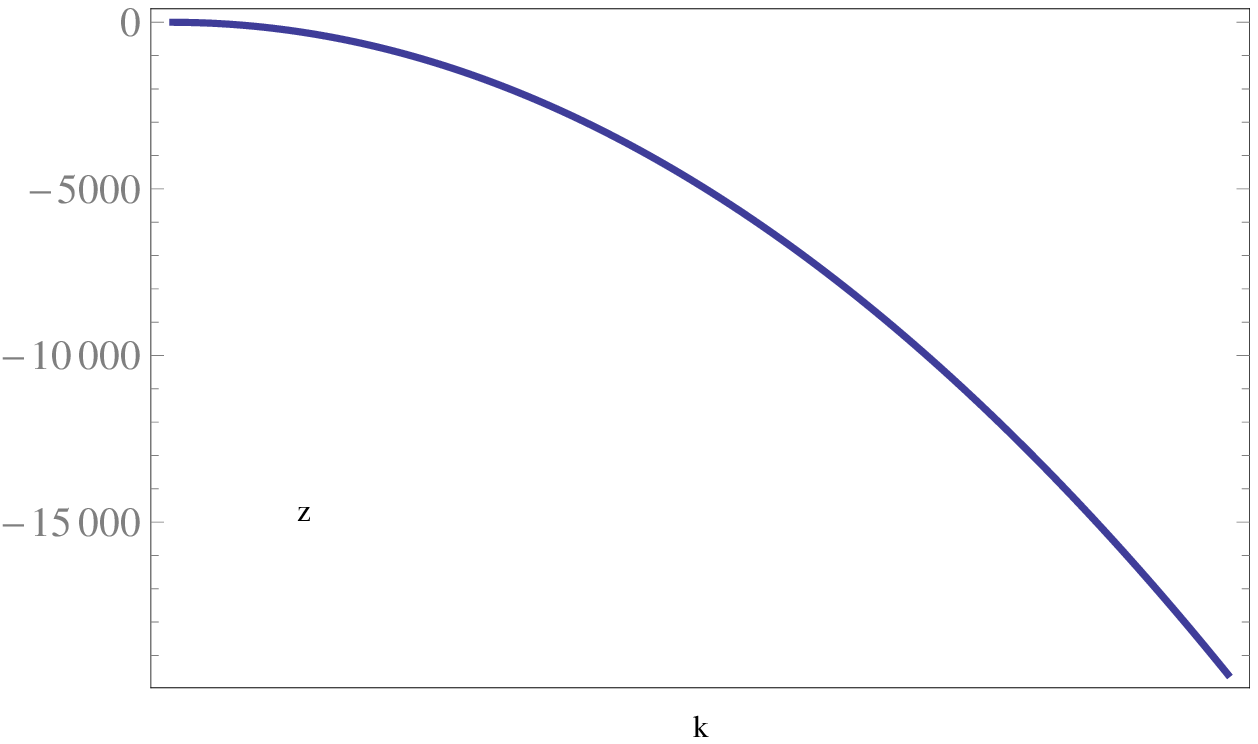}}\hspace{0.05\textwidth}
 \subfloat[Separatrix]{\label{fig:res4DkBT2}\includegraphics[width=0.40\textwidth]{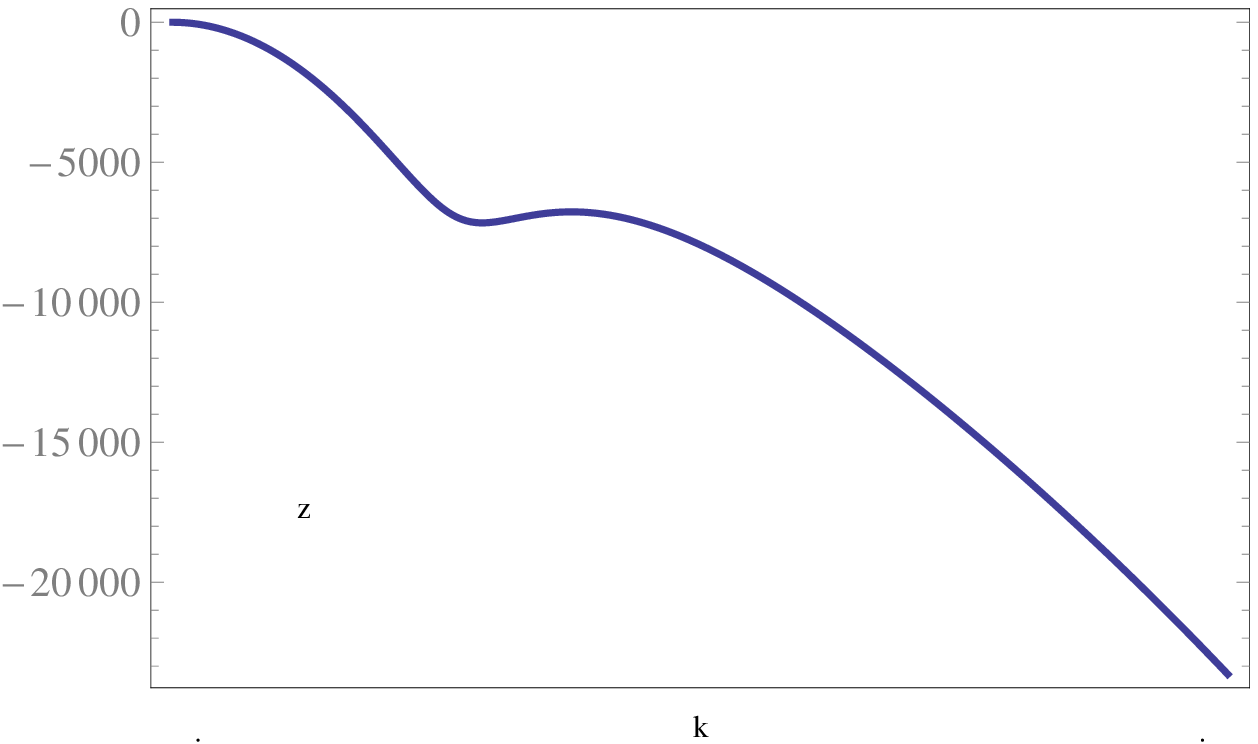}}\hspace{0.05\textwidth}
 \subfloat[Type (IIIa)$^{\dyn}$]{   \label{fig:res4DkBT3}\includegraphics[width=0.40\textwidth]{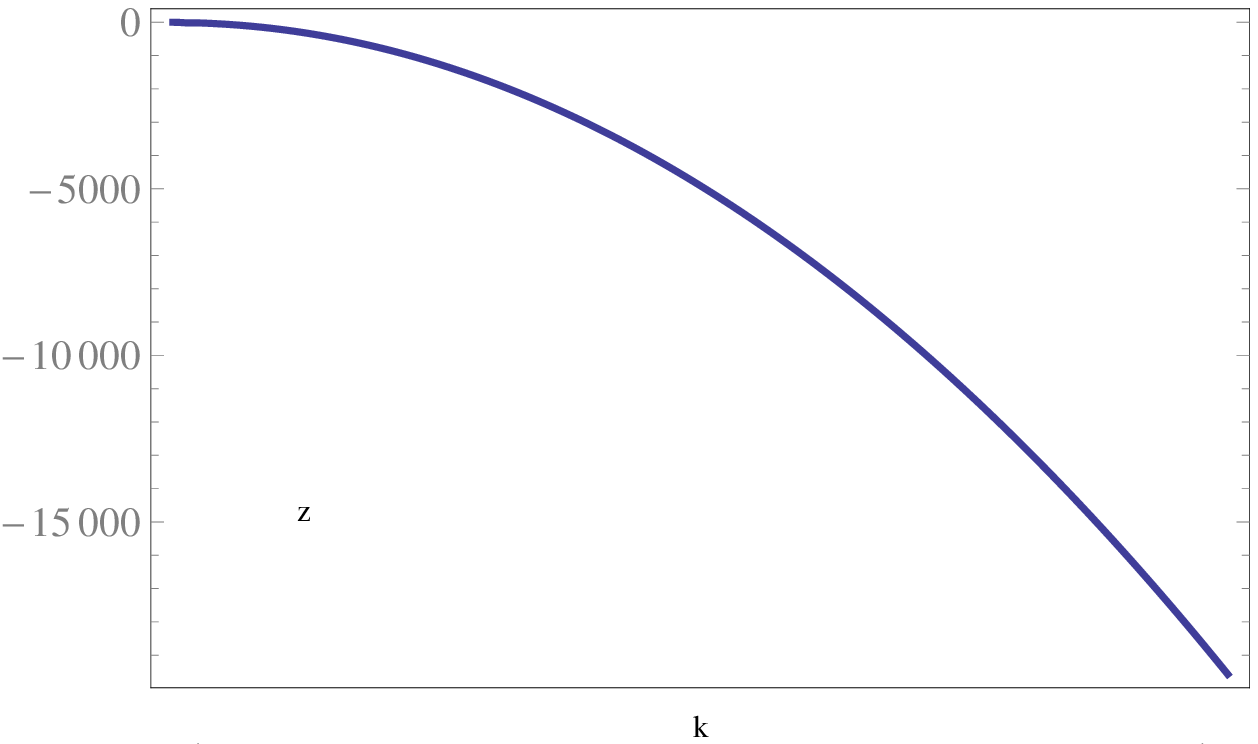}}
 \caption{The function $k\mapsto \int_{k_0}^k \,k^{\prime} \kB(k^{\prime})\md k^{\prime}$ for representative examples of the three types of trajectories. The function is seen to be negative throughout, demonstrating that the solution \eqref{eqn:res4D_022} is regular everywhere.}\label{fig:res4DkB}
\end{figure}

Let us demonstrate that {\it the solution \eqref{eqn:res4D_022} is  regular everywhere}. The denominator in eq. \eqref{eqn:res4D_022} does not vanish at any $k$. For this to happen the $k^{\prime}$-integral in \eqref{eqn:res4D_022} would have to be positive. It turns out however that $\kB(k)=\kB(\tg_k^{\dyn},\KkD_k)$ in the integrand is  positive in the small interval  $\KkD \in [0.213,0.5]$ only; furthermore the integral $\int_{k}^{k_0}\,k^{\prime} \kB(k^{\prime})\md k^{\prime} $ in \eqref{eqn:res4D_022}  remains negative even in the corresponding possibly `dangerous' regimes, as for example the IR branch for the type (IIIa)$^{\dyn}$ trajectories or the spirals in the vicinity of \fpnD-\fpL{}. Fig. \ref{fig:res4DkB} shows the decrease of $\int_{k_0}^k \,k^{\prime} \kB(k^{\prime})\md k^{\prime}$ as a function of $k$ towards the UV, and the fact that it is negative in the far IR, for the three different types of trajectories in the $\dyn$-sector. 
In fact, in the UV -- when the $\dyn$-trajectories spiral into \fpnD-\fpL{} -- the denominator in \eqref{eqn:res4D_022} becomes very large so that the $k^2_0$ term can be neglected. Then \eqref{eqn:res4D_022} approaches
\begin{align}
 \tg_k^{\background}&\,\stackrel{\text{UV}}{=}\, \frac{k^2}{-\int_{k_0}^k \,k^{\prime} \kB(k^{\prime})\md k^{\prime}} \,\, \xrightarrow{k\rightarrow \infty}\,\, 8.18 \,\equiv \tg_*^{\background}
 \label{eqn:res4D_023}
\end{align}
This limit equals precisely the \fpnB-\fpL{}-value of $\tg_k^{\background}$ for any  initial value $\tg_{k_0}^{\background}$. 
This shows that in the UV, and for any trajectory,
\begin{align}
\lim_{k^{\prime}\rightarrow\infty} \kB(k^{\prime})=C \quad \text{ with } |C|<\infty \text{ and } C<0 \label{eqn:res4D_022C}
\end{align}
 should hold true, where $C$ is some finite constant.

\paragraph{The `$\background$' cosmological constant.}
Upon inserting the solutions of the $\dyn$-sector and $\tg_k^{\background}$ we obtain from \eqref{eqn:res4D_021B} a single linear, inhomogeneous ODE with scale-dependent coefficients which determines $\KkB_k$:
\begin{align}
 \partial_t \KkB_k &= \kA(k)\,\tg^{\background}_k + \big[\kB(k)\,\tg_k^{\background } -2\big] \, \KkB_k \label{eqn:res4D_024}
\end{align}
The coefficient functions $\kA(k)$ and $\kB(k)$ are fixed once initial conditions are imposed on $\tg_k^{\dyn}$ and $\KkD_k$. 
The  solution to eq. \eqref{eqn:res4D_024} reads then
\begin{align}
\KkB_k &= \frac{\tg_k^{\background} }{\tg_{k_0}^{\background}} \left(\frac{k_0}{k}\right)^4 \Big[\KkB_{k_0} + \frac{\tg_{k_0}^{\background}}{k_0^4} \, \int_{k_0}^{k}\md k^{\prime}\, k^{\prime\,3} \kA(k^{\prime})\Big] \label{eqn:res4D_025}
\end{align}
where for $\tg_k^{\background}$ the expression \eqref{eqn:res4D_022} is to be inserted.
Inside the square brackets on the RHS of \eqref{eqn:res4D_025} we can distinguish two contributions, of first and of zeroth order in $\KkB_{k_0}$, respectively. 
\begin{figure}[!ht]
\centering
\psfrag{k}[br]{${\scriptscriptstyle  k}$}
\psfrag{j}[br]{${\scriptscriptstyle   \text{UV}}$}
\psfrag{i}[bl]{${\scriptscriptstyle \text{IR}}$}   
\psfrag{z}[bc]{$\scriptscriptstyle \propto \KkB_{k_0}$}  
 \subfloat[Type (Ia)$^{\dyn}$]{%
\psfrag{f}[l]{$\scriptscriptstyle  \propto \left(\KkB_{k_0}\right)^0$}  
 \label{fig:res4DkAT1}\includegraphics[width=0.45\textwidth]{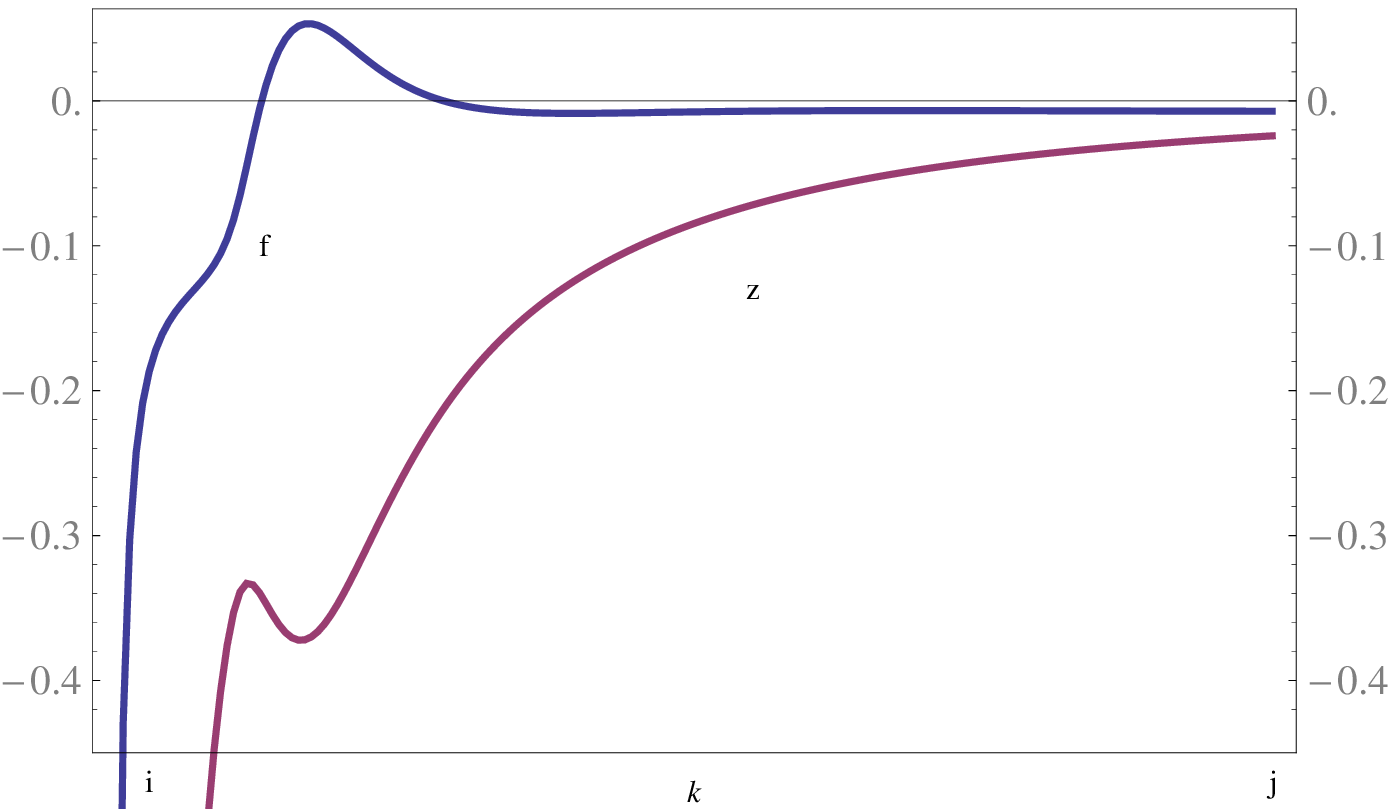}}
\hspace{0.05\textwidth}
 \subfloat[Separatrix]{%
\psfrag{f}[l]{$\scriptscriptstyle  \propto \left(\KkB_{k_0}\right)^0$}  
  \label{fig:res4DkAT2}\includegraphics[width=0.45\textwidth]{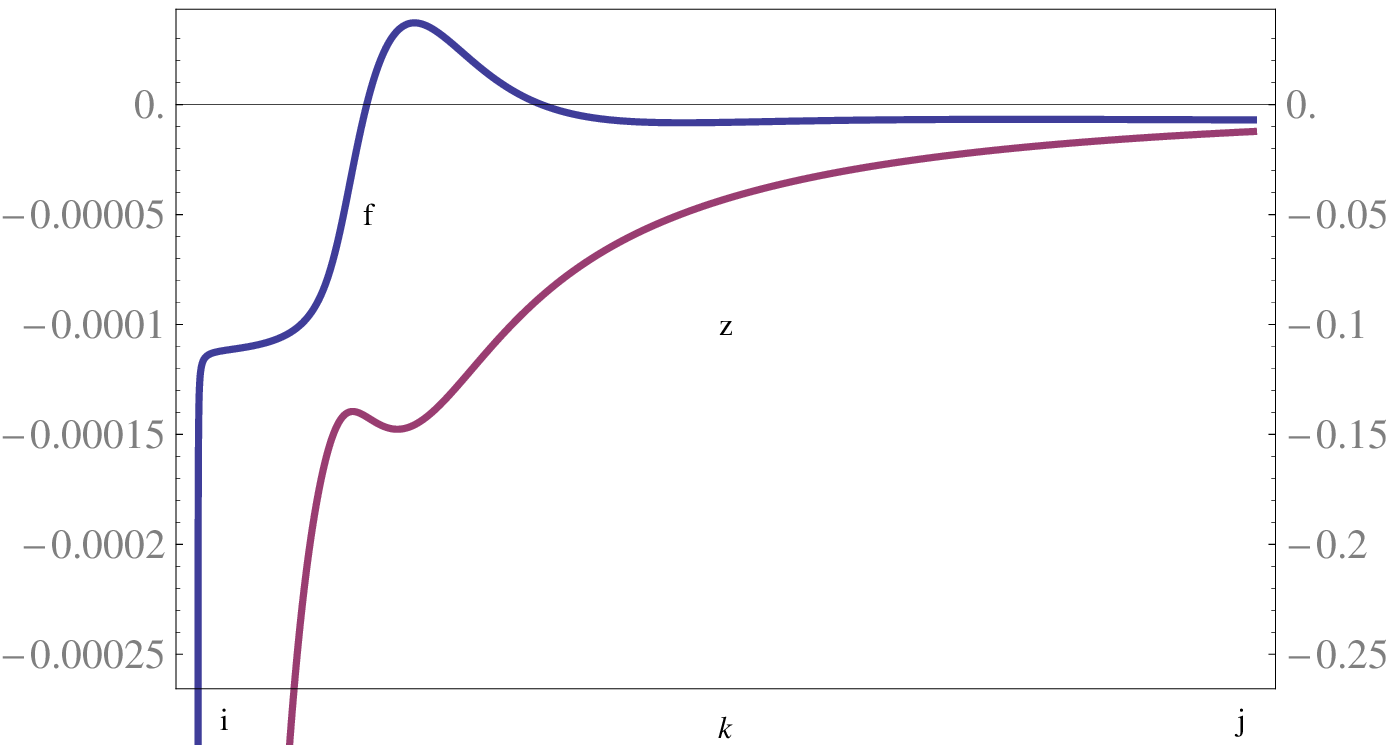}}
\hspace{0.05\textwidth}
 \subfloat[Type (IIIa)$^{\dyn}$]{%
\psfrag{f}[tc]{$\scriptscriptstyle  \propto \left(\KkB_{k_0}\right)^0$}  
\label{fig:res4DkAT3}\includegraphics[width=0.45\textwidth]{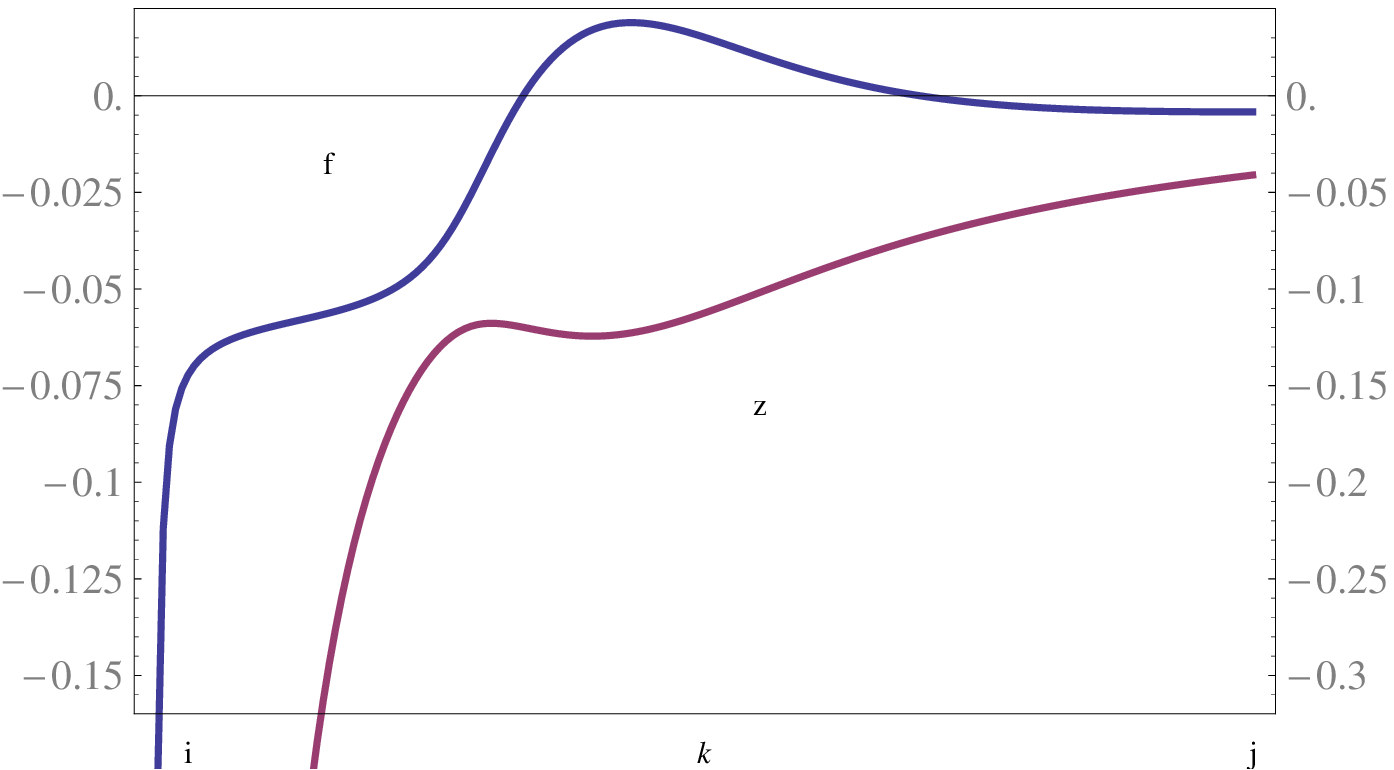}}
 \caption{The two contributions to the function $\KkB_k$ which are constant and linear in $\KkB_{k_0}$, respectively, see eq. \eqref{eqn:res4D_025}. Shown is the regime of scales when the denominator in \eqref{eqn:res4D_022} becomes large. The vertical scales on the LHS (RHS) of the diagrams correspond to the linear (constant) contribution.}\label{fig:res4DkA}
\end{figure}
The moment the denominator in \eqref{eqn:res4D_022} starts increasing rapidly, $\tg_k^{\background}$ becomes almost independent of the initial value $\tg_{k_0}^{\background}$. 
This renders the term in \eqref{eqn:res4D_025} which is of zeroth order in $\KkB_{k_0}$ completely independent of the initial data $\big(\KkB_{k_0},\,\tg_{k_0}^{\background}\big)$.

In Fig. \ref{fig:res4DkA} we display the $k$-dependence of both contributions separately. From the diagrams we conclude that the function $k\mapsto \KkB_k$ is indeed independent of the initial value $\tg_{k_0}^{\background}$ over a large range of scales. 
For those $k$-values the constant contribution, in the approximation of eq. \eqref{eqn:res4D_023} given by the fraction $ - \frac{\int k^3 \kA(k)}{k^2 \int { k \kB(k)}}$, starts dominating over the $\KkB_{k_0}$-linear term already at small scales $k$ and finally approaches the \fpnB\fpC\fpnD-\fpL{} value of $\KkB$ in the UV. 

\noindent {\bf Summary.}
Recalling that above the same qualitative property was found for $\tg_k^{\background}$ also, we see that {\it under upward evolution, all solutions $k\mapsto (\tg_k^{\background},\KkB_k)$ of the $\background$-sector `forget' their values at $k=k_0$ and converge to a single trajectory ultimately hitting \fpnB\fpC\fpnD-\fpL{} when moving towards the UV.}

This behavior is related to the observation that for $k\rightarrow\infty$ {\it all} solutions in the $\dyn$-sector spiral into \fpnD-\fpL{} (assuming, as always, $\tg_{k_0}^{\dyn}>0$). 
This fact is a global, and nonlinear extension of the above linear analysis, yielding 2 attractive directions at \fpnD-\fpL{}. 
Since all such solutions share the same fate in the UV, the differential equations for the $\background$-couplings, too, become independent of ($\tg_{k_0}^{\dyn}$, $\KkD_{k_0}$) in the UV. 
In principle the $\background$-couplings could still depend on their own initial values, ($\tg^{\background}_{k_0}$, $\KkB_{k_0}$). However, we saw that the influence of $\tg_{k_0}^{\background}$ and $\KkB_{k_0}$ is significant only in the IR. 
Hence, we discover that the UV attractivity of \fpnB\fpC\fpnD-\fpL{} in all 4 directions which previously was established on the basis of the critical exponents at the linearized level only, possesses a nonlinear extension: {\it The RG flow on the 4-dimensional theory space has the global property that all trajectories approach the doubly non-Gaussian fixed point under upward evolution:}
\begin{align}
 (\tg_k^{\dyn},\KkD_k,\tg_k^{\background},\KkB_k)\,\, \xrightarrow{k\rightarrow \infty}\,\, \text{\fpnB\fpC\fpnD-\fpL{}} \quad \text{ for all  } \, (\tg_{k_0}^{\dyn}>0,\,\KkD_{k_0},\,\tg_{k_0}^{\background}>0,\,\KkB_{k_0}) \nonumber 
\end{align}
To be precise, the fixed point's `basin of attraction' consists of all points with positive dynamical and background Newton constants.

\subsection[The running UV attractor in  the \texorpdfstring{$\background$}{B}-sector]{The running UV attractor in  the $\bm{\background}$-sector}
We saw that in the $\dyn$-sector the differential equations are autonomous: The beta-functions are invariant under translations in the RG-time $t$, and we obtain a { time independent} phase portrait  in the $\dyn$-sector, see Fig.  \ref{fig:flowD4Traj}.
However, after choosing initial values ($\tg_{k_0}^{\dyn}$, $\KkD_{k_0}$) and inserting the resulting Dyn-solution into the `$\background$'-equations, their translation invariance gets broken and the $\background$-system becomes  non-autonomous, depending explicitly on $k$. As a result, the 2 dimensional phase portrait on the $\tg^{\background}$-$\KkB$ plane is explicitly `time' dependent.

We shall nevertheless be able to deduce the essential qualitative properties of the phase portrait by analytical methods. 
Its structure is essentially determined by the {\it $k$-dependent analogue of fixed points} in the $\background$-system. 

\noindent {\bf (A) The `moving fixed point'.}
We consider the two equations \eqref{eqn:res4D_021} for $\tg_k^{\background}$ and $\KkB_k$ with given, externally prescribed coefficient functions $\kA(k)$ and $\kB(k)$ and search for $k$-dependent points $(\tg_{\attr}^{\background},\KkB_{\attr})$ on the $\tg^{\background}$-$\KkB$-plane at which both beta-functions vanish simultaneously:
\begin{align}
\beta_{\tg}^{\background}\big(\tg_{\attr}^{\background}(k),\KkB_{\attr}(k);k\big)=0 \nonumber \\
 \beta_{\Kk}^{\background}\big(\tg_{\attr}^{\background}(k),\KkB_{\attr}(k);k\big)=0 \label{eqn:res4D_026B}
\end{align}
As long as $\kB(k)\neq 0$, there exist two solutions to these equations.

\noindent {\bf (i)} The first one is trivial, $\tg^{\background}_{\attr}=0=\KkB_{\attr}$, and happens to be independent of $k$. As a consequence, we expect a  4-dimensional Gaussian fixed point $(\tg_*^{\dyn},\,\KkD_*,\,\tg_*^{\background},\,\KkB_*)=0$  to be present in all phase diagrams. 

\begin{figure}[!ht]
\centering
 \subfloat[Type (Ia)$^{\dyn}$]{%
\psfrag{j}[tr]{${\scriptscriptstyle  \text{UV}\rightarrow}$}
\psfrag{i}[tl]{${\scriptscriptstyle \leftarrow \text{IR}}$}   
\psfrag{z}[bc]{$\scriptscriptstyle \tg_{\attr}^{\background}(k)$}  
\psfrag{f}[bc]{$\scriptscriptstyle   \KkB_{\attr}(k)$}      
\label{fig:res4DattrBT1}\includegraphics[width=0.4\textwidth]{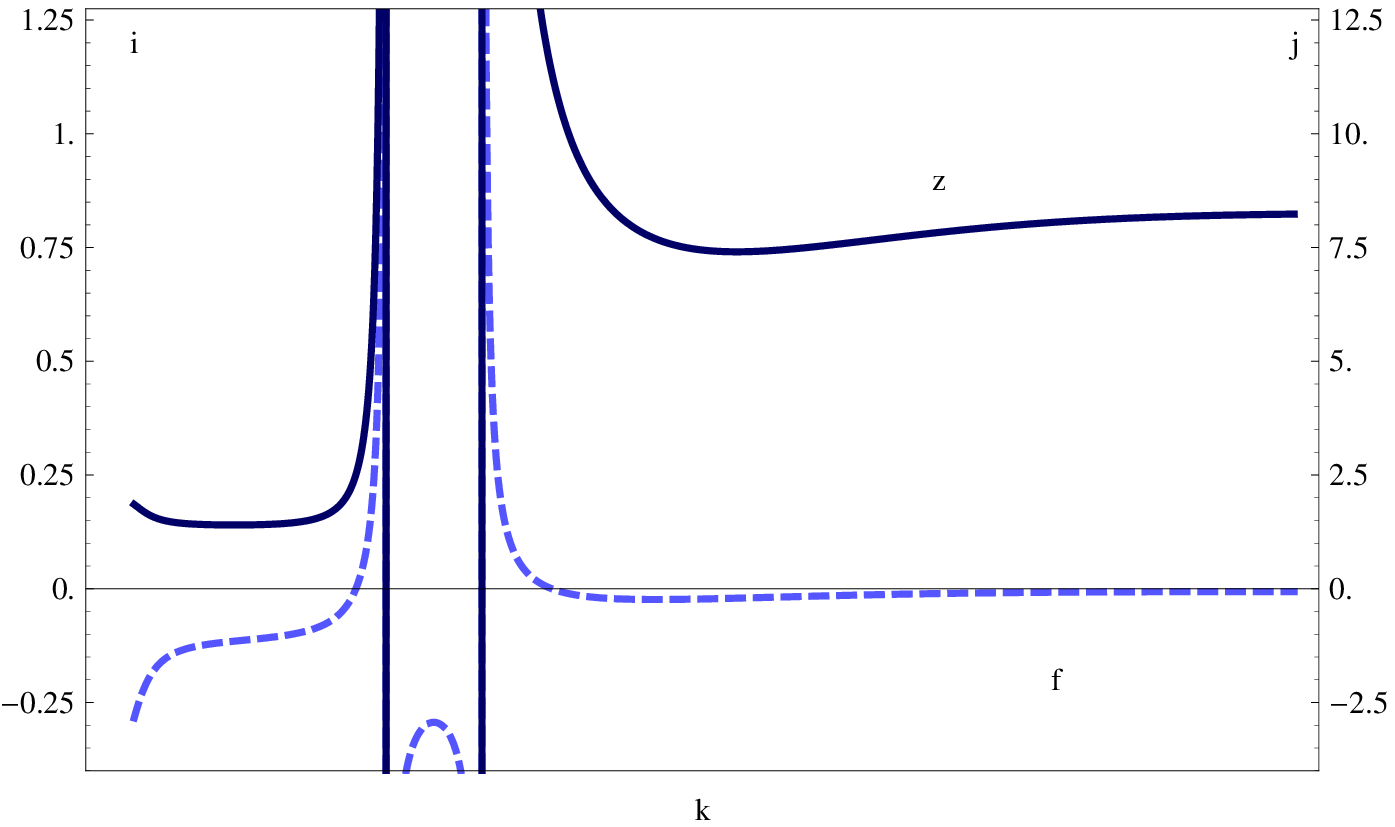}}
\hspace{0.05\textwidth}
 \subfloat[Separatrix]{%
\psfrag{j}[br]{${\scriptscriptstyle  \text{UV}\rightarrow}$}
\psfrag{i}[bl]{${\scriptscriptstyle \leftarrow \text{IR}}$}   
\psfrag{z}[bl]{$\scriptscriptstyle \tg_{\attr}^{\background}(k)$}  
\psfrag{f}[bl]{$\scriptscriptstyle  \KkB_{\attr}(k)$}      
\label{fig:res4DattrBT2}\includegraphics[width=0.4\textwidth]{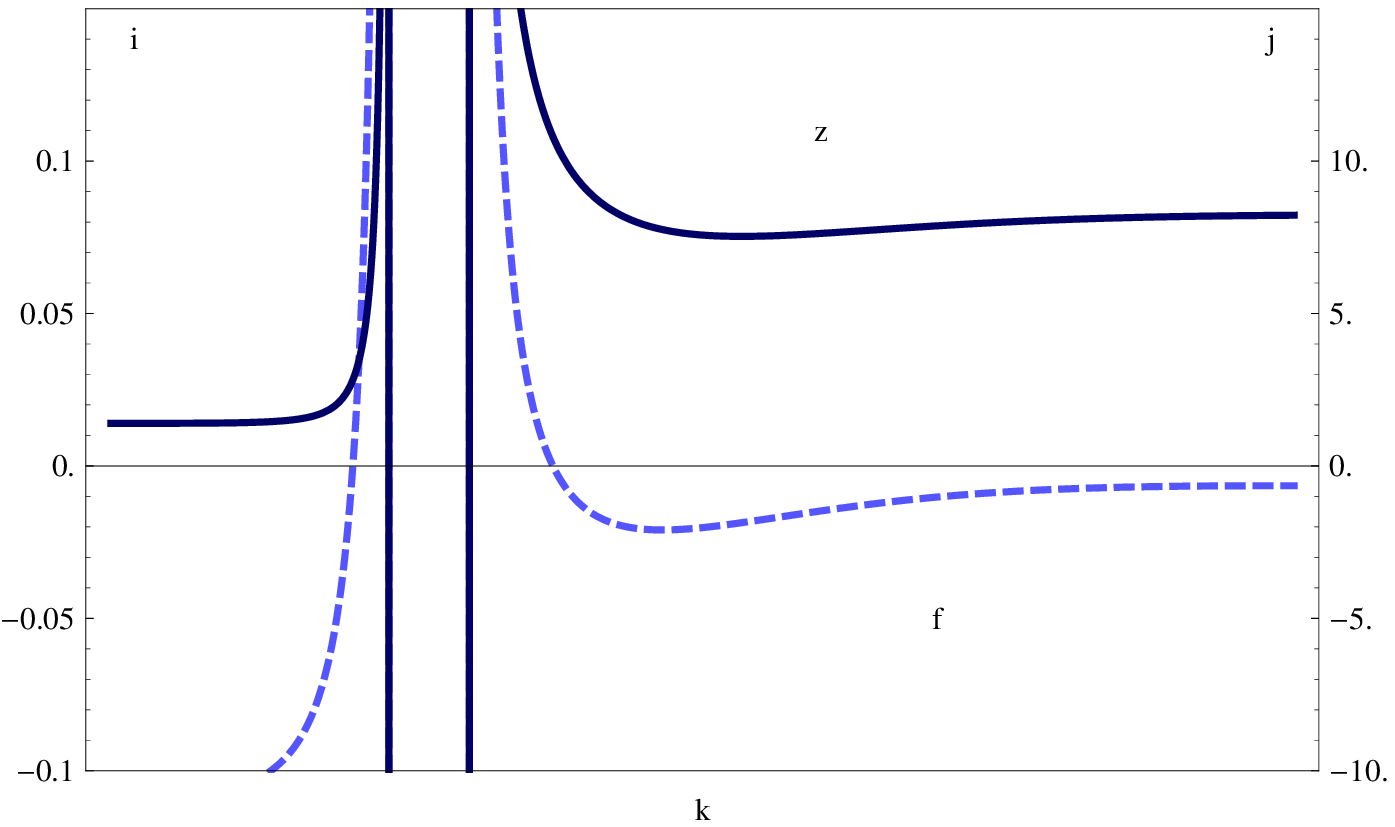}}
\hspace{0.05\textwidth}
 \subfloat[Type (IIIa)$^{\dyn}$]{%
\psfrag{j}[tr]{${\scriptscriptstyle   \text{UV}\rightarrow}$}
\psfrag{i}[tl]{${\scriptscriptstyle \leftarrow \text{IR}}$}   
\psfrag{z}[bc]{$\scriptscriptstyle \tg_{\attr}^{\background}(k)$}  
\psfrag{f}[bc]{$\scriptscriptstyle   \KkB_{\attr}(k)$}         
\label{fig:res4DattrB3}\includegraphics[width=0.4\textwidth]{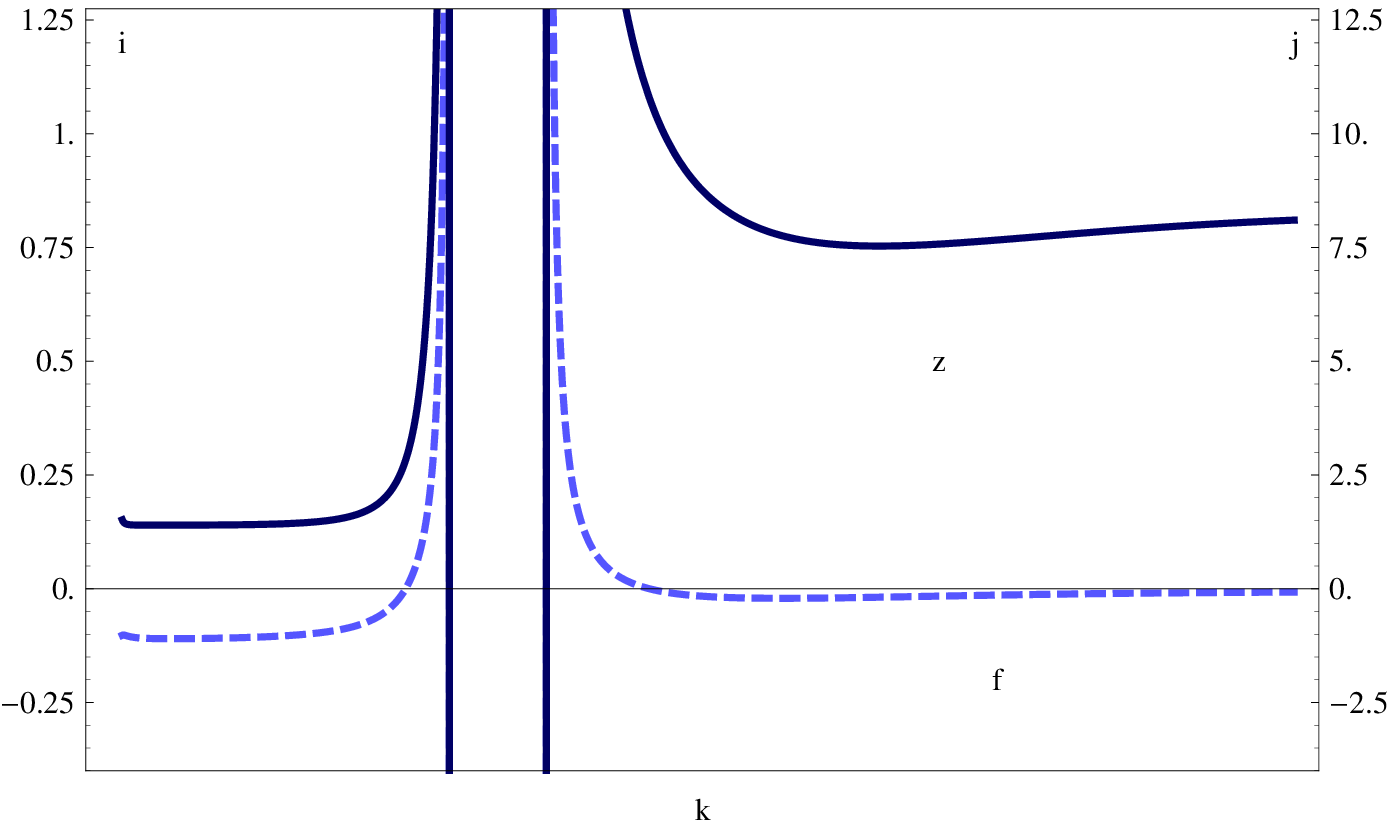}}
 \caption{The $k$-dependence of the running UV attractor $(\tg_{\attr}^{\background}(k),\KkB_{\attr}(k))$ for three representative $\dyn$ trajectories. The left (right) scale corresponds to $\KkB_{\attr}$ ($\tg_{\attr}^{\background}$). The running attractor $(\tg_{\attr}^{\background}(k),\KkB_{\attr}(k))$ approaches the \fpnB\fpC\fpnD-\fpL{} coordinates in the UV and finite, non-vanishing values in the IR. In between the running UV attractor touches the boundary of theory space as can be seen from the divergent coordinates. Notice that $(\tg_{\attr}^{\background}(k),\KkB_{\attr}(k))$ is the position of the `sink'  the inverse RG flow in the $\background$-sector is pointing to, but not a solution to the RG equations.}\label{fig:res4DattrB}
\end{figure}

\noindent {\bf (ii)} The second solution is non-trivial and explicitly $k$-dependent:
\begin{align}
\tg_{\attr}^{\background}(k)=-2 \slash \kB(k)    
\quad  \text{and} \quad 
 \KkB_{\attr}(k)= -\tfrac{1}{2} \kA(k)\,\slash \kB(k)  \label{eqn:res4D_027}
\end{align}
Its $k$-dependence is shown in Fig. \ref{fig:res4DattrB} for  three typical $\dyn$-trajectories, one of each type, which determine $\kA(k)$ and $\kB(k)$. 
The UV behavior of $(\tg_{\attr}^{\background}(k),\KkB_{\attr}(k))$ is seen to be the same for all underlying $\dyn$ trajectories $(\tg^{\dyn}_k, \KkD_k)$, namely
\begin{align}
\big(\tg_{\attr}^{\background}(k),\KkB_{\attr}(k)\big)\,\, \xrightarrow{k\rightarrow \infty}\,\, \text{\fpnB-\fpL}\,  \qquad \text{ for all }\, \, (\tg_{k_0}^{\dyn}>0,\KkD_{k_0},\tg_{k_0}^{\background}>0,\KkB_{k_0})
 \label{eqn:res4D_028}
\end{align}
Thus, for $k\rightarrow\infty$ the `running fixed point' $(\tg_{\attr}^{\background}(k),\KkB_{\attr}(k))$ approaches a true one, namely  \fpnB-\fpL{}.

Let us consider the imbedding of the running fixed point into the 4-dimensional theory space. It moves along a curve parametrized by
\begin{align}
 u_{\attr}(k)\equiv \big(\tg_k^{\dyn},\KkD_{k},\tg_{\attr}^{\background}(k),\KkB_{\attr}(k)\big) \label{eqn:res4D_028B}
\end{align}
Note that {\it the curve $k\mapsto u_{\attr}(k)$ is {\it not} an RG trajectory}. 
Since all $\dyn$ trajectories approach \fpnD-\fpL{} for $k\rightarrow\infty$, it is clear that $u_{\attr}(k)$ approaches \fpnB\fpC\fpnD-\fpL{} in the UV.
However, its  global properties depend significantly on the type of the $\dyn$-trajectory:

\noindent {\bf (i)}
The  type (IIa)$^{\dyn}$-trajectory, the separatrix, for instance, describes a cross-over:
\begin{align}
\text{\fpgD-\fpL{}} \xleftarrow{k\rightarrow 0}(\tg_k^{\dyn},\KkD_k) \xrightarrow{k\rightarrow\infty} \text{\fpnD-\fpL{}}
\end{align}
 Hence the resulting curve \eqref{eqn:res4D_028B} connects  two fixed points, namely the `doubly non-Gaussian' one in the UV, and the mixed  \fpnB\fpC\fpgD-\fpL{}  in the IR:
\begin{align}
 \text{\fpnB\fpC\fpgD-\fpL{}} \xleftarrow{k\rightarrow 0} u_{\attr}(k) \xrightarrow{k\rightarrow\infty} \text{\fpnB\fpC\fpnD-\fpL{}}
\end{align}

\noindent {\bf (ii)}
For type (Ia)$^{\dyn}$ and (IIIa)$^{\dyn}$ trajectories, where $\KkD$ for $k\rightarrow0$ goes to $-\infty$ and $+\infty$\footnote{For a moment we ignore the singularity in the beta-functions at $\KkD=1\slash2 $.}, respectively, the  point \eqref{eqn:res4D_027} approaches, in the IR,  
\begin{align}
\big(\tg^{\background}_{\attr},\, \KkB_{\attr}\big)\xrightarrow{\KkD\rightarrow \pm\infty}\left(\frac{3\pi}{5}\,, -\frac{2}{5} \right)
 \label{eqn:res4D_029}
\end{align}

At a certain intermediate scale $(0<k<\infty)$, the running attractor touches the boundary of theory space, i.e. it is pulled to infinity%
\footnote{Note that there is nothing wrong with diverging values of $\tg_{\attr}^{\background}$ and $\KkB_{\attr}$ at some $k$ since, as we stressed already, the curve followed by the `running fixed point' is not an RG trajectory. A divergent $\tg_{\attr}^{\background}$ and \slash or $\KkB_{\attr}$ simply means that at this special value of $k$ there exists no such fixed point.}
, then returns to the interior of theory space, and finally moves towards \fpnB-\fpL{}, under upward evolution. 
This behavior is clearly seen in the diagrams of Fig. \ref{fig:res4DattrB}.

\noindent {\bf (B) Stability of the `moving fixed point'.}
Let us linearize the $\background$-system \eqref{eqn:res4D_021} about $(\tg_{\attr}^{\background},\KkB_{\attr})$ and deduce a $k$-dependent analogue of a stability matrix, $\mathcal{B}(\tg_{\attr}^{\background},\KkB_{\attr};k)$. This matrix turns out to have two $k$-dependent eigenvectors $V_{\attr}^{(1\slash 2)}(k)$, associated to the `would-be critical exponents', i.e. its negative eigenvalues $\theta_{\attr}^{(1)}=2$ and $\theta_{\attr}^{(2)}=4$, respectively:
\begin{align}
\mathcal{B}(\tg_{\attr}^{\background},\KkB_{\attr};k)=
 \begin{pmatrix}
 -2 & 0 \\ \tfrac{1}{2}\kA(k) & -4 
 \end{pmatrix}\,,
 \quad V_{\attr}^{(1)}(k)= 4 \hat{e}_{\tg}^{\background} + \kA(k)\,\hat{e}_{\lambda}^{\background}\,,
 \quad V_{\attr}^{(2)}(k)=\hat{e}_{\lambda}^{\background}
 \label{eqn:res4D_030}
\end{align}
Here, $\hat{e}_{\tg}^{\background}$ and $\hat{e}_{\lambda}^{\background}$ are unit vectors in  the $\tg^{\background}$- and $\KkB$-direction, respectively.
As both $\theta^{(1)}_{\attr}$ and $\theta^{(2)}_{\attr}$ are found to be positive we may conclude that {\it at all scales the point $\big(\tg_{\attr}^{\background},\KkB_{\attr}\big)$ is UV-attractive in both $\background$-directions}. 
No matter which initial conditions we choose for the $\background$-couplings, under upward evolution the $\background$-trajectories $k\mapsto (\tg_{k}^{\background},\KkB_k)$ are always pulled towards this point for $k\rightarrow \infty$. Therefore we shall refer to it as the {\it running UV attractor} and denote it {\bf \AttrL} or {\bf \AttrL}$(k)$ in the following.

\noindent {\bf (C) Global structure of the $\background$-flow.}
For $k\rightarrow0$ the attractor property {\bf \AttrL} implies that all $\background$-trajectories converge to the values \eqref{eqn:res4D_029} in the (Ia)$^{\dyn}$ and (IIIa)$^{\dyn}$ cases, and to \fpnB\fpC\fpgD-\fpL{} in the case of the separatrix (IIa)$^{\dyn}$. 

Under upward evolution, when {\bf \AttrL}$(k)$ moves due to an increasing $k$, the $\background$-trajectories try to follow its motion until they all end up in the doubly non-Gaussian fixed point  \fpnB\fpC\fpnD-\fpL{} for $k\rightarrow \infty$. This behavior is shown  in the phase portraits of Figs. \ref{fig:res4DppT1a} - \ref{fig:res4DppT3a}  for representative type (Ia)$^{\dyn}$, (IIa)$^{\dyn}$, and (IIIa)$^{\dyn}$ trajectories, respectively. 
\afterpage{%
\clearpage
\thispagestyle{empty}
\begin{figure}[pt!]
\centering
\psfrag{N}[tc]{${\scriptscriptstyle  }$}
\psfrag{G}[tl]{${\scriptscriptstyle }$}   
\psfrag{b}[c]{${ \tg^{\background}  }$}
\psfrag{a}[m]{${ \KkB}$}  
\psfrag{s}[ml]{$\scriptscriptstyle \SolAttr(k)$}  
\psfrag{k}[ml]{$\scriptscriptstyle   \text{Sol}_{(0,0)}^{\background}(k)$}
\psfrag{t}[ml]{$\scriptscriptstyle  \text{RG-`time' } k$}    
 \subfloat{%
\label{fig:res4Dpp1T1a}\includegraphics[width=0.4\textwidth]{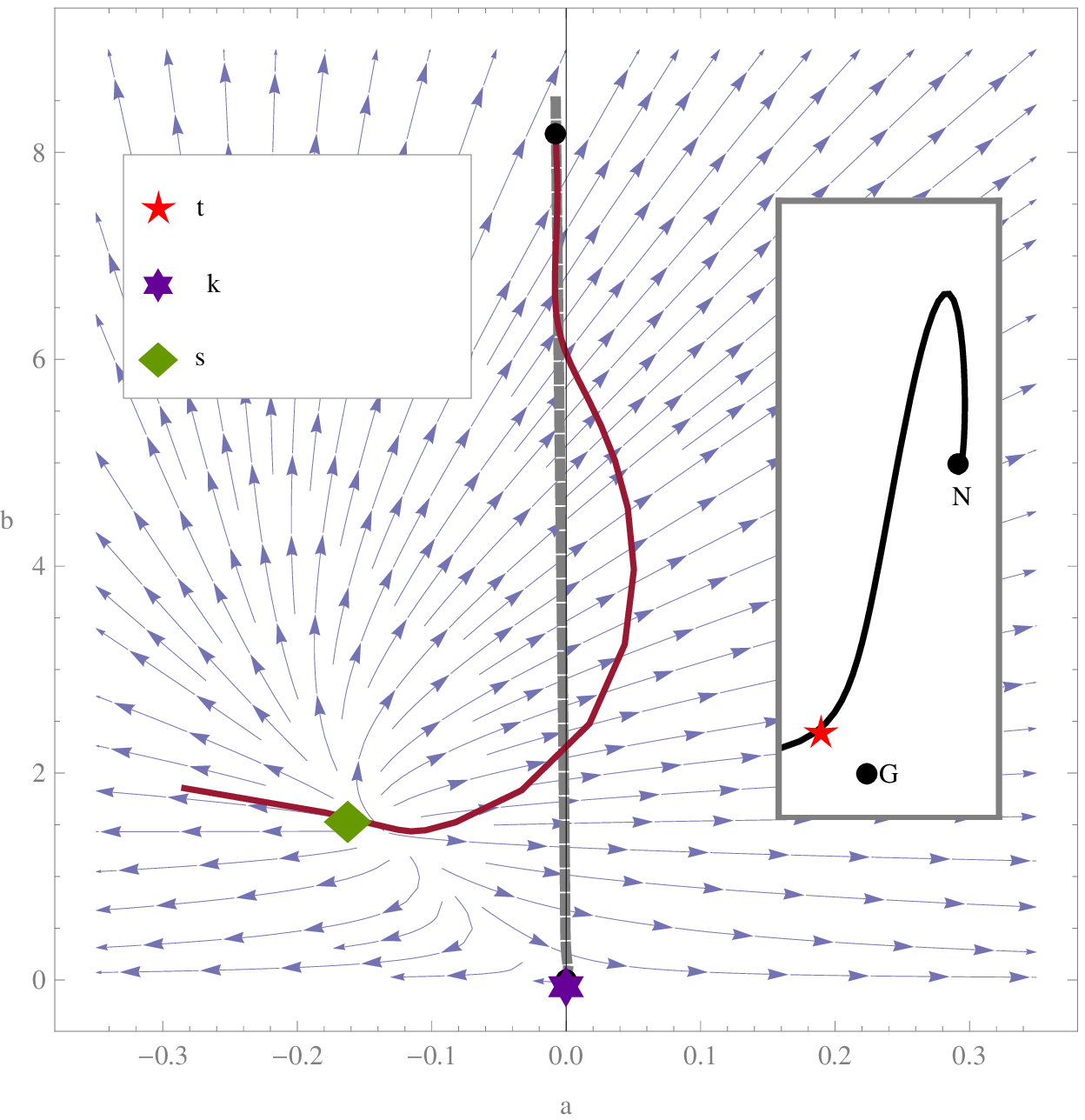}}
\hspace{0.05\textwidth}
 \subfloat{%
\label{fig:res4Dpp2T1a}\includegraphics[width=0.4\textwidth]{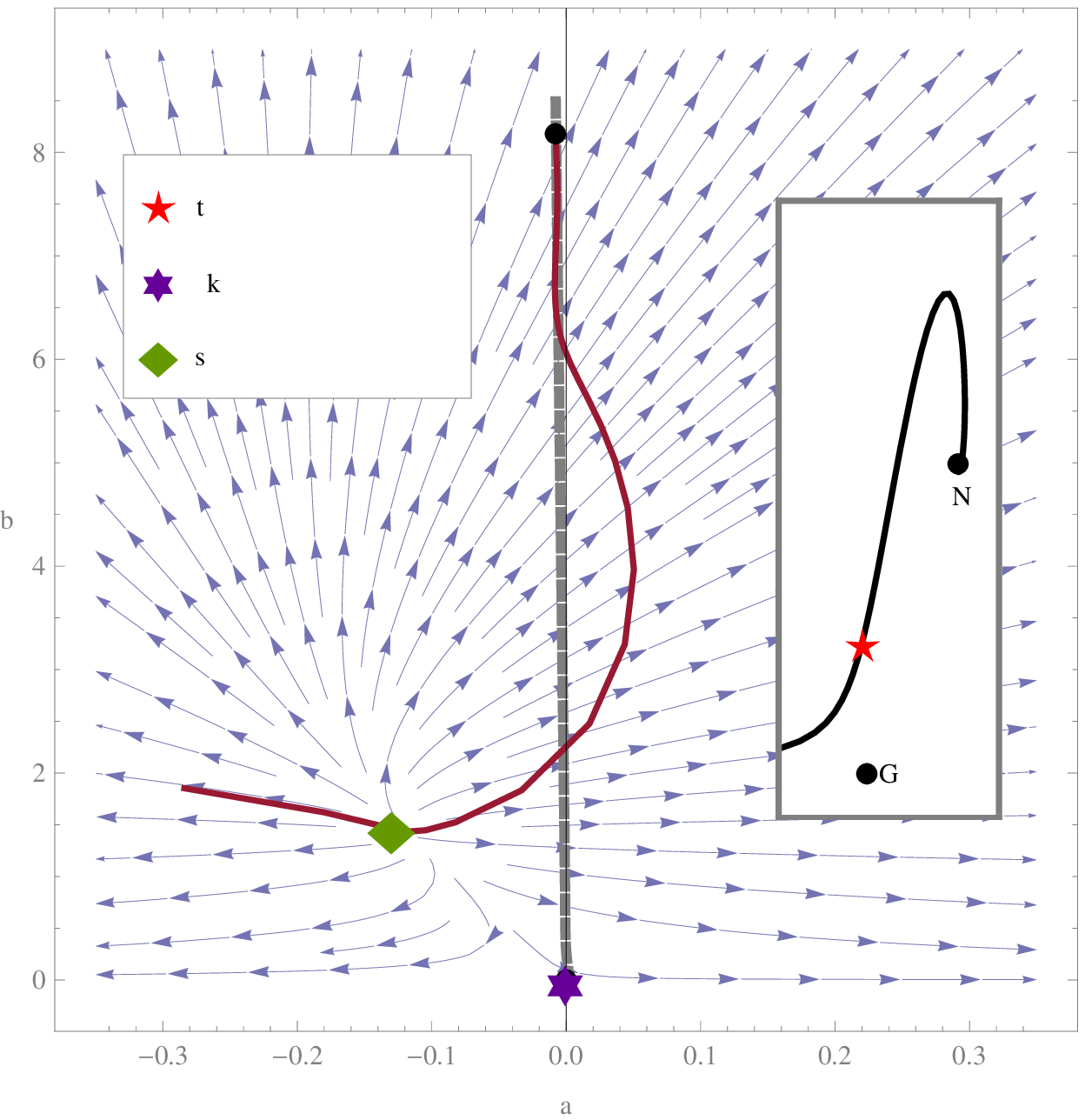}}
\hspace{0.05\textwidth}
 \subfloat{%
\label{fig:res4Dpp3T1a}\includegraphics[width=0.4\textwidth]{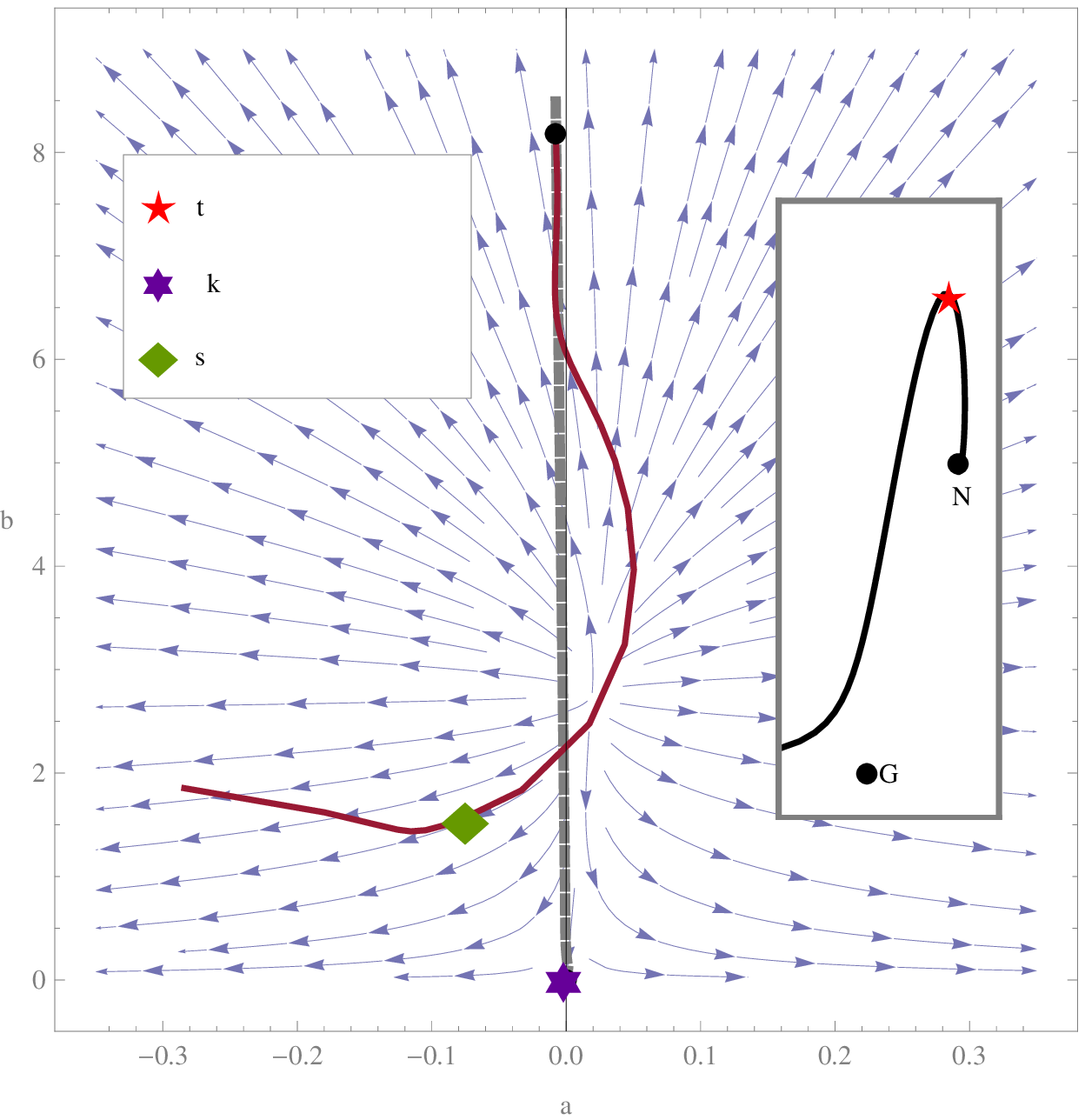}}
\hspace{0.05\textwidth}
 \subfloat{%
\label{fig:res4Dpp4T1a}\includegraphics[width=0.4\textwidth]{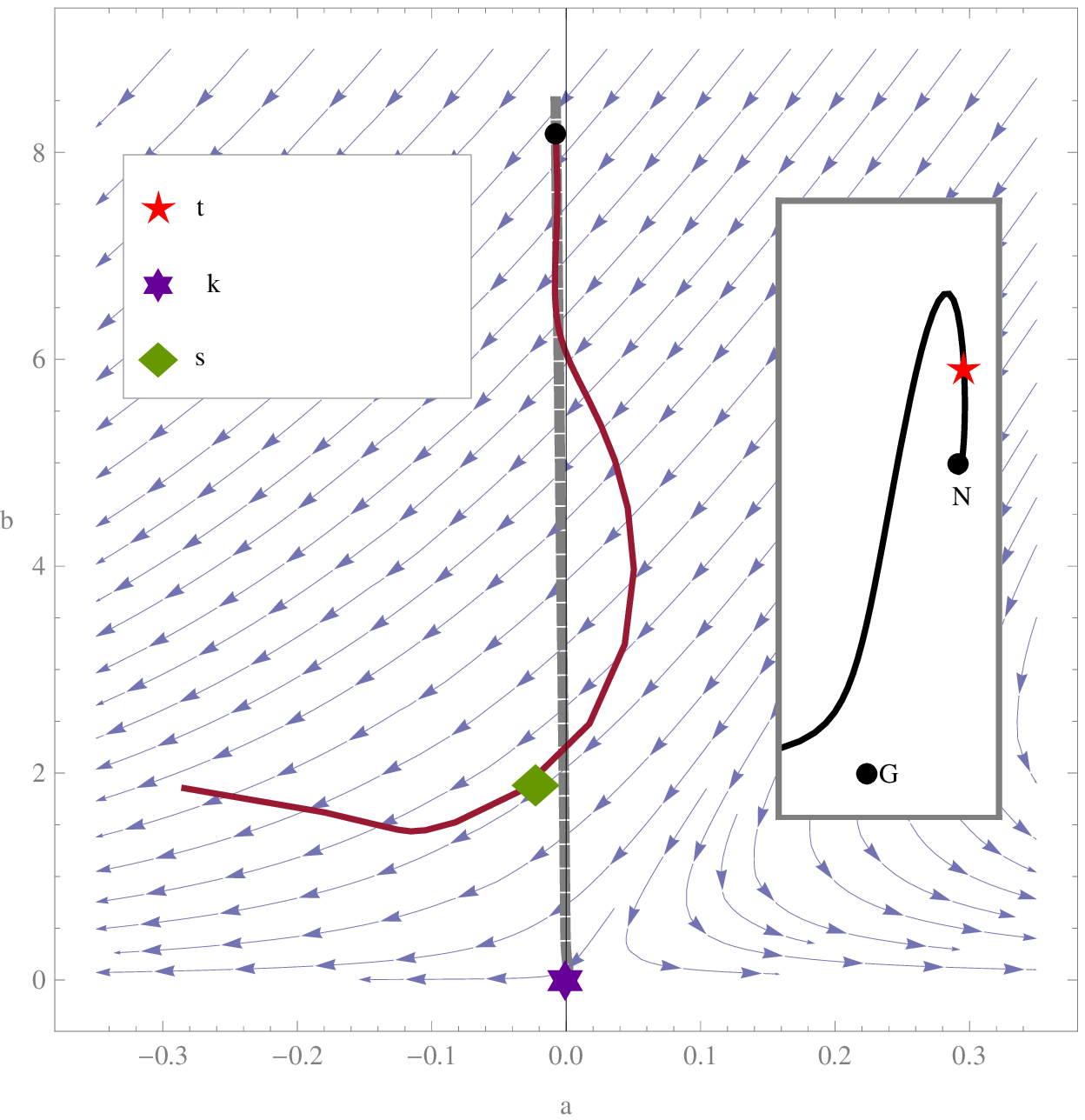}}
\hspace{0.05\textwidth}
 \subfloat{%
\label{fig:res4Dpp5T1a}\includegraphics[width=0.4\textwidth]{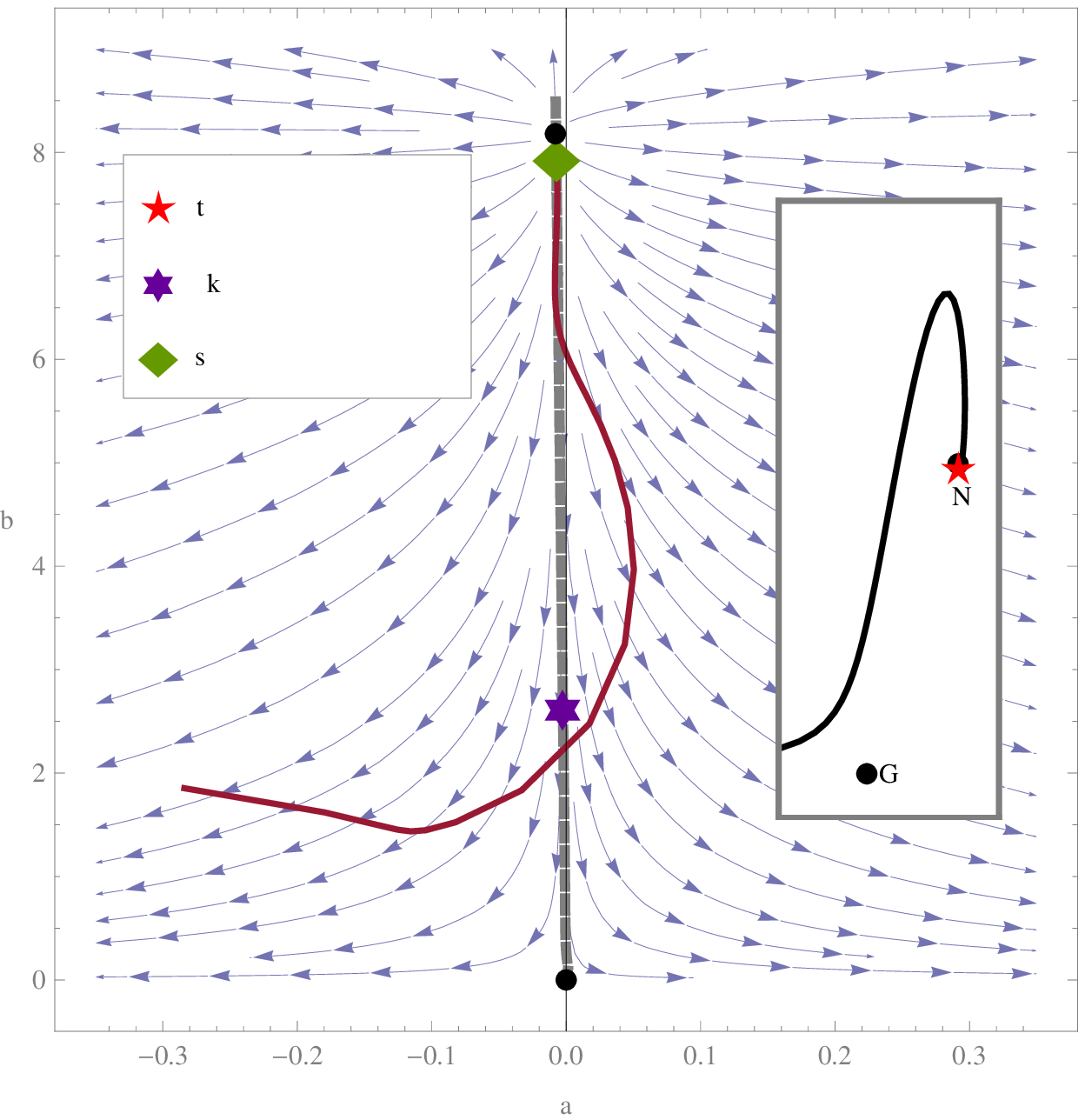}}
\hspace{0.05\textwidth}
 \subfloat{%
\label{fig:res4Dpp6T1a}\includegraphics[width=0.4\textwidth]{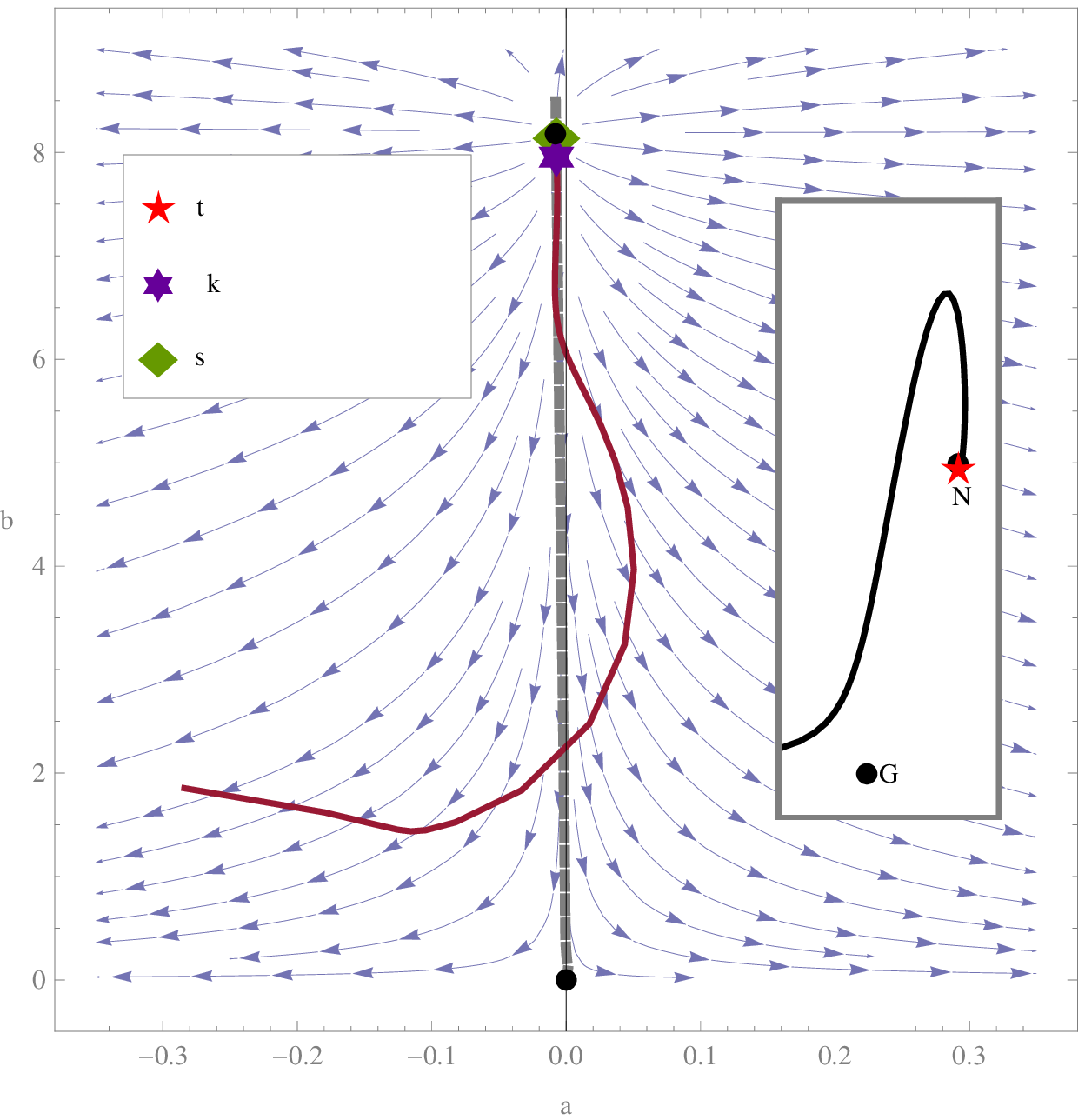}}
 \caption{The $\background$-phase portraits at increasing scales $k$. The underlying type (Ia)$^{\dyn}$ trajectory in the $\dyn$-sector is shown in the inset on the right, and the current RG time is marked with a star therein. The arrows point towards the IR and picture the instantaneous vector field in the $\background$-sector.  The (red) solid and the (gray) dashed curve highlight two important solutions in the $\background$-sector, namely $\SolAttr(k)$ and $\text{Sol}_{(0,0)}^{\background}(k)$, respectively. Their current position is indicated by the (green) diamond and the (violet) six-pointed star, respectively.}\label{fig:res4DppT1a}
\end{figure}
}
\afterpage{%
\clearpage
\thispagestyle{empty}
\begin{figure}[pt!]
\centering
\psfrag{N}[tc]{${\scriptscriptstyle  }$}
\psfrag{G}[tl]{${\scriptscriptstyle }$}   
\psfrag{b}[c]{${ \tg^{\background}  }$}
\psfrag{a}[m]{${ \KkB}$}  
\psfrag{s}[ml]{$\scriptscriptstyle \SolAttr(k)$}  
\psfrag{k}[ml]{$\scriptscriptstyle   \text{Sol}_{(0,0)}^{\background}(k)$}
\psfrag{t}[ml]{$\scriptscriptstyle  \text{RG-`time' } k$}    
 \subfloat{%
\label{fig:res4Dpp1S}\includegraphics[width=0.4\textwidth]{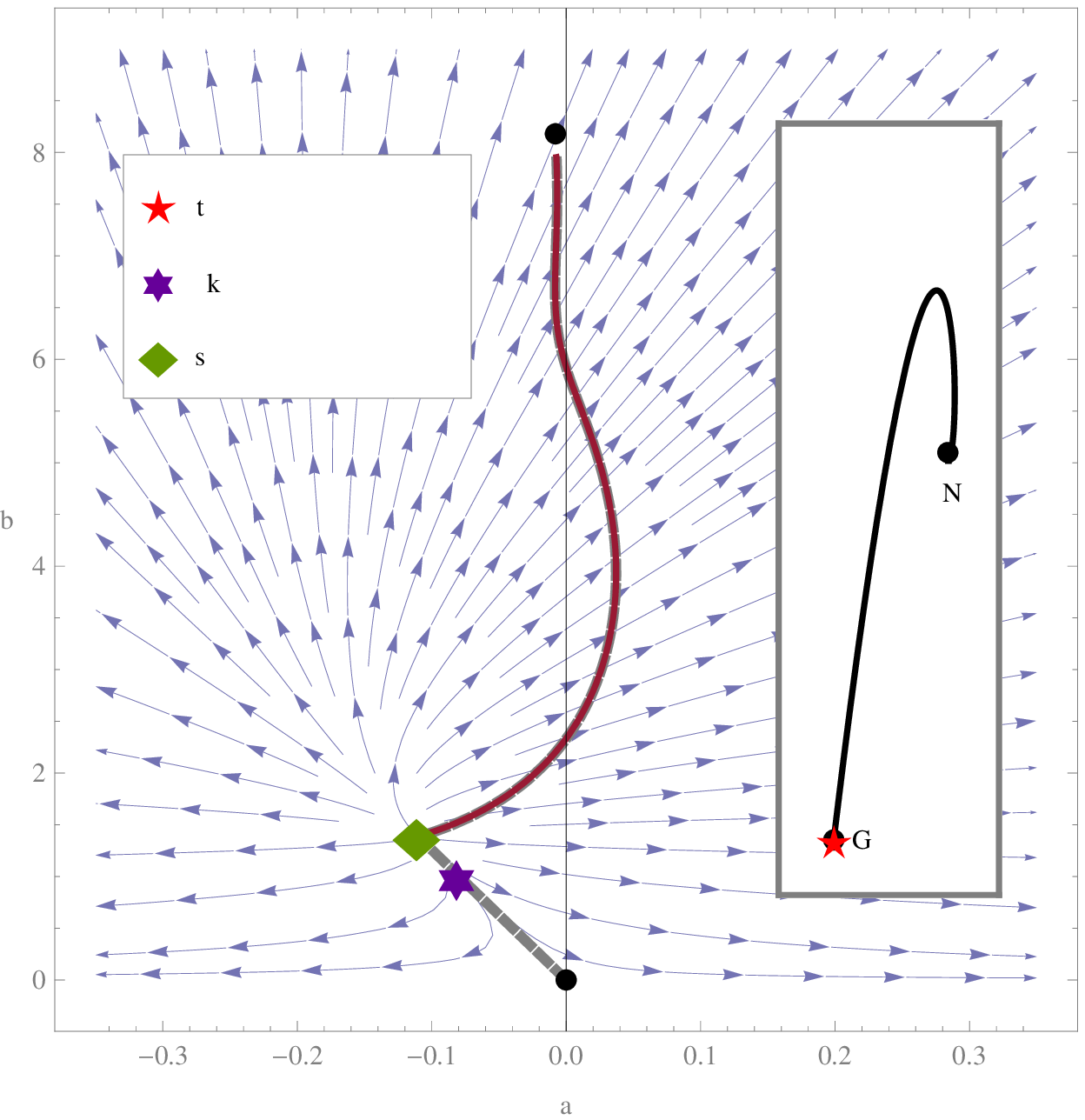}}
\hspace{0.05\textwidth}
 \subfloat{%
\label{fig:res4Dpp2S}\includegraphics[width=0.4\textwidth]{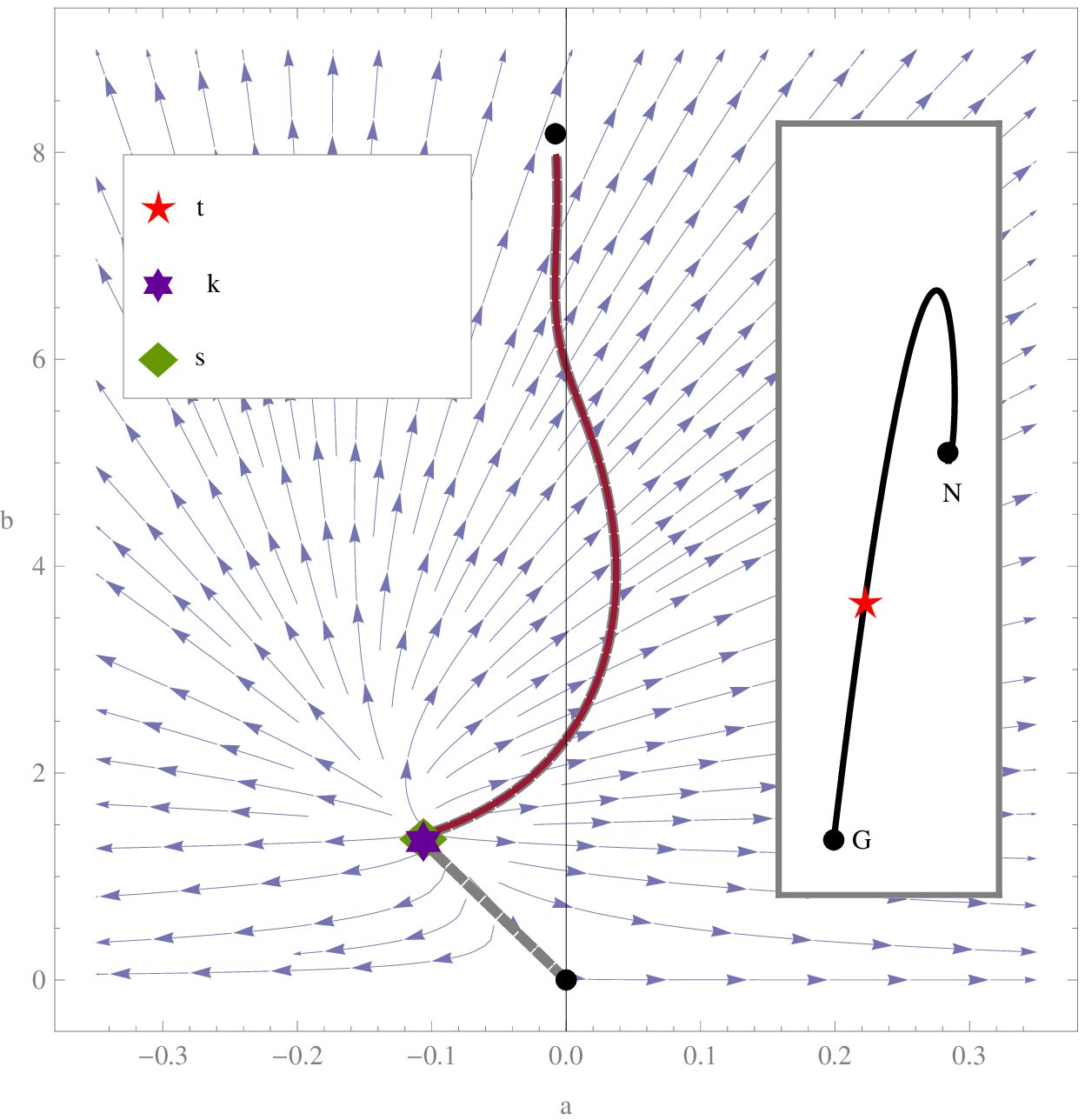}}
\hspace{0.05\textwidth}
 \subfloat{%
\label{fig:res4Dpp3S}\includegraphics[width=0.4\textwidth]{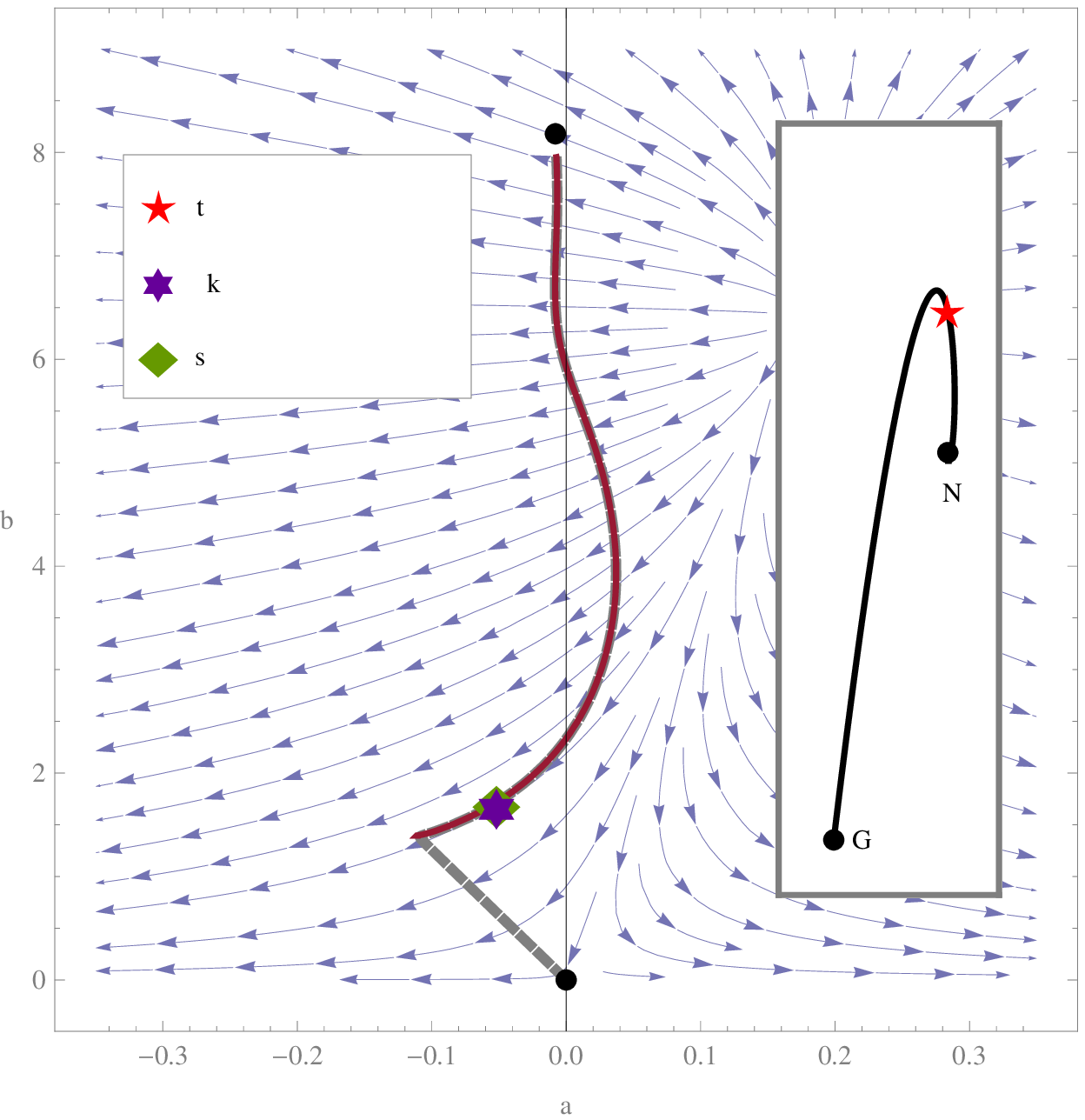}}
\hspace{0.05\textwidth}
 \subfloat{%
\label{fig:res4Dpp4S}\includegraphics[width=0.4\textwidth]{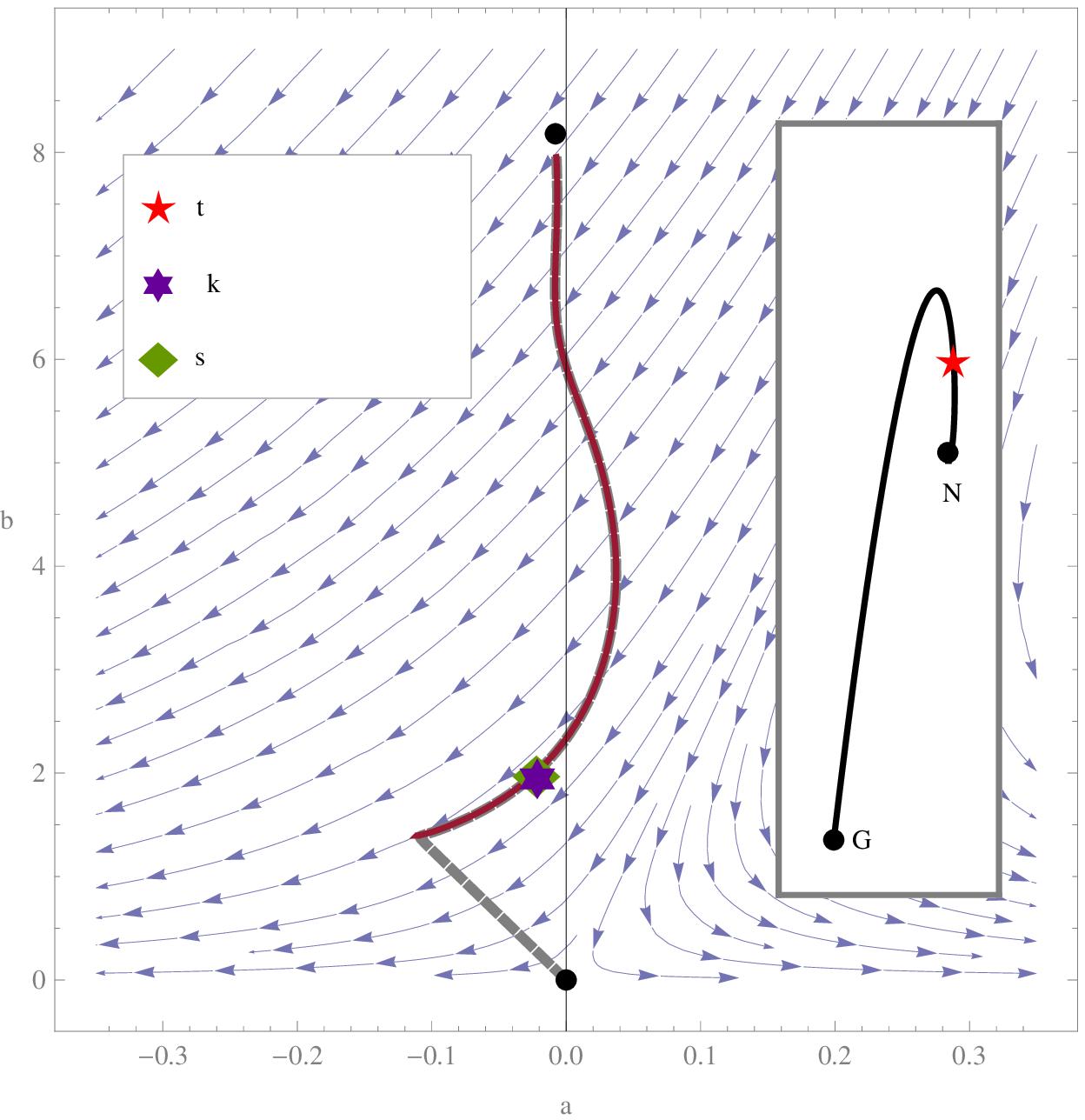}}
\hspace{0.05\textwidth}
 \subfloat{%
\label{fig:res4Dpp5S}\includegraphics[width=0.4\textwidth]{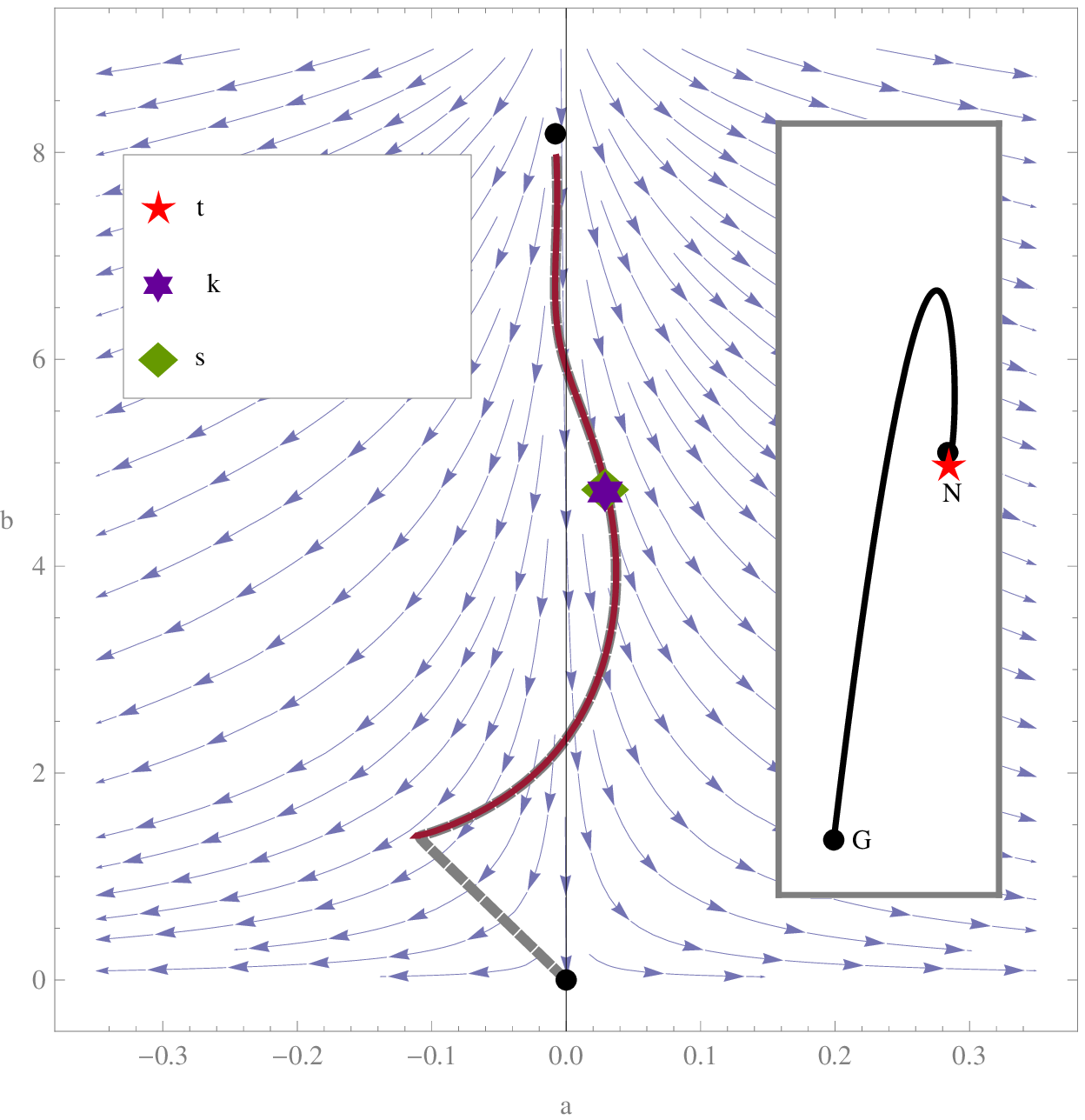}}
\hspace{0.05\textwidth}
 \subfloat{%
\label{fig:res4Dpp6S}\includegraphics[width=0.4\textwidth]{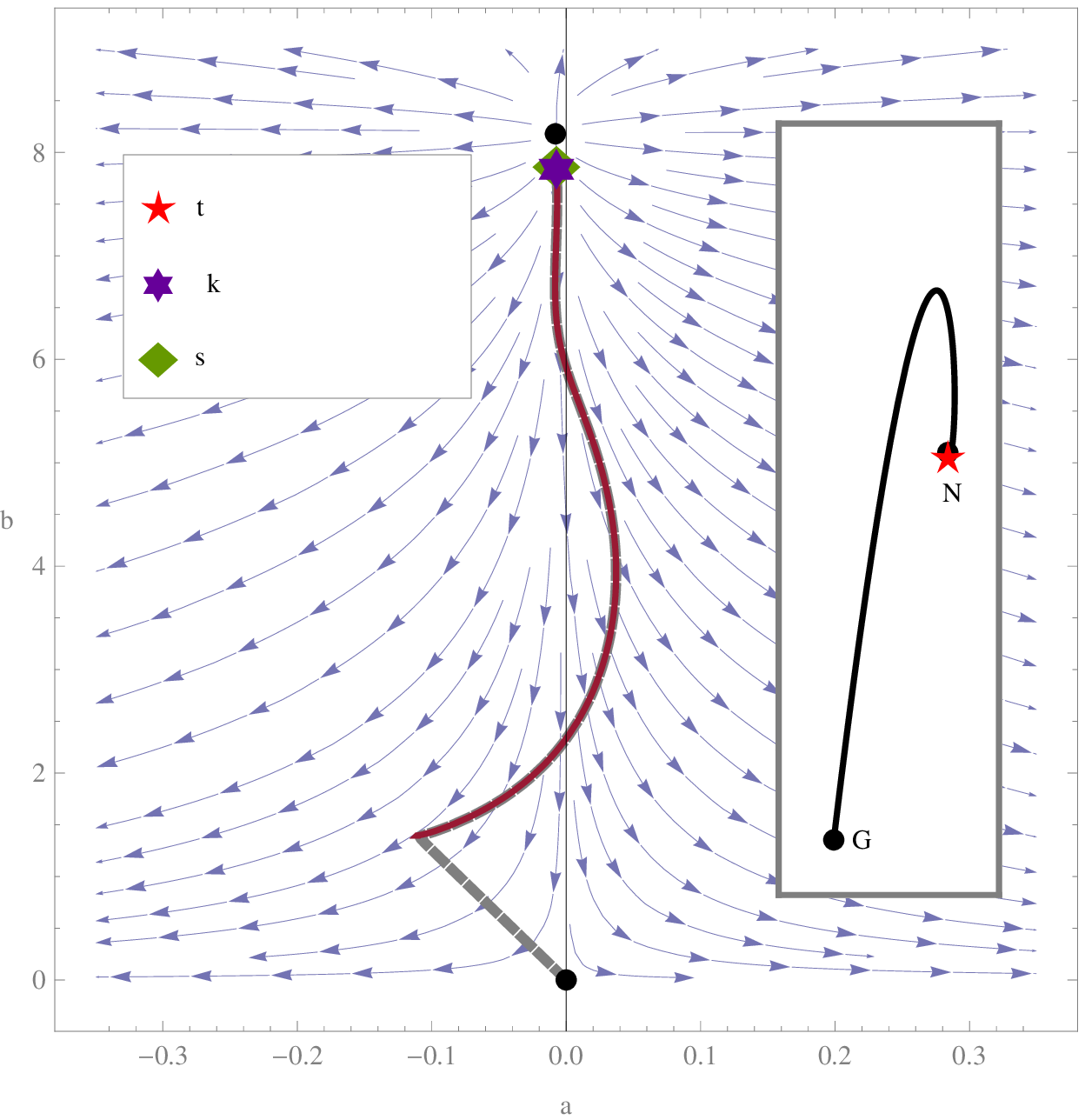}}
 \caption{The $\background$-phase portraits at increasing scales $k$. The underlying separatrix in the $\dyn$-sector is shown in the inset on the right, and the current RG time is marked with a star therein. The arrows point towards the IR and picture the instantaneous vector field in the $\background$-sector.  The (red) solid and the (gray) dashed curve highlight two important solutions in the $\background$-sector, namely $\SolAttr(k)$ and $\text{Sol}_{(0,0)}^{\background}(k)$, respectively. Their current position is indicated by the (green) diamond and the (violet) six-pointed star, respectively.}\label{fig:res4DppS}
\end{figure}
}
\afterpage{%
\clearpage
\thispagestyle{empty}
\begin{figure}[pt!]
\centering
\psfrag{N}[tc]{${\scriptscriptstyle  }$}
\psfrag{G}[tl]{${\scriptscriptstyle }$}   
\psfrag{b}[c]{${ \tg^{\background}  }$}
\psfrag{a}[m]{${ \KkB}$}  
\psfrag{s}[ml]{$\scriptscriptstyle \SolAttr(k)$}  
\psfrag{k}[ml]{$\scriptscriptstyle   \text{Sol}_{(0,0)}^{\background}(k)$}
\psfrag{t}[ml]{$\scriptscriptstyle  \text{RG-`time' } k$}    
 \subfloat{%
\label{fig:res4Dpp1T3a}\includegraphics[width=0.4\textwidth]{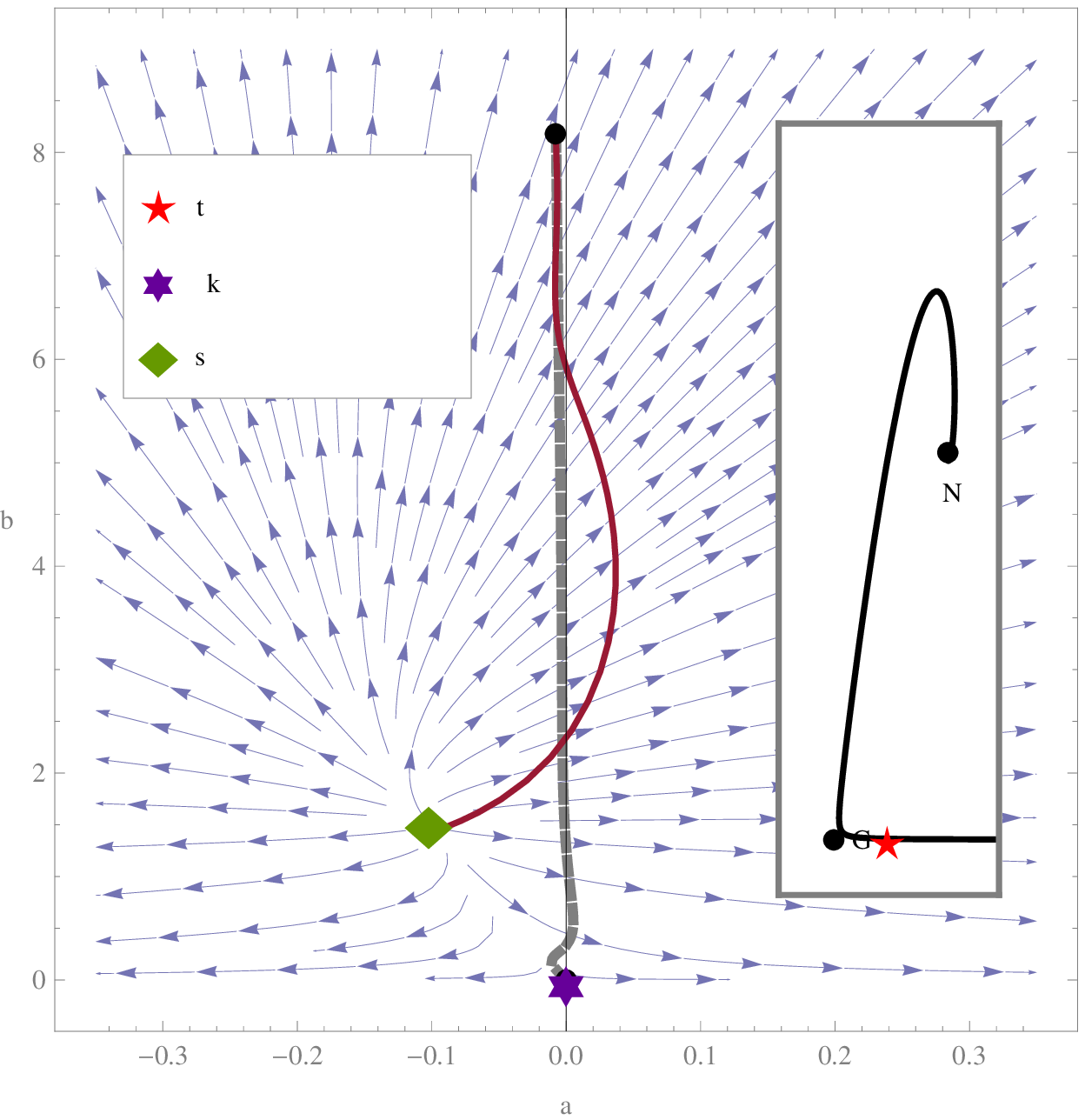}}
\hspace{0.05\textwidth}
 \subfloat{%
\label{fig:res4Dpp2T3a}\includegraphics[width=0.4\textwidth]{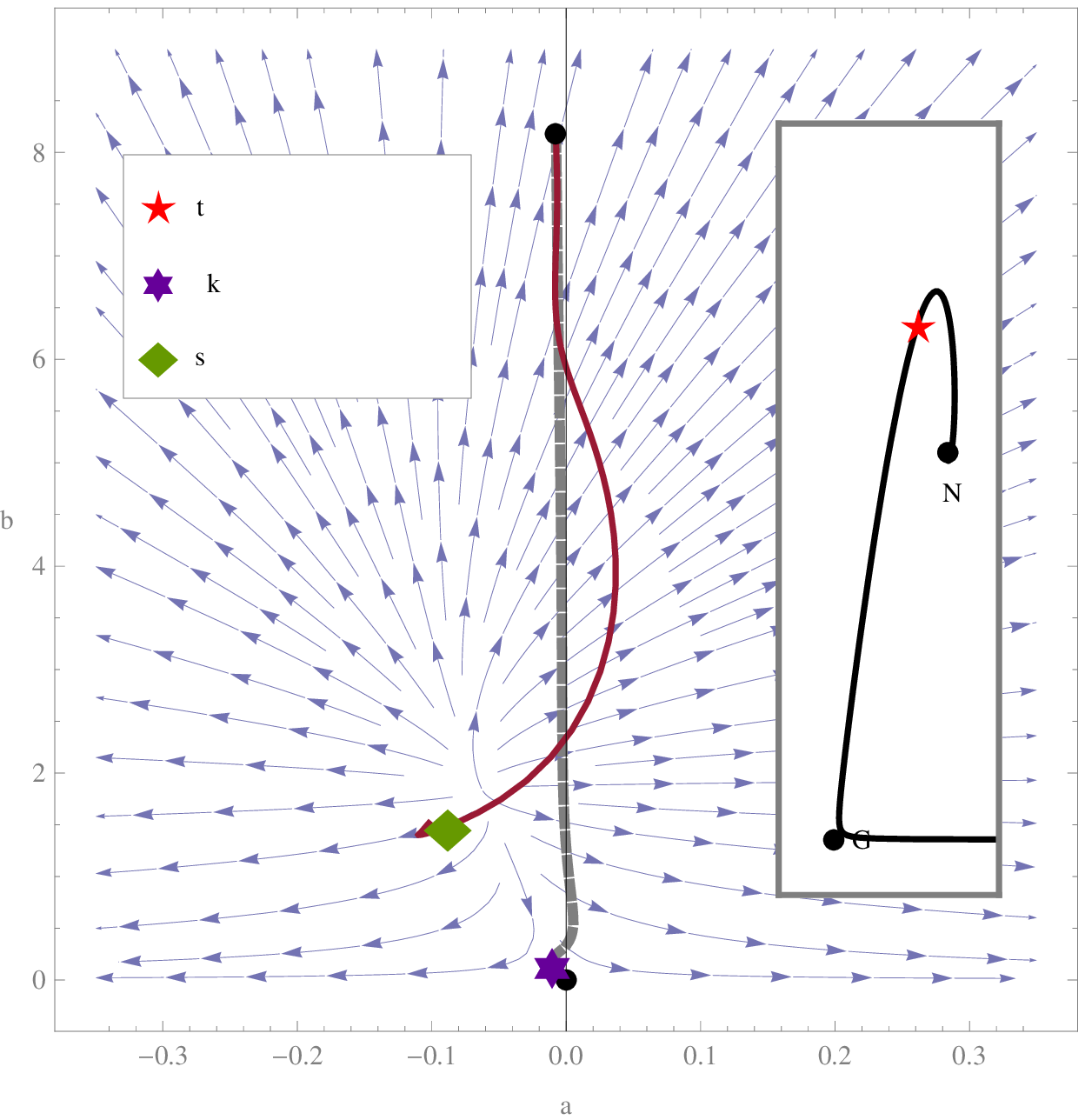}}
\hspace{0.05\textwidth}
 \subfloat{%
\label{fig:res4Dpp3T3a}\includegraphics[width=0.4\textwidth]{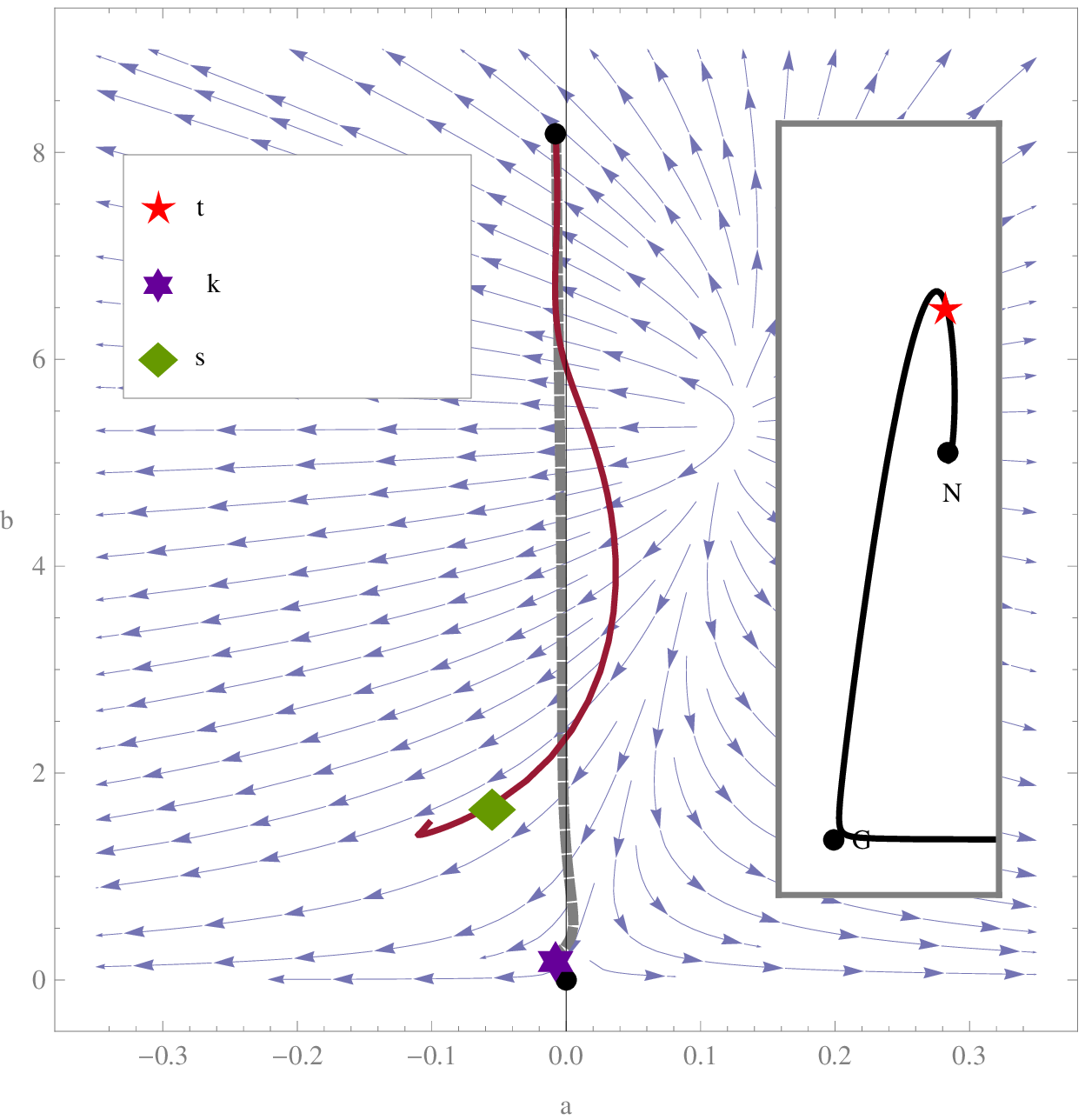}}
\hspace{0.05\textwidth}
 \subfloat{%
\label{fig:res4Dpp4T3a}\includegraphics[width=0.4\textwidth]{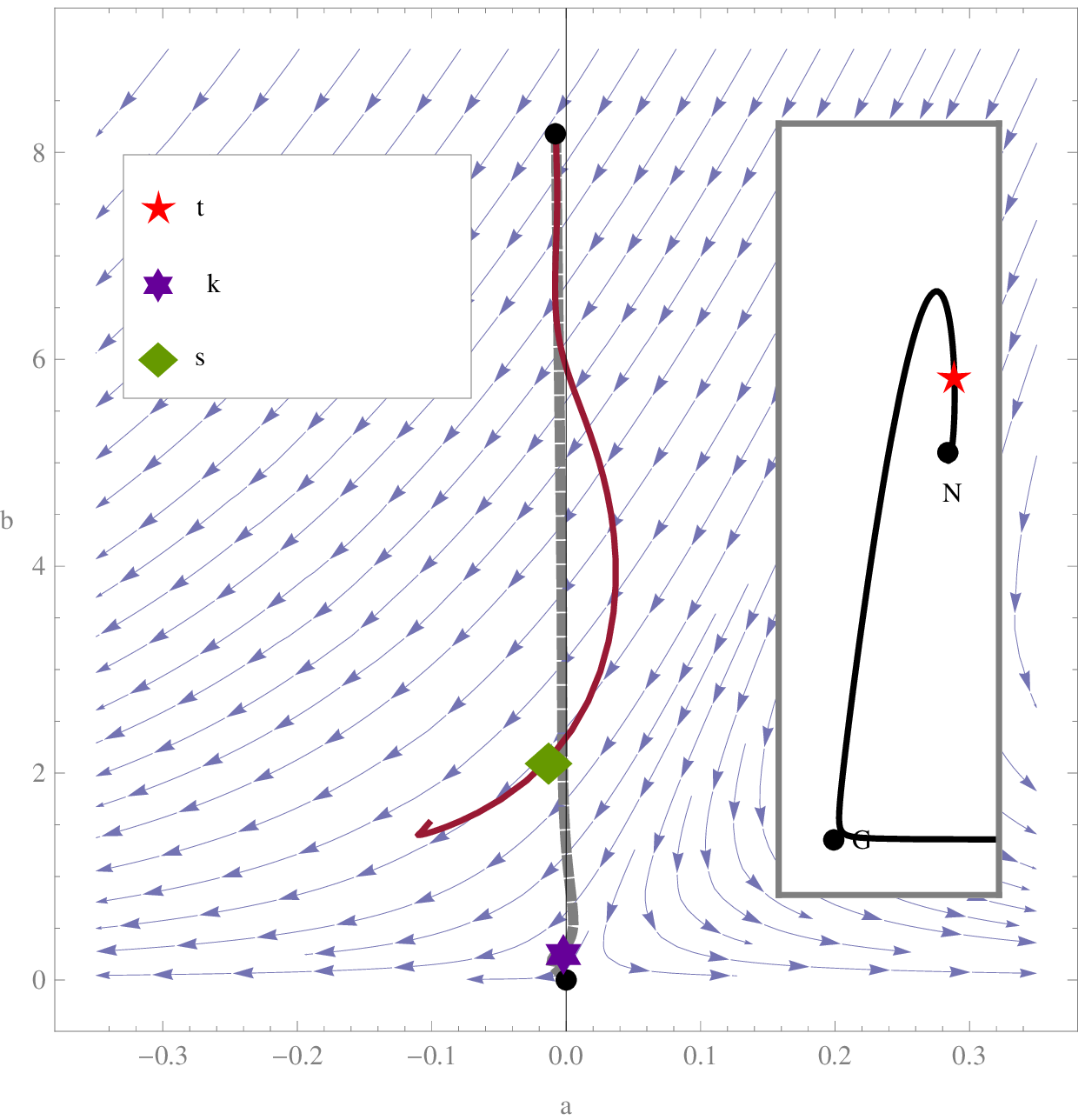}}
\hspace{0.05\textwidth}
 \subfloat{%
\label{fig:res4Dpp5T3a}\includegraphics[width=0.4\textwidth]{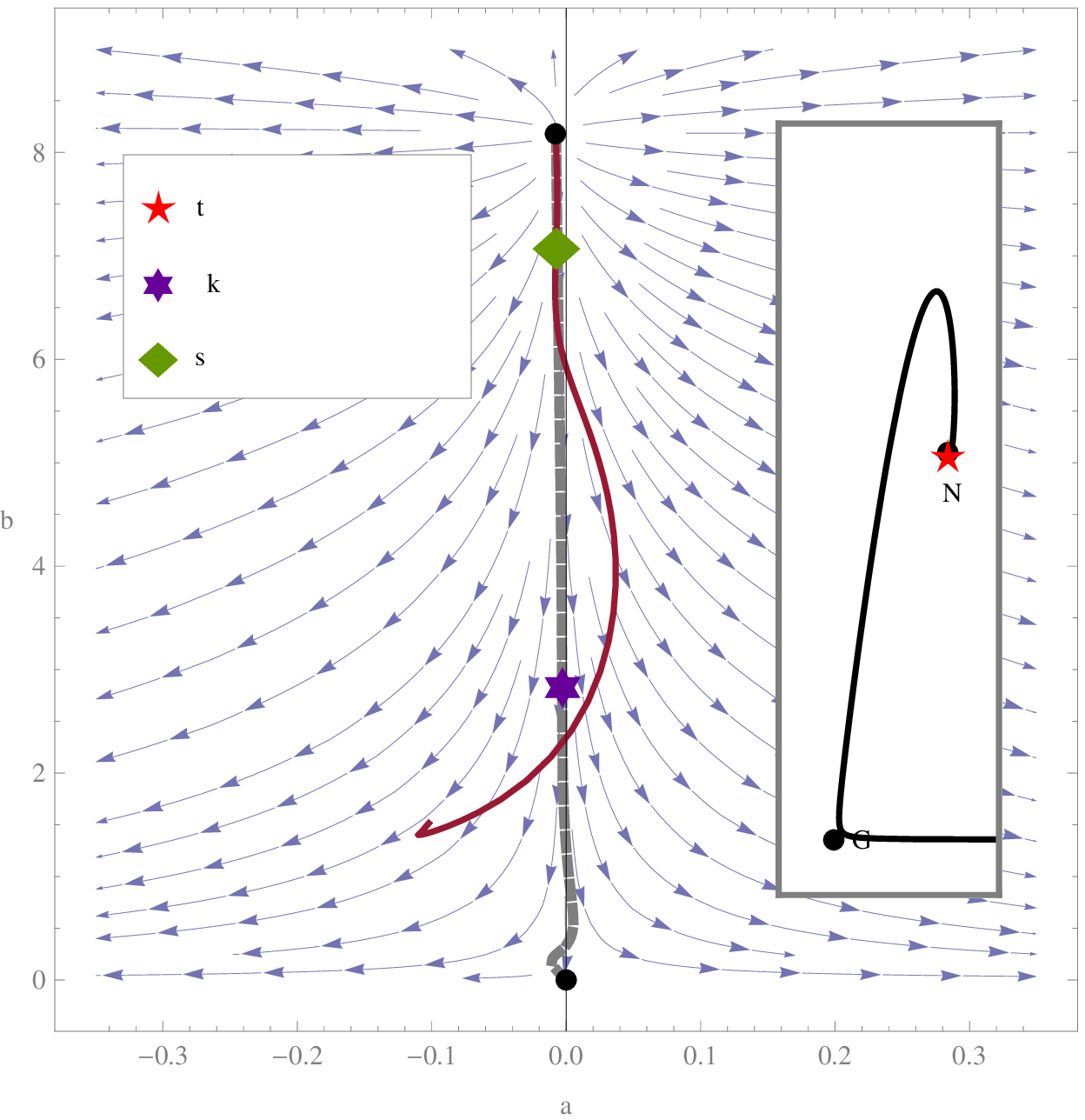}}
\hspace{0.05\textwidth}
 \subfloat{%
\label{fig:res4Dpp6T3a}\includegraphics[width=0.4\textwidth]{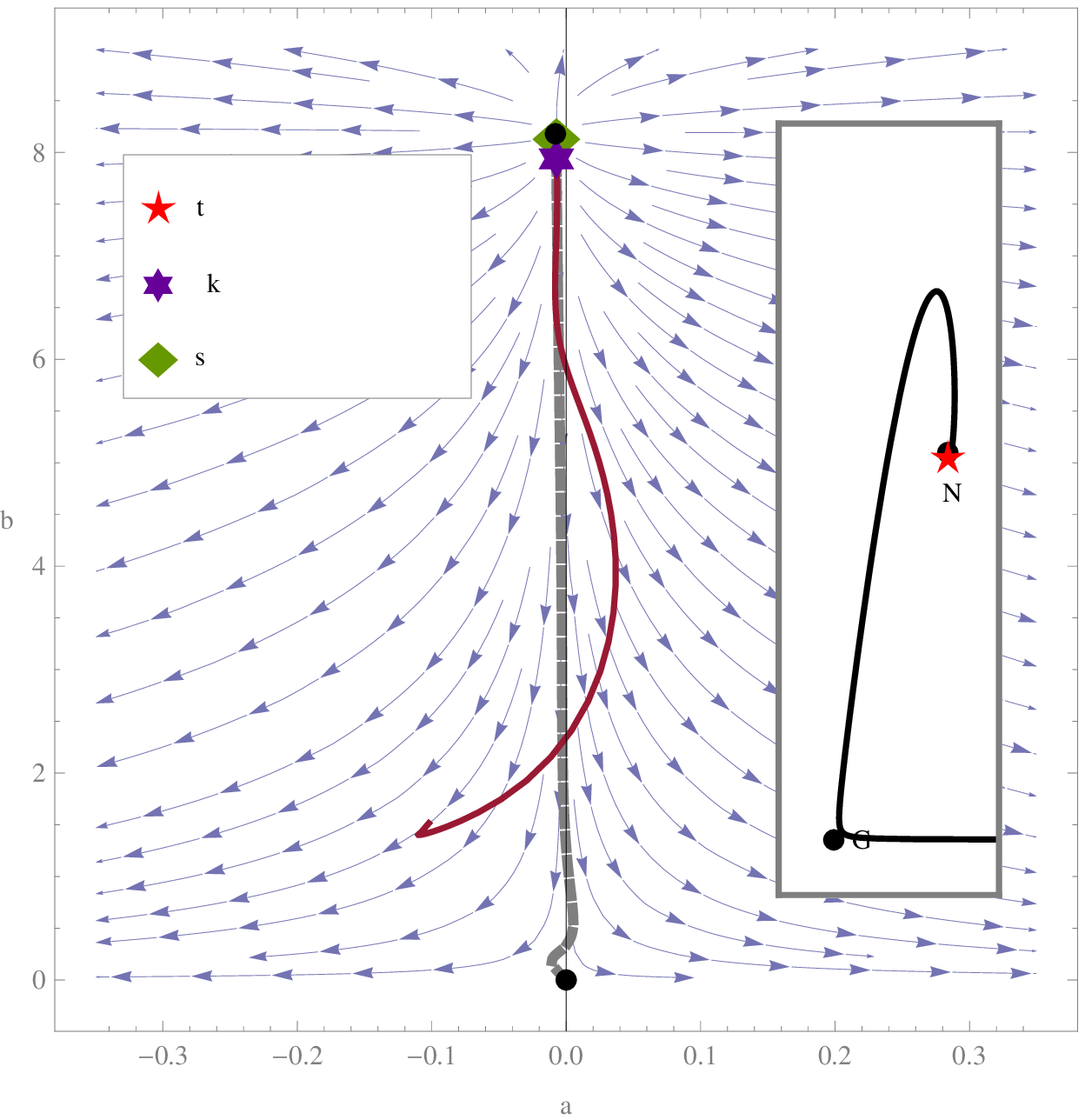}}
 \caption{The $\background$-phase portraits at increasing scales $k$. The underlying type (IIIa)$^{\dyn}$ trajectory in the $\dyn$-sector is shown in the inset on the right, and the current RG time is marked with a star therein. The arrows point towards the IR and picture the instantaneous vector field in the $\background$-sector.  The (red) solid and the (gray) dashed curve highlight two important solutions in the $\background$-sector, namely $\SolAttr(k)$ and $\text{Sol}_{(0,0)}^{\background}(k)$, respectively. Their current position is indicated by the (green) diamond and the (violet) six-pointed star, respectively.}\label{fig:res4DppT3a}
\end{figure}
\clearpage
}

Each one of the `snapshots' displayed in Figs. \ref{fig:res4DppT1a} - \ref{fig:res4DppT3a}  is structured as follows. 
The `current RG-time' can be inferred from the position of the five-pointed star on the underlying $\dyn$-trajectory; it is sketched inside the small  box on the right of the phase portrait.
The arrows in this larger diagram represent the 2-component vector field $\vec{\beta}_{\background}\equiv\big(\beta_{\tg}^{\background},\beta_{\Kk}^{\background}\big)$ on the $\tg^{\background}$-$\KkB$-plane at this particular instant of time. 
The instantaneous integral curves of this vector field are shown, too; because of its time dependence, those integral curves are no RG trajectories, however.

Furthermore, information about a single, especially interesting RG trajectory is provided by the time dependent location of the six-pointed star which can be seen in all snap shots. 
It  indicates the current position of the solution $\big(\tg_k^{\background},\,\KkB_k\big)$ of the RG equations which is fixed by the `final condition' $\lim_{k\rightarrow0} (\tg_k^{\background},\KkB_k)=(0,0)$. 
This solution is denoted Sol$^{\background}_{(0,0)}(k)$ in the diagrams.
The dashed curve, for clarity shown in the plots at any time, is the set of points visited by Sol$^{\background}_{(0,0)}(k)$ for $0\leq k < \infty$. It is a true RG trajectory, i.e. a solution to the eqs. \eqref{eqn:res4D_021}.

Similarly, we included in all phase portraits a (red) curve that shows another genuine RG trajectory, denoted $\SolAttr(k)$.
It is the solution picked by the `final condition' 
$\lim_{k\rightarrow0} \big(\tg_k^{\background},\,\KkB_k\big) = \lim_{k\rightarrow0} \big(\tg_{\attr}^{\background}(k),\, \KkB_{\attr}(k)\big)$.
In other words, under upward evolution it grows out of the UV attractor, with which it coincides at $k=0$.

As time elapses from $k=0$ to $k=\infty$, we expect $\SolAttr(k)$ and $\text{\bf $\AttrL$}(k)$ to follow different paths, since the former curve is an RG trajectory, while the latter is not.
This characteristic feature can be clearly observed in the diagrams.
The trajectory $\SolAttr(k)$ ultimately gets pulled into the doubly non-Gaussian fixed point for $k\rightarrow\infty$:
\begin{align*} 
\text{\bf $\AttrL$}(0)\oplus (\cdots)^{\dyn} \xleftarrow{k\rightarrow 0}   \SolAttr(k) \xrightarrow{k\rightarrow \infty} \text{\fpnB\fpC\fpnD-\fpL{}}
\end{align*}
Here $(\cdots)^{\dyn}$ stands for the various possible IR regimes of the underlying dynamical trajectory.
The RG-trajectory $\SolAttr(k)$  is an especially important one since {\it all trajectories  converge towards $\SolAttr(k)$  when $k$ is increased.} 

In the snapshots of Fig. \ref{fig:res4DppS}, obtained from  the $\dyn$-separatrix, this convergence is clearly seen to occur for the trajectory which, under upward evolution, begins in the doubly Gaussian fixed point \fpgB\fpC\fpgD-\fpL{} at $k=0$. 
This solution  coincides with the attractor's position already at a rather low RG-scale where $(\tg^{\background}_{\attr},\KkB_{\attr})$ has not yet moved much and  seems to be $k$-independent. 
Only after both trajectories have merged, $(\tg^{\background}_{\attr},\KkB_{\attr})$ actually starts running rapidly, and the remaining evolution towards $k\rightarrow\infty$ can be entirely described by the red curve that sits on top of the dashed one. 

In the plots the attraction towards $(\tg^{\background}_{\attr},\KkB_{\attr})$ is well visible since the running UV attractor is first heading for  infinity in the vertical direction, then returns, moves to $\KkD\approx0$, and lowers its $\tg^{\dyn}$ value until it ultimately reaches \fpnB-\fpL{}. This motion of the running attractor reflects itself in the bow of the trajectories.

In Fig. \ref{fig:res4DppT1a}, pertaining to a (Ia)$^{\dyn}$ trajectory, this feature is less pronounced: The $\SolAttr$-solution remains close to the Gaussian fixed point value for a long period of RG-time, and the confluence of both curves  takes place in the  far UV only.

Finally, Fig. \ref{fig:res4DppT3a} for the (IIIa)$^{\dyn}$ case looks very similar to the type (Ia)$^{\dyn}$ result and most of the significant running in the $\background$-sector  takes place in the UV only. In the IR we went down only to RG-scales which are such that $\KkD_k$ is still  well separated from the singular boundary of theory space at $\KkD=1\slash2$.

\begin{figure}[!ht]
\centering
\psfrag{g}[l]{$\tg^{\background}$}
\psfrag{l}{$\KkB$}      
\psfrag{a}[t]{$\scriptscriptstyle \AttrL(k=0)$}
\psfrag{b}[r]{$\scriptstyle \partial(\tg^{\background},\,\KkB)$}
\psfrag{n}[r]{$\scriptscriptstyle {\text{NGFP}}$}
\includegraphics[width=0.4\textwidth]{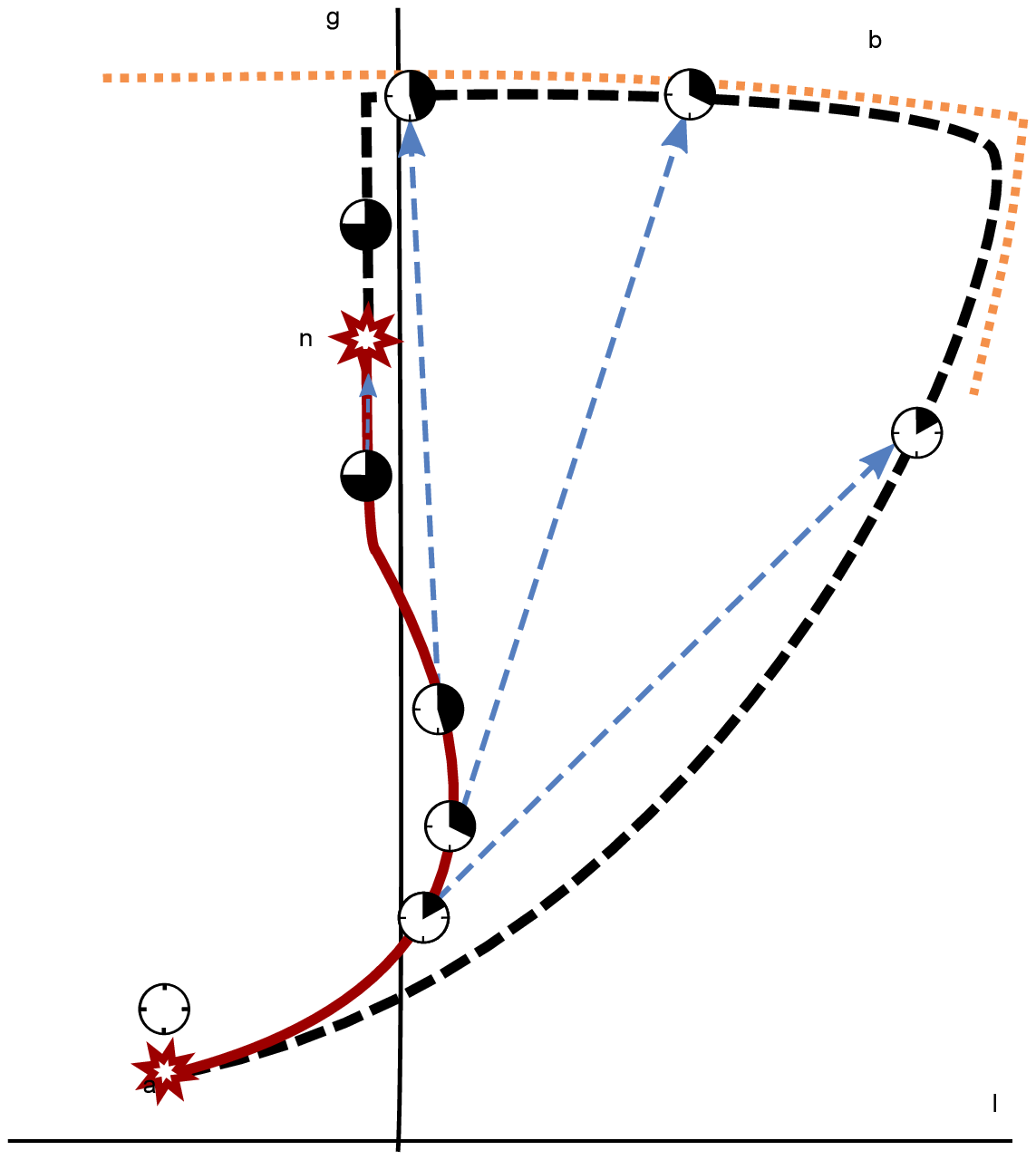}
\caption{The dashed (black) and solid (red) curve show the $k$-dependent positions of {\bf \AttrL}$(k)$ and $\SolAttr(k)$ on the $\tg^{\background}$-$\KkB$-plane. 
Recall that $\SolAttr(k)$ is an RG trajectory while {\bf \AttrL}$(k)$ is not.
The (orange) dotted curve indicates the boundary `$\partial(\tg^{\background},\,\KkB)$' where the $\background$-couplings diverge. 
The clocks mark equal-time  positions on the curves; the black filling indicates the elapsed RG time for upward evolution.
The dashed (blue) arrows indicate  the direction in which $\SolAttr(k^{\prime})$ is pulled at `time' $k^{\prime}$, namely the position of the attractor at this instant of time, {\bf \AttrL}$(k^{\prime})$.
While initially, at $k=0$, the trajectory $\SolAttr(0)$ coincides with {\bf \AttrL}$(0)$ the curves depart due to their different velocities. They meet again at  \fpnB\fpC\fpnD-\fpL{} for $k\rightarrow\infty$.
However, the motion of {\bf \AttrL}$(k)$ at intermediate scales is encoded in the indentation of the $\SolAttr(k)$ curve before it approaches  \fpnB\fpC\fpnD-\fpL{}.} \label{fig:runnAttr}
\end{figure}

\noindent {\bf (D) Summary of the attractor mechanism.}
In Fig. \ref{fig:runnAttr} we give a schematic description of the attractor mechanism which we uncovered in the `snapshots'. It is sufficient to focus on the (upward!) evolution of {\bf \AttrL}$(k)$ and its most interesting `follower' the RG trajectory $\SolAttr(k)$.
The dashed (black) and solid (red) curves describe the footprints of {\bf \AttrL}$(k)$ and $\SolAttr(k)$, respectively,  during their full evolution from $k=0$ to $k\rightarrow \infty$.
In the upper right part of the plot the (orange) dotted curve adumbrates the boundary of  $(\tg^{\background},\,\KkB)$-space where the $\background$-couplings diverge. 
The clocks mark equal-time positions on both curves; their black filling indicates the RG time elapsed  since they left their common initial point, {\bf \AttrL}($k=0$). 
The attraction of $\SolAttr(k^{\prime})$ to the current position of {\bf \AttrL}$(k^{\prime})$ is indicated by the dashed (blue) arrows. 

Notice that the two tracks meet only in the IR,  at{\bf \AttrL}$(0)$ (red star on the bottom), and at \fpnB\fpC\fpnD-\fpL{} in the UV. 
This is due to the following fact. 
While $\SolAttr(k^{\prime})$ is an RG trajectory whose velocity is determined by the beta-functions,  {\bf \AttrL}$(k)$ is a $k$-dependent solution to a `non-evolution' equation. 
The latter moves to the boundary of $(\tg^{\background},\,\KkB)$-space for some intermediate scales, but rapidly turns back to finite values of $\tg^{\background}$ and $\KkB$, and then  slowly approaches \fpnB\fpC\fpnD-\fpL{} from above.
On the other hand, the RG trajectories, in particular $\SolAttr(k)$, have a smaller velocity and thus only see the `taillamp' of  {\bf \AttrL}$(k)$ with which they try to catch up. 
The journey of {\bf \AttrL}$(k)$ to the boundary and back is reflected in the indentation of the $\SolAttr(k)$ curve before it approaches  \fpnB\fpC\fpnD-\fpL{}. 

\noindent {\bf Summary:}
We have seen that the RG-evolution in the $\background$-sector is crucially determined by the scale dependent UV attractor {\bf $\AttrL$}$(k)$. 
Whereas, in the UV, the RG-trajectories have no other choice but ultimately run into \fpnB\fpC\fpnD-\fpL{} for $k\rightarrow\infty$, they differ strongly in their IR behavior, in particular in the way they approach the physical point $k=0$. 
Following the trajectory backward, i.e. for increasing $k$, the dependence on their `initial' point $\big(\tg_{k_0}^{\dyn\slash \background},\,\Kk_{k_0}^{\dyn \slash \background}\big)$ reduces the more the closer $\big(\tg^{\background}_k,\KkB_k\big)$ gets to the running UV attractor $\big(\tg_{\attr}^{\background}(k),\KkB_{\attr}(k)\big)$.

\subsection[Split symmetry restoration in the physical limit\texorpdfstring{ $k\rightarrow0$}{}]{Split symmetry restoration in the physical limit $\bm{k\rightarrow0}$}\label{subsec:splitsym}
We will now search for RG trajectories which comply with  the requirement of split-symmetry restoration in the IR, i.e. when $k$ is lowered towards the physical point $k=0$.

Recall that split-symmetry is a property of idealized solutions of the FRGE where $\EAA_k[\flcb;\bg]$ reduces to a functional of a single field,  $g_{\mu\nu}=\bg_{\mu\nu}+\flcb_{\mu\nu}$. 
If fully intact in the truncation ansatz \eqref{eqn:trA07}, its $G_k^{(p)}$'s and $\Kkbar_k^{(p)}$'s are the same then at all levels, $p=0,1,2,\cdots$. 
In $\background$-$\dyn$ language, this is tantamount to saying that $\EAA^{\text{grav}}_k[g,\bg]$ looses its `extra $\bg$-dependence' so that, for $p=1,2,3,\cdots$:
\begin{subequations}
\begin{align}
\frac{1}{G_k^{(p)}}\equiv \frac{1}{G_k^{\dyn}}&\stackrel{!}{=}\frac{1}{G_k^{(0)}}\equiv\frac{1}{G_k^{\dyn}}+\frac{1}{G_k^{\background}} \quad \Leftrightarrow \quad \,\frac{1}{G_k^{\background}}\equiv\frac{k^2}{\tg_k^{\background}}\stackrel{!}{=}0
\label{eqn:res4D_019A} \\
\frac{\Kkbar^{(p)}}{G_k^{(p)}}\equiv\frac{\KkbarD}{G_k^{\dyn}}&\stackrel{!}{=}\frac{\Kkbar_k^{(0)}}{G_k^{(0)}}\equiv\frac{\KkbarD}{G_k^{\dyn}}+\frac{\Kkbar_k^{\background}}{G_k^{\background}} \quad \Leftrightarrow \quad \frac{\Kkbar_k^{\background}}{G_k^{\background}} \equiv k^4\,\frac{\KkB_k}{\tg_k^{\background}}\stackrel{!}{=}0
\label{eqn:res4D_019B}
\end{align}
\label{eqn:res4D_019}
\end{subequations}
Clearly we cannot expect those conditions to hold everywhere along a trajectory, at best, 
and only approximately, in a restricted regime of scales. 
After all, split-symmetry is broken explicitly both by the gauge fixing and the cutoff term. 
It is desirable, however, to base the construction of QEG on an asymptotically safe trajectory which reinstalls split-symmetry as exactly as possible%
\footnote{Which is not to say, {\it fully}. The gauge fixing dependent contents of $\EAA_k$ which never makes its way into observables may remain split-symmetry violating. Note that in the present truncation this `gauge fixing dependent contents' is the $\EAA^{\text{gf}}+\EAA^{\text{gh}}$ part of the EAA which never becomes split-symmetric, of course. Our discussion concerns only the $\EAA^{\text{grav}}$-part of the EAA ansatz.}
 for $k\rightarrow0$: 
In this limit the EAA approaches the standard effective action whose $n$-point functions, taken on-shell, are related to observable $S$-matrix elements.

By eqs. \eqref{eqn:res4D_019}, approximate split-symmetry demands $G_k^{\background}$ to be very `large', and $\KkbarB$ to be very `small', in an appropriate sense. Note that these are conditions on the $\background$-couplings only, the $\dyn$ ones are left unconstrained.

If the relations \eqref{eqn:res4D_019} indeed hold true for all $k$ in some interval $(k_1,k_2)$, the $k$-differentiated relations are satisfied, too. They require the beta-functions of the dimensionful $\background$-couplings to vanish; from \eqref{eqn:res4D_021dim}: 
\begin{subequations}
\begin{align}
&\partial_t\big(1\slash G_k^{\background}\big)=-k^2 \kB(\KkD_k,\tg_k^{\dyn})\stackrel{!}{=}0
 \label{eqn:res4D_020C} \\
 &\partial_t\big(\Kkbar_k^{\background}\slash G_k^{\background}\big)=k^4 \kA(\KkD_k,\tg_k^{\dyn})\stackrel{!}{=}0 \label{eqn:res4D_020D}
\end{align}
\label{eqn:res4D_020}
\end{subequations}
The conditions \eqref{eqn:res4D_020} guarantee the stability of \eqref{eqn:res4D_019} under the RG-evolution.
Notice that, by them, the running of the $\background$-couplings is not explicitly restricted,  rather they put constraints on the $\dyn$ quantities $\KkD_k$ and $\tg_k^{\dyn}$. This is just opposite as above. 

When we try to solve \eqref{eqn:res4D_020} by finding simultaneous zeros of $\kB\stackrel{!}{=}0$ and $\kA\stackrel{!}{=}0$ we find that (on the $\tg^{\dyn}>0$ half-plane) there exists {\it only one} such zero, namely the point $(\tg^{\dyn}_{\text{zero}},\, \KkD_{\text{zero}}) \approx (0.708,\,0.207)$. This shows clearly that we have to abandon the idea of finding a {\it full} trajectory that preserves split-symmetry, but rather look for RG-trajectories that  restore split-symmetry in the physical limit $k\rightarrow0$ at least, which is perfectly sufficient. 

Nevertheless, it is quite remarkable that the point $(\tg^{\dyn}_{\text{zero}},\, \KkD_{\text{zero}})$ is strikingly close to the \fpnD-\fpL{} fixed point which we located at $(\tg_*^{\dyn},\,\KkD_*)\approx(0.703,\,0.207)$. 
There is no obvious general reason for this `miracle' to happen.

\subsubsection[Split symmetric `final conditions' in the \texorpdfstring{$\background$}{B}-sector]{Split symmetric `final conditions' in the $\bm{\background}$-sector}
From now on we shall be modest and try to establish split-symmetry at $k=0$ only. 
In order to explore the implications of \eqref{eqn:res4D_019} for this case let us assume we have solved the  differential equations of the $\dyn$ sector, found all trajectories $k\mapsto \big(\tg_k^{\dyn},\,\KkD_k\big)$, and labeled them by their position $\big(\tg_{k_0}^{\dyn},\,\KkD_{k_0}\big)$ at some intermediate scale, $0<k_0<\infty$.
Then, inserting the $\dyn$-trajectories into the flow eqs. \eqref{eqn:res4D_021dimB} for $\tg_k^{\background}$ and $\KkB_k$,  
our task is  to identify those initial, or more appropriately, final conditions for the $\background$ couplings that lead to intact split-symmetry in the physical limit $k\rightarrow 0$.

Taking advantage of the explicit solution to the two $\background$-equations given in  \eqref{eqn:res4D_022} and \eqref{eqn:res4D_025} the requirement of split-symmetry, at some $k$ which is still arbitrary, assumes the form
\begin{subequations}
\begin{align}
 \frac{1}{G_k^{\background}}&=k^2 \frac{1}{\tg_k^{\background}}=
\frac{k^2_0}{\tg_{k_0}^{\background}} -  \int_{k_0}^k \,k^{\prime} \kB(k^{\prime})\md k^{\prime} \stackrel{!}{=} 0 	\label{eqn:res4D_046}
\end{align}
\begin{align}
\frac{\KkbarB}{G_k^{\background}}&=k^4 \frac{\KkB_k}{\tg_k^{\background}} = k_0^4 \frac{\KkB_{k_0}}{\tg_{k_0}^{\background}}  +  \int_{k_0}^{k}\md k^{\prime}\, k^{\prime\,3} \kA(k^{\prime})  \stackrel{!}{=} 0	\label{eqn:res4D_047}
\end{align}\label{eqn:res4D_047F}
\end{subequations}
We want split-symmetry to be intact at $k=0$, so we now let $k\rightarrow0$ in the conditions \eqref{eqn:res4D_047F}, while keeping $k_0$ strictly nonzero throughout.
Then these conditions {\it uniquely} fix  `initial' values $\big(G_{k_0}^{\background},\,\Kkbar^{\background}_{k_0}\big)$ for the dimensionful%
\footnote{Note that because of the explicit factors of $k^2$ and $k^4$, respectively, which relate dimensionless and dimensionful quantities in eqs. \eqref{eqn:res4D_047F} the implications of split-symmetry in the limit $k\rightarrow0$ are discussed most easily in {\it dimensionful} terms.} 
background couplings at $k=k_0$:
\begin{subequations}
\begin{align}
 G_{k_0}^{\background} &=\tg_{k_0}^{\background}\slash k_0^2=  -\big( \int^{k_0}_0 \,k^{\prime} \kB(k^{\prime})\md k^{\prime} \big)^{-1} \label{eqn:res4D_048A} \\
 \Kkbar^{\background}_{k_0}&=\KkB_{k_0}\,k_0^2= G_{k_0}^{\background} \int^{k_0}_0 \md k^{\prime }\, k^{3\,\prime} \,\kA(k^{\prime})\label{eqn:res4D_048B}
\end{align}\label{eqn:res4D_048}
\end{subequations}
Here $\kB(k)\equiv \kB(\tg_k^{\dyn},\,\KkD_k)$ and $\kA(k)\equiv \kA(\tg_k^{\dyn},\,\KkD_k)$ depend manifestly on the $\dyn$ trajectory under consideration.

This result is good news for the Asymptotic Safety program in a twofold way:
First, there does indeed exist a trajectory in the $\background$-sector which complies with the requirement of split-symmetry at $k=0$, but second, there is {\it only  one} such trajectory; as a consequence, when solving the RG equations for the $\background$-couplings there are no constants of integration that could be chosen freely, and this increases the predictivity of the theory.

Another remarkable feature of the initial conditions \eqref{eqn:res4D_048} is their close relationship to the running attractor $\text{\bf $\AttrL$}(k)$, to which we turn next.

\subsubsection[The \texorpdfstring{$k\rightarrow0$}{IR} asymptotics of the \texorpdfstring{$\dyn$}{Dyn} trajectories]{The $\bm{k\rightarrow0}$ asymptotics of the $\bm{\dyn}$ trajectories}
As a necessary preparation for  the exploration of the split-symmetry restoration of the full 4-dimensional system and to demonstrate the role played by the running attractor we first summarize the IR behavior of ($\tg_k^{\dyn},\KkD_k$) along  trajectories of the three types,  (Ia)$^{\dyn}$, (IIa)$^{\dyn}$, and (IIIa)$^{\dyn}$, respectively. 

\paragraph{Type (Ia)$^{\dyn}$:}
This type of trajectories has the defining property that $\KkD_k \xrightarrow{k\rightarrow 0} -\infty$. 
If we consider the $\dyn$-system for very large negative (positive) values of $\KkD_k$ we obtain the following asymptotic solutions (for $ \tg_{k_0}^{\dyn}>0$):
\begin{align}
 \tg_k^{\dyn}&=\frac{2\pi \tg_{k_0}^{\dyn} k^2  }{\tg_{k_0}^{\dyn} \left(k_0^2-k^2\right)+2\pi k_0^2  }\, \xrightarrow{k\rightarrow 0} 0 \\
\KkD_k &= \frac{ \tg_{k_0}^{\dyn} \left(k_0^4-k^4\right)-6\pi k_0^4  \KkD_{k_0}}{3 k^2 \left( \tg_{k_0}^{\dyn}\left(k^2-k_0^2\right)-2\pi k_0^2  \right)} \,\xrightarrow{k\rightarrow 0} \pm \infty  \label{eqn:res4D_032}
\end{align} 
Here the sign for the limit of $\KkD_0$ agrees with that of  $\KkD_{k_0}$ $\text{sign}\big(6\pi \KkD_{k_0}-\tg_{k_0}^{\dyn}\big)$. 
For type (Ia)$^{\dyn}$ trajectories the initial point $(\tg_{k_0}^{\dyn},\KkD_{k_0})$ is chosen such that the IR-limit of $\KkD_k$ is negative. (The positive sign in \eqref{eqn:res4D_032} applies to the type (IIIa)$^{\dyn}$  we will discuss below.) 
The corresponding {\it dimensionful}  $\dyn$-couplings have the following asymptotic behavior:
\begin{subequations}
\begin{align}
\frac{1}{ G_k^{\dyn}}&= \frac{\tg_{k_0}^{\dyn} \left(k_0^2-k^2\right)+2\pi k_0^2  }{2\pi \tg_{k_0}^{\dyn} }\, && \xrightarrow{k\rightarrow 0} \quad \frac{1}{G_{k_0}^{\dyn}}+\frac{k_0^2}{2\pi}   \label{eqn:res4D_033A}\\
\frac{\Kkbar_k^{\dyn}}{ G_k^{\dyn}}&= k_0^4 \, \big[\KkD_{k_0}\slash  \tg_{k_0}^{\dyn} - \tfrac{1}{6\pi} \big(1 -(k\slash k_0)^4\big)\big]\,&& \xrightarrow{k\rightarrow 0}\quad  \,\frac{\Kkbar^{\dyn}_{k_0}}{  G_{k_0}^{\dyn}} - \frac{k_0^4}{6\pi}\,  \label{eqn:res4D_033B}
\end{align} \label{eqn:res4D_033}
\end{subequations}
The IR-limit of $G^{\dyn}_k$ in eq. \eqref{eqn:res4D_033A} will later on be used to define the Planck mass for the class of type (Ia)$^{\dyn}$ trajectories.

\paragraph{Type (IIa)$^{\dyn}$:}
Along the  separatrix   the dimensionless cosmological and Newton's constant approach zero in the IR: $\KkD_k \xrightarrow{k\rightarrow 0}0$ and $\tg_k^{\dyn} \xrightarrow{k\rightarrow 0}0$. 
Linearizing around the  values $\tg^{\dyn}=0=\KkD$ we find 
\begin{subequations}
\begin{align}
 \tg_k^{\dyn}&=\tg_{k_0}^{\dyn} \, (k\slash k_0)^2\, \xrightarrow{k\rightarrow 0} 0 
&& 1\slash G_k^{\dyn}= \frac{k_0^2}{\tg_{k_0}^{\dyn} }\,  \xrightarrow{k\rightarrow 0} \quad \frac{k_0^2}{\tg_{k_0}^{\dyn}} \label{eqn:res4D_035A}\\
\KkD_k &= \frac{3}{8\pi} \tg_{k_0}^{\dyn} (k\slash k_0)^2  \,\xrightarrow{k\rightarrow 0} \pm 0 
&& \Kkbar_k^{\dyn}\slash G_k^{\dyn}= \frac{3}{8\pi}\,k^4 \,\, \xrightarrow{k\rightarrow 0}\quad 0 \label{eqn:res4D_035B}
\end{align} \label{eqn:res4D_035}
\end{subequations}
The initial values $\KkD_{k_0}$ and $\tg^{\dyn}_{k_0}$, imposed near the Gaussian fixed point, are constrained by the `separatrix condition' $\KkD_{k_0}= (3\slash 8\pi )\tg_{k_0}^{\dyn}$.

\paragraph{Type (IIIa)$^{\dyn}$:}
The type (IIIa)$^{\dyn}$ trajectories suffer from the well known problem that they run into a divergence of the beta-functions and terminate at some low but nonzero scale $k_{\text{term}}$ when $\KkD_k$ reaches the value $1\slash2$: 
Neither within the single- nor the bi-metric Einstein-Hilbert truncation their full extension to $k=0$ can be computed. Nevertheless, the situation is fairly clear for the trajectories of interest, namely those that before hitting the singularity enjoy a long classical regime \cite{h1-2,entropy} in which $G_k^{\dyn}\equiv G_{\text{cl.reg.}}^{\dyn}$ and $\KkbarD\equiv\Kkbart_{\text{cl.reg.}}^{\dyn}$ are approximately constant. Their dimensionless counterparts are then given by, for $k\gg k_{\text{term}}$ in the classical regime (`$\text{cl.reg.}$'),
\begin{subequations}
\begin{align}
 \tg_k^{\dyn}&=G_{\text{cl.reg.}}^{\dyn} k^2 \label{eqn:res4D_038A}\\
\KkD_k &= \Kkbart_{\text{cl.reg.}}^{\dyn}\slash k^2  \label{eqn:res4D_038B}
\end{align}  \label{eqn:res4D_038}
\end{subequations}
For the purposes of the present paper we hypothesize that in reality  the classical regime  even extends to scales $k\rightarrow 0$. 
The running \eqref{eqn:res4D_038} corresponds to that found in \eqref{eqn:res4D_032} then.

\subsubsection{The attractor mechanism of split-symmetry restoration}
Next we are going to insert the various types of asymptotic (for $k\rightarrow0$) solutions  $k\mapsto \big(\tg_k^{\dyn},\,\KkD_k\big)$ into $\kB(k)\equiv \kB(\tg_k^{\dyn},\,\KkD_k)$ and $\kA(k)\equiv \kA(\tg_k^{\dyn},\,\KkD_k)$.
Then we use the resulting functions in eqs. \eqref{eqn:res4D_048} for the initial values $\big(G_{k_0}^{\background},\,\Kkbar^{\background}_{k_0}\big)$ whereby we can perform the two $k^{\prime}$-integrals analytically. 
For the asymptotic formulae to be sufficiently good  approximations in the integrands we must choose $k_0$ very close to zero (more precisely, much smaller than the Planck scale, $k_0\ll\left(G_0^{\dyn}\right)^{-1\slash2}$, see below).
To convince yourself that the use of the asymptotic solutions is indeed permissible then, and to see what it entails, notice also the following facts:

\noindent{\bf (i)} Since $\tg_k^{\dyn}$ approaches zero in the IR for all three classes of $\dyn$ trajectories, contributions to $\kB$ and $\kA$ containing the anomalous dimension $\aDz\propto \tg^{\dyn}$ produce only terms of subleading order in the asymptotic expansion, which we may neglect. 
This implies in particular $\qA{p}{n}{-2\KkD_k}\approx \ThrfA{p}{n}{-2\KkD_k}$ for $\tg_k^{\dyn}\rightarrow 0$.

\noindent{\bf (ii)} For the large values of $\KkD_k$ which occur in  IR  of type (Ia)$^{\dyn}$ and (IIIa)$^{\dyn}$ trajectories we may exploit that $\lim_{\KkD_k\rightarrow \pm\infty}\ThrfA{p}{n}{-2\KkD_k}=0$ for $n\geq1$. 
Only the ghost terms contribute to $\kA$ and $\kB$. 
For the separatrix we instead use $\lim_{\KkD_k\rightarrow 0}\ThrfA{p}{n}{-2\KkD_k}=\ThrfA{p}{n}{0}$ in lowest order.

\noindent{\bf (iii)} The ghost contributions to both $\kB$ and $\kA$ are unaffected by any approximation in the $\dyn$ sector, simply because they do not depend on the $\dyn$ couplings at all and are thus $k$-independent and independent of the $\dyn$ initial conditions.

Going through the explicit formulae for $\kB$ and $\kA$ it is now easy to check that, as a consequence of these three simplifying properties, we are entitled to perform the integrals \eqref{eqn:res4D_048} in the far IR by substituting {\it constant} functions $\kB(k)\approx\kB(0)$ and $\kA(k)\approx\kA(0)$. 
In this manner we find that  the initial values of the $\background$ couplings that lead to a fully intact split-symmetry at $k=0$, are given by
\begin{subequations}
\begin{align}
 G_{k_0}^{\background} &=   \big( -2\slash \, \kB(0) \big) \slash k_0^2 = \tg^{\background}_{\attr}(0) \slash k_0^2 \label{eqn:res4D_049A} \\
 \Kkbar^{\background}_{k_0}&= \big(- \tfrac{1}{2}\kA(0)\slash  \kB(0) \big) k_0^2 = \Kk_{\attr}^{\background}(0)\,k_0^2\label{eqn:res4D_049B}
\end{align}\label{eqn:res4D_049}
\end{subequations}
This is an important result, and various comments are in order here:

\noindent{\bf(A)} By comparing the initial values \eqref{eqn:res4D_049} to the coordinates  \eqref{eqn:res4D_027} of $\AttrL(k)$ in the limit $\KkD\rightarrow\pm\infty$ or $\KkD\rightarrow 0$, respectively, we find that {\it the split-symmetry restoring initial point $(\tg_{k_0}^{\background},\KkB_{k_0})$ coincides exactly with the location of the running UV attractor $(\tg_{\attr}^{\background},\KkB_{\attr})$ for $k=0$.} 

\noindent{\bf(B)} In the first place, the result demonstrates that {\it there exists indeed a  fully extended RG trajectory, well behaved at all scales between zero and infinity.} 
It defines an asymptotically safe theory, hitting a NGFP for $k\rightarrow\infty$, and at the same time restores split-symmetry in the physical limit $k\rightarrow0$ when all fluctuations are integrated out. At least numerically, we can compute this trajectory for all $k\in[0,\infty)$. In this way we have verified  that the trajectory lies indeed on the UV critical hypersurface of the doubly non-Gaussian fixed point, \fpnB\fpC\fpnD-\fpL{}.

\noindent{\bf(C)} {\it There exists one,  and only one set of initial values in the $\background$-sector, $(\tg_{k_0}^{\background},\KkB_{k_0})$, that leads to split-symmetry at $k=0$.} (The uniqueness follows from the fact that the running UV attractor is IR repulsive in all directions.) This has the positive side effect that the number of free parameters that characterize the asymptotically safe quantum theories one can construct does not increase when we generalize the 2-parameter single-metric Einstein-Hilbert truncation to the 4-parameter bi-metric one.
In fact, the newly introduced parameters immediately get `eaten up' by the necessity to turn the split-symmetry violations to zero at  $k=0$.

\noindent{\bf(D)} For type (Ia)$^{\dyn}$ and (IIIa)$^{\dyn}$ trajectories the presence of the ghost contributions ($\rhoG=1$) is found to be essential for the existence of the symmetry restoring point. 
Making $\rhoG$ explicit, the corresponding IR solutions assume the form
\begin{subequations}
\begin{align}
  \frac{1}{G_k^{\background}} &\quad \xrightarrow{k\rightarrow 0} \quad  \frac{1}{G_{k_0}^{\background}}-\frac{5\rhoG}{3\pi}k_0^2  \label{eqn:res4D_034A}\\
\frac{\KkbarB}{G_k^{\background}} &\quad  \xrightarrow{k\rightarrow 0}  \quad \frac{\Kkbar_{k_0}}{G_{k_0}^{\background}}+\frac{2\rhoG}{3\pi}k_0^4  \label{eqn:res4D_034B}
\end{align}  \label{eqn:res4D_034}
\end{subequations}
From here  the split-symmetry restoring initial values $G_{k_0}^{\background}=3\pi\slash (5k_0^2\,\rhoG)$ and $\Kkbar^{\background}_{k_0}=-2\slash5$ can be deduced.
Omitting the ghost terms ($\rhoG=0$) split-symmetry restoration in the IR would require $G_{k_0}^{\background}\rightarrow \infty$ at a nonzero $k_0$! 
Note also that the corresponding dimensionless initial data, for $\rhoG=1$, are $(\tg_{k_0}^{\background},\,\KkB_{k_0})=(3\pi\slash5,\,-2\slash5)$ which, by \eqref{eqn:res4D_030}, is exactly the IR position of the attractor, $\AttrL(k\rightarrow0)$.

\noindent{\bf(E)} For the  separatrix  there are additional graviton contributions shifting the initial data towards smaller values:
\begin{subequations}
\begin{align}
  \frac{1}{G_k^{\background}}& \quad \xrightarrow{k\rightarrow 0} \quad \frac{1}{G_{k_0}^{\background}}-\frac{(7+20\rhoG)}{12\pi}\, k_0^2   \label{eqn:res4D_037A} \\
\frac{\KkbarB}{G_k^{\background}}& \quad \xrightarrow{k\rightarrow 0} \quad \frac{\Kkbar^{\background}_{k_0}}{G_{k_0}^{\background}}-\frac{(5-8\rhoG)}{12\pi}\, k_0^4   \label{eqn:res4D_037B}
\end{align}  \label{eqn:res4D_037}
\end{subequations}
`Switching on' the ghosts ($\rhoG=1$) the split-symmetry restoration happens at $\big(\tg_{k_0}^{\background},\, \KkB_{k_0}\big)=\big(4\pi\slash 9, \,-1\slash9\big)$. 
These are precisely the $\background$-coordinates of the fixed point \fpnB\fpC\fpgD-\fpL{}, which in turn equals  the $k\rightarrow0$ limit of $\AttrL(k)$ for the type (IIa)$^{\dyn}$ trajectory.

\noindent {\bf Summary:} 
For every RG trajectory of the $\dyn$ sector there exists precisely one associated trajectory of the $\background$ couplings which restores split-symmetry for $k\rightarrow0$. 
In the IR, the $\background$ trajectory approaches the (UV attractive, i.e., IR repulsive) running attractor $\AttrL(k)$.
For $k\rightarrow\infty$, the combined 4-dimensional trajectory is asymptotically safe and runs into the fixed point \fpnB\fpC\fpnD-\fpL{}.

\subsection{All classes of split-symmetry restoring trajectories}\label{subsec:classDiagr}
In this subsection we provide a survey of all classes of RG trajectories with restored split-symmetry in the IR, and we present explicit examples.  
In the $\dyn$-sector, we do have the freedom of choosing initial data $\tg^{\dyn}_{k_0}$ and $\KkD_{k_0}$ at some $k=k_0$, and so we will select a typical representative of each  type (Ia)$^{\dyn}$, (IIa)$^{\dyn}$, and (IIIa)$^{\dyn}$, respectively. 
Once such a solution is picked we have no further freedom: the values of  $\tg^{\background}_{k_0}$ and $\KkB_{k_0}$ are then uniquely determined by the requirement of split-symmetry at $k=0$, and so there is a unique `lift' of the 2-dimensional $\dyn$ trajectory to the 4-dimensional theory space.
The resulting trajectories will be referred to as of type (Ia)$^{\dyn}$-\AttrL, (IIa)$^{\dyn}$-\AttrL, and (IIIa)$^{\dyn}$-\AttrL, respectively.
Clearly the last type is the most interesting one since it is closest to real Nature, presumably.

In the following we shall always express the RG-scale $k$ and the dimensionful couplings in units of the Planck mass defined by the dynamical Newton constant: $m_{\text{Pl}}=[G^{\dyn}_{k_{\text{IR}}}]^{-1\slash2}$. 
For the infrared normalization scale $k_{\text{IR}}$ we choose $k_{\text{IR}}=0$ for the type (Ia)$^{\dyn}$ and (IIa)$^{\dyn}$ trajectories. In the (IIIa)$^{\dyn}$ case were we cannot follow the evolution down to $k=0$ we choose for $k_{\text{IR}}$ a value in the (semi-)classical regime where $G_k^{\dyn}$ and $\Kkbar_k^{\dyn}$ are approximately constant (lower horizontal branch of the trajectory).
In principle the notion of a  `Planck mass'  depends on the level of the Newton constant used to define it.  
In our case there is no ambiguity  since we enforce split-symmetry in the IR. Hence all $G^{(p)}_k$ give rise to the same Planck mass:
$ m_{\text{Pl}}= \lim_{k\rightarrow0} \big(1\slash \sqrt{G_k^{(p)}}\big) \equiv 1\slash \sqrt{G_0^{(0)}} \equiv 1\slash \sqrt{G_0^{\dyn}}$. 
This definition depends on the chosen trajectory, however.

\begin{figure}[!ht]
\centering
\psfrag{k}{${\scriptstyle   k \, \slash m_{\text{Pl}}}$}
\psfrag{l}[bc][bc][1][90]{$\scriptstyle    \Kk^{\cix}_k $}
\psfrag{m}[bc][bc][1][90]{$\scriptstyle   \tg^{\cix}_k\,\Kk^{\cix}_k  $}
\psfrag{L}[bc][bc][1][90]{$\scriptstyle    \Kkbar^{\cix}_k \, {\scriptscriptstyle [m_{\text{Pl}}]}$}
\psfrag{g}[bc][bc][1][90]{$\scriptstyle   \tg^{\cix}_k $}
\psfrag{G}[bc][bc][1][90]{$\scriptstyle   G^{\cix}_k  \, {\scriptscriptstyle [m_{\text{Pl}}]}$}
 \subfloat{%
\label{fig:res4Dpp1T1aAA}\includegraphics[width=0.4\textwidth]{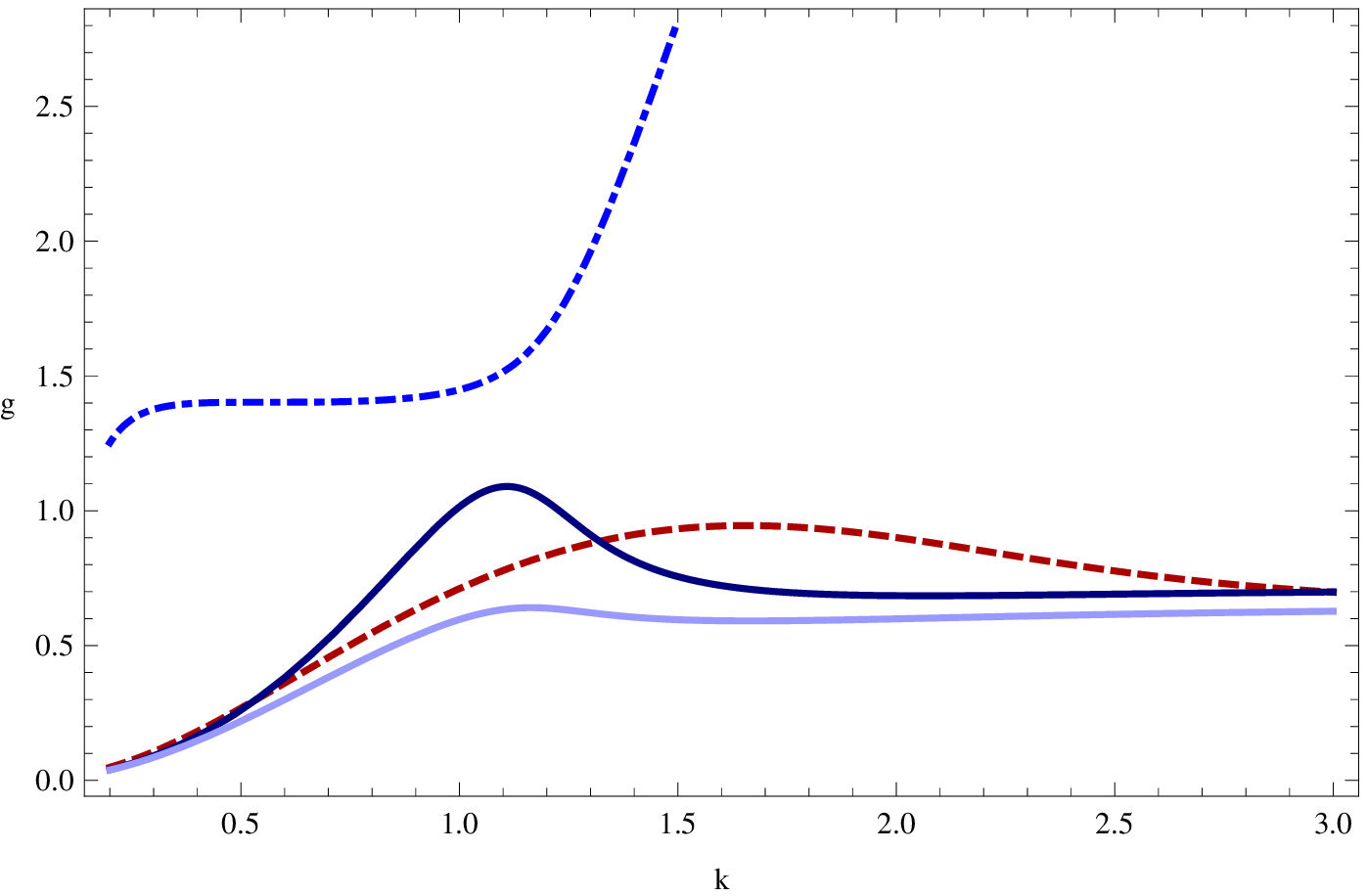}}
\hspace{0.05\textwidth}
 \subfloat{%
\label{fig:res4Dpp1T1aAB}\includegraphics[width=0.4\textwidth]{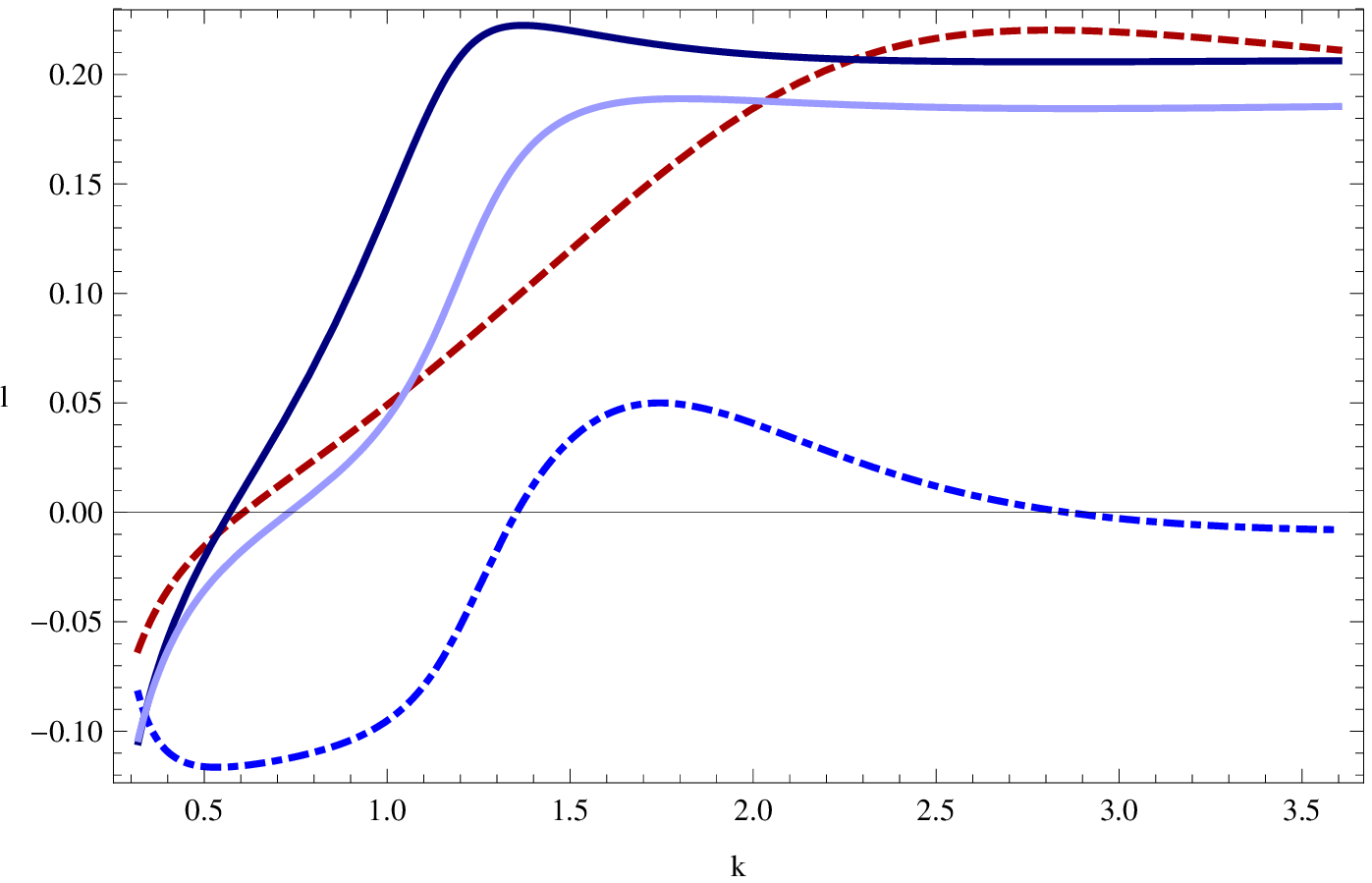}}
 \caption{Type (Ia)$^{\dyn}$-\AttrL trajectory: dimensionless couplings.}\label{fig:res4DcpT1aA}
\end{figure}

\begin{figure}[!ht]
\centering
\psfrag{k}{${\scriptstyle   k \, \slash m_{\text{Pl}}}$}
\psfrag{l}[bc][bc][1][90]{$\scriptstyle    \Kk^{\cix}_k $}
\psfrag{m}[bc][bc][1][90]{$\scriptstyle   \tg^{\cix}_k\,\Kk^{\cix}_k  $}
\psfrag{L}[bc][bc][1][90]{$\scriptstyle    \Kkbar^{\cix}_k \, {\scriptscriptstyle [m_{\text{Pl}}]}$}
\psfrag{g}[bc][bc][1][90]{$\scriptstyle   \tg^{\cix}_k $}
\psfrag{G}[bc][bc][1][90]{$\scriptstyle   G^{\cix}_k  \, {\scriptscriptstyle [m_{\text{Pl}}]}$}
 \subfloat{%
\label{fig:res4Dpp3T1aCA}\includegraphics[width=0.4\textwidth]{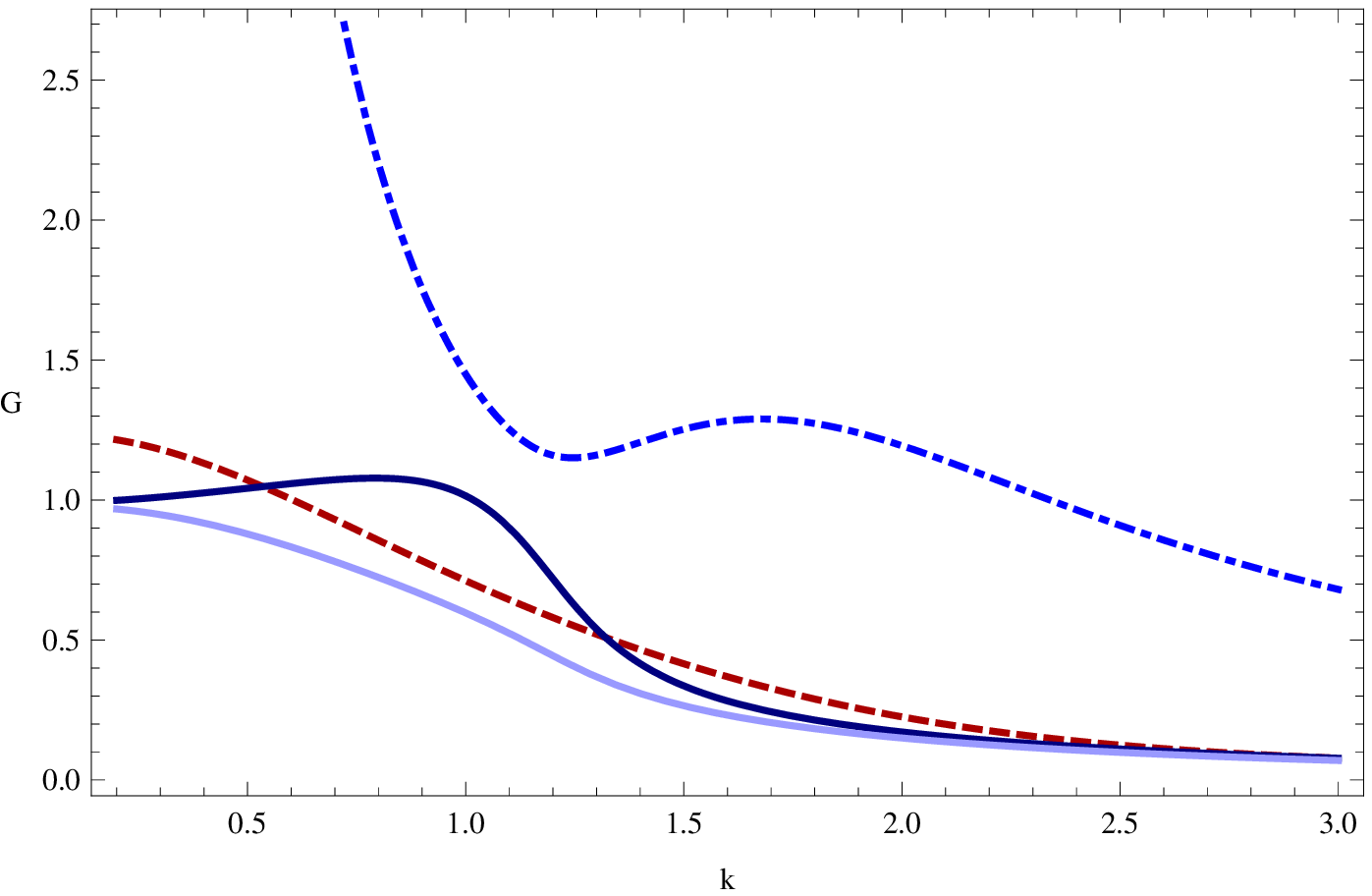}}
\hspace{0.05\textwidth}
 \subfloat{%
\label{fig:res4Dpp3T1aCB}\includegraphics[width=0.4\textwidth]{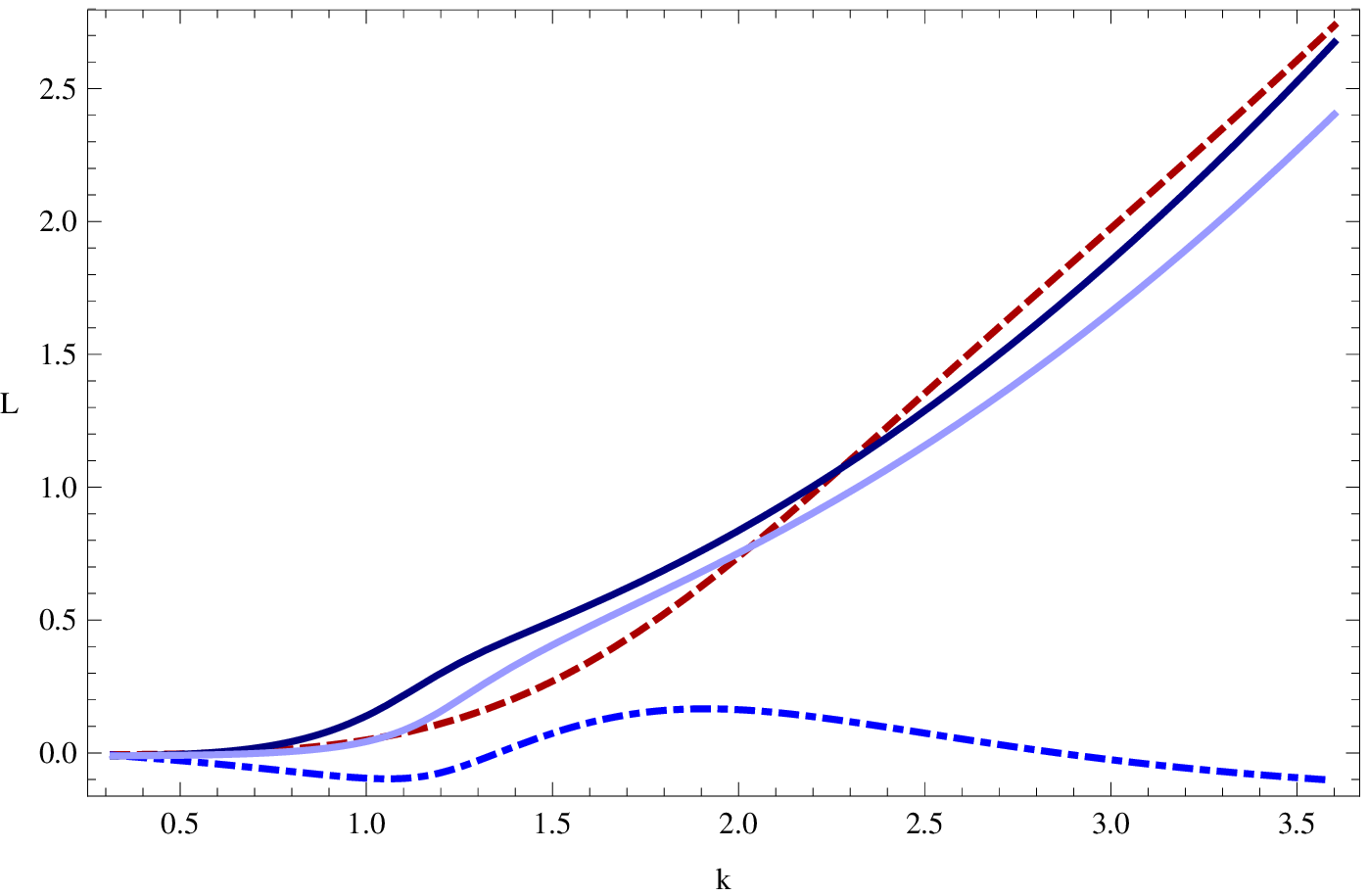}}
 \caption{Type (Ia)$^{\dyn}$-\AttrL trajectory: dimensionful couplings.}\label{fig:res4DcpT1aC}
\end{figure}

\begin{figure}[!ht]
\centering
\psfrag{k}{${\scriptstyle   k \, \slash m_{\text{Pl}}}$}
\psfrag{l}[bc][bc][1][90]{$\scriptstyle    \Kk^{\cix}_k $}
\psfrag{m}[bc][bc][1][90]{$\scriptstyle   \tg^{\cix}_k\,\Kk^{\cix}_k  $}
\psfrag{L}[bc][bc][1][90]{$\scriptstyle    \Kkbar^{\cix}_k\slash G^{\cix}_k \, {\scriptscriptstyle [m_{\text{Pl}}]}$}
\psfrag{g}[bc][bc][1][90]{$\scriptstyle   \tg^{\cix}_k $}
\psfrag{G}[bc][bc][1][90]{$\scriptstyle   1\slash G^{\cix}_k  \, {\scriptscriptstyle [m_{\text{Pl}}]}$}
 \subfloat{%
\label{fig:res4Dpp3T1aFA}\includegraphics[width=0.4\textwidth]{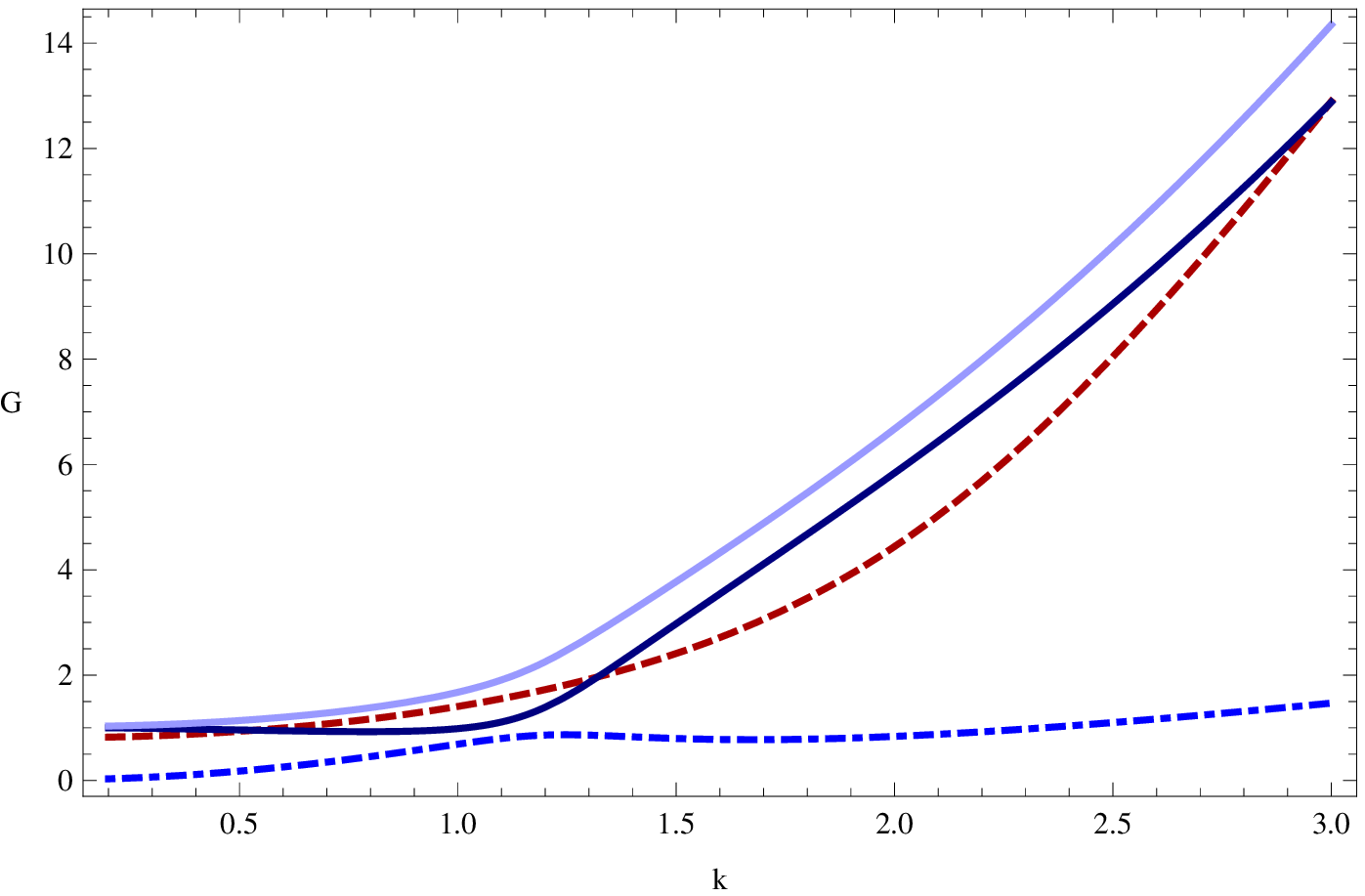}}
\hspace{0.05\textwidth}
 \subfloat{%
\label{fig:res4Dpp3T1aFB}\includegraphics[width=0.4\textwidth]{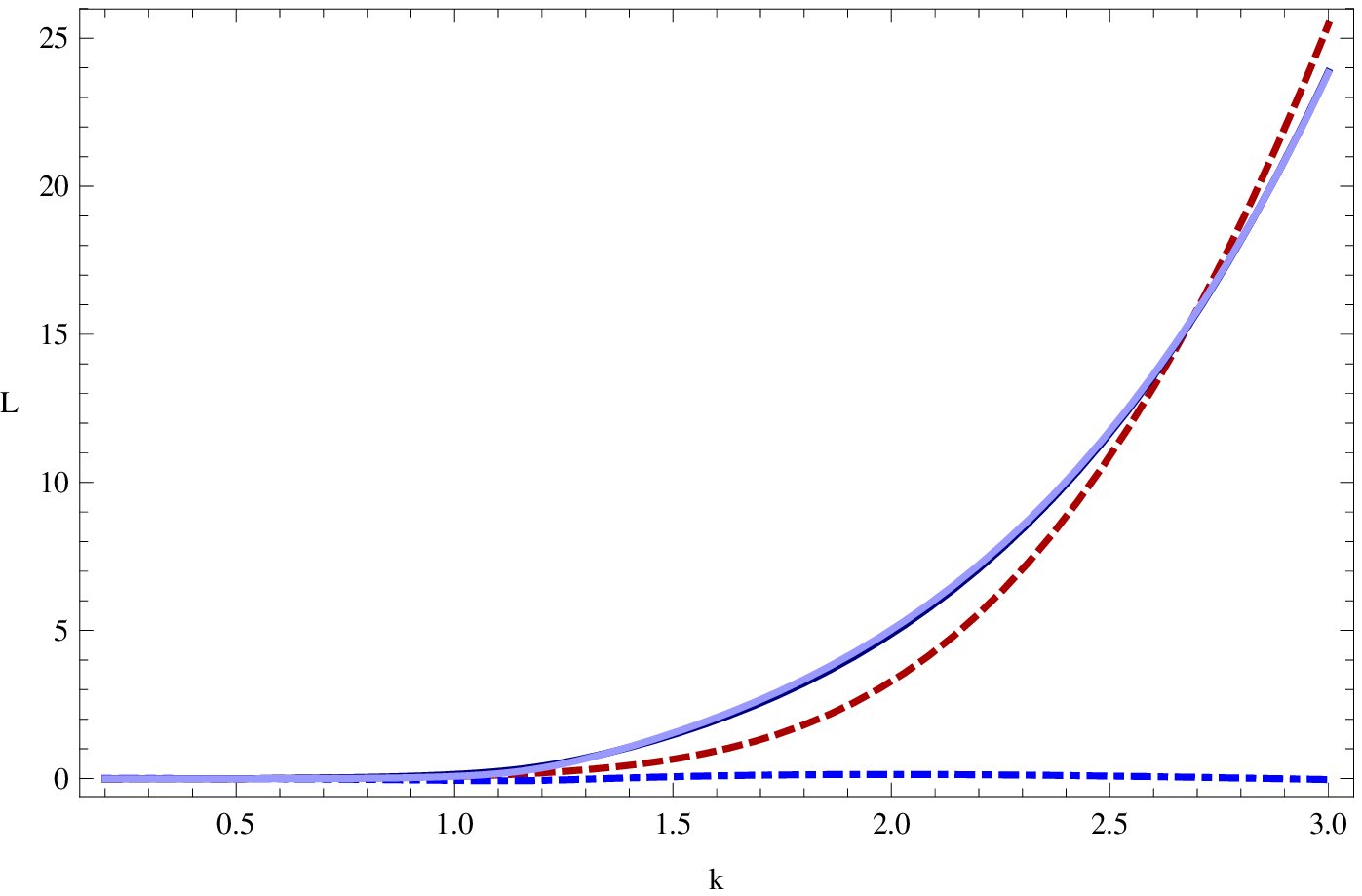}}
 \caption{Type (Ia)$^{\dyn}$-\AttrL trajectory: the coefficients as they appear in the EAA.
Note the perfect split-symmetry restoration in the IR: $1\slash G_k^{\background}$ and $\KkbarB\slash G_k^{\background}$ vanish for $k\rightarrow0$, implying that $\EAA_k^{\text{grav}}$ looses its extra $\bg_{\mu\nu}$ dependence.
}\label{fig:res4DcpT1aF}
\end{figure}

\begin{figure}[!ht]
\centering
\psfrag{k}{${\scriptstyle   k \, \slash m_{\text{Pl}}}$}
\psfrag{l}[bc][bc][1][90]{$\scriptstyle    \Kk^{\cix}_k $}
\psfrag{m}[bc][bc][1][90]{$\scriptstyle   \tg^{\cix}_k\,\Kk^{\cix}_k  $}
\psfrag{L}[bc][bc][1][90]{$\scriptstyle    \Kkbar^{\cix}_k \, {\scriptscriptstyle [m_{\text{Pl}}]}$}
\psfrag{g}[bc][bc][1][90]{$\scriptstyle   \tg^{\cix}_k $}
\psfrag{G}[bc][bc][1][90]{$\scriptstyle   G^{\cix}_k  \, {\scriptscriptstyle [m_{\text{Pl}}]}$}
 \subfloat{%
\label{fig:res4Dpp1T2aAA}\includegraphics[width=0.4\textwidth]{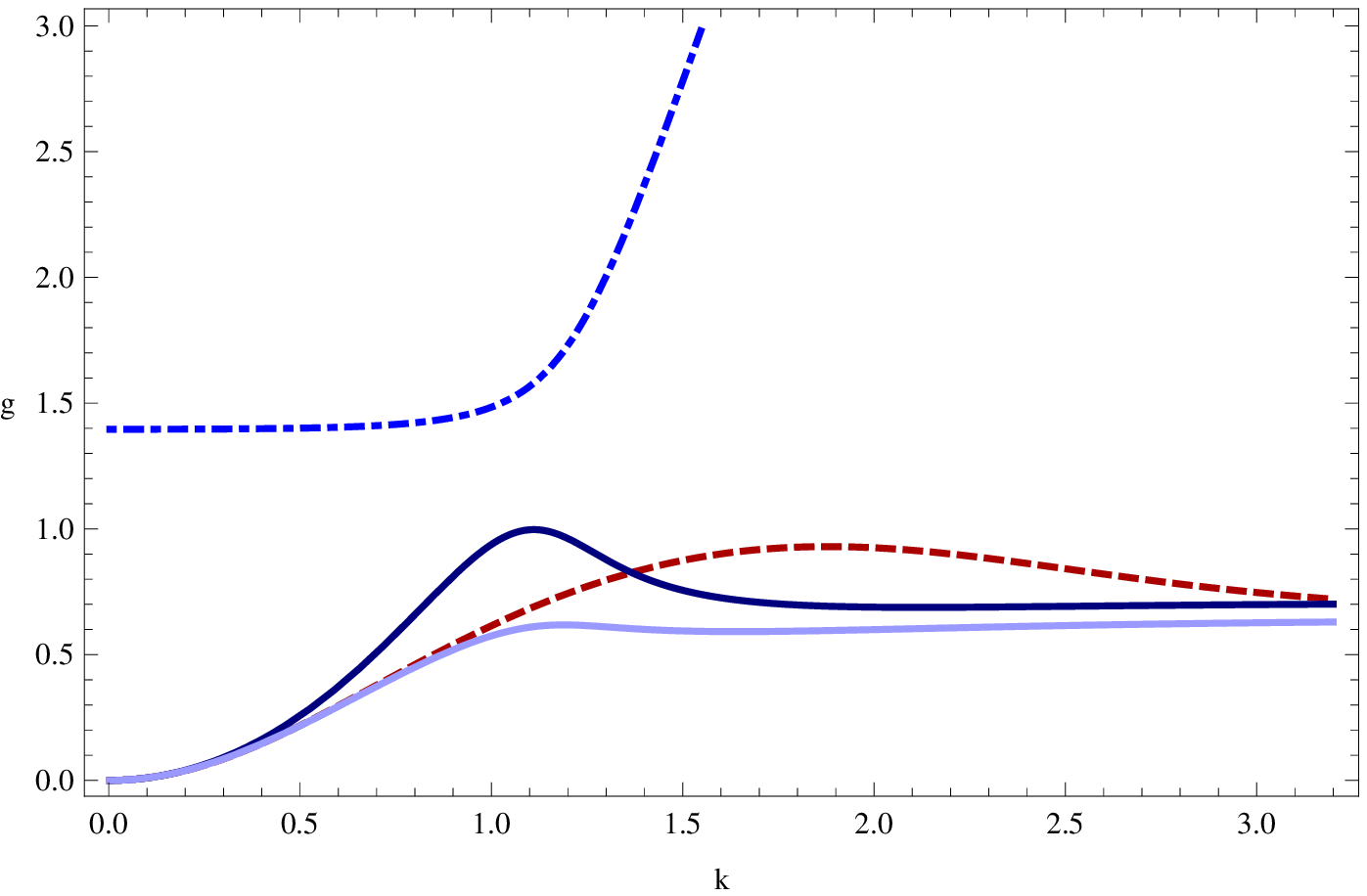}}
\hspace{0.05\textwidth}
 \subfloat{%
\label{fig:res4Dpp1T2aAB}\includegraphics[width=0.4\textwidth]{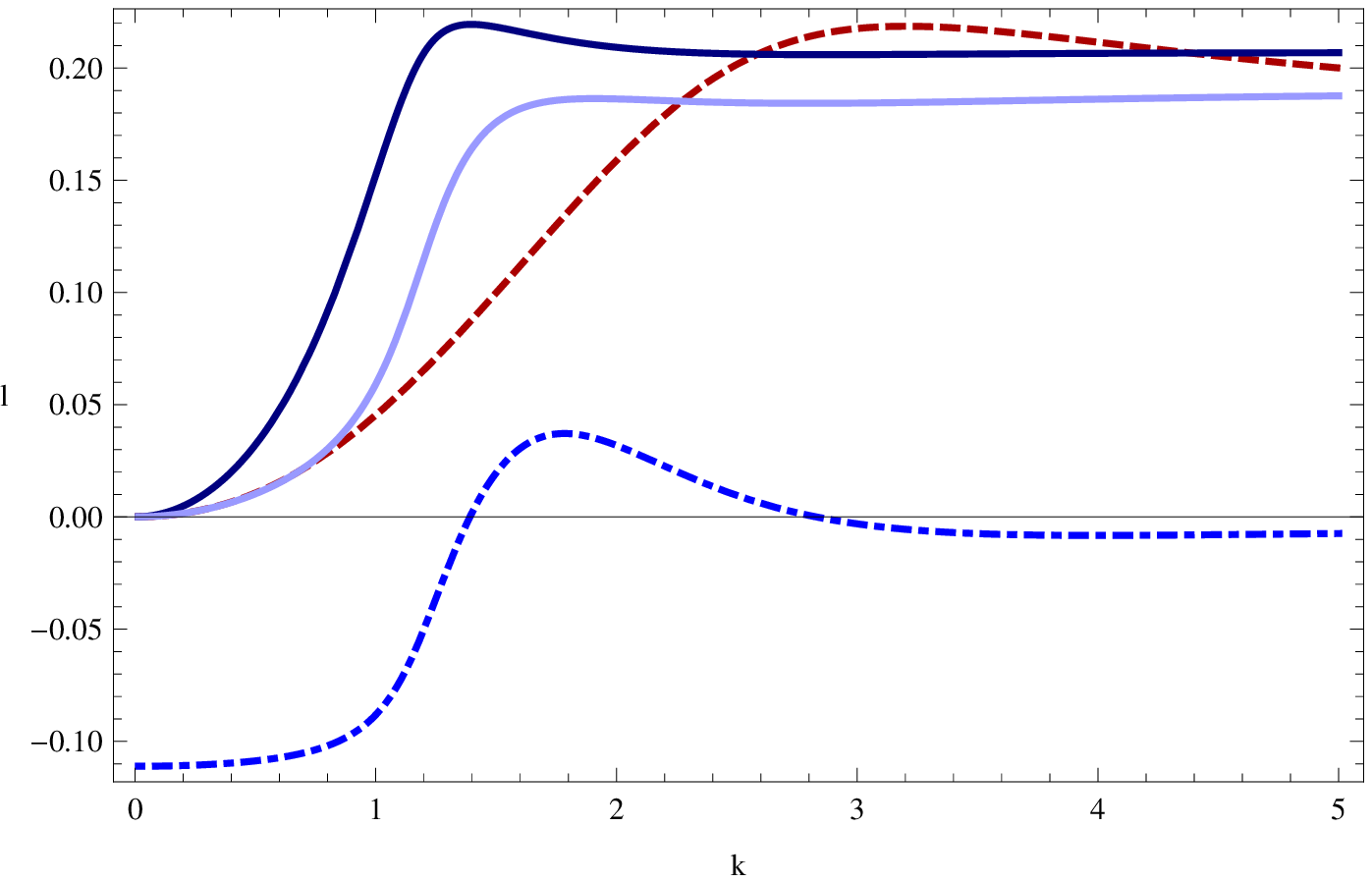}}
 \caption{Type (IIa)$^{\dyn}$-\AttrL trajectory: dimensionless couplings.}\label{fig:res4DcpT2aA}
\end{figure}

\begin{figure}[!ht]
\centering
\psfrag{k}{${\scriptstyle   k \, \slash m_{\text{Pl}}}$}
\psfrag{l}[bc][bc][1][90]{$\scriptstyle    \Kk^{\cix}_k $}
\psfrag{m}[bc][bc][1][90]{$\scriptstyle   \tg^{\cix}_k\,\Kk^{\cix}_k  $}
\psfrag{L}[bc][bc][1][90]{$\scriptstyle    \Kkbar^{\cix}_k \, {\scriptscriptstyle [m_{\text{Pl}}]}$}
\psfrag{g}[bc][bc][1][90]{$\scriptstyle   \tg^{\cix}_k $}
\psfrag{G}[bc][bc][1][90]{$\scriptstyle   G^{\cix}_k  \, {\scriptscriptstyle [m_{\text{Pl}}]}$}
 \subfloat{%
\label{fig:res4Dpp3T2aCA}\includegraphics[width=0.4\textwidth]{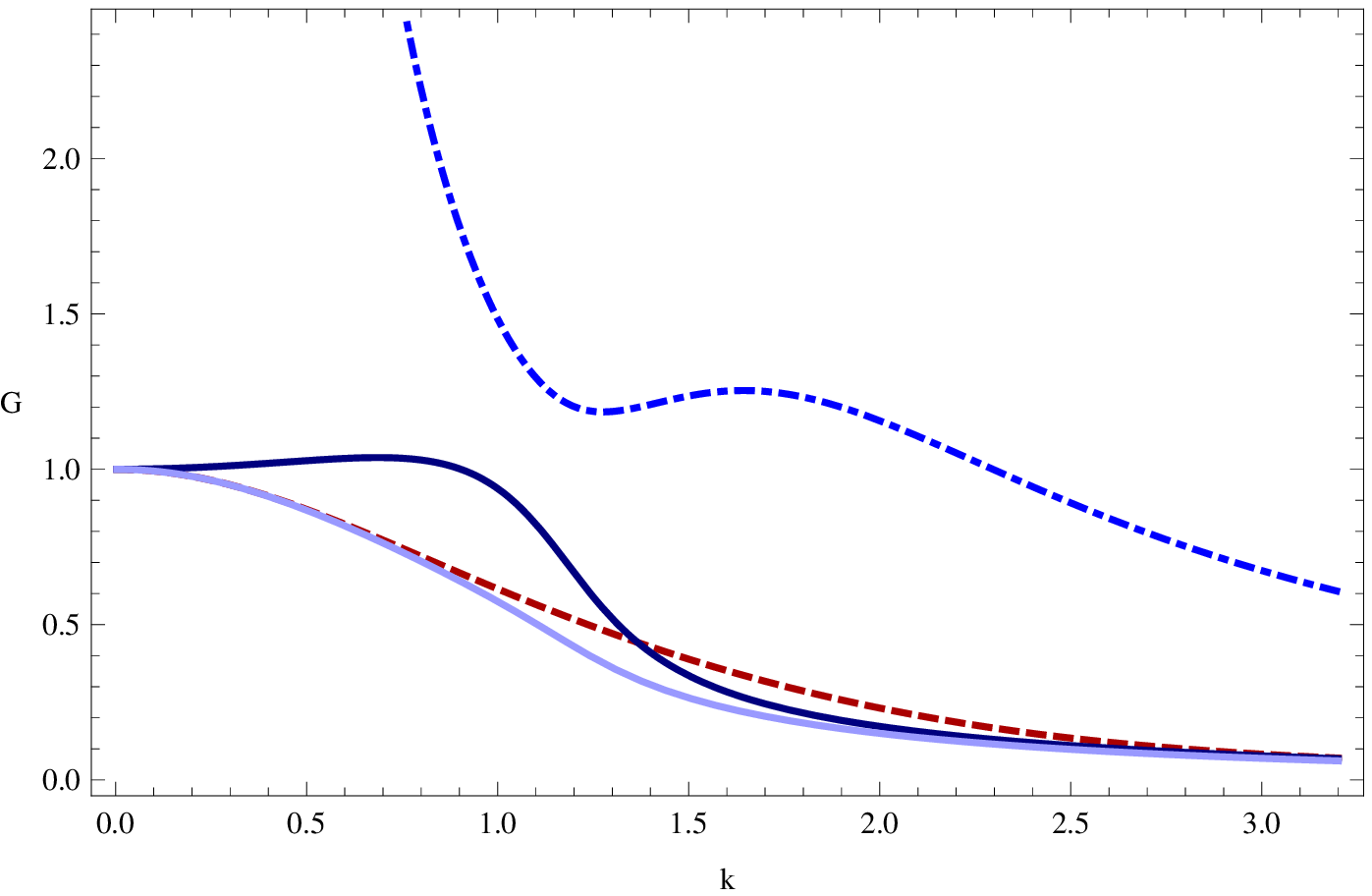}}
\hspace{0.05\textwidth}
 \subfloat{%
\label{fig:res4Dpp3T2aCB}\includegraphics[width=0.4\textwidth]{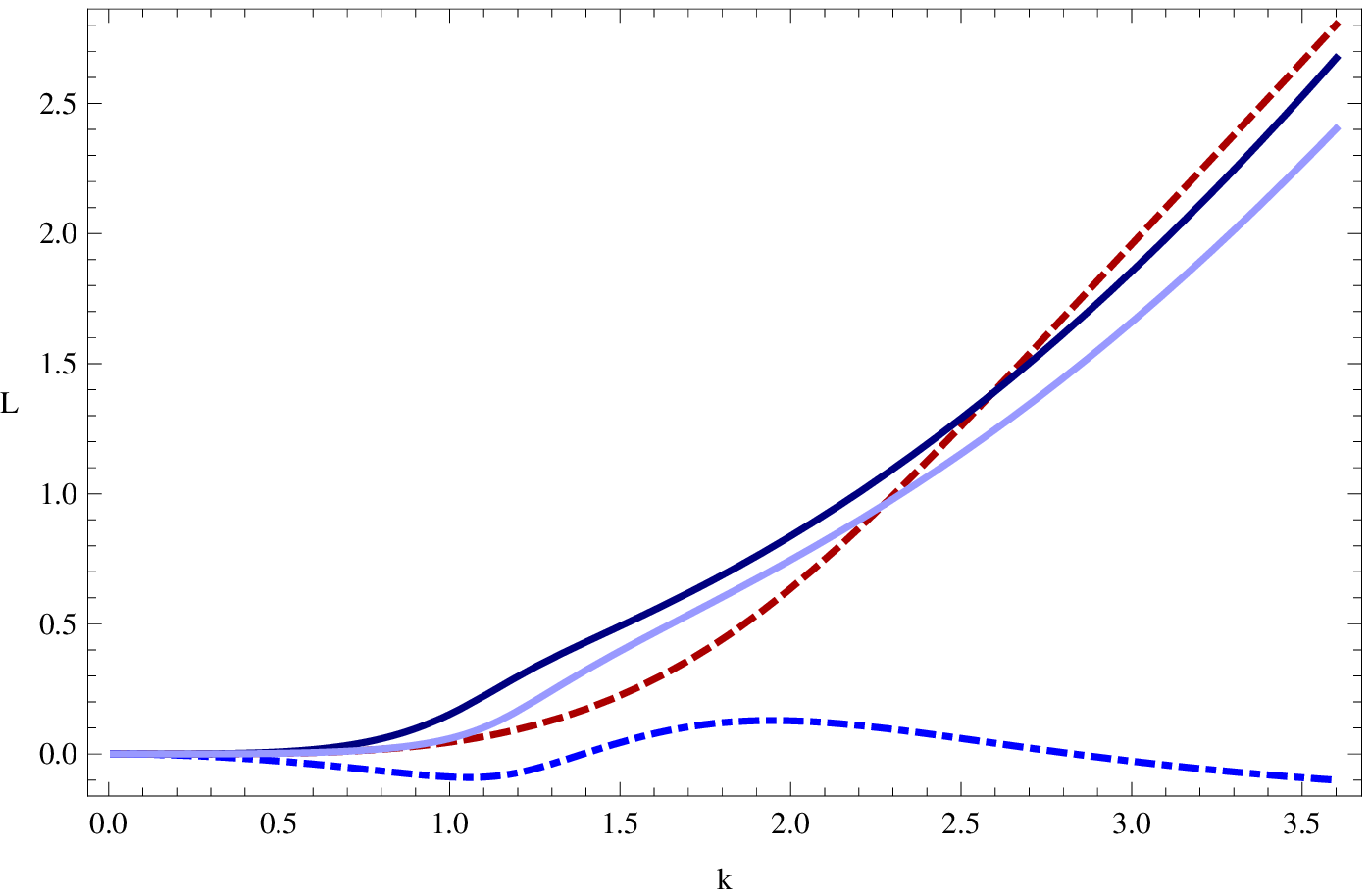}}
 \caption{Type (IIa)$^{\dyn}$-\AttrL trajectory: dimensionful couplings.}\label{fig:res4DcpT2aC}
\end{figure}

\begin{figure}[!ht]
\centering
\psfrag{k}{${\scriptstyle   k \, \slash m_{\text{Pl}}}$}
\psfrag{l}[bc][bc][1][90]{$\scriptstyle    \Kk^{\cix}_k $}
\psfrag{m}[bc][bc][1][90]{$\scriptstyle   \tg^{\cix}_k\,\Kk^{\cix}_k  $}
\psfrag{L}[bc][bc][1][90]{$\scriptstyle    \Kkbar^{\cix}_k\slash G^{\cix}_k \, {\scriptscriptstyle [m_{\text{Pl}}]}$}
\psfrag{g}[bc][bc][1][90]{$\scriptstyle   \tg^{\cix}_k $}
\psfrag{G}[bc][bc][1][90]{$\scriptstyle   1\slash G^{\cix}_k  \, {\scriptscriptstyle [m_{\text{Pl}}]}$}
 \subfloat{%
\label{fig:res4Dpp3T2aFA}\includegraphics[width=0.4\textwidth]{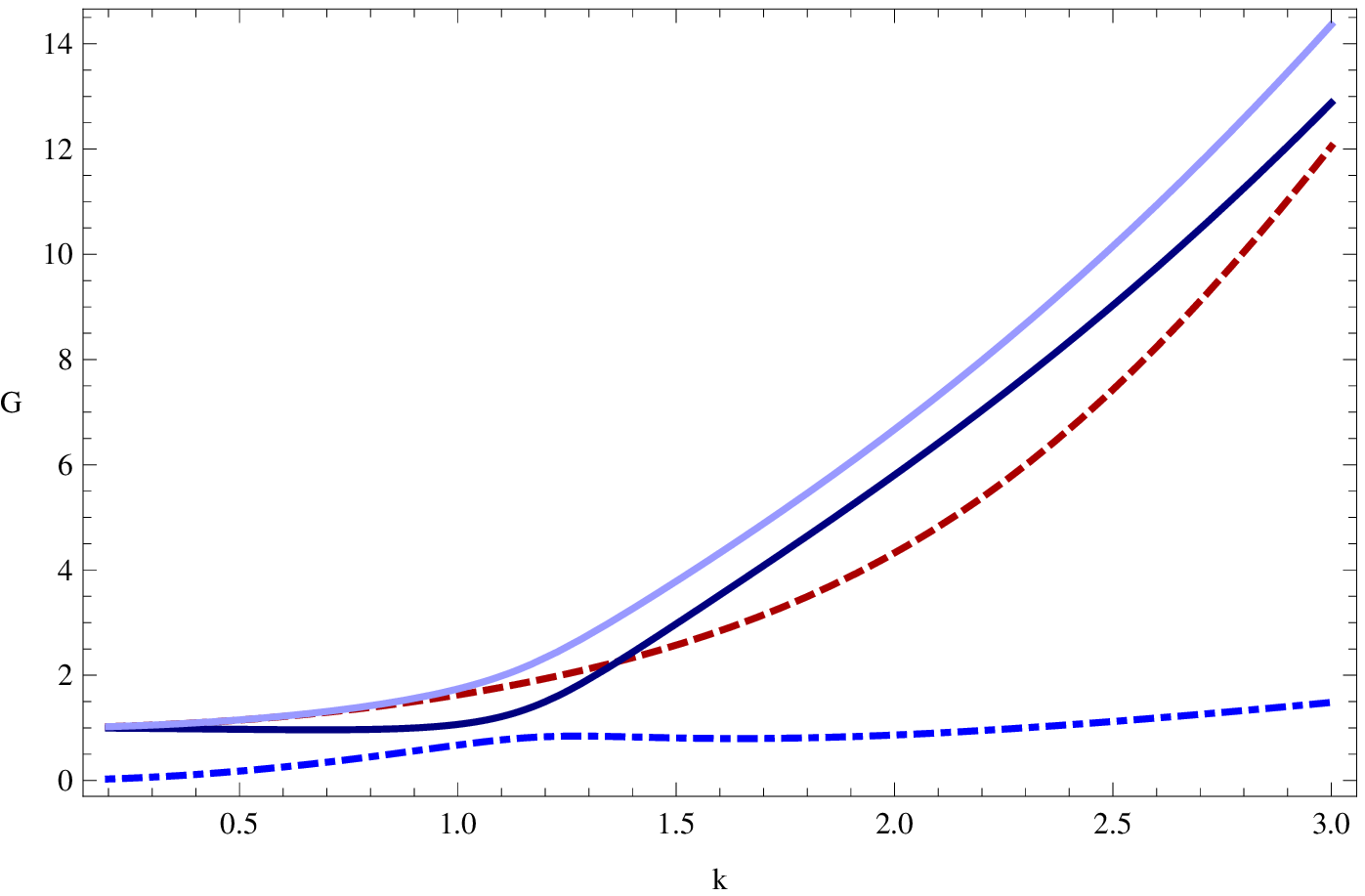}}
\hspace{0.05\textwidth}
 \subfloat{%
\label{fig:res4Dpp3T2aFB}\includegraphics[width=0.4\textwidth]{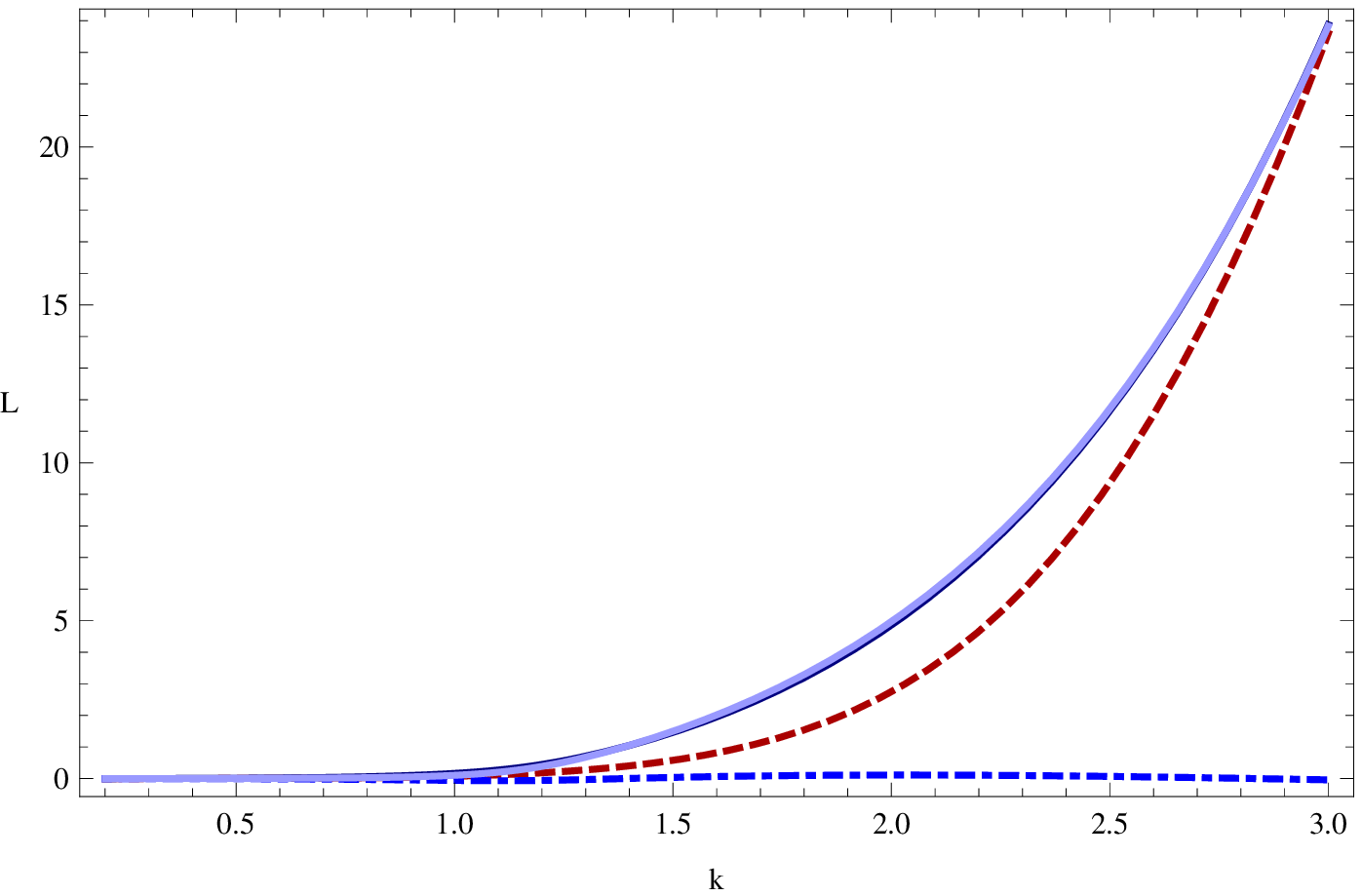}}
 \caption{Type (IIa)$^{\dyn}$-\AttrL trajectory: the coefficients as they appear in the EAA. 
Note the split-symmetry restoration for $k\rightarrow0$.}\label{fig:res4DcpT2aF}
\end{figure}

\begin{figure}[!ht]
\centering
\psfrag{k}{${\scriptstyle   k \, \slash m_{\text{Pl}}}$}
\psfrag{l}[bc][bc][1][90]{$\scriptstyle    \Kk^{\cix}_k $}
\psfrag{m}[bc][bc][1][90]{$\scriptstyle   \tg^{\cix}_k\,\Kk^{\cix}_k  $}
\psfrag{L}[bc][bc][1][90]{$\scriptstyle    \Kkbar^{\cix}_k \, {\scriptscriptstyle [m_{\text{Pl}}]}$}
\psfrag{g}[bc][bc][1][90]{$\scriptstyle   \tg^{\cix}_k $}
\psfrag{G}[bc][bc][1][90]{$\scriptstyle   G^{\cix}_k  \, {\scriptscriptstyle [m_{\text{Pl}}]}$}

 \subfloat{%

\label{fig:res4Dpp1T3aAA}\includegraphics[width=0.4\textwidth]{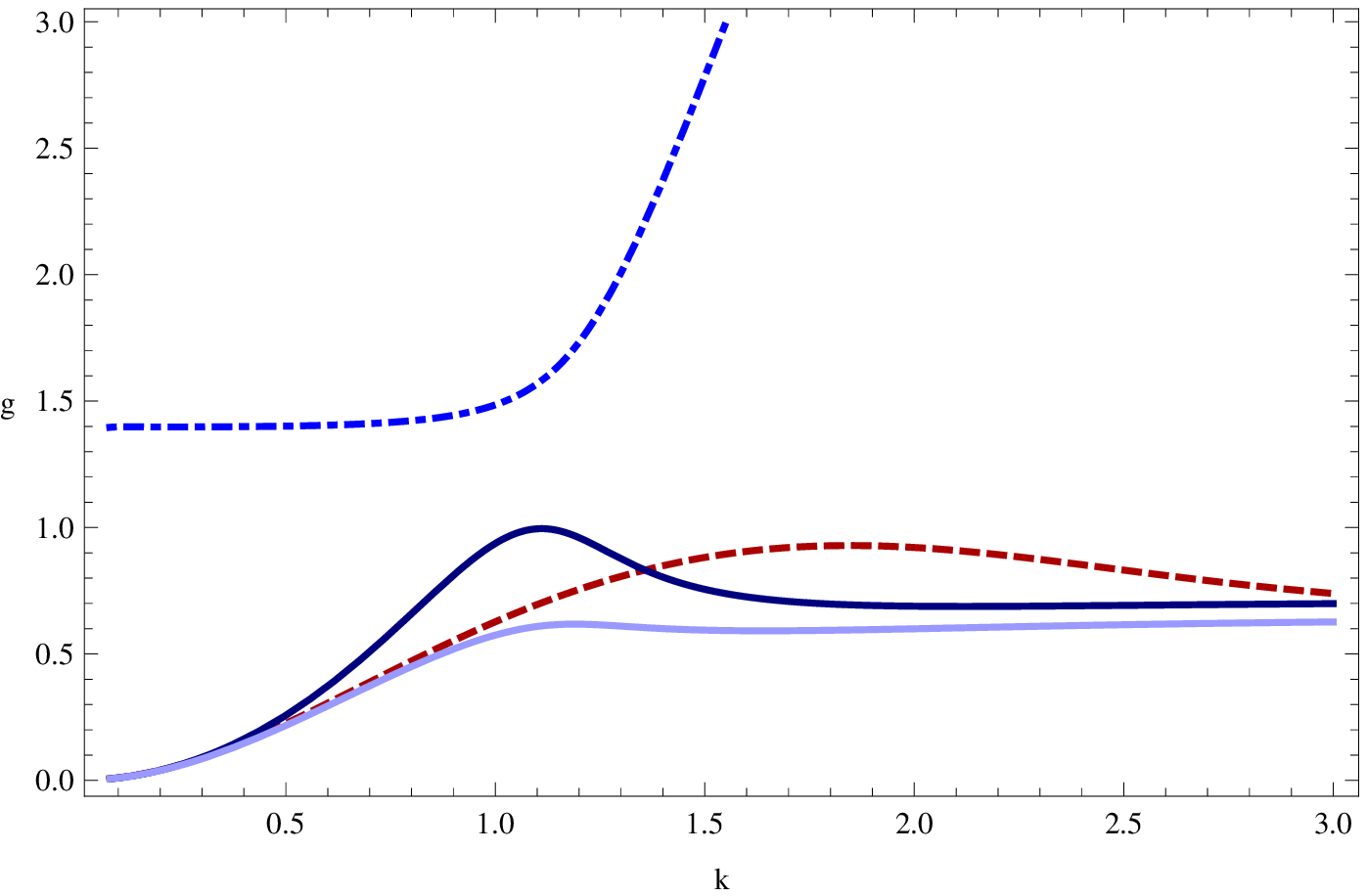}}
\hspace{0.05\textwidth}
 \subfloat{%
\label{fig:res4Dpp1T3aAB}\includegraphics[width=0.4\textwidth]{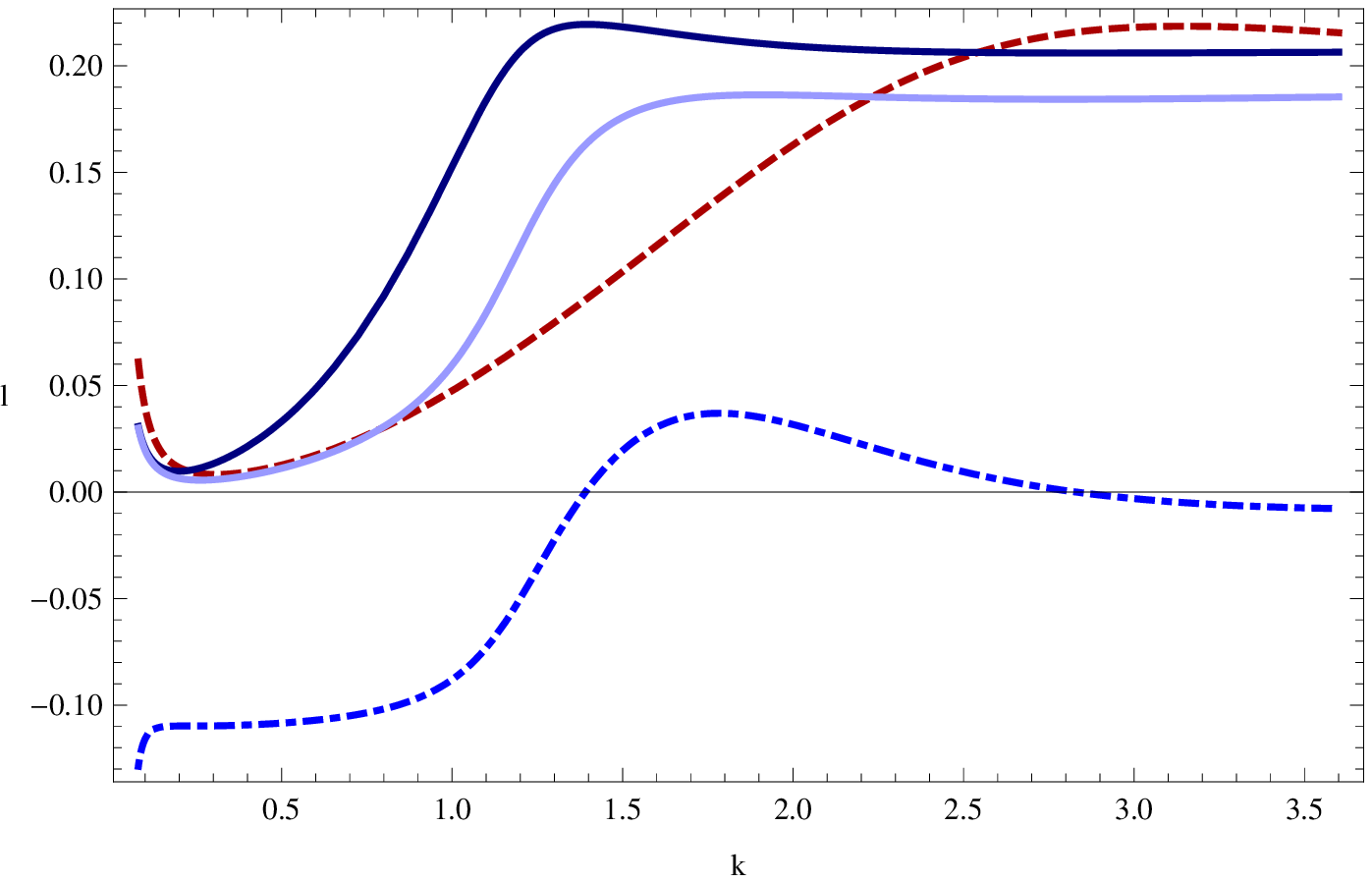}}
 \caption{Type (IIIa)$^{\dyn}$-\AttrL trajectory: dimensionless couplings.}\label{fig:res4DcpT3aA}
\end{figure}

\begin{figure}[!ht]
\centering
\psfrag{k}{${\scriptstyle   k \, \slash m_{\text{Pl}}}$}
\psfrag{l}[bc][bc][1][90]{$\scriptstyle    \Kk^{\cix}_k $}
\psfrag{m}[bc][bc][1][90]{$\scriptstyle   \tg^{\cix}_k\,\Kk^{\cix}_k  $}
\psfrag{L}[bc][bc][1][90]{$\scriptstyle    \Kkbar^{\cix}_k \, {\scriptscriptstyle [m_{\text{Pl}}]}$}
\psfrag{g}[bc][bc][1][90]{$\scriptstyle   \tg^{\cix}_k $}
\psfrag{G}[bc][bc][1][90]{$\scriptstyle   G^{\cix}_k  \, {\scriptscriptstyle [m_{\text{Pl}}]}$}
 \subfloat{%
\label{fig:res4Dpp3T3aCA}\includegraphics[width=0.4\textwidth]{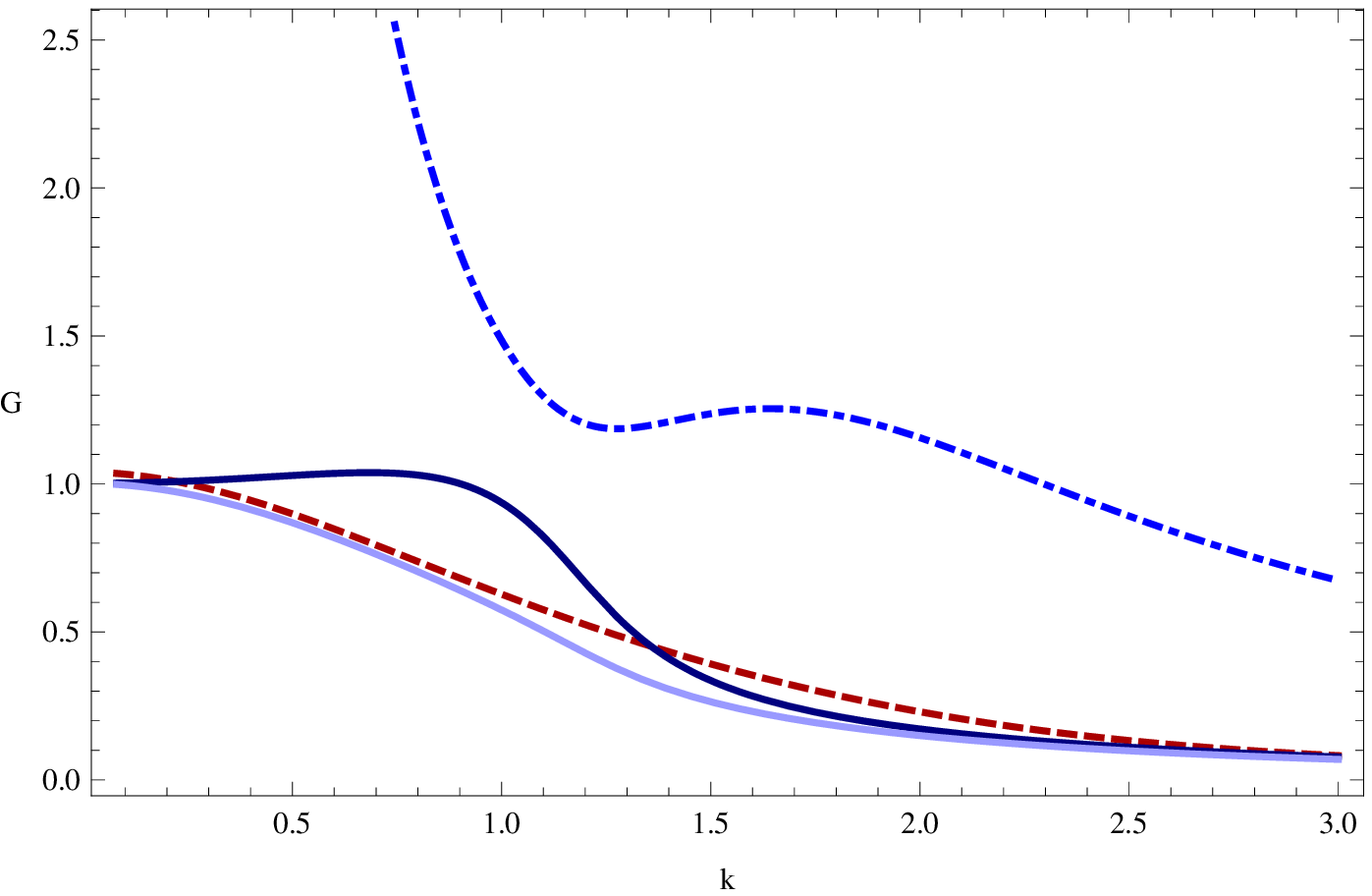}}
\hspace{0.05\textwidth}
 \subfloat{%
\label{fig:res4Dpp3T3aCB}\includegraphics[width=0.4\textwidth]{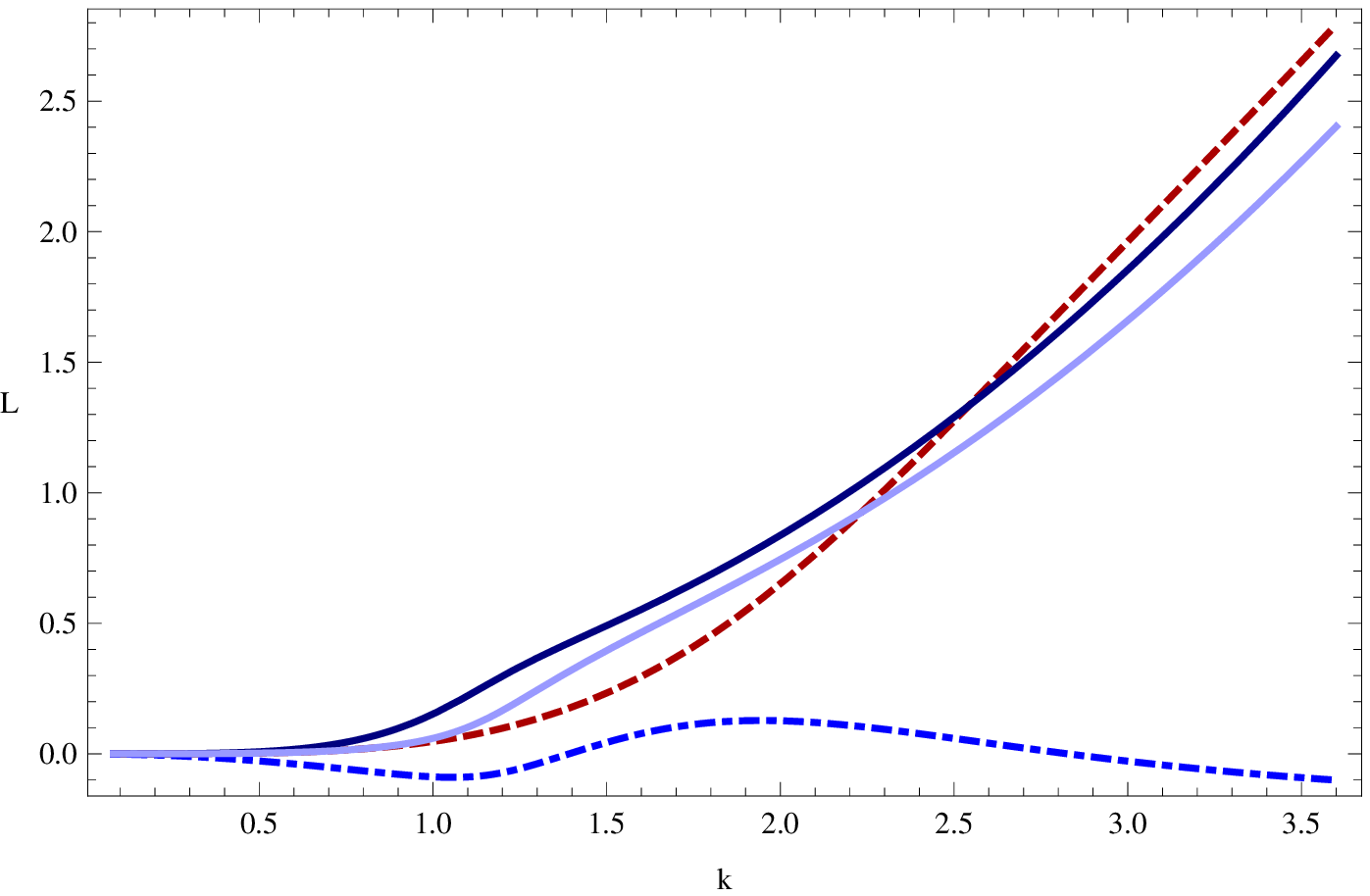}}
 \caption{Type (IIIa)$^{\dyn}$-\AttrL trajectory: dimensionful couplings.}\label{fig:res4DcpT3aC}
\end{figure}

\begin{figure}[!ht]
\centering
\psfrag{k}{${\scriptstyle   k \, \slash m_{\text{Pl}}}$}
\psfrag{l}[bc][bc][1][90]{$\scriptstyle    \Kk^{\cix}_k $}
\psfrag{m}[bc][bc][1][90]{$\scriptstyle   \tg^{\cix}_k\,\Kk^{\cix}_k  $}
\psfrag{L}[bc][bc][1][90]{$\scriptstyle    \Kkbar^{\cix}_k\slash G^{\cix}_k \, {\scriptscriptstyle [m_{\text{Pl}}]}$}
\psfrag{g}[bc][bc][1][90]{$\scriptstyle   \tg^{\cix}_k $}
\psfrag{G}[bc][bc][1][90]{$\scriptstyle   1\slash G^{\cix}_k  \, {\scriptscriptstyle [m_{\text{Pl}}]}$}
 \subfloat{%
\label{fig:res4Dpp3T3aFA}\includegraphics[width=0.4\textwidth]{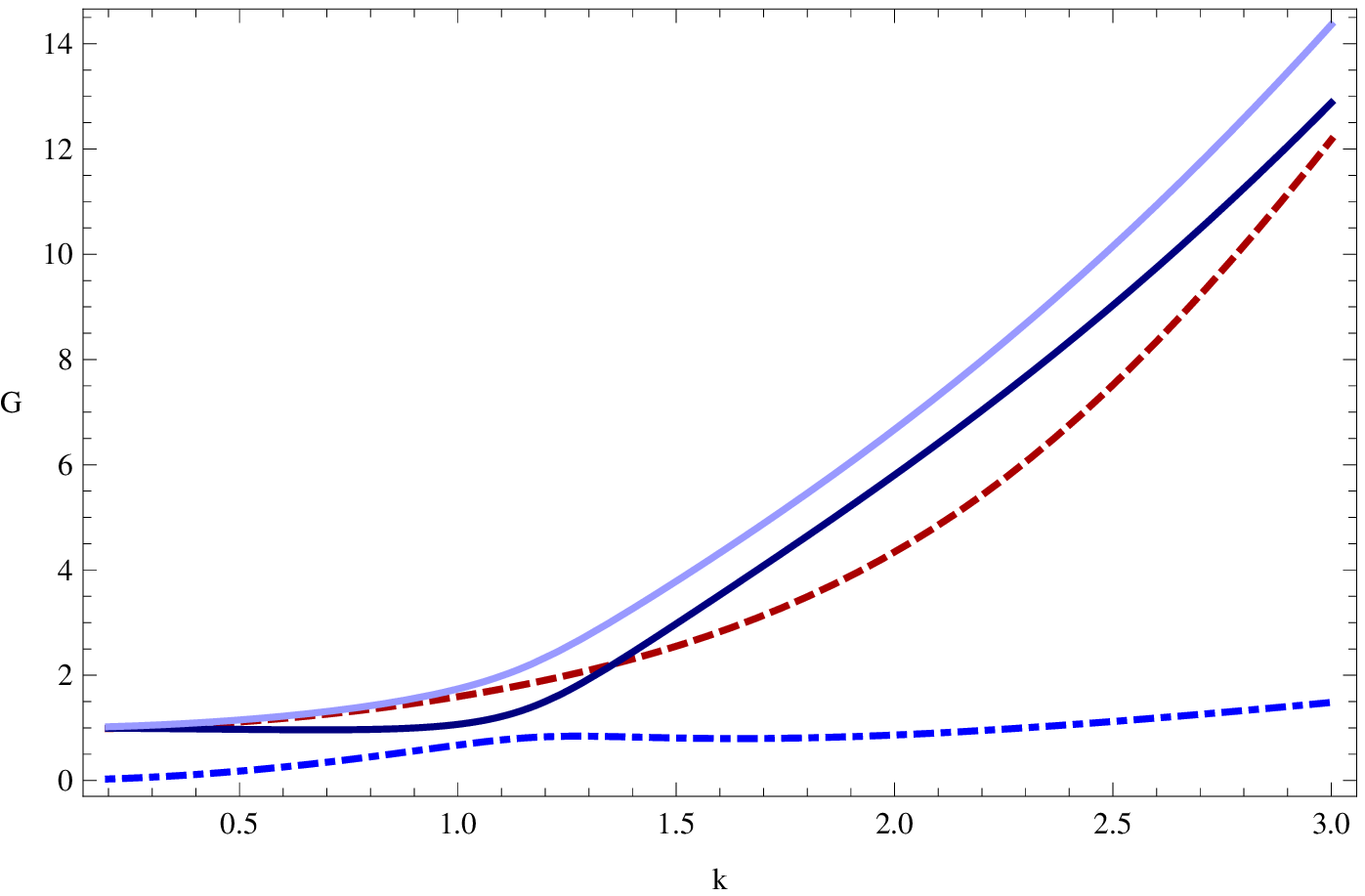}}
\hspace{0.05\textwidth}
 \subfloat{%
\label{fig:res4Dpp3T3aFB}\includegraphics[width=0.4\textwidth]{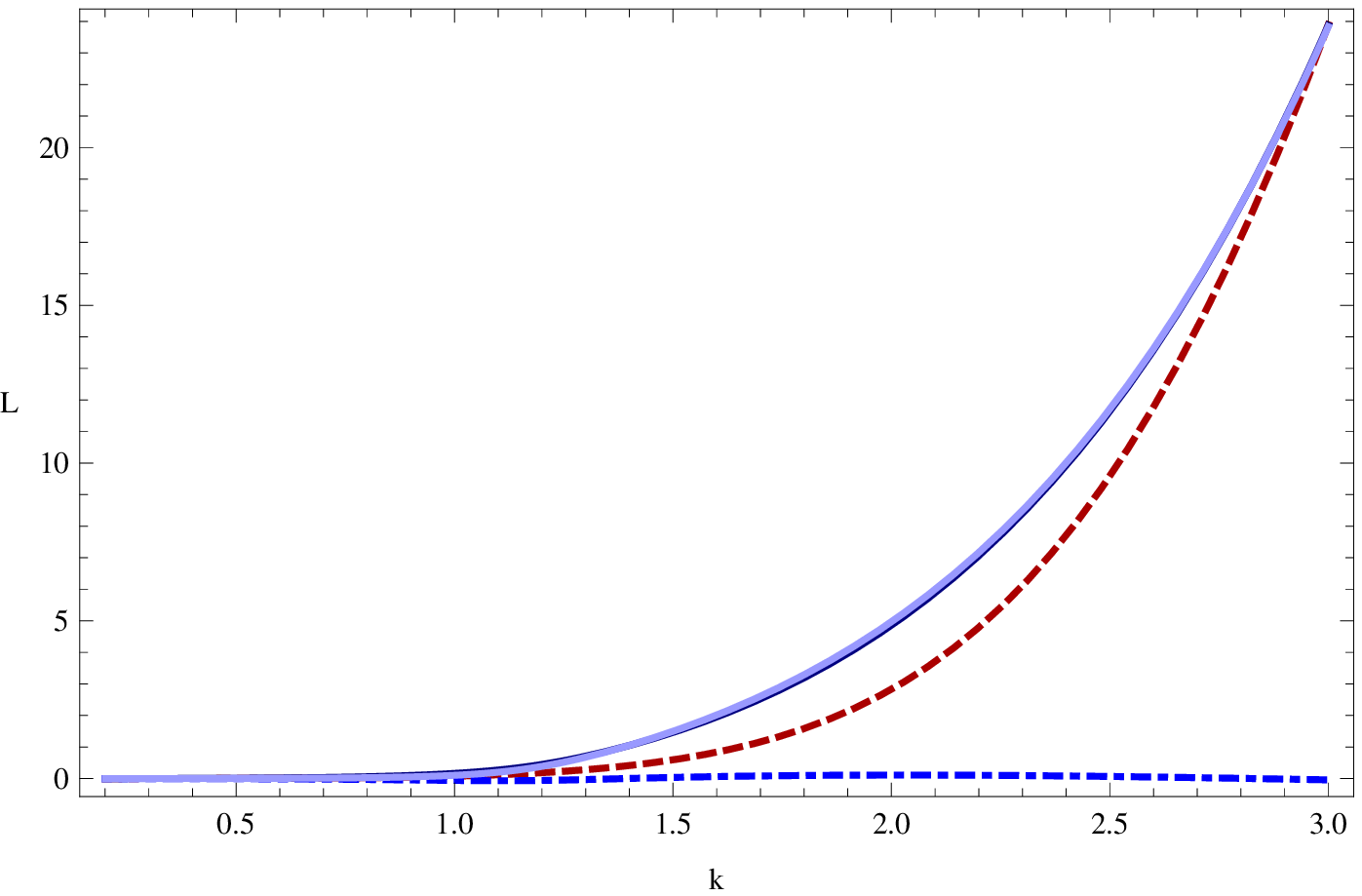}}
 \caption{Type (IIIa)$^{\dyn}$-\AttrL trajectory: the coefficients as they appear in the EAA. Note again the vanishing $1\slash G_k^{\background}$ and $\KkbarB\slash G_k^{\background}$, indicative of split-symmetry restoration in the limit $k\rightarrow0$.}\label{fig:res4DcpT3aF}
\end{figure}

Next we present the $k$-dependence of all running couplings,  both in dimensionless and dimensionful form, for one representative numerical solution in each class of trajectories.%
\footnote{For the `initial' conditions at the scale $k_0=1$ we set $(\tg_{k_0}^{\dyn},\KkD_{k_0})=(0.3,\, -0.01)$, $(\tg_{k_0}^{\dyn},\KkD_{k_0})=(\frac{3\pi}{8}\,10^{-4},\, 10^{-4})$, and $(\tg_{k_0}^{\dyn},\KkD_{k_0})=(0.05,\, 0.01)$ for the representative of type (Ia)$^{\dyn}$, (IIa)$^{\dyn}$, and (IIIa)$^{\dyn}$, respectively.} 
For comparison  we include the single-metric result for the Einstein-Hilbert truncation \cite{mr} using the same initial data as for the $\dyn$ couplings.
The diagrams in Figs. \ref{fig:res4DcpT1aA}-\ref{fig:res4DcpT1aF},  \ref{fig:res4DcpT2aA}-\ref{fig:res4DcpT2aF}, and  \ref{fig:res4DcpT3aA}-\ref{fig:res4DcpT3aF}, respectively, are devoted to the  (Ia)$^{\dyn}$-,  (IIa)$^{\dyn}$-, and  (IIIa)$^{\dyn}$-\AttrL trajectories.

In all plots we employ the following color $\slash$ style-coding to distinguish the  couplings $\tg^{\cix}$, $G^{\cix}$, $\Kk^{\cix}$, and $\Kkbar^{\cix}$, for $\cix\in\{\dyn,\,\background,\, (0),\, \sm \}$:
\begin{align*}
& \text{ { dashed (red): }} && {\color{darkred}\hdashrule[0.5ex]{4cm}{2pt}{3mm 3pt}} &&\quad  \cix=\sm \,\text{ (single-metric)} \\
 &\text{ { solid (dark-blue): }} && {\color{darkblue}\rule[0.5ex]{3.5cm}{2pt}} &&\quad \cix=\dyn\equiv (p) \text{ for } p\geq1\\
 &\text{  { solid (light-blue): }} && {\color{lightblue}\rule[0.5ex]{3.5cm}{2pt}} &&\quad  \cix=(0)\\
 &\text{ { dot-dashed (blue): }} && {\color{blue}\hdashrule[0.5ex]{4cm}{2pt}{3mm 3pt 1mm 3pt}} && \quad \cix=\background
\end{align*}

\noindent {\bf (A)}
Let us first consider the running {\it dimensionless} couplings   shown in Figs. \ref{fig:res4DcpT1aA}, \ref{fig:res4DcpT2aA}, and \ref{fig:res4DcpT3aA}.
The following features are shared by all three types of trajectories:
 
The solutions for the single-metric, the level-$(0)$ and the $\dyn$ couplings do not agree in any approximate sense, but  differ quite significantly for most $k$. 
In the IR, we imposed the requirement of split-symmetry and this is clearly seen even in the results for the dimensionless quantities: For $k\rightarrow0$, in the classical regimes, the $p=0$ and $p\geq1$ curves overlap basically.
 Likewise, in the UV, we observe that, consistent with the analysis in Section \ref{sec:sm_bm}, there is a remarkable similarity of the single- and bi-metric curves in the vicinity of their non-Gaussian fixed points. 
All plots confirm this numerical `miracle' which, as we emphasized already, is not due to any general principle. (But it is highly welcome of course.)
At intermediate scales the single-metric and the bi-metric solutions are found to be rather different, even qualitatively. This is precisely the symptom of the broken split-symmetry. 

In conclusion we can say in comparison with the bi-metric truncation, the single-metric treatment seems to be a good approximation in the far IR and UV, but at the quantitative level it does  not account for what happens in between.
It must be said that the single-metric results convey the correct qualitative picture, nevertheless.

\noindent {\bf (B)} Next we take a look at  the {\it dimensionful} couplings, expressed in units of the Planck mass. Figs.  \ref{fig:res4DcpT1aC},  \ref{fig:res4DcpT2aC}, and \ref{fig:res4DcpT3aC} show the results for the Newton  and  cosmological constants for the three classes.
As for the asymptotic $k$-dependence of the Newton constants $G_k^{\cix}$, $\cix\in\{\dyn,\,\background,\,(0),\,\sm\}$ we observe that all of them {\it vanish} for $k\rightarrow\infty$.

Thus {\it we recover gravitational anti-screening in the bi-metric setting, but only at a high (Planckian) scale.}
In fact, it is quite impressive to see that, for $k$ below the Planck scale,  the dynamical Newton constant $G_k^{\dyn}$ actually {\it increases} with $k$ then, assumes a maximum near $k\approx m_{\text{Pl}}$, and finally decreases for $k\gtrsim m_{\text{Pl}}$. 
On the other hand, the single-metric Newton constant $G_k^{\sm}$ decreases with $k$ at all scales $k\geq0$.
Clearly this behavior of $G_k^{\dyn}$ is a consequence of the sign-flip of $\kAD(\KkD)$, and therefore  $\aDz$, at $\KkD=\KkD_{\text{crit}}$.

\noindent {\bf (C)}
For $k\rightarrow0$ the background Newton constant $G_k^{\background}$ diverges, and $\Kkbar_k^{\background}$ vanishes, exactly as it should be in order to make $1\slash G_k^{\background}$ vanish in this limit, which is necessary for split-symmetry.
This is best seen in Figs. \ref{fig:res4DcpT1aF}, \ref{fig:res4DcpT2aF}, and \ref{fig:res4DcpT3aF}. There the dependence of the prefactors of the $\background$-type invariants in the truncation ansatz on $k$ is shown, namely $1\slash G_k^{\background}$ and $\KkbarB\slash G_k^{\background}$, respectively. 
Remember that the (blue) dot-dashed line is related to the $\background$-sector, it is very impressive to see how close to zero it stays in the IR for all three types of trajectories. 
These plots confirm that we were indeed successful in combining Background Independence with Asymptotic Safety.

Notice that for moderately large values of $k$ the $\background$-prefactors increase. 
Their deviation from zero is relatively small when compared with the $\dyn$- or level-($0$) sector, and this  implies that split-symmetry is intact at least approximately. 
For intermediate scales  we again find considerable violation of split-symmetry, which manifests itself by $\background$-coefficients which are now of the same order as the $\dyn$- and level-($0$) ones.
Therefore the single-metric (red, dashed line) only converges to the bi-metric curves for $k\rightarrow0$ and $k\rightarrow\infty$.

From the differences between the $p=0$ and $p\geq1$ curves, too, we see again that for all three trajectory types
 split-symmetry is apparently well restored in the IR and  the UV, but in between it suffers from a considerable breaking; 
the $\dyn$ couplings show a more pronounced $k$-dependence than the level-$(0)$ couplings, whereas the single-metric functions  are monotone.

\section{Single-metric vs. bi-metric truncation:\\a confrontation}\label{sec:sm_bm}
In this section we perform an in-depth analysis of the differences between the bi-metric Einstein-Hilbert truncation and its single-metric approximation.
We shall describe how precisely their results are interrelated on general grounds, and what can be learned from the numerical comparison about the validity of the single-metric truncation. 
As we shall see, its degree of reliability varies considerably over the theory space.
For future work it will be important to know of course where, and to what extent it can be trusted.
In particular we shall also understand why in the past it has always been notoriously difficult to obtain accurate and stable results for the critical exponents.

In this section, in subsection \ref{subsec:compMRS2}, we shall also critically examine how our new method based on the `deformed $\alpha=1$ gauge' compares to the bimetric calculation in ref. \cite{MRS2} which employed  the transverse-traceless approach.

This section is of a somewhat technical nature, and can be skipped by the reader who is mostly interested in the results.

\subsection{Collapsed level hierarchies}
Let us write the level-expanded EAA symbolically as 
$\EAA_k[\flcb;\bg]=\sum_{p=0}^{\infty} F_k^{(p)}[\bg]\, (\flcb_{\mu\nu})^p $ where the $F_k^{(p)}$'s depend on the background metric only. When we insert this expansion into the FRGE and project on a fixed level $p$ we see that $\partial_t F_k^{(p)}$ which appears on its LHS gets equated to an expression exclusively involving $\{ F_k^{(q)} \, | \, q = 2,\cdots,p+2 \}$. Hence the scale derivative of all (dimensionful) level-($p$) couplings is given by a beta-function depending on the level-($q$) couplings, with $q=2,\cdots,p+2$, only.%
\footnote{The {\it dimensionless} couplings also contain trivial canonical terms in their beta-functions, of course. They play no role in this discussion and are ignored here.}
This is the generic situation when no special restrictions on the form of the EAA are assumed: the FRGE amounts to an infinite hierarchy of equations $\partial_t F_k^{(p)}=\cdots$ for $p=0,1,2,3,\cdots$ which does not terminate at any finite level and couples all levels therefore.
Only if it was possible to realize split-symmetry exactly this tower of equations collapses to a single equation that governs all levels.

The bi-Einstein-Hilbert truncation used in the present paper involves the assumption that split-symmetry is broken only weakly, and that differences among the `higher' levels $p=1,2,3,\cdots$ are sufficiently small so that they may be ignored.
The lowest level, $p=0$, however is dealt with separately and is allowed to show a RG behavior different from $p\geq1$. 
As always, the $p=0$ couplings have beta-functions which depend on the $p=2$ couplings. 
For the present truncation the latter happen to be equal to  those at all non-zero levels $p\geq1$.

Stated more abstractly, what reduced the infinite hierarchy of RG equations to just 2 equations was an {\it additional hypothesis about the RG flow}, namely that the split-symmetry breaking is such that it lifts only the degeneracy between levels with $p=0$ and $p>0$, while those with $p>0$ remain degenerate among themselves.

The logical status of the familiar single-metric truncations can be characterized analogously. 
Here the additional hypothesis invoked is even stronger: one pretends that the solutions to the FRGE exhibit {\it exact} split-symmetry so that {\it all} levels $p=0,1,2,3,\cdots$ undergo an equivalent RG evolution. 
It is sufficient then to retain the RG equations for the lowest level at $p=0$ to fix the $k$-dependence of all running couplings. 

\subsection{Relating single- and bi-metric beta-functions}
Let us return to the concrete example of the bi-Einstein-Hilbert truncation with its 4 independent couplings $\{\tg^{(0)},\Kk^{(0)},\tg^{(1)},\Kk^{(1)}\}$ and let us see in which way precisely its beta-functions are related to those of the standard single-metric Einstein-Hilbert truncation with only 2 running couplings.
Recall that after the conformal projection $g_{\mu\nu}=e^{2\cf}\bg_{\mu\nu}$ the EAA of the former equals that of the latter for $\cf=0$. In the bi-metric case,  the flow equation is expanded in powers of $\cf$, whereby the zeroth and first orders in $\cf$ correspond to the level-(0) and level-(1) couplings, respectively. 
Structurally  the single-metric beta-functions $\beta_{\tg}^{\sm}$ and $\beta_{\Kk}^{\sm}$ thus coincide with the beta-functions of the level-(0) couplings, however only {\it  after we have identified all couplings of different orders}.
 
Even though this sounds trivial it changes the form of the beta-functions quite significantly so that the new differential equations are of a rather different type, with qualitatively new properties. For example, the anomalous dimension $\aDz$ that (contrary to $\eta^{\background}$!) can appear on the RHS of the bi-metric flow equation is no longer related to an independent coupling $\tg^{(2)}$ then,  but in fact to the $\eta^{\background}$-related $\tg^{(0)}$. This transition changes the simple, Bernoulli-type differential equation \eqref{eqn:res4D_021} into a much more complicated non-polynomial (but autonomous) one. 

In detail, we have to apply the following identifications:
\begin{subequations}
\begin{align}
 &\beta_{\tg\slash \Kk}^{\sm}(\tg^{\sm},\Kk^{\sm};d)\equiv \beta_{\tg\slash \Kk}^{(0)}(\{\tg^{(q)}=\tg^{\sm},\Kk^{(q)}=\Kk^{\sm}\};d)
\big|_{ \eta^{(p)}\equiv \eta^{\sm} }\qquad p\,\in\, \{0,1,2,\cdots\}  \label{eqn:res4D_051A}\\
&\tg^{(0)}=\tg^{(1)}=\cdots=\tg^{\sm}\,,\qquad \Kk^{(0)}=\Kk^{(1)}=\cdots = \Kk^{\sm}\,.  \label{eqn:res4D_051B}
\end{align}  \label{eqn:res4D_051}
\end{subequations}
The running of all couplings at higher levels is pretended to be described by the two beta-functions  from level-(0). 

\subsection{Conditions for the reliability of a single-metric calculation}
To check whether the single-metric truncation is a good approximation to the bi-metric one, we must study the  beta-functions of the higher levels,  the conditions \eqref{eqn:res4D_051}, and their implications:
\begin{subequations}
\begin{align}
&\tg^{\sm}_k\stackrel{!}{=}\tg^{(0)}_k=\tg^{(1)}_k=\cdots\,,\qquad \Kk^{\sm}_k \stackrel{!}{=} \Kk_k^{(0)}=\Kk_k^{(1)}=\cdots \,,
\label{eqn:res4D_052A}\\
& \beta_{\tg\slash \Kk}^{\sm}(\tg^{\sm},\Kk^{\sm};d)\stackrel{!}{=} \beta_{\tg\slash \Kk}^{(p)}(\{\tg^{(q)}=\tg^{\sm},\Kk^{(q)}=\Kk^{\sm}\};d)
\qquad p,q\,\in\, \{0,1,2,\cdots\} \label{eqn:res4D_052B}
\end{align} \label{eqn:res4D_052}
\end{subequations}
Notice that contrary to split-symmetry requirement where it was more natural to consider {\it dimensionful} couplings, the requirements  \eqref{eqn:res4D_052}  are constraints on the {\it dimensionless} couplings. 
But of course as long as we are not taking the $k\rightarrow0$ or $k\rightarrow\infty$ limit we can simply strip off the explicit $k$-factors from the split-symmetry condition and end up with the conditions \eqref{eqn:res4D_052}: 
\begin{subequations}
\begin{align}
 G_k^{\sm}=k^{-(d-2)} \tg_k^{\sm} \stackrel{!}{=}k^{-(d-2)}  \tg_k^{(0)} =k^{-(d-2)}  \tg_k^{(1)}=\cdots=G_k^{(p)} \\
\Kkbar_k^{\sm}= k^{2} \Kk^{\sm}_k \stackrel{!}{=} k^{2} \Kk^{(0)}_k  = k^{2} \Kk^{(1)}_k=\cdots = \Kkbar_k^{(p)}
\end{align}
\end{subequations}
Though in principle there is the possibility that  the explicit $k$-dependence of the dimensionful couplings gives rise to split-symmetry  for $k=0$ or $k\rightarrow\infty$ only, we never relied on this possibility, and we shall never do it in what follows.  
This puts the respective requirements for intact split-symmetry and a valid single-metric approximation on an equal footing.

\subsection{The anomalous dimensions: structural differences}
In the remainder of this section we  restrict the discussion to  the Newton couplings. 
This  covers already all subtleties that arise in concretely working out the conditions \eqref{eqn:res4D_052} and their generalization to the full 4-dimensional theory space.

The beta-functions of the Newton constants in the level-description were found in eqs. \eqref{eqn:res4D_004} and \eqref{eqn:res4D_003}, respectively:
\begin{subequations}
\begin{align}
&\beta_{\tg}^{(0)}(\tg^{\dyn},\KkD,\tg^{(0)};d)= \Big[d-2+ \big(B_1^{(0)}(\KkD;d)+ \aDz\, B_2^{(0)}(\KkD;d)\big)\tg^{(0)} \Big]\tg^{(0)} \label{eqn:res4D_053A} \\
&\beta_{\tg}^{\dyn}(\tg^{\dyn},\KkD;d)= \Big[d-2+ \big(B_1^{\dyn}(\KkD;d)+ \aDz\, B_2^{\dyn}(\KkD;d)\big)\tg^{\dyn} \Big]\tg^{\dyn} \label{eqn:res4D_053B}
\end{align}\label{eqn:res4D_053}
\end{subequations}
Due to the back-reaction of the $\dyn$-couplings, i.e. those with $p\geq1$, especially via $\aDz$, eqs. \eqref{eqn:res4D_053A} and \eqref{eqn:res4D_053B} amount to structurally quite different expressions for the anomalous dimensions:
\begin{subequations}
\begin{align}
 \eta^{(0)}(\tg^{\dyn},\KkD,\tg^{(0)};d)&= \big[B_1^{(0)}(\KkD;d)+ \aDz\, B_2^{(0)}(\KkD;d)\big]\tg^{(0)}\label{eqn:res4D_054A} \\
\aDz(\tg^{\dyn},\KkD;d)&=\frac{B_1^{\dyn}(\KkD;d)\tg^{\dyn}}{1- B_2^{\dyn}(\KkD;d)\tg^{\dyn}} \label{eqn:res4D_054B}
\end{align}\label{eqn:res4D_054}
\end{subequations}
The `confusion' of  different levels by the single-metric truncation yields  a formula for its anomalous dimension $\eta^{\sm}$ that combines the non-polynomial $\tg^{\dyn}$ dependence  \eqref{eqn:res4D_054B} at the levels $p\geq1$, with the  dependence on the cosmological constant from level-($0$), the latter given by $B_{1\slash2}^{(0)}(\Kk;d)$.
Explicitly, we can extract the single-metric beta-functions for Newton's coupling by applying \eqref{eqn:res4D_051} to eq. \eqref{eqn:res4D_053}:
\begin{align}
 \beta_{\tg}^{\sm}(\tg^{\dyn},\KkD,\tg^{(0)};d)&=\beta_{\tg}^{(0)}(\tg^{\sm},\Kk^{\sm},\tg^{\sm};d)\big|_{\aDz=\eta^{\sm}}\nonumber\\
&= \Big[d-2+ \big(B_1^{(0)}(\Kk^{\sm};d)+ \eta^{\sm}\, B_2^{(0)}(\Kk^{\sm};d)\big)\tg^{\sm} \Big]\tg^{\sm}
\label{eqn:res4D_055}
\end{align}
The identification of $\aDz$ with $\eta^{\sm}$ leads to an implicit equation for $\eta^{\sm}$ from which we obtain
\begin{align}
 \eta^{\sm}(\tg^{\sm},\Kk^{\sm};d)&=\frac{B_1^{(0)}(\Kk^{\sm};d)\tg^{\sm}}{1- B_2^{(0)}(\Kk^{\sm};d)\tg^{\sm}}
\label{eqn:res4D_056}
\end{align}
Eq. \eqref{eqn:res4D_056} highlights the limitations of the single-metric formula in approximating the full bi-metric RG flow: 
Even though $\eta^{\sm}$ inherits the dynamical $\tg$-dependence and thus reproduces all findings based on its non-polynomial form, it completely looses any information on the dynamical $\Kk$-dependence. 
It is thus rather non-trivial that our bi-metric results, even at the numerical level, in many cases stayed very close to the single-metric ones. 

\subsection{(Un-)Reliable portions of the single-metric theory space}
Since intact split-symmetry is closely related to the reliability of the single-metric truncation, we expect the IR regime of the trajectories --- where we explicitly restored split-symmetry --- to be well approximated  by the trajectories of $\tg^{\sm}$ and $\Kk^{\sm}$. 
The plots presented in Section \ref{subsec:classDiagr} actually confirm this expectation, see Figs. \ref{fig:res4DcpT1aA},  \ref{fig:res4DcpT2aA},  and \ref{fig:res4DcpT3aA}. 


Nevertheless, we already pointed out that there exists no fully extended solution $\{\EAA_k,\, k\in[0,\infty)\}$ along which split-symmetry would be intact everywhere. 
As can be seen in the diagrams of Section \ref{subsec:classDiagr}, only in the extreme IR and UV, that is {\it only when $k\slash m_{\text{\rm Pl}}\ll1$ or $k\slash m_{\text{\rm Pl}}\gg1$ the dimensionless bi-metric couplings are satisfactorily approximated by the single-metric ones,  in between the curves disagree even qualitatively.}

Instead of focusing on a single trajectory only, we next investigate the validity of the single-metric approximation in an extended region of the bi-metric theory space. 
For this purpose we determine the set $\mathfrak{R}$ of all pairs $(g,\Kk)$ with $g>0$ such that at the points of theory space associated to them via $(\tg^{(0)},\Kk^{(0)},\tg^{\dyn},\KkD)=(\tg,\Kk,\tg,\Kk)\in \mathcal{T}$ the `single-metric condition' \eqref{eqn:res4D_052} is satisfied. Those points form a subset of the 4-dimensional theory space, denoted 
$$\mathcal{T}_{\mathfrak{R}}=\{(g,\Kk,g,\Kk)\,\mid \, \, (g,\Kk)\in\mathfrak{R}\}\subset \mathcal{T}$$ 
We may think of this submanifold as set of initial points for RG trajectories at which the single-metric approximation is exact. 
The crucial question is to what extent this submanifold is invariant under the $4$-dimensional RG flow. 

If we start from a point with $(\tg^{(0)},\Kk^{(0)})=(\tg,\Kk)=(\tg^{\dyn},\KkD)$, and insist that this equality is preserved under an infinitesimal RG transformation, we must require that the corresponding anomalous dimensions agree: $\eta^{(0)}(\tg,\Kk,\tg;d)\stackrel{!}{=}\aDz(\tg,\Kk;d)$. Using \eqref{eqn:res4D_054} this entails a constraint for the pair $(\tg,\Kk)$ which involves the $B_{1\slash2}$-functions:
\begin{align}
 0\stackrel{!}{=} \tg \Big[B_1^{(0)}(\Kk;d)- B_1^{\dyn}(\Kk;d)\Big] +\tg^2\,\Big[B_1^{\dyn}(\Kk;d)B_2^{(0)}(\Kk;d)- B_1^{(0)}(\Kk;d)B_2^{\dyn}(\Kk;d)\Big]  \, 
\label{eqn:res4D_057}
\end{align}
This equation is a necessary condition for $(\tg,\Kk)\in\mathfrak{R}$.

Besides the requirement $\eta^{(0)}\stackrel{!}{=}\aDz$, leading to \eqref{eqn:res4D_057}, the `single-metric validity conditions' \eqref{eqn:res4D_052} also include the constraint  arising from $ \eta^{\sm}(\tg,\Kk;d)\stackrel{!}{=}\aDz(\tg,\Kk;d)$.
Expressed in terms of the $B_{1\slash2}$-functions it is found to be exactly identical with eq. \eqref{eqn:res4D_057}.%
\footnote{The coincidence of the $\eta^{(0)}=\aDz$ and $\eta^{\sm}=\aDz$ conditions is owed to the fact that $B^{\sm}_{1\slash2}\equiv B^{(0)}_{1\slash2}$,
which reflects the general relationship between the single-metric and the level-$(0)$ conformally projected bi-metric results.} 
As a result, {\it equation \eqref{eqn:res4D_057} is  a sufficient condition for the validity of the single-metric truncation within $\mathfrak{R}$.}

Let us now solve eq. \eqref{eqn:res4D_057} explicitly to determine the pairs $(g,\Kk)\in\mathfrak{R}$.

\noindent {\bf (i)} A first class of solutions consists of points $(\tg,\Kk_{\mathfrak{R}})$ with $\tg$ arbitrary and $\Kk_{\mathfrak{R}}$ satisfying both
$B_1^{(0)}(\Kk;d)=B_1^{\dyn}(\Kk;d)$ and $B_2^{(0)}(\Kk;d)=B_2^{\dyn}(\Kk;d)$. 

\begin{figure}[!ht]
\psfrag{d}[bc][bc][1][90]{${\scriptstyle B_1^{\dyn}(\Kk;4)-B_1^{(0)}(\Kk;4)}$}
\psfrag{a}{${\scriptstyle  \Kk}$}
 \psfrag{c}{${\scriptstyle }$}
\psfrag{b}[bc][bc][1][90]{${\scriptstyle B_1^{\dyn\slash(0)}(\Kk;4)}$}
\centering
\includegraphics[width=0.7\textwidth]{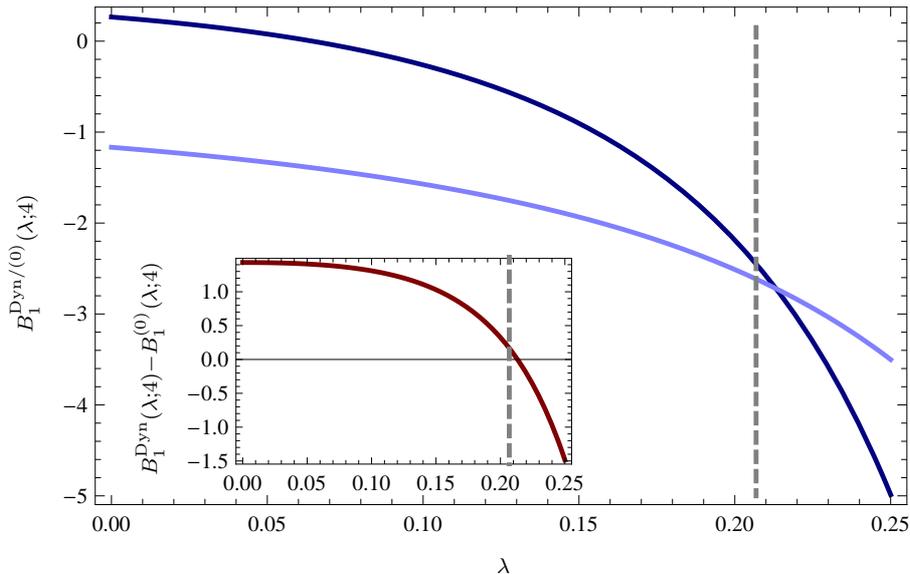}
\caption{This figure depicts the dependence of $\kAD(\Kk;4)$ and $B_1^{(0)}(\Kk;4)$ on their argument $\Kk$ by the dark, respectively light, blue line. The difference of the functions is shown as the (red) curve inserted in the lower left  corner. In both plots the fixed point value  $\KkD_*$ is marked with the dashed vertical line. The smaller the difference of $\kAD(\Kk;4)$ and $B_1^{(0)}(\Kk;4)$ the better is the single-metric approximation to the bi-metric truncation. While in the vicinity of the Gaussian fixed point there is a large discrepancy, the point of  best approximation is seen to be (miraculously) close to the NGFP value $\KkD_*$.}  \label{fig:b1db1sm}
\end{figure}

\begin{figure}[!ht]
\psfrag{d}[bc][bc][1][90]{${\scriptstyle B_2^{\dyn}(\Kk;4)-B_2^{(0)}(\Kk;4)}$}
\psfrag{a}{${\scriptstyle  \Kk}$}
 \psfrag{c}{${\scriptstyle  }$}
\psfrag{b}[bc][bc][1][90]{${\scriptstyle B_2^{\dyn\slash(0)}(\Kk;4)}$}
\centering
\includegraphics[width=0.7\textwidth]{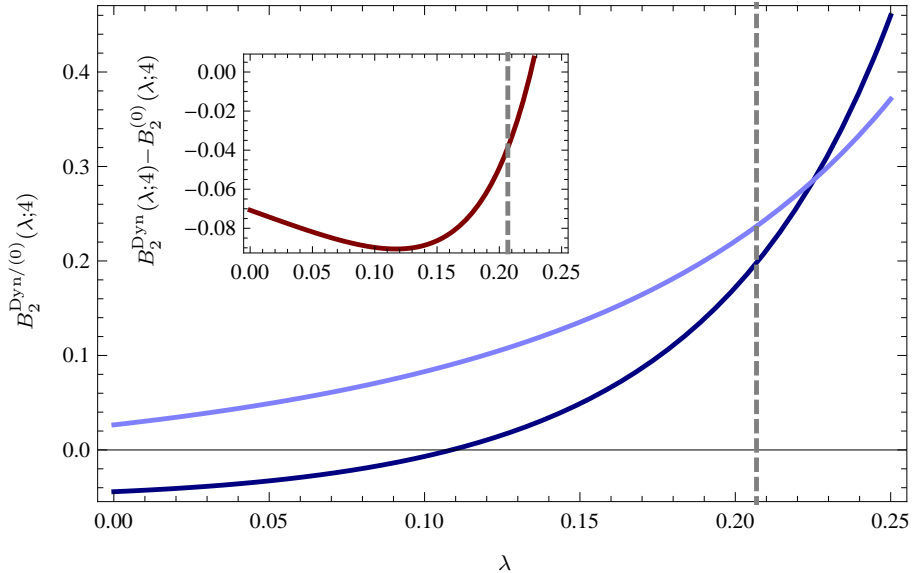}
\caption{The second condition for the fulfillment of eq. \eqref{eqn:res4D_057} is the agreement of $\kBD(\Kk;4)$ with $B_2^{(0)}(\Kk;4)$. Their dependence on $\Kk$ is given by the dark, respectively light, blue line. The difference of the functions is shown in the upper left  corner. The fixed point value $\KkD_*$ is marked with the dashed vertical line. Notice that the agreement of $\kBD(\Kk;4)$ with $B_2^{(0)}(\Kk;4)$ is best in the vicinity of the NGFP. There,  the single-metric truncation is  a good approximation to the bi-metric one. Away from the NGFP its quality deteriorates considerably.}  \label{fig:b2db2sm}
\end{figure}

Are there $\Kk$-values for which these conditions are satisfied, or satisfied approximately at least?
The explicit comparison of $B_{1\slash2}^{(0)}$ and $B_{1\slash2}^{\dyn}$ with respect to their $\Kk$-dependence in  $d=4$ is depicted in Fig. \ref{fig:b1db1sm} and Fig. \ref{fig:b2db2sm}, respectively.
The dashed vertical line marks the fixed point value of the dynamical cosmological constant, $\KkD_*$. The dark line gives the dependence of $B^{\dyn}_{1\slash2}$ on $\Kk$, whereas the light  one shows the functions $B^{(0)}_{1\slash2}$.
In the inset pictures we plotted the difference between the $\dyn$- and level-$(0)$ functions. 

We observe that while $B_{1\slash2}^{(0)}(\Kk)$ and  $B_{1\slash2}^{\dyn}(\Kk)$ have a qualitatively similar $\Kk$-dependence, the exact equality $B_{1\slash2}^{(0)}(\Kk)=B_{1\slash2}^{\dyn}(\Kk)$ holds only at a single value of $\Kk$. However, what comes as a real surprise is that this distinguished $\Kk$-value at which the single-metric truncation performs best is impressively close to the $\Kk$-coordinate of \fpnD-\fpL{}, $\KkD_*$.
This `miracle' is not explained by any general principle. Its implication is clear though:
{\it The single-metric approximation to the bi-Einstein-Hilbert truncation is most reliable precisely in that region of theory space where the non-Gaussian fixed point is located}.
This discovery may be seen as an a posteriori justification of the single-metric investigations of the Asymptotic Safety conjecture.

\begin{figure}[!ht]
\psfrag{g}{${\scriptstyle \tg^{\dyn}}$}
\psfrag{l}{${\scriptstyle  \KkD}$}
\centering
\includegraphics[width=0.6\textwidth]{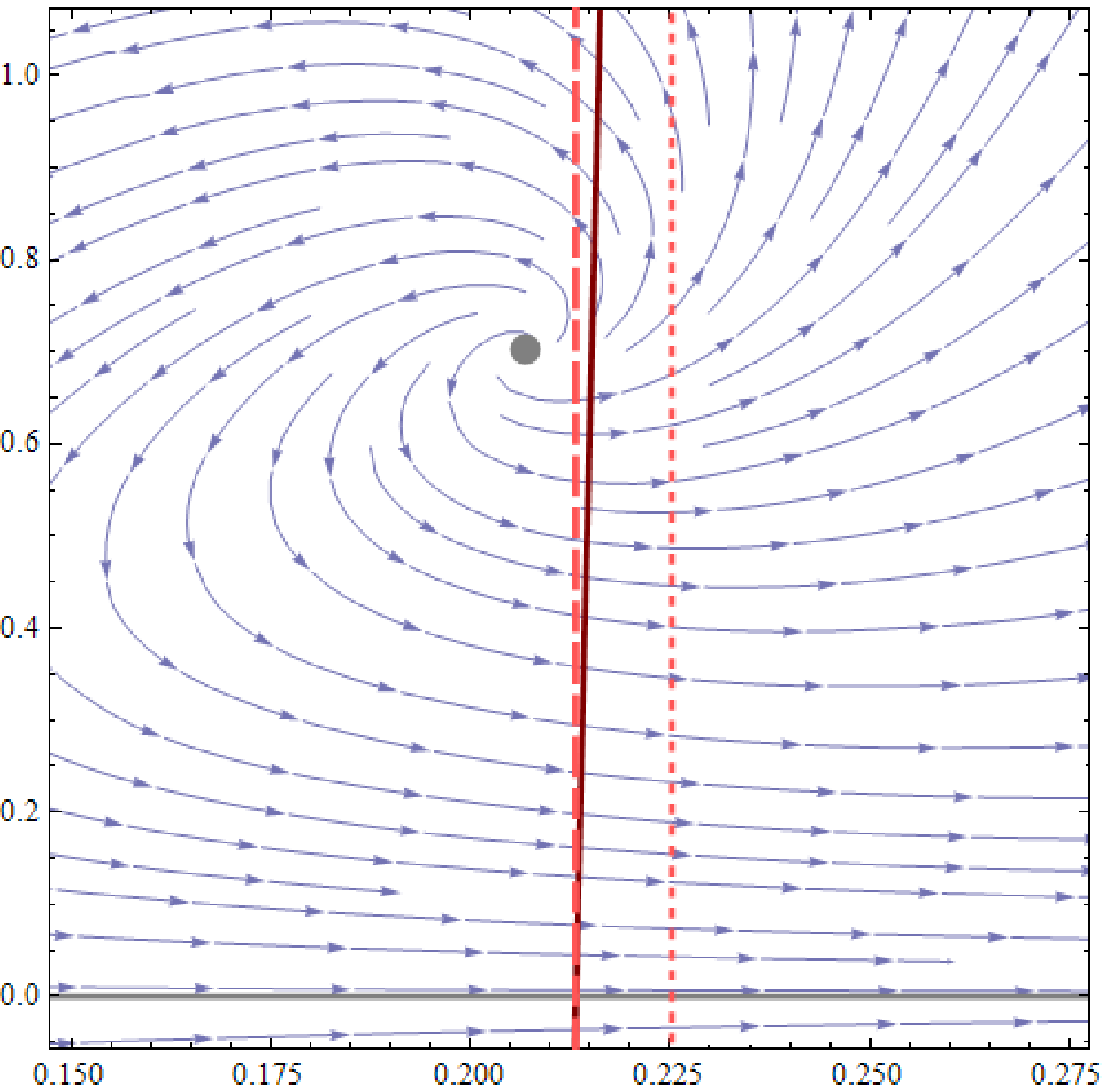}
\caption{The two solutions to the split-symmetry condition \eqref{eqn:res4D_057}, given by pairs $(\tg,\Kk_{\mathfrak{R}})$ and $(\tg_{\mathfrak{R}}(\Kk),\Kk)$, respectively, are shown  by the two light, and the dark (red) curve. They are superimposed on the phase portrait of the `$\dyn$' sector. In the vicinity of these curves split-symmetry is approximately intact. The `miraculous' result is that both of them, the almost linear (dark red) curve, and the (light red) exactly vertical line, pass very close to the NGFP (gray disc). This justifies the use of the single-metric truncation in the  vicinity of the NGFP. On the other hand, it is also apparent that every RG trajectories stays only briefly within a neighborhood of $(\tg,\Kk_{\mathfrak{R}})$ or $(\tg_{\mathfrak{R}}(\Kk),\Kk)$. Away from these regions the single-metric truncation might become problematic.}  \label{fig:b1db1smSolg}
\end{figure}

\noindent {\bf (ii)} Up to now we solved the condition \eqref{eqn:res4D_057} by setting to zero the coefficients of $\tg$ and $\tg^2$ separately.
There is a second type of solutions consisting of pairs $(\tg_{\mathfrak{R}}(\Kk),\Kk)$ where
\begin{align}
 \tg_{\mathfrak{R}}(\Kk)&\equiv \frac{B_1^{(0)}(\Kk;d)-B_1^{\dyn}(\Kk;d)}{B_1^{(0)}(\Kk;d)B_2^{\dyn}(\Kk;d)-B_1^{\dyn}(\Kk;d)B_2^{(0)}(\Kk;d)}
\end{align}
In Fig. \ref{fig:b1db1smSolg} the function $\tg_{\mathfrak{R}}(\Kk)$, along with the  above $(\tg,\Kk_{\mathfrak{R}})$ solution, is superimposed on the phase portrait of the `$\dyn$' sector which, as we know, is qualitatively similar to the `$\sm$' one.

The good news we learn from this plot is that {\it the NGFP is situated very close to the curve $(\tg_{\mathfrak{R}}(\Kk),\Kk)$}. 
This is a second `miracle', again un-explained by any general argument, and independent of the first one.
So the single-metric approximation seems indeed most reliable when it comes to locating the NGFP and exploring its properties.

The bad news is that there does not exist a single RG trajectory that would stay on, or close to the $(\tg_{\mathfrak{R}}(\Kk),\Kk)$-line for {\it all} scales; the trajectories intersect it at most once or twice.
The  consequence is that every `$\sm$' trajectory unavoidably contains segments where it differs substantially from its bi-metric, i.e. `$\dyn$' analogue.

\noindent {\bf Summary:} The general picture which emerges is a as follows. 
In the extreme UV, all RG trajectories start out from initial points which are infinitesimally close to the non-Gaussian fixed point which is at the heart of the Asymptotic Safety construction. 
In this region, the bi-metric description is well approximated by the single-metric truncation.
Once the trajectories escape from the non-Gaussian fixed point regime and enter an intermediate region where $\Kk\lesssim0.15$, say,  split-symmetry is significantly violated, and the single-metric approximation becomes unreliable. This can be seen in the Figs. \ref{fig:b1db1sm}, \ref{fig:b2db2sm}, and the explicit RG trajectories given in the diagrams of Section \ref{subsec:classDiagr}.

\subsection{The (un-)reliability of the \\ critical exponent calculations}
What remains to be investigated is the precise relation  between the single- and bi-metric results concerning the non-Gaussian fixed point, in particular its location and critical exponents.
Figs. \ref{fig:b1db1sm} and  \ref{fig:b2db2sm} already shed some light on this question since `miraculously'  the best agreement of single-metric and bi-metric results was found to be close to $\KkD_*$. 
In  this subsection we concentrate on the implications of this observation and, most importantly, try to understand why the predictions for the critical exponents differ so much in the two truncations.

\noindent{\bf (A)} Consider the single-metric and the bi-metric non-Gaussian fixed point regimes. 
Exactly at the NGFP all anomalous dimensions are $\eta_*^{\cix}=-(d-2)$. 
We first turn our attention to the bi-metric Newton couplings at the fixed point:
\begin{subequations}
\begin{align}
\tg_*^{\dyn}&= \frac{(d-2)}{B_2^{\dyn}(\KkD_*;d)(d-2)- B_1^{\dyn}(\KkD_*;d)} \\ 
\tg_*^{(0)}&= \frac{(d-2)}{B_2^{(0)}(\KkD_*;d)(d-2)- B_1^{(0)}(\KkD_*;d)} 
\end{align}
\end{subequations}
It is obvious that the better $B_{1\slash2}^{(0)}(\KkD_*;d)\approx B_{1\slash2}^{\dyn}(\KkD_*;d)$ is satisfied the closer are $\tg_*^{(0)}$ and $\tg_*^{\dyn}=\tg_*^{(p)}$, $p\geq1$, and the better is the split-symmetry. 
At the \fpnD-\fpL{} the deviation is small but non-zero; we expect $\tg^{(0)}_*$ to be at best approximately equal to $\tg^{\dyn}_*$, which in fact was found in Section \ref{sec:fps}. 
The main reason for the differing fixed point values is the difference in the $B_1^{(0)\slash\dyn}$-functions. 
Both, $B_1^{\dyn}(\Kk;d)$ and $B_1^{(0)}(\Kk;d)$ decrease for increasing $\Kk$, and so does their difference. 
For $\Kk<\KkD_*$ the difference $B_1^{\dyn}(\Kk;d)-B_1^{(0)}(\Kk;d)$ is positive, and thus $\tg^{(0)}_*<\tg^{\dyn}_*$.

\noindent{\bf (B)} The relation of $\tg_*^{(0)}$ and $\tg_*^{\dyn}$ to their single-metric cousin $\tg_*^{\sm}$ is more involved:
\begin{align}
\tg_*^{\sm}&= \frac{(d-2)}{B_2^{(0)}(\Kk_*^{\sm};d)(d-2)- B_1^{(0)}(\Kk_*^{\sm};d)} 
\end{align}
The difference of $\tg_*^{\sm}$ and $\tg_*^{(0)}$ is a consequence of the differing fixed point values of the cosmological constants, in particular we find $\Kk_*^{\sm}<\KkD_*$. This discrepancy has a balancing effect such that $\tg^{\dyn}_*\approx \tg_*^{\sm}$ while $\tg_*^{(0)}<\tg_*^{\sm}$ owed to the fact that  $B_1^{(0)}(\Kk_*^{\sm};d)>B_1^{(0)}(\KkD_*;d)$ for $\Kk_*^{\sm}<\KkD_*$.

\noindent{\bf (C)} Moving away from the NGFPs the first question that arises is whether the respective linearized flows in their vicinity are similar. The solution for the Newton constants in the bi-metric truncation reads
\begin{align}
 \tg^{\dyn}(k)&= g_*^{\dyn} + 2c_1 V_{1}^{(1)}\,\left(\frac{k_0}{k}\right)^{\theta^{\prime}}\!\!\cos\big(\vartheta_c+\vartheta^{(1)}+ \theta^{\prime\prime} \ln(k_0\slash k)\big)  \\
\tg^{(0)}(k)&= g_*^{(0)} + 2c_1 V_{3}^{(1)}\,\left(\frac{k_0}{k}\right)^{\theta^{\prime}}\!\!\cos\big(\vartheta_c+\vartheta^{(1)}+ \theta^{\prime\prime} \ln(k_0\slash k)\big) + c_3 V_{3}^{(3)}\, \left(\frac{k_0}{k}\right)^{\theta_3}
\end{align}
Here, $c_j$, $\vartheta_c$ are real constants of integration, $V_{r}^{(j)}$ is the $r^{\text{th}}$ component of the $j^{\text{th}}$ (real) eigenvector, and $\vartheta^{(j)}$ is its phase. The set of critical exponents consists of the complex pair 
$\theta_{1\slash2}= \theta^{\prime} \pm \Ii\,\theta^{\prime\prime}$, where $\theta^{\prime}>0$, and the purely real, positive critical exponent $\theta_3$. in addition to the spiral motion into the non-Gaussian fixed point present for the dynamical coupling, there is an additional $k^2$-term ($\theta_3=-2$) that contributes to the running of $\tg^{(0)}(k)$.

In the single-metric truncation the linearization yields instead
\begin{align}
  \tg^{\sm}(k)&= g_*^{\sm} + 2c^{\sm}_1 V_{1}^{(1;\sm)}\,\left(\frac{k_0}{k}\right)^{\theta^{\prime}_{\sm}}\!\!\cos\big(\vartheta^{\sm}_c+\vartheta^{(1;\sm)}+ \theta_{\sm}^{\prime\prime} \ln(k_0\slash k)\big) 
\end{align}
We see that qualitatively $\tg^{\sm}(k)$  has the same kind of running as $\tg^{\dyn}(k)$. Again, there is a pair of complex conjugate critical exponents with a positive real part $\theta^{\prime}_{\sm}$ and a non-vanishing imaginary part $\theta^{\prime\prime}_{\sm}$ that produces spirals.
From a quantitative perspective the results differ  considerably, however, and it is instructive to understand why. 

\noindent {\bf (D)}
Despite the small difference of the respective fixed point coordinates, {\bf the critical exponents disagree substantially} in the single- and bi-metric truncation, as we discuss next. 
Considering the single-metric and the dynamical bi-metric sector the two critical exponents are given by
\begin{align}
 &\theta_{1\slash2}=-\frac{1}{2} \big( \bmA+\bmD \big)\mp \frac{1}{2} \, \sqrt{4\bmB\bmC + (\bmA-\bmD)^2} 
\label{eqn:critExpD}
\end{align}
where $\mathcal{B}_{rs}$ is the $r$-$s$ entry of the respective stability matrix:
\begin{align}
 &\mathcal{B}=\begin{pmatrix}
\partial_{\tg}\beta_{\tg} & \partial_{\Kk}\beta_{\tg} \\
\partial_{\tg}\beta_{\Kk} & \partial_{\Kk}\beta_{\Kk}
\end{pmatrix} \, \Rightarrow \quad 
  \mathcal{B}^{\dyn}= 
\begin{pmatrix}
-2.32 & -27.8 \\
0.72 & -4.9
\end{pmatrix} \, , \quad 
\mathcal{B}^{\sm}= 
\begin{pmatrix}
-2.34 & -10.4 \\
0.96 & -0.61
\end{pmatrix} 
\label{eqn:bmatricesD}
\end{align}
Comparing $\kAD$ to $\kB$ we observe that their first columns agree quite well, but the second columns are rather different.
The main difference between the matrices $\mathcal{B}^{\dyn}$ and $\mathcal{B}^{\sm}$ is the way they depend on the cosmological constant. 
Whereas a change in the Newton couplings has, even quantitatively, a similar impact on $\kAD$ and $B^{\sm}$, a change in the cosmological constant affects the bi-metric system much more strongly than the `$\sm$' one. 
 Since in both cases the product $4\bmB\bmC$ is negative and much larger than $ (\bmA-\bmD)^2$, the square root in \eqref{eqn:critExpD} is purely imaginary, yielding
\begin{align}
 &\theta^{\prime}= -\frac{1}{2} \big( \bmA+\bmD \big)\, , \qquad \theta^{\prime\prime}=\frac{1}{2} \, \sqrt{-4\bmB\bmC - (\bmA-\bmD)^2} 
\end{align}
The differences of the single-metric and dynamical bi-metric critical exponents are therefore approximately 
\begin{subequations}
\begin{align}
&\theta^{\prime}-\theta^{\prime}_{\sm}\approx -\frac{1}{2} \big( \bmD^{\dyn}-\bmD^{\sm} \big)= 2.2 \,,\\
&\theta^{\prime\prime}-\theta^{\prime\prime}_{\sm}\approx  \sqrt{-\bmB^{\dyn}\bmC^{\dyn}}- \sqrt{-\bmB^{\sm}\bmC^{\sm}}=1.3\,.
\end{align} \label{eqn:xxx}
\end{subequations}
So, numerically the error due to the single-metric approximation is indeed considerable, more than two units (one unit) in the real (imaginary) part.

Taking the explicit structure of the beta-functions into account it thus becomes clear that it is the {\it slope} of the functions $B_{1\slash2}(\Kk;d)$ and $A_{1\slash2}(\Kk;d)$ at $\Kk=\Kk_*$ that is mainly responsible for the  differences. 
It is apparent from Figs. \ref{fig:b1db1sm} and \ref{fig:b2db2sm} that these slopes  are indeed quite different for the `$\dyn$' and the `$\sm$' case, even at the intersection points where the functions themselves agree. 
These $\Kk$-derivatives are the main reason for the quantitative differences

\noindent {\bf Summary:} 
In conclusion we can say that the vicinity of the non-Gaussian fixed point is sufficiently well described within the single-metric approximation if we are satisfied with `semi-quantitative' results.
It correctly captures all qualitative properties of the flow.
Our experience with the present truncation suggests however that it will hardly be possible to perform precision calculations of critical exponents in a single-metric truncation, not even with a very general ansatz for $\EAA_k$.

\subsection[Comparison with the TT-based approach\texorpdfstring{ of  ref. \cite{MRS2}}{}]{Comparison with the TT-based approach of  ref. \cite{MRS2}} \label{subsec:compMRS2}
At this point it is worthwhile to check how our present bi-metric results compare to those obtained in ref. \cite{MRS2} where a similar truncation ansatz including two Einstein-Hilbert actions for $g_{\mu\nu}$ and $\bg_{\mu\nu}$ was used, and the same running couplings were investigated.
Both calculations rely on the conformal projection technique, actually first employed in \cite{MRS2}, however with  two main differences:

\noindent {\bf (i)} The gauge choice: While in ref. \cite{MRS2} the `anharmonic gauge', $\varpi=1\slash d$, with gauge parameter $\alpha\rightarrow0$ was chosen, the present calculation uses the  harmonic gauge fixing condition, $\varpi=1\slash2$, with gauge parameter $\alpha=1-(d-6)\cf+\Order{\cf^2}$. The dependence of the `$\sm$' results on the gauge fixing parameter had already been investigated in ref. \cite{oliver1,oliver2,oliver3}, and the changes between $\alpha=0$ and $\alpha=1$ were found to be of the order of a few percent only, so we expect the choice of $\alpha$ to be of minor importance.  

\noindent {\bf (ii)} Uncontracted derivatives: In the present calculation the uncontracted derivative terms cancel due to the gauge choice. 
Therefore the heat kernel expansion becomes straightforward. 
Instead, in  ref. \cite{MRS2}, one had to deal with contracted as well as uncontracted derivative terms, and it was necessary to apply a  TT-decomposition to project the traces in the FRGE onto the truncated theory space. This led to a set of new field variables (irreducible component fields) and made the heat kernel expansion by far more involved and lengthy.

Recall that a truncation of theory space is not specified by the ansatz for the EAA alone, but  in addition by a prescription for the projection on the field monomials. Variation of either ingredient  may alter the results. 
The difference of our present calculation to \cite{MRS2} can be understood as stemming from the $\EAA_k$-ansatz only, namely in the gauge fixing part of the EAA.
Furthermore, the cutoff action $\Delta S_k$ is slightly different in the two cases, since in \cite{MRS2} it is formulated in terms of the irreducible (TT) components of $\flcb_{\mu\nu}$, while in our new approach simply in terms of the undecomposed $\flcb_{\mu\nu}$.
The comparison can help distinguishing between artifacts of the specific truncation and robust, truncation-independent, results.
 
In what follows we  focus on  the properties of the  doubly non-Gaussian fixed point   in $d=4$, and compare the results of the TT-based bi-metric calculation in \cite{MRS2} with our present one, as well as with the single-metric approximation. 

\subsubsection{Existence and location of  non-Gaussian fixed points}
In ref. \cite{MRS2} the same number  of fixed points,  in the same six categories as in our present approach has been found, namely three  different $\dyn$ fixed points, and for each of them two $\background$ fixed points. 
The most important one for Asymptotic Safety is \fpnD-\fpL{}, having $\tg_*^{\dyn}>0$. In Tab. \ref{tab:MRS2_tab01} we summarize its properties.
\begin{table}\centering
\begin{tabular}[!ht]{|c|c|c|}\hline
 Bi-metric \cite{MRS2} & Bi-metric (present)	&	Single-metric \cite{mr} \\[2.2ex] \hline\hline
&& \\[0pt] 
$\substack{\text{{\bf NG$^{\background}_{-}$}\fpC\fpnD-\fpL{}}\\ (\tg_*^{\dyn}= 1.05,\,\KkD_*= 0.22) \\ (\tg_*^{\background}= -41.6,\,\KkB_*= 0.58)}$
&$\substack{\text{\fpnB\fpC\fpnD-\fpL{}}\\ (\tg_*^{\dyn}= 0.70,\,\KkD_*= 0.21) \\(\tg_*^{\background}= 8.2,\,\KkB_*= -0.01)}$
 &  $\substack{\text{{\bf NG}-\fpL{}}\\ (\tg_*^{\sm}= 0.71,\,\Kk^{\sm}_*= 0.19)}$			
 \\[2.2ex] \hlinewd{0.2pt}
$\substack{\text{\fpnO\fpC\fpnD-\fpL{}}\\ (\tg_*^{\dyn}= 1.05,\,\KkD_*= 0.22) \\ (\tg_*^{(0)}= 1.08,\,\Kk^{(0)}_*= 0.21)}$
&$\substack{\text{\fpnO\fpC\fpnD-\fpL{}}\\ (\tg_*^{\dyn}= 0.70,\,\KkD_*= 0.21) \\(\tg_*^{(0)}= 0.65,\,\Kk^{(0)}_*= 0.19)}$
&  	$\substack{\text{{\bf NG}-\fpL{}}\\ (\tg_*^{\sm}= 0.71,\,\Kk^{\sm}_*= 0.19)}$		
 \\[2.2ex] \hlinewd{0.2pt}
$\theta_{\pm}=4.5 \pm 4.2 \Ii $
 &  $\theta_{\pm}=3.6 \pm 4.3\Ii $
& $\theta_{\pm}=1.5  \pm 3.0\Ii $
 \\[0pt] \hlinewd{0.2pt}
 $s_{\text{UV}}=2_{\dyn}+2_{(0)}=4$
 &  $s_{\text{UV}}=2_{\dyn}+2_{(0)}=4$
& $s_{\text{UV}}=2$
\\\hline
\end{tabular}	\caption{The properties of the doubly non-Gaussian fixed points are listed for the bi-metric calculation  in \cite{MRS2}, for the present one, and for the single-metric approximation \cite{mr}. The dynamical fixed point properties obtained by the two bi-metric approaches are seen to be qualitatively equivalent, and also well approximated by the single-metric truncation. In the background sector, the results for $\tg_*^{\background}$ and $\KkB$ differ by sign, but only their combination with the `$\dyn$' parameters yielding the level-($0$)  couplings is numerically meaningful, and those are indeed qualitatively similar. 
In addition the critical exponents for the dynamical sector are given. They are complex, with a positive real part.}\label{tab:MRS2_tab01}
\end{table}
Notice that the new and the old bi-metric results in the $\dyn$ sector and their single-metric counterparts share the same qualitative  as well as semi-quantitative features, the $\background$-sector differs in that the two approaches lead to $\KkB_*$ and $\tg_*^{\background}$ values, with the opposite signs even. 
However, the $\background$-couplings describe only differences between level-$(0)$ and the $\dyn$ or $p\geq1$ couplings and thus their  precise values and signs are not meaningful as such; they are an indication for the degree of split-symmetry, however.
(The negative $\tg_*^{\background}$ found in \cite{MRS2} only tells us that in this case $\tg_*^{(0)}>\tg_*^{\dyn}$.) 

\subsubsection[Impact of \texorpdfstring{$\alpha$}{alpha} and \texorpdfstring{${\varpi}$}{the gauge condition} on the NGFPs]{Impact of $\bm{\alpha}$ and $\bm{\varpi}$ on the NGFPs}
The impact of changing the functional form of the  gauge fixing condition, concretely the parameter $\varpi$, and the parameter $\alpha$ in its prefactor can be observed in the fixed point coordinates. 
The single-metric, and the present bi-metric truncation employ the same   harmonic gauge fixing and $\alpha=1$ choices; in Table \ref{tab:MRS2_tab01} we see that their fixed point values almost coincide. 
The TT-based bi-metric calculation \cite{MRS2} which uses the  `anharmonic' choice for $\varpi$ together with $\alpha=0$ leads to a somewhat different fixed point value $\tg_*^{\dyn}>1$, with about the same $\KkD_*$. 
This is a small, but visible effect due to the different gauge choices.

\begin{figure}[!ht]
\centering
\psfrag{r}{$\rhoG$}
\psfrag{g}{$\tg^{\dyn}_*$}
\psfrag{l}{$\KkD_*$}      
\psfrag{f}{$\scriptscriptstyle {\text{\fpgD-\fpL{}}}$}
\psfrag{m}{$\scriptscriptstyle {\text{\fpnDn-\fpL{}}}$}
\psfrag{n}{$\scriptscriptstyle {\text{\fpnD-\fpL{}}}$}
 \subfloat[The  dependence of $\tg_*^{\dyn}$ on $\rhoG$.]{\label{fig:rhoGMRS2A}\includegraphics[width=0.4\textwidth]{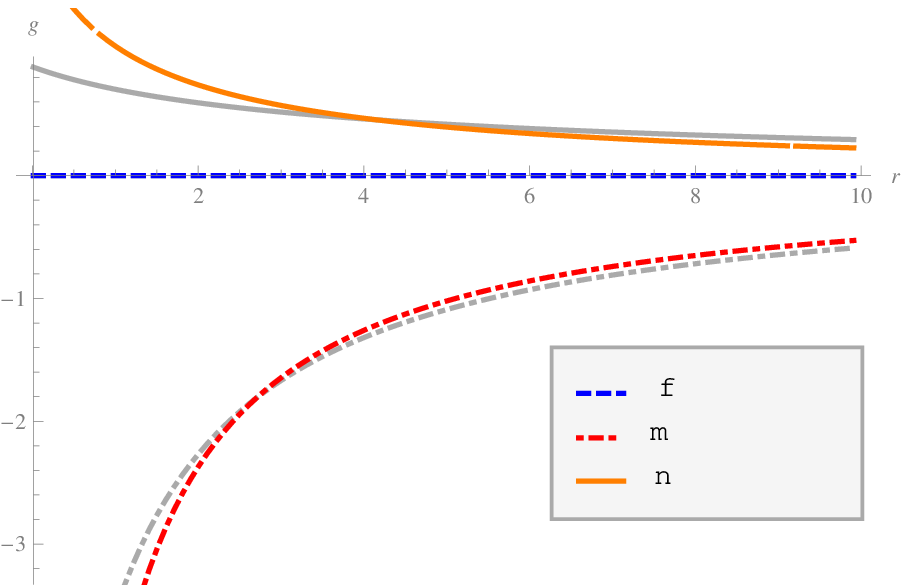}}\hspace{0.1\textwidth}
 \subfloat[The dependence of $\KkD_*$ on $\rhoG$.]{\label{fig:rhoGMRS2B}\psfrag{g}{$\KkD_*$}  \includegraphics[width=0.4\textwidth]{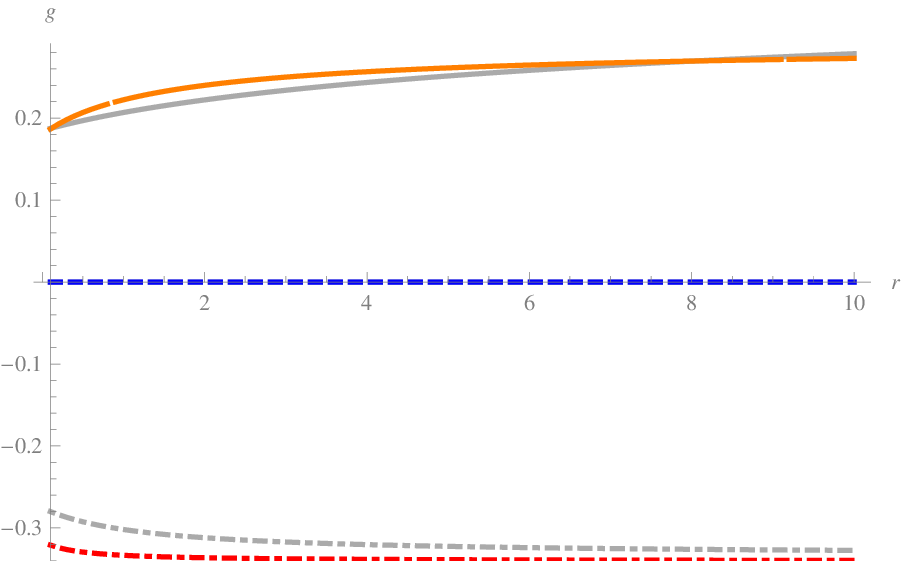}}
\caption{The  dependence of $\tg_*^{\dyn}$ and $\KkD_*$ on $\rhoG$ for the three fixed points, \fpgD-\fpL{}, \fpnDn-\fpL{}, and \fpnD-\fpL{} according to the approach of \cite{MRS2} (dark lines). 
The gray lines indicate the corresponding results of the present calculation. The qualitative agreement of these results for all  fixed points is obvious. Except for very small values of $\rhoG$, even quantitatively the results are seen to be almost equal, indicating that the impact of $\varpi$ on the ghost sector is relatively small, and is probably negligible compared to the other sources of uncertainty.} \label{fig:rhoGMRS2}
\end{figure}

For the design of future, more advanced truncations it is also instructive to monitor how the gauge choice influences the ghost sector and the beta-functions.
In Fig. \ref{fig:rhoGMRS2} we therefore plot the $\rhoG$-dependence of the fixed point coordinates obtained with the old approach \cite{MRS2}. 
For comparison our findings from Fig. \ref{fig:rhoG} are indicated as light gray lines. 
The diagrams show basically overlapping curves for most coordinates. 
If $\rhoG\lesssim2$ the positive $\tg_*^{\dyn}$ values, though qualitatively displaying the same increasing behavior for decreasing $\rhoG$, differ quantitatively, but by much less than one order of magnitude. The standard choice $\rhoG=1$ is within this regime and thus reflects the small, but visible difference in the fixed point values of Tab. \ref{tab:MRS2_tab01}.
 At large $\rhoG$, the influence of $\varpi$ in the ghost sector becomes completely negligible, at least in the non-Gaussian fixed point regime.

We can get an indication for the quality of the split-symmetry  in the vicinity of the NGFP by the size of its  $\tg_*^{\background}$ value, or more appropriately, by its inverse.
The smaller  $1\slash \tg_*^{\background}$, the better is the coincidence of $\tg_*^{(0)}$ and $\tg_*^{\dyn}$.
\begin{figure}[!ht]
\centering
\psfrag{r}{$\rhoG$}
\psfrag{g}{$1\slash \tg^{\background}_*$}
\psfrag{l}{$\KkB_*$}      
\psfrag{f}{$\scriptscriptstyle {\text{\fpgB\fpC($\cdots$)$^{\dyn}$-\fpL{}}}$}
\psfrag{m}{$\scriptscriptstyle {\text{\fpnB\fpC\fpnDn-\fpL{}}}$}
\psfrag{n}{$\scriptscriptstyle {\text{\fpnB\fpC\fpnD-\fpL{}}}$}
\psfrag{o}{$\scriptscriptstyle {\text{\fpnB\fpC\fpgD-\fpL{}}}$}
 \subfloat[The  dependence of $1\slash g^{\background}_*$ on $\rhoG$ in \cite{MRS2}.]{
\psfrag{m}{$\scriptscriptstyle {\text{{\bf NG$^{\background}_{-}$}\fpC\fpnDn-\fpL{}}}$}
\psfrag{n}{$\scriptscriptstyle {\text{{\bf NG$^{\background}_{-}$}\fpC\fpnD-\fpL{}}}$}
\psfrag{o}{$\scriptscriptstyle {\text{{\bf NG$^{\background}_{-}$}\fpC\fpgD-\fpL{}}}$}
\label{fig:rhoGMRS2C}\includegraphics[width=0.45\textwidth]{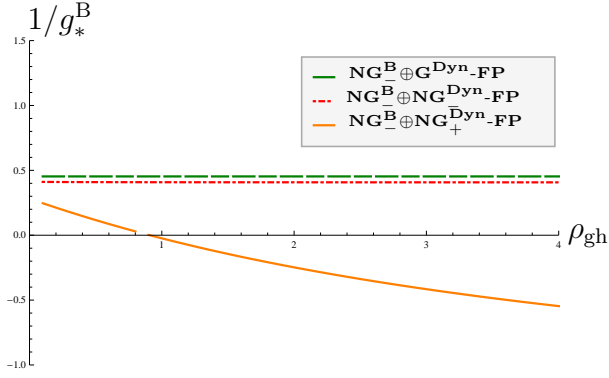}}\hfill
 \subfloat[The dependence of $1\slash g^{\background}_*$ on $\rhoG$ in the present calculation.]{\label{fig:rhoGMRS2D}    \includegraphics[width=0.45\textwidth]{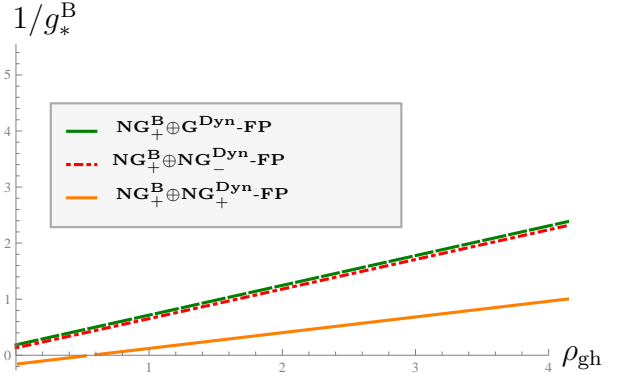}}
\caption{The dependence of $1\slash \tg_*^{\background}$  on $\rhoG$ is shown for the three `$\background$' non-Gaussian fixed points.  The smaller $1\slash \tg_*^{\background}$, the better split-symmetry is realized at the respective non-Gaussian fixed point. We see that the curves for the physically most relevant one, based upon \fpnD-\fpL{}, displays a zero which in  both calculations is located very close to $\rhoG=1$.}\label{fig:rhoGMRS2n}
\end{figure}
In Fig. \ref{fig:rhoGMRS2n} the dependence of  $1\slash \tg_*^{\background}$  on $\rhoG$ is shown for both bi-metric calculations, that is \cite{MRS2} and the present one.
As already pointed out, they yield different signs for $\tg_*^{\background}$, but this is irrelevant. 

The good news we learn from  Fig. \ref{fig:rhoGMRS2n} is that $1\slash\tg_*^{\background}$ possesses a zero, and that this zero occurs  in both calculations very close to $\rhoG=1$, that is, to  the actually implemented value of the ghost normalization!
There, the magnitude of $\tg_*^{\background}$ diverges, and this in turn forces $\tg_*^{(0)}$ and $\tg_*^{\dyn}=\tg_*^{(p)}$, $p\geq1$ to be equal.
The privileged status of a choice near $\rhoP=1$ seems to be at least one of the reasons for the `miraculously' good agreement of the single- and bi-metric truncation in the vicinity of the NGFP.

\subsubsection{Critical exponents and UV-critical hypersurface}
Turning next to the linearized flow near the (doubly) non-Gaussian fixed point, Tab. \ref{tab:MRS2_tab01} shows that the critical exponents for the two bi-metric calculations are more similar among themselves than in comparison to the single-metric results. 
In fact, for the spiral motion they predict almost  the same frequency.

Still there is a difference between the bi-metric truncations, the origin of which can again be traced back to the $\KkD$-dependence of the beta-functions. The stability matrix implied by the differential equations in \cite{MRS2} reads:
\begin{align}
  \mathcal{B}= 
\begin{pmatrix}
-2.37 & -46.0 \\
0.49 & -6.6
\end{pmatrix}  \qquad \text{(TT-approach)}
\end{align}
While the first column, corresponding to the derivatives with respect to $\tg^{\dyn}$, shows only relatively small deviations from $\kAD$ in  eq. \eqref{eqn:bmatricesD}, the source of the quantitative change in the critical exponents again originates in the $\KkD$-derivatives in the second column, which in turn are governed by the $\KkD$-dependence of $\kAD(\KkD)$ in the vicinity of the non-Gaussian fixed point.

The agreement of the dimensionality of the critical hypersurface as predicted by  both bi-metric calculations is an additional point in support of  the following general  picture:
The results of the present, new bi-metric approach and the earlier TT-based one in \cite{MRS2} are in close analogy, at least in the vicinity of the non-Gaussian fixed point. 
Deviations in the UV are of a minor numerical  kind, but most importantly all qualitative results for physically essential quantities agree among the two approaches.

 \section{An application: the running spectral dimension}\label{sec:spctDim}
It has been observed very early on that the effective spacetimes described by the EAA, in particular those along asymptotically safe RG trajectories display self-similar properties reminiscent of fractals, with a $k$-dependent effective dimensionality $d_{\text{eff}}\equiv 4+\eta_{\text{N}}$ which interpolates between $d_{\text{eff}}=4$ macroscopically and $d_{\text{eff}}=2$ microscopically \cite{oliver1, oliver2}. This observation gave rise to the development of a general scale dependent analog of Riemannian geometry for those spacetimes \cite{jan1, jan2}, and the discovery of a dynamically generated minimum length, a notion that turned out surprisingly subtle \cite{jan1}. 

Computing the spectral dimension $\mathcal{D}_s$ of those spacetimes \cite{oliverfrac} revealed the same crossover from $4$ dimensions in the IR to $2$ in the UV which was observed on the basis of $d_{\text{eff}}$. It is an exact, truncation independent prediction of asymptotically safe gravity, with or without matter.

More recently \cite{frankfrac} the scale dependence of $\mathcal{D}_s$ was reconsidered under the more restrictive assumptions of (i) pure gravity, (ii) the validity of the (single-metric) Einstein-Hilbert truncation, and (iii) the choice of an RG trajectory which admits a long classical regime \cite{h3, entropy}. 
Under these conditions, the EAA predicts in addition an extended semiclassical regime at intermediate scales in which the spectral dimension assumes the rational value $\mathcal{D}_s=4\slash3$. It has been argued \cite{frankfrac} and substantiated by a detailed comparison with Monte Carlo data that the dimensional reduction observed in numerical CDT simulations \cite{janCDT, Benedetti-Henson} actually originates in this semiclassical regime rather than the asymptotic scaling region of a fixed point. 
(See \cite{frankfrac, NJP, Naxos-book} for further details, and \cite{stefan-frankfrac, gianluca-astrid-frankfrac} for extensions.)

The scale dependent spectral dimension $\mathcal{D}_s(k)\equiv \mathcal{D}_s(\tg_k,\Kk_k)$ derived in \cite{frankfrac}, under the above conditions, reads (for $d=4$)
\begin{align}
 \mathcal{D}_s(\tg,\Kk)&= \frac{8}{4  + \Kk^{-1} \beta_{\Kk}(\tg,\Kk)}	\label{eqn:res4DspD_001}
\end{align}
where $\Kk$ was the dimensionless cosmological constant of the single-metric truncation, $\Kk^{\sm}\equiv \Kkbar^{\sm}\slash k^2$.
From the EAA-based derivation of eq. \eqref{eqn:res4DspD_001} it is obvious \cite{oliverfrac} that $\Kk_k$ enters this formula via the (contracted) Einstein-equation $\SR(\langle g\rangle_k)= 4 \Kkbar_k^{\sm}=4k^2 \Kk_k^{\sm}$ which describes how the effective metric responds to the scale dependence of the cosmological constant.

When we now go over from the single- to the bi-metric Einstein-Hilbert truncation we interpret $\langle g \rangle_k$ as a self-consistent background metric. It is given by the tadpole equation%
\footnote{See Section 4 of ref. \cite{MRS1} for a detailed discussion.}
$\big(\delta \EAA_k \slash \delta \flcb\big)\big|_{\flcb=0}=0$ which, for the present truncation, has again the same structure as the classical vacuum Einstein equation, but now containing the running {\it level-$(1)$ cosmological constant}:
$\bar{G}_{\mu\nu}=-\Kkbar_k^{(1)}\bg_{\mu\nu}$. Going through its derivation \cite{frankfrac} it is therefore easy to see that the above formula for the spectral dimension remains correct for the bi-metric truncation of the present paper provided we set $\Kk\equiv \Kk^{(1)}$ in eq. \eqref{eqn:res4DspD_001}, and interpret $\beta_{\Kk}$ as the beta-function of $\Kk^{(1)}$, depending on the level-$(1)$ couplings:%
\footnote{In a truncation that distinguishes between $\tg^{(p)}$ for $p\geq1$ the spectral dimension inherits an implicit dependence on higher level couplings since those appear in $\beta^{(1)}_{\Kk}$.}
\begin{align}
 \mathcal{D}_s(\tg^{(1)},\Kk^{(1)})&= \frac{8}{4  + (\Kk^{(1)})^{-1} \beta^{(1)}_{\Kk}(\tg^{(1)},\Kk^{(1)})}	\label{eqn:res4DspD_002}
\end{align}
Note that $\mathcal{D}_s$ is a scalar function on the $\tg^{(1)}$-$\Kk^{(1)}$ theory space: it depends only on the value of the couplings, but not the scale at which the RG trajectory passes there. It  is well-defined if the denominator on the RHS of \eqref{eqn:res4DspD_002} is always positive. This is indeed the case for the separatrix and all type (IIIa)$^{\dyn}$ trajectories, but not for those of type (Ia)$^{\dyn}$. They pass through a point where the assumptions behind \eqref{eqn:res4DspD_001} and \eqref{eqn:res4DspD_002} do not apply \cite{mr} and $\mathcal{D}_s$ diverges.

It is instructive to insert solutions of the RG equations, $k\mapsto(\tg_k^{(1)},\,\Kk^{(1)}_k)= (\tg_k^{\dyn},\,\KkD_k)$, into eq. \eqref{eqn:res4DspD_002} and determine the resulting scale-dependence of $\mathcal{D}_s$. For each class of trajectories we take one representative and depict the result in the following. 
While $\mathcal{D}_s$ is actually independent of the $\background$-couplings, it is most natural to think of these trajectories as 4-dimensional ones whose $\background$-sector is chosen in the split symmetry-restoring way.

\paragraph{Type (Ia)$^{\dyn}$ trajectories.}
For this class of trajectories the cosmological constant turns negative in the IR. 
\begin{figure}[h!]
\psfrag{d}[bc][bc][1][90]{$\mathcal{D}_s$}
\psfrag{k}{${\scriptstyle   k \, \slash m^{\dyn}_{\text{Pl}}}$}
\centering
\includegraphics[width=0.7\textwidth]{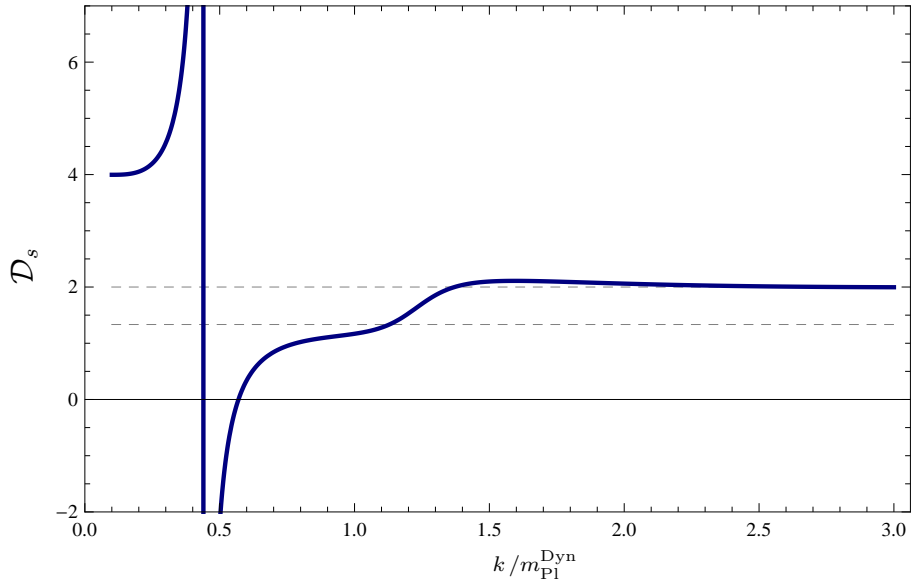}
\caption{The spectral dimension along a typical (Ia)$^{\dyn}$ trajectory. The interpretation as a spectral dimension is lost near the singularity where  the cosmological constant turns negative. Besides a short semiclassical plateau at $\mathcal{D}_s=4\slash 3$, disturbed by the divergence, the IR and the UV limits at $4$ and $2$, respectively, are well visible.}  \label{fig:specDimTypIa}
\end{figure}
This entails a divergence of $\mathcal{D}_s$ at a certain scale where the denominator of eq. \eqref{eqn:res4DspD_001} vanishes. In the vicinity of this scale, the function $\mathcal{D}_s$ admits no physical interpretation as a `$k$-dependent spectral dimension', since this would require a very slow (`adiabatic') dependence on $k$. 
Apart from that, we find in the corresponding Fig. \ref{fig:specDimTypIa} qualitatively precisely  the same $k$-dependence of $\mathcal{D}_s$, which had  been obtained within the single-metric approximation \cite{frankfrac,stefan-frankfrac}: 
In the IR the running spectral dimension approaches $\mathcal{D}_s=4$, there exists a semiclassical  plateau at $\mathcal{D}_s=4\slash 3$, here relatively short because it is disturbed by the divergence, and in the UV the running $\mathcal{D}_s$ converges to its fixed point value, $2$.

\begin{figure}[h!]
\psfrag{d}[bc][bc][1][90]{$\mathcal{D}_s$}
\psfrag{k}{${\scriptstyle   k \, \slash m^{\dyn}_{\text{Pl}}}$}
\centering
\includegraphics[width=0.7\textwidth]{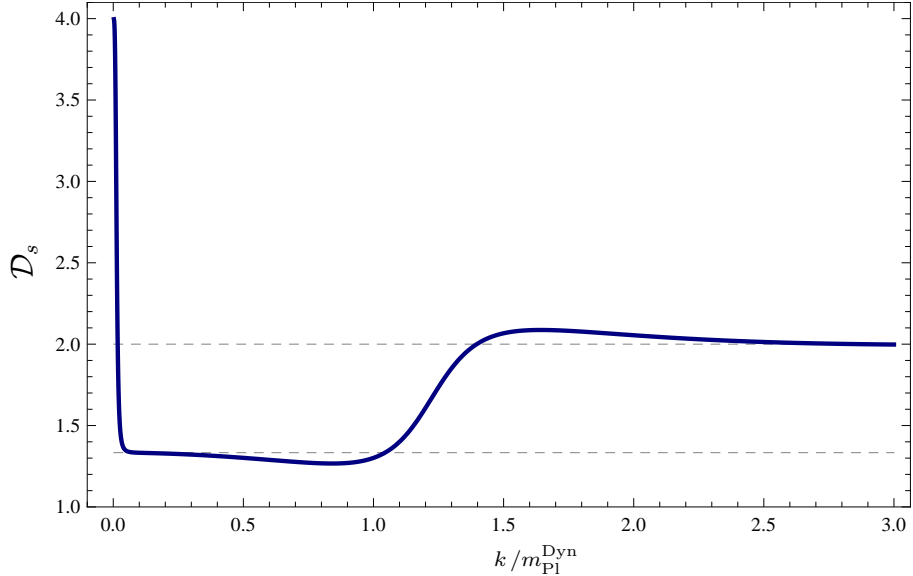}
\caption{Spectral dimension along the  (IIa)$^{\dyn}$ trajectory, the separatrix in the dynamical sector. The jump at $k=0$ is a computational artifact; in reality the semiclassical regime with $\mathcal{D}_s=4\slash 3$ extends down to $k=0$.}  \label{fig:specDimTypIIa}
\end{figure}

\paragraph{Type (IIa)$^{\dyn}$ trajectory.}
The separatrix  in the $\dyn$-sector gives rise to the scale-dependent  $\mathcal{D}_s$ shown in Fig. \ref{fig:specDimTypIIa}.
The semiclassical plateau at $\mathcal{D}_s=4\slash3$ is quite pronounced in this case, and it seems to terminate in a sudden jump to $\mathcal{D}_s=4$ for $k\rightarrow0$.
Actually this jump is a computational artifact since in a numerical calculation one is never able to find the separatrix exactly; rather, almost always the computer will generate  a type (Ia) or (IIIa) trajectory.
Hence it is ultimately pushed away from the GFP along the $\Kk$-direction, at some very low scale $k\slash m_{\text{Pl}} \ll1$, and this is exactly what caused the apparent jump to $\mathcal{D}_s=4$ in Fig. \ref{fig:specDimTypIIa}.
In reality, for the perfect separatrix solution (and in absence of matter!) the semiclassical regime with $\mathcal{D}_s=4\slash3$ extends down to $k=0$; there exists no genuinely classical regime with $\mathcal{D}_s=4$ \cite{frankfrac}.
 Towards the UV, we find a smooth cross over of the semiclassical plateau to $\mathcal{D}_s=2$, as expected in the NGFP regime.

\begin{figure}[h!]
\psfrag{d}[bc][bc][1][90]{$\mathcal{D}_s$}
\psfrag{y}[bc][bc][1][90]{$\mathcal{D}_s$}
\psfrag{k}{${\scriptstyle   k \, \slash m^{\dyn}_{\text{Pl}}}$}
\centering
\includegraphics[width=0.7\textwidth]{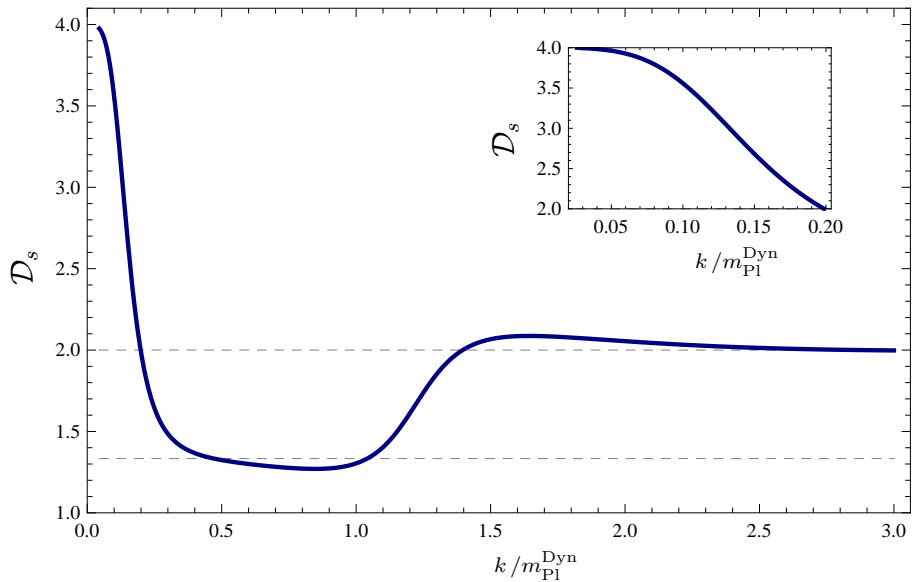}
\caption{The spectral dimension along a type (IIIa)$^{\dyn}$ trajectory. All three plateaus are well developed: we find  $\mathcal{D}_s=2$ in the UV, $\mathcal{D}_s=4$ in the IR, and in between a semiclassical regime with $\mathcal{D}_s=4\slash3$. The insert shows the classical regime at $k\ll m^{\dyn}_{\text{Pl}}$.}  \label{fig:specDimTypIIIa}
\end{figure}

\paragraph{Type (IIIa)$^{\dyn}$ trajectories.}
For this type of trajectories,  Fig. \ref{fig:specDimTypIIIa} shows
the running spectral dimension along a typical example. All three plateaus  are well visible here, with  $\mathcal{D}_s=2$ in the UV,    $\mathcal{D}_s=4$ in the IR, and  an intermediate  plateau, well below the Planck scale, with the semiclassical value $\mathcal{D}_s=4\slash3$. This, again, is in accord with  the results obtained in \cite{frankfrac} by means of a single-metric truncation.

Summarizing this subsection we can say that as far as the running spectral dimension is concerned, the single-metric Einstein-Hilbert truncation is a fully reliable approximation to its bi-metric generalization. 
By its very definition, a $k$-dependent spectral dimension makes sense only if $\mathcal{D}_s(k)$ changes with $k$ at most `adiabatically' \cite{frankfrac}.
Basically $\mathcal{D}_s$ can be interpreted meaningfully only when it develops a plateau.
Yet, for all three types of trajectories, the single- and bi-metric truncations agree on the respective plateau structures, and on the values which $\mathcal{D}_s$ assumes there.

 \section[A brief look at \texorpdfstring{$\bm{d=2+\epsilon}$}{d = 2+epsilon} and \texorpdfstring{$\bm{d=3}$}{d = 3}]{A brief look at $\bm{d=2+\epsilon}$ and $\bm{d=3}$} \label{sec:dOther}
Gravity in, or near two dimensions has always been an important theoretical laboratory for quantum gravity. 
In particular, the Asymptotic Safety scenario was first proposed in the $2+\epsilon$ dimensional Einstein-Hilbert theory.
In this section we re-analyze this theory in the bi-metric setting.
Because of its universality properties, absent in higher dimensions, our findings about the relation between the single- and bi-metric treatment are particularly clearcut, and in fact quite striking.

Also three dimensions are of special interest since in $d=3$ the metric and ghost fluctuations compensate exactly {\it on shell}.
For the EAA which is  a typical off shell object this implies by no means that there is no RG flow in $d=3$ (as is often believed wrongly).
Rather, while certain characteristic terms indeed disappear from the beta-functions, there is still a non-trivial RG running which needs to be taken seriously, for instance, when one uses the EAA to construct the continuum limit of a regularized functional integral for gravity. 
In $d=3$ we have the advantage that this characteristic `off-shell running' can be studied {\it in isolation}.

In the following two subsections we discuss $d=2+\epsilon$ and $d=3$ in turn; a complete list of the pertinent beta-functions can be found in the Appendix.

\subsection{Near dimension two}
The case of $d=2+\epsilon$ dimensions is special in that all Newton constants become dimensionless for $\epsilon\rightarrow 0$. In the lowest nontrivial order in $\epsilon$, and for vanishing (dynamical) cosmological constant, all anomalous dimensions have the structure
\begin{align}
 \eta^{\cix}= -b^{\cix}\,\tg^{\cix}+ \Order{\epsilon} \,,\qquad \cix\in \{\dyn,\,\background,\,(p)\} \label{eqn:dOther_001}
\end{align}
with certain constants $b^{\cix}$.
For $\epsilon\searrow0$, the leading term in $\eta^{\cix}$ is of order $\epsilon^0$. 
Within our approximation of retaining in all formulae only the lowest nontrivial order the condition for a non-Gaussian fixed point $(\epsilon+\eta^{\cix}=0)$ has a solution  which is linear in $\epsilon$, namely $\tg_*^{\cix}=\epsilon\slash b^{\cix}$. 
Note also that in the approximation \eqref{eqn:dOther_001} the general relationship \eqref{eqn:res4D_016} connecting the level- to the $\dyn$-$\background$-language boils down to the simple statement
\begin{align}
 b^{(0)}=b^{\background}+b^{\dyn} \quad\text{ and }\quad b^{(p)}=b^{\dyn}\text{ for }p\geq1. \label{eqn:dOther_001B}
\end{align}

In the single-metric Einstein-Hilbert truncation \cite{mr}, too, the anomalous dimension is well known to have the structure \eqref{eqn:dOther_001}. In its dia- vs. para-magnetic decomposed form, the pertinent coefficient $b^{\cix}\equiv b^{\sm}$ was found to be \cite{andi1}
\begin{align}
 b^{\sm}=\frac{2}{3}\Big[(-3)_{\text{dia}} +(4)_{\text{gh-dia}}+ (6)_{\text{para}} + (12)_{\text{gh-para}} \Big]= \frac{2}{3}\times [+19] =\frac{38}{3} 
\label{eqn:dOther_004}
\end{align}
All four contributions to $b^{\sm}$ are separately universal, i.e. independent of the cutoff shape function $R^{(0)}$.

Guided by our experience from 4 dimensions we might expect that the result of the single-metric calculation approximately agrees at some level of accuracy with that in the $\dyn$ sector of the bi-metric computation. 
In particular $b^{\sm}$ of \eqref{eqn:dOther_004} should not be very different from $b^{\dyn}$, the crucial coefficient in the anomalous dimension $\aDz$ of the dynamical Newton constant $G^{\dyn}_k$. What we actually find reads as follows:
\begin{align}
b^{\dyn}&=
 \frac{2}{3}\Big[-2\big(3+2\rhoG+15\rhoP\big)  \, \ThrfA{2}{1}{0}
+12\big(7 -2\rhoG \big)\rhoP\,
     \, \ThrfA{3}{2}{0} 
 \Big] \nonumber\\
&= \frac{2}{3}\Big[-6-4\rhoG+12\rhoP -12\rhoG\rhoP
    \Big]
\label{eqn:dOther_005}
\end{align}
The dia-magnetic contributions are obtained from this expression by setting $\rhoP=0$, and the graviton part by letting $\rhoG=0$; furthermore, the paramagnetic contributions are those proportional to $\rhoP$, and the ghost part is  proportional to $\rhoG$. 
Hence, when written in the style of the single-metric result, eq. \eqref{eqn:dOther_005} reads
\begin{align}
 b^{\dyn}=\frac{2}{3}\Big[(-6)_{\text{dia}} +(-4)_{\text{gh-dia}}+ (12)_{\text{para}} + (-12)_{\text{gh-para}} \Big]= \frac{2}{3}\times [-10]=-\frac{20}{3}
\label{eqn:dOther_006}
\end{align}
Obviously $b^{\sm}$ and $b^{\dyn}$ are quite different, not even their signs are in agreement so that screening and anti-screening behavior get interchanged. 
While the single-metric calculation predicts $b^{\sm}>0$, hence a NGFP at $\tg_*^{\sm}>0$ (for $\epsilon>0$), the bi-metric analogue has $b^{\dyn}<0$ with a corresponding fixed point at a {\it negative} Newton constant, something one normally considers unphysical. 
This clash is particularly striking since, like in $b^{\sm}$, {\it all 4 separate contributions appearing in the dia$\slash$para, metric$\slash$ghost decomposition of $b^{\dyn}$ are separately scheme independent, but none of them agrees with its single-metric analogue.}
Indeed, in the second line of \eqref{eqn:dOther_005} we exploited that all threshold functions of the type $\ThrfA{n+1}{n}{{\textstyle0}}$, for vanishing argument, assume the universal, i.e. $R^{(0)}$-independent values $\ThrfA{n+1}{n}{0}=1\slash \Gamma(n+1)$, $n\geq0$. 
In this sense, both the single-  and the bi-metric results can be considered particularly robust and `clean'.

Turning to the  anomalous dimension of the $\background$-sector we find likewise
\begin{align}
b^{\background}&=
\frac{2}{3}\Big[(4\rhoG-3)\ThrfA{1}{0}{0}+2\big(3+2\rhoG+6(3+\rhoG)\rhoP\big)\ThrfA{2}{1}{0}-12\big(7-2\rhoG)\rhoP \ThrfA{3}{2}{0}	 \Big] \nonumber\\
&=\frac{2}{3}\Big[	3 +8\rhoG -6\rhoP +24\rhoG\rhoP\Big]\,. \label{eqn:dOther_007}
\end{align}
This expression, again, contains only universal values of the threshold functions. Casting \eqref{eqn:dOther_007} in a more instructive way, we obtain
\begin{align}
 b^{\background}=\frac{2}{3}\Big[(3)_{\text{dia}} +(8)_{\text{gh-dia}}+ (-6)_{\text{para}} + (24)_{\text{gh-para}} \Big]\equiv \frac{2}{3}\times [+29]=\frac{58}{3}
\label{eqn:dOther_008}
\end{align}
Notice  that in total $b^{\background}$   is {\it positive} while $b^{\dyn}$ was negative. 

What makes these findings particularly alarming, or at least puzzling at first sight is that they show a considerable degree of internal consistency.
To see this, let us add up the $b$-coefficients of the background and the dynamical sector, thereby maintaining the dia$\slash$para, metric$\slash$ghost decomposition:
\begin{align}
 b^{\background}+b^{\dyn}=\frac{2}{3}\Big[ (3-6)_{\text{dia}} +(8-4)_{\text{gh-dia}}+ (-6+12)_{\text{para}} + (24-12)_{\text{gh-para}}	\Big]=b^{\sm}
\label{eqn:dOther_008B}
\end{align}
Remarkably enough, not only does the sum $b^{\background}+b^{\dyn}$ exactly agree with the old single-metric result, even all 4 terms of the decomposition separately do so.
The `miracle' behind \eqref{eqn:dOther_008B} finds its explanation when we evaluate the level-$(0)$ anomalous dimension, the corresponding coefficient $b^{(0)}$ being
\begin{align}
b^{(0)}&=
\frac{2}{3}\Big[(4\rhoG-3)\ThrfA{1}{0}{0}+6\big(1+2\rhoG\big)\rhoP\ThrfA{2}{1}{0} \Big] \nonumber\\
&=\frac{2}{3}\Big[	-3 +4\rhoG +6\rhoP +12\rhoG\rhoP\Big]
\label{eqn:dOther_010}
\end{align}
This result coincides, not only as a sum but even term by term, exactly with  the single-metric result  \eqref{eqn:dOther_004}:
\begin{align}
 b^{(0)}=\frac{2}{3}\Big[(-3)_{\text{dia}} +(4)_{\text{gh-dia}}+ (6)_{\text{para}} + (12)_{\text{gh-para}} \Big]=b^{\sm}
\label{eqn:dOther_011}
\end{align}
From our discussion in Section \ref{sec:sm_bm} of the relation between single- and bi-metric beta-functions the equality $b^{(0)}=b^{\sm}$ was indeed to be expected; there we demonstrated that quite generally the single-metric RG equations are closely related to the level-($0$) ones in the bi-metric computation. It is reassuring to see this rule at work here, despite potential subtleties related to the limit $\epsilon\rightarrow0$.

On the other hand, from \eqref{eqn:dOther_001B} we know that at the level of the $b$-coefficients the translation rule from the $\dyn$-$\background$ to the level description is simply $b^{(0)}=b^{\background}+b^{\dyn}$. Combining this with $b^{(0)}=b^{\sm}$ from \eqref{eqn:dOther_011} we obtain $b^{\background}+b^{\dyn}=b^{\sm}$, and this is precisely what we found \eqref{eqn:dOther_008B}!

\noindent At this point several important remarks are in order:

\noindent {\bf (A)} The main message conveyed by the above universal numbers is that {\it the leading order in $\epsilon$ is suffering from a significant violation of split-symmetry}, as is testified by the values  $b^{(0)}=b^{\sm}=+\frac{38}{3}$ at level zero, and $b^{(1,2,3,\cdots)}=b^{\dyn}=-\frac{20}{3}$ at the higher levels. 
Equivalent to that, there is a corresponding disagreement between the single- and bi-metric results, even at the level of signs.

\noindent {\bf (B)} In order to judge to what extent these somewhat disconcerting findings might carry over to higher dimensions, $d=4$ in particular, the (perhaps) worried reader should recall the following facts. 

\noindent {\bf (i)} In general dimensions, $\eta^{\dyn}(\tg^{\dyn},\KkD)$ is given by eq. \eqref{eqn:appx_BetaD_003}, and it involves two functions of the dynamical cosmological constant, $\kAD$ and $\kBD$. 
Expanding this anomalous dimension in $\tg^{\dyn}$, and retaining the leading, i.e. linear term only, it reads
\begin{align}
 \aDz(\tg^{\dyn},\KkD;d)= \kAD(\KkD;d) \, \tg^{\dyn}+ \Order{\tg^{\dyn\, 2}}
\label{eqn_dOther_eta1}
\end{align}
Upon an additional expansion of $\kAD$ for small $\KkD$ the approximation for $\aDz$ boils down to 
\begin{align}
 \aDz(\tg^{\dyn},\KkD;d)=-(d-2)\omega_d^{\dyn} \, \tg^{\dyn}+ \Order{\tg^{\dyn\, 2}}+\Order{\KkD}
\label{eqn_dOther_eta2}
\end{align}
It is characterized by a single coefficient only, $\omega_d^{\dyn}\equiv - \kAD(0;d)\slash (d-2)$.

\noindent {\bf (ii)} If we set  $d\equiv2+\epsilon$ and then expand $\aDz$ of eq. \eqref{eqn:appx_BetaD_003} in powers of $\epsilon$ we obtain $\aDz=-b^{\dyn}\,\tg^{\dyn}+\Order{\epsilon}$ for the lowest non-trivial order in $\epsilon$, with $b^{\dyn}=\lim_{\epsilon\rightarrow0} \big(\epsilon\,\omega_{2+\epsilon}^{\dyn}\big)$
a well defined, and non-zero number.
Notice that this simple equation for $\aDz$ has the same structure as the doubly expanded (in $\tg^{\dyn}$ and $\KkD$) general result \eqref{eqn_dOther_eta2} valid for all $d$. 
Here, however, it results from nothing more than the $\epsilon\rightarrow0$ limit alone \cite{mr}; no separate assumption about the smallness of the couplings and, based on that, expansions with respect to $\tg^{\dyn}$, $\KkD$ are invoked. 
Stated differently, if we retain only at the lowest order in $\epsilon$, it is unavoidable that $\aDz$ becomes linear in $\tg^{\dyn}$ and, more importantly, that the coefficient function $\kAD(\KkD)$ automatically always gets evaluated at the point $\KkD=0$ only.
So, the essential conclusion for the present discussion is that {\it in the limit $\epsilon\rightarrow0$, to leading order in $\epsilon$, the RG flow completely `forgets' about how a non-zero value of $\KkD$ affects the RG running of $\tg^{\dyn}$.}

\noindent
{\bf (iii)} In accord with this last remark, even the NGFP at $\tg_*^{\dyn}=\epsilon\slash b^{\dyn}$ is fully determined by $b^{\dyn}$, i.e. by $\kAD(\KkD=0)$. 
For $\epsilon\searrow0$, the very same coefficient which for general $d$ controls the semiclassical regime near the GFP only also decides about whether or not there exists the desired NGFP on the $\tg^{\dyn}>0$ half-space. 
We saw that this led to a clash between the single- and bi-metric calculation, since $b^{\sm}>0$, but $b^{\dyn}<0$. 
However, in Subsection \ref{subsec:paramagnetic} we also saw already that in $d=4$ the situation is different in a crucial way. There $\kAD(\KkD)$ changes its sign between $\KkD=0$ and $\KkD=\KkD_*$; as a result, the semiclassical regime on the $\tg^{\dyn}>0$ half-space exhibits gravitational screening ($\aDz>0$), as in $d=2+\epsilon$, while for larger values of $\KkD$ the sign of $\aDz$ turns negative, ultimately approaches $\aDz_*=-2$, and a NGFP forms at a {\it positive} value of Newton's constant, exactly where we would like it to be.

\noindent {\bf (C)}
Summarizing the above remarks, we can say that {\it the most problematic property of the $(2+\epsilon)$-dimensional theory, the dynamical Newton constant having a negative fixed point value%
\footnote{We assume $\epsilon>0$ here, as always.}, 
is an artifact of the $\epsilon$-expansion.}
In $d=4$, instead, screening at small $\KkD$ can coexist with anti-screening and a NGFP at larger values of the cosmological constant thanks to the $\KkD$-dependence of $\aDz$, a property the leading order of the $\epsilon$-expansion is completely insensitive to.

\subsection{Three dimensions}
Next, let us turn our attention to spacetimes of dimension $d=3$. 
The full set of results is given in Appendix \ref{app:BetaD3}. 
First we focus on the anomalous dimensions in the semiclassical regime which, for all $d$, read $\eta^{\cix}=-(d-2)\,\omega_d^{\cix}\,\tg^{\cix}+ \Order{\tg^{\cix\,2},\KkD}$. 
In $d=2+\epsilon$ the coefficients $\omega_d^{\cix}$ were universal, in $d=3$ they explicitly depend on the cutoff $R^{(0)}$ instead.
To make our main point as clear as possible, it is instructive to write down the dia$\slash$para and metric$\slash$ghost decomposed form of the coefficients appearing in $\eta^{\cix}=-\omega^{\cix}_3\, \tg^{\cix}+\Order{\tg^{\cix\,2},\KkD}$ for $\cix=\{\dyn,\,\background,\, (0)\}$. In the $\dyn$ and $\background$ sector we find, respectively:
\begin{align}
&\omega_3^{\dyn}=
  - \frac{1}{\sqrt{\pi}}\,\Big[\big(3+2\rhoG+2\rhoP(9-4\rhoG)\big) \ThrfA{2}{3\slash 2}{0}
-6\rhoP \big(9 -4\rhoG\big) \ThrfA{3}{5\slash2}{0} 
 \Big]
\label{eqn:dOther_013}
\\
& \omega_3^{\background}=  
 \frac{1}{\sqrt{\pi}}\, \Big[
2(\rhoG-1) \ThrfA{1}{1\slash 2}{0} 
+ \big(3+2\rhoG +4\rhoP(6-\rhoG)\big)\, \ThrfA{2}{3\slash 2}{0}
 -6\,  \rhoP\,\big(9-4\rhoG\big)
   \ThrfA{3}{5\slash2}{0} 
\Big]\nonumber
\end{align}
Notice that $\omega_3^{\dyn}$ and $\omega_3^{\background}$ are perfectly generic, in the sense that they receive contributions of both paramagnetic and diamagnetic origin, from both ghost and metric fluctuations.

The expected `magic' happens only in the case of  the level-$(0)$ coefficient: 
From the general identity \eqref{eqn:trA27D} we  infer $\omega^{(0)}_3=\omega_3^{\dyn}+\omega_3^{\background}$ in the present approximation, yielding
\begin{align}
 \omega_3^{(0)}&=
 \frac{2}{\sqrt{\pi}}\, \Big[ (\rhoG-1)\,\ThrfA{1}{1\slash 2}{0} + \rhoP (3+2\rhoG ) \ThrfA{2}{3\slash 2}{0}\Big]\label{eqn:dOther_015}
\end{align}
Upon `switching on' the ghosts by setting $\rhoG=1$, the entire coefficient $\omega_3^{(0)}$ is seen to be proportional to $\rhoP$.
Hence {\it the level-(0) anomalous dimension $\eta^{(0)} \propto \omega_3^{(0)}$ is of purely paramagnetic origin}. 

The non-zero diamagnetic contributions from $\omega^{\dyn}_3$ and $\omega^{\background}_3$ have canceled precisely.
At the level of the bi-metric EAA, for generic arguments (`off shell' in particular), this cancellation is the only immediate reflection of the fact that classical Einstein-Hilbert gravity has no propagating modes in $d=3$.
There is no comparable compensation of metric and ghost modes at higher orders.

The observed perfect cancellation of all diamagnetic terms in $\omega_3^{(0)}$ is fully consistent with earlier results on the single-metric case \cite{andi1} where the same cancellation was found to occur in the analogous coefficient $\omega_3^{\sm}$. Indeed, it can be verified again that $\omega_3^{(0)}=\omega_3^{\sm}$, as it should be for general reasons.

Finally, let us leave the semiclassical regime and  consider the possibility of non-Gaussian fixed points in $3$ dimensions. 
Using the full fledged beta-functions of  Appendix \ref{app:BetaD3} we do indeed find such fixed points. 
The results can be summarized as  follows:
\begin{align}
\begin{tabular}[h!]{|l||c|c|c|}
\hline
	d=3			  & $\substack{\text{\fpnD-\fpL{}}\\ (\tg_*^{\dyn}= 0.13,\,\KkD_*= 0.2)}$ & $\substack{\text{\fpnDn-\fpL{}}\\ (\tg_*^{\dyn}= -0.38,\,\KkD_*=-0.25)}$	&	$\substack{\text{\fpgD-\fpL{}}\\ (\tg_*^{\dyn}= 0,\,\KkD_*= 0)}$ \\[2.2ex] \hline \hline
&&& \\[0pt] 
\text{\fpnB-\fpL{}}	 &  	$\substack{\text{\fpnB\fpC\fpnD-\fpL{}}\\ (\tg_*^{\background}= 1.3,\,\KkB_*= -0.97)}$		&$\substack{\text{\fpnB\fpC\fpnDn-\fpL{}}\\ (\tg_*^{\background}= 0.17,\,\KkB_*= -0.14)}$ &$\substack{\text{\fpnB\fpC\fpgD-\fpL{}}\\ (\tg_*^{\background}= 0.18,\,\KkB_*= -0.15)}$\\[2.2ex] \hlinewd{0.2pt}
&&& \\[1pt] 
\text{\fpgB-\fpL{}} 	 &  $\substack{\text{\fpgB\fpC\fpnD-\fpL{}}\\ (\tg_*^{\background}= 0,\,\KkB_*=0)}$			& $\substack{\text{\fpgB\fpC\fpnDn-\fpL{}}\\ (\tg_*^{\background}=0,\,\KkB_*=0)}$ & $\substack{\text{\fpgB\fpC\fpgD-\fpL{}}\\ (\tg_*^{\background}= 0,\,\KkB_*= 0)}$\\[2.2ex]
\hline
\end{tabular}	\label{eqn:res3D_017B}
\end{align}
We find, as in $d=4$, a total of six fixed points, five of which are non-Gaussian. 
Remarkably enough, the qualitative picture is exactly the same as in  the four dimensional case, displaying a non-Gaussian fixed point both in the upper and lower half-plane of $\tg^{\dyn}$, as well as a Gaussian one at $\tg_*^{\dyn}=0$. 
Instead, the background coordinate value $\tg^{\background}_*$ ($\KkB_*$) is found to be positive (negative). 
Furthermore, a relatively large hierarchy $\tg_*^{\dyn}\approx 10\, \tg^{\background}_*$ for the  doubly non-Gaussian \fpnB\fpC\fpnD-\fpL{}, indicating approximate split-symmetry, is found also in $d=3$.
 \section{Summary and conclusion} \label{sec:conclusion}
In this section we start with a summary of  our work by means of a brief and concise list displaying the main results.
We then close with a number of general conclusions and the essential lessons for future investigations, which we learned here.

\subsection{Summary of the main results}

\noindent {\bf (A)}
On the technical side, we employed and tested a new method for dealing with operator traces involving uncontracted covariant derivatives which is much simpler than those used before.
We found that, within the expected truncation uncertainty and cutoff dependence, the resulting RG flow matches the results obtained with the conventional method based on a York decomposition, even though, mathematically, the pertinent beta-functions are quite different.
We hope that similar methods will be helpful also in  future bi-metric calculations.

\noindent {\bf (B)}
On the 4-dimensional theory space $\mathcal{T}\equiv \big\{\big(\tg^{\dyn},\,\KkD,\,\tg^{\background},\,\KkB\big)\big\}$ the RG flow was found to decompose hierarchically according to 
$ \big(\tg^{\dyn},\KkD\big) \rightarrow \tg^{\background} \rightarrow \KkB$.
This allows us to compute the flow on the dynamical subspace $\mathcal{T}_{\dyn}\equiv \big\{\big(\tg^{\dyn},\,\KkD\big)\big\}$ without reference to the background couplings.

\noindent {\bf (C)}
The main characteristics of the 2-dimensional flow on $\mathcal{T}_{\dyn}$ are as follows:

 {\bf (c.1)}
There exist two non-Gaussian and one Gaussian fixed point on $\mathcal{T}_{\dyn}$, namely \fpnDn-\fpL{}, \fpnD-\fpL{}, and \fpgD-\fpL{}.
They are located at a negative, positive, and vanishing coordinate $\tg_*^{\dyn}$, respectively.

 {\bf (c.2)}
Reliability analyses reveal that \fpnD-\fpL{} is very robust, while \fpnDn-\fpL{} is likely to be a truncation artifact.

 {\bf (c.3)}
The critical exponents of \fpnD-\fpL{} are given by a complex conjugate pair, with a non-zero imaginary part leading to spiral-shaped trajectories.
The fixed point is UV attractive in both directions.

 {\bf (c.4)}
There are no RG trajectories crossing the $\tg^{\dyn}=0$ line, in particular there exists no cross-over trajectory connecting \fpnDn-\fpL{} to \fpnD-\fpL{}.
We may therefore restrict $\mathcal{T}_{\dyn}$ to the half-plane with $\tg^{\dyn}>0$ which is invariant under the flow.

 {\bf (c.5)}
The phase portrait on $\mathcal{T}_{\dyn}$ is very similar to the one obtained with the single-metric truncation.
In particular the properties of \fpnD-\fpL{} are numerically similar to those of the single-metric fixed point, and the RG trajectories admit exactly the same 
(`type Ia, IIa, IIIa') classification that has been introduced for the single-metric case.

 {\bf (c.6)}
In contrast to all single-metric based predictions, gravitational anti-screening is lost near the Gaussian fixed point \fpgD-\fpL{}.

\noindent {\bf (D)}
Each RG trajectory on $\mathcal{T}_{\dyn}$, together with initial conditions for $\tg^{\background}$ and $\KkB$, gives rise to a 4-dimensional trajectory, and we analyzed the corresponding flow on the complete theory space $\mathcal{T}$.

 {\bf (d.1)}
On $\mathcal{T}$, there exists a total of 6 fixed points, each one of the three $\dyn$-fixed points above can be combined with both a Gaussian and a non-Gaussian fixed point of the background couplings.

 {\bf (d.2)}
The doubly non-Gaussian fixed point on $\mathcal{T}$, \fpnB\fpC\fpnD-\fpL{}, is the natural candidate for the construction of an asymptotically safe infinite cutoff limit.
It possesses 4 relevant directions, so the UV critical hypersurface associated to it, $\cUV$, has maximum dimension within the present truncation.

\noindent {\bf (E)}
In order to define a quantum field theory, we need an RG trajectory which, at `$k=\infty$', starts out infinitesimally close to the fixed point, always runs on $\cUV$, and thereby gradually becomes split-symmetric, at the very least in the physical limit when $k$ approaches zero.
This is the crucial requirement of Background Independence.
Within the bi-Einstein-Hilbert truncation it requires $\EAA_k^{\text{grav}}$ to loose its `extra $\bg$-dependence', that is, $1\slash G_k^{\background}$ and $\KkbarB\slash G_k^{\background}$ must vanish in the IR limit of  low scales $k$ approaching zero.

 {\bf (e.1)}
For the first time, it was possible to demonstrate explicitly there do indeed exist trajectories which meet both of the two key requirements, Asymptotic Safety and Background Independence, simultaneously.
Those trajectories are labelled by only {\it two} free parameters. 
Thus the theory's predictivity is actually higher than  expected on the basis of a 4-dimensional $\cUV$.

 {\bf (e.2)}
The RG trajectories which restore split-symmetry in the IR (at $k=k_{\text{IR}}$) were found to be precisely those which merge with the `running UV attractor' at low scales, {\bf \AttrL}($k$).

\noindent {\bf (F)}
The relevance of the running UV attractor {\bf \AttrL}($k$) to the problem of split-symmetry restoration was uncovered by the following sequence of results and observations.

 {\bf (f.1)}
For every fixed trajectory $U^{\dyn}$ on $\mathcal{T}_{\dyn}$, i.e. $k\mapsto \big(\tg_k^{\dyn},\,\KkD_k\big)\equiv U^{\dyn}(k)$ the background couplings $\tg_k^{\background}$ and $\KkB_k$ are governed by a non-autonomous, i.e. explicitly RG-time dependent system of differential equations on the 2-dimensional subspace $\mathcal{T}_{\background}\equiv \big\{\big(\tg^{\background},\,\KkB\big)\big\}$.
It can be visualized as a time dependent vector field $\vec{\beta}_{\background}(k)$ on $\mathcal{T}_{\background}$, which also depends on the $\dyn$-trajectory chosen.

 {\bf (f.2)}
For every trajectory $U^{\dyn}$, and at every fixed RG time $k$, the vector field $\vec{\beta}_{\background}(k)$ has an (instantaneous) zero at $\big(\tg^{\background}_{\attr}(k),\,\KkB_{\attr}(k)\big)\in\mathcal{T}_{\background}$.
The (likewise instantaneous) stability analysis reveals that it is UV attractive in both $\background$-directions, hence the curve $k\mapsto \AttrL(k)\equiv \big(\tg_k^{\dyn},\,\KkD_k,\,\tg_{\attr}^{\background}(k),\,\KkB_{\attr}(k)\big)$ acts as a `running UV attractor'.
It is {\it not} an RG trajectory by itself.

 {\bf (f.3)}
The UV limit of the running attractor is precisely the doubly non-Gaussian fixed point: $\lim_{k\rightarrow\infty}\AttrL(k)=\text{\fpnB\fpC\fpnD-\fpL{}}$.

 {\bf (f.4)}
Picking a $\dyn$-trajectory $U^{\dyn}$ we are equipped with a $k$-dependent vector field $\vec{\beta}_{\background}$ on $\mathcal{T}_{\background}$.  
Integral curves  of $\vec{\beta}_{\background}(k^{\prime})$ at any fixed moment of `time', $k^{\prime}$, are {\it not} projections onto $\mathcal{T}_{\background}$ of an RG trajectory on the full theory space in general; they correspond to `snapshots' of the phase portrait on $\mathcal{T}_{\background}$ at this very moment.

 {\bf (f.5)}
For every `initial' point $\big(\tg^{\background}_{\text{in}},\,\KkB_{\text{in}}\big)$ specified on $\mathcal{T}_{\background}$ at the time $k_{\text{in}}$ there exists a unique solution of the RG equations through this point, $k\mapsto U^{\background}(k)$, with $U^{\background}(k_{\text{in}})=\big(\tg^{\background}_{\text{in}},\,\KkB_{\text{in}}\big)$.
It is obtained by integrating the (now explicitly $k$ dependent) equation $\partial_t U^{\background}(k)=\vec{\beta}_{\background}(U^{\background}(k);k)$ both upward and downward.
Making all input data  explicit, we denote this solution as 
\mbox{$U^{\background}\{U^{\dyn};k_{\text{in}}; \tg^{\background}_{\text{in}},\, \KkB_{\text{in}}\}(k)$.} 
It gives rise to an RG trajectory on the full theory space: $U(k)=\big(U^{\dyn}(k),U^{\background}(k)\big)$.

 {\bf (f.6)}
We find that  in the limit $k\gg k_{\text{in}}$ of a long upward evolution, $U^{\background}\{U^{\dyn};k_{\text{in}}; \tg^{\background}_{\text{in}},\, \KkB_{\text{in}}\}$ looses its memory of the initial point $\big(\tg^{\background}_{\text{in}},\,\KkB_{\text{in}}\big)$, and a universal limit curve is obtained:
$\lim_{ k\gg k_{\text{in}}} U^{\background}\{U^{\dyn};k_{\text{in}}; \tg^{\background}_{\text{in}},\, \KkB_{\text{in}}\}(k) = U^{\background}_{\attr}(k) $.

 {\bf (f.7)}
Every  $\dyn$ trajectory $U^{\dyn}$ implies a specific limit curve $U^{\background}_{\attr}(k)$, and together they define an RG trajectory on the full theory space: $k\mapsto U_{\attr}(k)= \big( U^{\dyn}(k),U^{\background}_{\attr}(k)\big))$. 
This trajectory is  precisely the one which, in the IR, ends on the running attractor $\AttrL(k)$.%
\footnote{Note that numerically constructing the trajectory reaching $\AttrL(k_{\text{IR}})$ involves fine-tuning, in the sense that the desired final point is IR repulsive in both $\background$-directions. From this perspective the UV attractor is better called an `IR repeller'.} 
It is determined by the `final condition' $\big(\tg^{\background}_{k_{\text{IR}}},\,\KkB_{k_{\text{IR}}}\big)=\big(\tg^{\background}_{\attr}(k_{\text{IR}}),\,\KkB_{\attr}(k_{\text{IR}})\big)$ with $k_{\text{IR}}\rightarrow0$.
In short, $U_{\attr}(k_{\text{IR}})= \AttrL(k_{\text{IR}})$, while in the opposite limit $U_{\attr}(k)$ approaches the doubly non-Gaussian fixed point: $U_{\attr}(k\rightarrow\infty)=\text{\fpnB\fpC\fpnD-\fpL{}}$.

 {\bf (f.8)}
We found that the class of all trajectories $U_{\attr}(k)$, obtained by varying the underlying $U^{\dyn}(k)$, are of special importance:
The trajectories in this class, and only those, restore split-symmetry in the IR, being at the same time asymptotically safe.
This explains the relevance of the running UV attractor to the problem of split-symmetry restoration.

\noindent {\bf (G)}
Performing a detailed comparison of the single- and the bi-metric Einstein-Hilbert truncation we found that, quite unexpectedly, the former is a rather precise approximation to the latter  in the vicinity of the non-Gaussian fixed point. 
In the far IR the split-symmetry restoring trajectories, by construction, give rise to another regime in which the two truncations agree well.  
However, in between there are strong qualitative differences, for instance with respect to the sign of the dynamical anomalous dimension $\aDz$. 
Furthermore, the quantitative differences of the critical exponents in both settings are quite significant, despite the `miraculous' precision of the single-metric truncation near the NGFP. They clearly show the limitations of the single-metric approximation. Results from an independent calculation using the  TT-decomposition techniques \cite{MRS2} support these findings.

\noindent {\bf (H)}
As a concrete application of the bi-metric flow in $d=4$ we computed the running spectral dimension $\mathcal{D}_s(k)$ of the emergent fractal spacetimes according to the definition proposed in \cite{frankfrac}.
We found that the bi- and single-metric results agree on all universal predictions this definition can give rise to, namely the formation of plateaus on which $\mathcal{D}_s(k)$ assumes the values $\mathcal{D}_s=4$, $\mathcal{D}_s=4\slash3$, and $\mathcal{D}_s=2$, respectively.

\noindent {\bf (I)}
We saw that in $d=3$ the bi-metric flow is very similar to the one in $d=4$, even though the classical Einstein-Hilbert theory in $d=3$ has no propagating modes.
This makes it particularly clear  that, in any dimension, due to the off-shell nature of the EAA not only the `radiative' field modes but also the `coulombic' ones play an important role in driving the RG evolution. 
This is in accord with the physical picture of `paramagnetic dominance' put forward in \cite{andi1, andi-MG}.

\noindent {\bf (J)}
The most striking and even qualitatively essential discrepancies between the single- and the bi-metric predictions we encountered in the leading order of the $\epsilon$-expansion about two dimension ($d=2+\epsilon$, $\epsilon>0$).
Disentangling metric \slash ghost and dia- \slash paramagnetic contributions, both the anomalous dimension $\aDz$ and its single-metric approximation $\eta^{\sm}$ are characterized by 4 {\it separately universal} coefficients.
All 4 of them were found to {\it disagree} between $\aDz$ and $\eta^{\sm}$.
Thus, contrary to the single-metric prediction, the bi-metric RG flow possesses {\it no NGFP at positive Newton constant}.
We were able to show that this discrepancy and the related strong breaking  of split-symmetry are an artifact of the leading order $\epsilon$-expansion, which does not generalize to higher dimensions.

\subsection{Conclusions and outlook}
The main message of the present investigations for future work on the Asymptotic Safety program is quite clear:
From a certain degree of precession onward it seems to make little sense to keep including further invariants in the truncation ansatz that are built from the dynamical metric alone. 
Rather, to the same extent we allow the effective average action to depend on $g_{\mu\nu}$ in a more complicated way, also its dependence on the background metric $\bg_{\mu\nu}$ must be generalized.
While the picture conveyed by the single-metric truncations is qualitatively correct often, they are certainly insufficient for quantitatively precise calculations, the determination of critical exponents for example.

Taking the bi-metric structure of the EAA seriously we find ourselves in a situation which is very widespread in quantum field theory:
One starts from a classical field theory with a certain symmetry, tries to quantize it, thereby discovers that some sort of regularization is needed, then picks a regulator which spoils the symmetry, perhaps because there exists no invariant regularization, and finally after the quantization one  tries to re-establish this symmetry at the observable level, or close to it.
A well known example of this situation is the quantization of the electromagnetic field in a scheme which violates gauge invariance.
Computing radiative corrections  one then might encounter a non-zero photon mass at the intermediate steps of the calculation.
At the end, however, it must always be possible to choose the bare parameters in such a way that the renormalized, physical photon mass turns out exactly zero.

In quantum gravity, the background quantum split-symmetry should be seen as analogous to gauge invariance in this example.
Split-symmetry is a formal way to express the arbitrariness of the background on which we quantize the metric fluctuations, and unbroken split-symmetry is tantamount to Background Independence.
The effective average action is not invariant per se, but the space of RG trajectories contains special solutions which re-establish split-symmetry at the physical level.

In this paper we showed explicitly that the restoration of split-symmetry is indeed possible, and how it can be achieved in practice.
In future bi-metric analyses of Quantum Einstein Gravity this will always be a central and important step.
The analogy with the photon mass makes it very clear that the implementation of split-symmetry deserves considerable attention since otherwise we never can be sure to deal with the right `universality class'. 
After all, comparing QEG to the familiar matter field theories on flat space, its most momentous distinguishing feature is Background Independence; it  implies in particular the necessity of an ab initio derivation of the arena all non-gravitational physics takes place in, namely spacetime.
Clearly this has much more profound consequences for the general structure of the theory than its notorious perturbative non-renormalizability, for example, which shows up at a secondary technical level only.

\paragraph{Acknowledgment:} We are grateful to Andreas Nink for a careful reading of the manuscript.
\clearpage
\begin{appendix}
\section*{Appendix}
\section{Beta-functions for all spacetime dimensions} \label{sec:BetaD}
In this appendix, we list the beta-functions of all dimensionless couplings for general spacetime dimension $d$. The conversion rules relating  dimensionless couplings to  dimensionfull ones are 
$ \tg^{\cix}_k = G_k^{\cix}\, k^{d-2}$, $\Kk^{\cix}_k = \Kkbar^{\cix}_k\, k^{-2}$ for all $\cix\in\{\dyn,\,\background,\, (0),\,(1),\,(2),\,\cdots\}$.
They entail the following relations between their scale derivatives:
\begin{subequations}
\begin{align}
 \partial_t \tg^{\cix}_k &= (d-2)k^{(d-2)}G^{\cix}_k + k^{d-2}\partial_t G^{\cix}_k=\big(d-2+\eta^{\cix}_k\big)\tg^{\cix}_k \equiv \beta_{\tg}^{\cix}\,,
\label{eqn:appx_BetaD_001}
\end{align}
\begin{align}
\partial_t \Kk^{\cix}_k&= k^{-2}\partial_t \Kkbar^{\cix}_k -2 k^{-2}\Kkbar^{\cix}_k =  k^{-2}\partial_t \Kkbar^{\cix}_k - 2 \Kk^{\cix}_k \equiv \beta_{\Kk}^{\cix}
\label{eqn:appx_BetaD_002}
\end{align}
\end{subequations}
In the following we will list the dimensionless beta-functions in their explicit, $d$-dependent form first in the $\{\text{$\dyn$,$\background$}\}$, then in the level-description.
The threshold functions $\ThrfA{p}{n}{w}$ and $\ThrfB{p}{n}{w}$ which they contain are defined as in ref. \cite{mr}. To identify the paramagnetic graviton and ghost contributions, respectively, the corresponding terms are multiplied by the factors $\rhoP$ and $\rhoG$ which should be put to unity if this information is not needed. In the first subsection we allow for arbitrary cutoff shape functions, in the second we specialize for the `optimized' one.

\subsection{Arbitrary cutoff shape funtion} \label{app:BetaDGT}

\paragraph{The anomalous dimension related to $\tg^{\dyn}$:}
The additional term in $\beta_{\tg}^{\dyn}$ that adds to the canonical $k$-dependence in eq. \eqref{eqn:appx_BetaD_001} is the anomalous dimension $\aDz$ that encapsulates all the non-trivial contributions generated by the RG flow. In case of the anomalous dimension for `$\dyn$'-Newton constant, we have to solve an implicit equation to obtain the corresponding  $\aDz$, since it also appears on the RHS of the flow equation. It  yields
\begin{align}
 \aDz(\tg^{\dyn},\KkD;d)&= \frac{\kAD(\KkD;d)\,\tg^{\dyn}}{1-\kBD(\KkD;d)\,\tg^{\dyn}}
\label{eqn:appx_BetaD_003}
\end{align}
Here, $\kAD(\KkD;d)$ and $\kBD(\KkD;d)$ are functions depending on the {\it dynamical} cosmological constant $\KkD$ only, as well as parametrically on the spacetime dimension $d$. Their explicit forms are, for the function in the numerator,
\begin{align}
\kAD(\KkD;d)&= \frac{d}{(4\pi)^{\frac{d}{2}-1}}\,\Big\{-\tfrac{(d-6)(d+1)}{6}  \, \ThrfA{2}{(d\slash 2)}{-2\KkD}
\, \nonumber \\
&\phantom{= \frac{d}{(4\pi)^{\frac{d}{2}-1}}\,\Big\{}+\frac{2}{3}\tfrac{(d-4)(d+1)}{(d-2)}\, \KkD \, \ThrfA{2}{(d-2)\slash 2}{-2\KkD}
\,  \nonumber \\
&\phantom{= \frac{d}{(4\pi)^{\frac{d}{2}-1}}\,\Big\{}
+2\tfrac{(d-6)(d-1)}{(d-2)}   \rhoP\,
   \big[  d \, \ThrfA{3}{(d+2)\slash2}{-2\KkD} 
-  \ThrfA{2}{(d\slash 2)}{-2\KkD} \big] \nonumber \\
&\phantom{= \frac{d}{(4\pi)^{\frac{d}{2}-1}}\,\Big\{}
-8\tfrac{(d-4)(d-1)}{(d-2)}\rhoP\, \KkD\, \ThrfA{3}{(d\slash 2)}{-2\KkD}
 \, \Big\} \nonumber \\ &\quad
+\frac{4  }{(4\pi)^{\frac{d}{2}-1}} \rhoG \,\Big\{ \big(\tfrac{d}{3}-\tfrac{4}{(d-2)}\rhoP\big) \ThrfA{2}{(d\slash 2)}{0} + \tfrac{4d}{(d-2)} \rhoP\ThrfA{3}{(d+2)\slash 2}{0} \Big\}\,
\label{eqn:appx_BetaD_004}
\end{align}
Likewise the function in the denominator reads:
\begin{align}
\kBD(\KkD;d)&= \frac{d}{2(4\pi)^{\frac{d}{2}-1}}\,\Big\{\tfrac{(d-6)(d+1)}{6}  \, \ThrfB{2}{(d\slash 2)}{-2\KkD}
\, \nonumber \\
&\quad \phantom{\quad + (4\pi)^{-\frac{d}{2}}\epsilon\,\, }-\frac{2}{3}\tfrac{(d-4)(d+1)}{(d-2)}\, \KkD \, \ThrfB{2}{(d-2)\slash 2}{-2\KkD}
\,  \nonumber \\
&\quad \phantom{\quad + (4\pi)^{-\frac{d}{2}}\epsilon\,\, }
-2\tfrac{(d-6)(d-1)}{(d-2)}   \rhoP\,
   \big[  d \, \ThrfB{3}{(d+2)\slash2}{-2\KkD} 
-  \ThrfB{2}{(d\slash 2)}{-2\KkD} \big] \nonumber \\
&\quad \phantom{\quad + (4\pi)^{-\frac{d}{2}}\epsilon\,\, }
+8\tfrac{(d-4)(d-1)}{(d-2)}\rhoP\, \KkD\, \ThrfB{3}{(d\slash 2)}{-2\KkD}
 \, \Big\} \label{eqn:appx_BetaD_005}
\end{align}
 Notice that in the present truncation the dimensionalities $d=4$ and $d=6$ play a special role: in these cases the graviton contributions to $\kAD$ and $\kBD$ are either all given by terms with an extra factor of $\KkD$ multiplying the threshold functions, in $d=6$, or precisely those terms are all absent, in $d=4$. This pattern is not found in the ghost contributions (which can be identified by their $\rhoG$ factor).

\paragraph{The beta-function of $\KkD$:}
The running of the dynamical cosmological constant is governed by the beta-function
\begin{align}
\beta_{\Kk}^{\dyn}(\tg^{\dyn},\KkD;d)
&= (\aDz-2)\,\KkD +  \tg^{\dyn} \frac{4d\,\rhoG}{(4\pi)^{\frac{d}{2}-1}} \,\ThrfA {2}{(d+2)\slash 2}{0} 
 \label{eqn:appx_BetaD_006} \\
&\quad + \tg^{\dyn} \frac{(d+1)}{(4\pi)^{\frac{d}{2}-1}} \,
 \Big\{ 2(d-4)\, \KkD \,\qA{2}{(d\slash 2)}{-2\KkD}
-\tfrac{(d-6) d}{2}\qA{2}{(d+2)\slash 2}{-2\KkD}  
\Big\} \nonumber
\end{align}
Note that the non-canonical graviton contributions on the RHS of \eqref{eqn:appx_BetaD_006} show the same property as discussed above: they are proportional to {\it either} $(d-4)\KkD$ {\it or} $(d-6)$.

Together with eq. \eqref{eqn:appx_BetaD_005}, eq. \eqref{eqn:appx_BetaD_006} gives rise to a closed system of two coupled differential equations which can be solved independently of the other equations.

\paragraph{The anomalous dimension related to $\tg^{\background}$:}
The non-canonical $k$-dependence of the background Newton constant is given by the anomalous dimension
\begin{align}
&\eta^{\background}(\tg^{\dyn},\KkD,\tg^{\background};d)= \label{eqn:appx_BetaD_007}\\ \nonumber
&\phantom{\eta^{\background}}=  
 \frac{d}{(4\pi)^{\frac{d}{2}-1}}\, \Big\{
\frac{(d+1)}{3} \qA{1}{(d-2)\slash 2}{-2\KkD}
-\frac{2}{3}\tfrac{(d-4)(d+1)}{(d-2)}\, \KkD \, \qA{2}{(d-2)\slash 2}{-2\KkD}
\,   \\
&\phantom{\eta^{\background}}\quad \phantom{\quad + (4\pi)^{-\frac{d}{2}}\epsilon\,\, }
+ \big(\tfrac{(d-6)(d+1)}{6} -8\tfrac{(d-1)}{(d-2)}\rhoP\big)\, \qA{2}{(d\slash 2)}{-2\KkD}\nonumber \\
%
&\phantom{\eta^{\background}}\quad \phantom{\quad + (4\pi)^{-\frac{d}{2}}\epsilon\,\, }+8\tfrac{(d-4)(d-1)}{(d-2)}\rhoP\, \KkD \, \qA{3}{(d\slash 2)}{-2\KkD}
 \nonumber \\
&\phantom{\eta^{\background}}\quad \phantom{\quad + (4\pi)^{-\frac{d}{2}}\epsilon\,\, } -2\tfrac{(d-6)(d-1) d}{(d-2)}\,  \rhoP\,
   \qA{3}{(d+2)\slash2}{-2\KkD} \qquad
 \,\,\, \Big\}\,\,\tg^{\background} \nonumber \\ &\quad
-\frac{4}{(4\pi)^{\frac{d}{2}-1}} \rhoG \,\Big\{ \frac{d}{3}  \ThrfA{1}{(d-2)\slash 2}{0}    + \big(\tfrac{d}{3}+\tfrac{2(d-4)}{(d-2)}\rhoP\big) \ThrfA{2}{(d\slash 2)}{0} + \tfrac{4d}{(d-2)} \rhoP\ThrfA{3}{(d+2)\slash 2}{0} \Big\}\,\tg^{\background} \nonumber
\end{align}
The property found of $\aDz$ concerning its $\KkD$-dependence in $d=4$ and $d=6$ is only partially shared by $\eta^{\background}$; there are contributions from the graviton sector that neither drop out for $d=4$ nor for $d=6$. 

\paragraph{The beta-function of $\KkB$:}
The last member in the hierarchicy of the differential system is the cosmological constant $\KkB$, which does not influence - but in turn shows a dependence on - all the other 3 couplings. Its beta-function is given by
\begin{align}
& \beta_{\Kk}^{\background}(\tg^{\dyn},\KkD,\tg^{\background},\KkB;d) = (\eta^{\background}-2)\KkB 
- \frac{4  d}{(4\pi)^{\frac{d}{2}-1}} \rhoG \Big\{\ThrfA {1}{(d\slash 2)}{0} +  \,\ThrfA {2}{(d+2)\slash 2}{0}  \Big\}\, \tg^{\background} 
\nonumber \\
&\qquad 
+ \frac{(d+1)}{(4\pi)^{\frac{d}{2}-1}}
\Big\{		
d\, \qA{1}{(d\slash 2)}{-2\KkD}
-2(d-4)\, \KkD \, \qA{2}{(d\slash 2)}{-2\KkD} 
+(d-6) \frac{d}{2}  \qA{2}{(d+2)\slash 2}{-2\KkD}
\Big\} \,\tg^{\background}
\label{eqn:appx_BetaD_008}
\end{align}

At this point all 4 RG equations of the $\dyn$-$\background$ system are fully specified.

In the level language, the $\dyn$ beta-functions provide the beta-functions at all higher levels $p=1,2,3,\cdots$
\begin{align*}
\eta^{(p)}=\eta^{\dyn} \quad\text{ and } \quad \beta_{\Kk}^{(p)}=\beta_{\Kk}^{\dyn}\quad \text{ for all } p\geq1.
\end{align*}
What is special are the level-$(0)$ couplings. They are governed by combinations of the `$\background$' and `$\dyn$' beta-functions.

\paragraph{The anomalous dimension related to $\tg^{(0)}$:}
The anomalous dimension of the level-($0$) Newton coupling fulfills 
\begin{align*}
  \frac{\eta^{(0)}}{\tg^{(0)}} =\frac{ \eta^{\dyn}}{\tg^{\dyn}} + \frac{\eta^{\background}}{\tg^{\background}},
\end{align*}
 from which,  using eqs. \eqref{eqn:appx_BetaD_005} and \eqref{eqn:appx_BetaD_007}, it follows that
\begin{align}
 \eta^{(0)}&=
+ \frac{2\, d}{(4\pi)^{\frac{d}{2}-1}}\, \Big\{\frac{(d+1)}{6} \qA{1}{(d-2)\slash 2}{-2\KkD}
- (d-1)\rhoP\, \qA{2}{(d\slash 2)}{-2\KkD}\Big\}\tg^{(0)} \nonumber \\&\quad
-\frac{4 }{(4\pi)^{\frac{d}{2}-1}} \rhoG \Big\{  \frac{d}{3}  \ThrfA{1}{(d-2)\slash 2}{0}    +2\rhoP \,\ThrfA{2}{(d\slash 2)}{0}\Big\}\tg^{(0)}
\label{eqn:appx_BetaD_009}
\end{align}
For notational consistency, one should identify $\KkD=\Kk^{(1)}=\Kk^{(2)}=\cdots$ on the RHS of this equation.

As a check on our calculation, we mention that if instead we identify $\aDz=\eta^{(0)}$ and $\KkD=\Kk^{(0)}$ in the $\qA{p}{n}{-2\KkD}$  functions we reobtain precisely the anomalous dimension of the single-metric truncation found in \cite{mr},  $\eta^{(0)}=\eta^{\sm}$, as it should be. (See also Section \ref{sec:sm_bm})

\paragraph{The beta-function of $\Kk^{(0)}$:}
For the cosmological constant  at level-$(0)$ one finds
\begin{align}
\beta_{\Kk}^{(0)}
&= (\eta^{(0)}-2)\Kk^{(0)}+  \tg^{(0)}\,  \frac{d(d+1)}{(4\pi)^{\frac{d}{2}-1}} 
 \, \qA{1}{(d\slash 2)}{-2\KkD}  -\tg^{(0)}\frac{4d}{(4\pi)^{\frac{d}{2}-1}}  \rhoG \ThrfA {1}{(d\slash 2)}{0} 
\label{eqn:appx_BetaD_010}
\end{align}
where $\KkD=\Kk^{(1)}=\Kk^{(2)}=\cdots$ in the present truncation. Again, it can be checked that $\beta_{\Kk}^{(0)}$ gives rise  to precisely the single-metric beta-function  $\beta_{\Kk}^{\sm}$, obtained in \cite{mr} if instead we make the identification   $\aDz=\eta^{(0)}$ and $\KkD=\Kk^{(0)}$.


\subsection{Optimized cutoff shape function}

The `optimized' shape function \cite{litimPRL} is given by $R^{(0)}(z)=(1-z)\theta(1-z)$ and allows for an explicit evaluation of the treshold-functions:
\begin{align}
 \ThrfA{p}{n}{w}= \frac{1}{\Gamma(n+1)}\big(1+w\big)^{-p}\,, \qquad \text{ and }\qquad \ThrfB{p}{n}{w}= \frac{1}{\Gamma(n+2)}\big(1+w\big)^{-p}
\label{eqn:appx_BetaD_020}
\end{align}
In the following we list the beta-functions in general spacetime dimensions $d$ within this optimized scheme.

\paragraph{The anomalous dimension related to $\tg^{\dyn}$:}
The key ingredients for the anomalous dimension $\aDz$ in the $\dyn$-sector, eq. \eqref{eqn:appx_BetaD_003}, are the functions $\kAD(\KkD;d)$ and $\kBD(\KkD;d)$ given by \eqref{eqn:appx_BetaD_004} and \eqref{eqn:appx_BetaD_005}, respectively. 
In the optimized scheme they are obtained upon inserting  \eqref{eqn:appx_BetaD_020}  into \eqref{eqn:appx_BetaD_004}:
\begin{subequations}
\begin{align}
\kAD(\KkD;d)&= \frac{ (1-2\KkD)^{-2} }{(4\pi)^{\frac{d}{2}-1}\,\Gamma(d\slash2)}\,\Big\{-\tfrac{(d+1)}{3} \big((d-6) -2\tfrac{(d-4)d}{d-2}\KkD\big)  
 \,  \nonumber \\
&\qquad\quad \phantom{\quad + (4\pi)^{-\frac{d}{2}}\epsilon\,\, }
+4\frac{(d-1)}{(d+2)} \frac{\big((6-d) + 2(d+2)\KkD\big)}{(1-2\KkD)}\rhoP 
 \, \Big\} \nonumber \\ &\quad
+\frac{8\Big(\frac{1}{3}+\frac{4}{d(d+2)}\rhoP  \Big)  }{(4\pi)^{\frac{d}{2}-1}\Gamma(d\slash2)} \rhoG 
\label{eqn:appx_BetaD_021}
\end{align}
\begin{align}
\kBD(\KkD;d)&= \frac{d(1-2\KkD)^{-2}}{(4\pi)^{\frac{d}{2}-1}\Gamma(d\slash2+1)}\,\Big\{\tfrac{(d-6)(d+1)}{6(d+2)}  
-\frac{1}{3}\tfrac{(d-4)(d+1)}{(d-2)}\, \KkD \, 
+2\tfrac{(d-6)(d-1)}{(d-2)(d +2)}   \rhoP \nonumber \\
&\qquad \phantom{\quad + (4\pi)^{-\frac{d}{2}}\epsilon\,\, }
-\frac{4(d-1)}{(d-2)(d+2)}\, \frac{ \big(\frac{(d-6)d}{(d+4)}-2 (d-4)\KkD\big)}{(1-2\KkD)} \,\rhoP
 \, \Big\} \label{eqn:appx_BetaD_022}
\end{align}
\end{subequations}
If we are not interested in distinguishing the  paramagnetic from diamagnetic terms, and ghost from metric contributions, we may set $\rhoG=1=\rhoP$. This yields the following simplified expressions:
\begin{subequations}
\begin{align}
\kAD(\KkD;d)&= \frac{ (1-2\KkD)^{-3} }{3\,(4\pi)^{\frac{d}{2}-1}\,\Gamma(d\slash2)}\,\Big\{
-\frac{(d-7)(d-6)(d-2)}{(d+2)}
+ \frac{4(d^3-11d^2+18d-6)}{(d-2)}\KkD \nonumber \\
&\phantom{ \frac{ (1-2\KkD)^{-3} }{3\,(4\pi)^{\frac{d}{2}-1}\,\Gamma(d\slash2)}\, \quad}
-\frac{4d(d-4)(d+1)}{(d-2)}\KkD\,\!^2
 \Big\} 
+\frac{8\Big(\frac{1}{3}+\frac{4}{d(d+2)}  \Big)  }{(4\pi)^{\frac{d}{2}-1}\Gamma(d\slash2)}
\label{eqn:appx_BetaD_021B}
\end{align}
\begin{align}
\kBD(\KkD;d)&= \frac{d(1-2\KkD)^{-3}}{3\,(4\pi)^{\frac{d}{2}-1}\Gamma(d\slash2+1)}\,\Big\{
\frac{(d-6)(d^3-9d^2+54d -56)}{2(d^2-4)(d+4)} \label{eqn:appx_BetaD_022B} \\
& \phantom{\frac{d(1-2\KkD)^{-2}}{3\,(4\pi)^{\frac{d}{2}-1}\Gamma(d\slash2+1)}}
-\frac{2(d^3-10d^2+15d -10)}{(d^2-4)}\KkD
+\frac{2(d-4)(d+1)}{(d-2)}\KkD\,\!^2
\Big\} \nonumber
\end{align}
\end{subequations}

\paragraph{The beta-function of $\KkD$:}
The beta-function of eq. \eqref{eqn:appx_BetaD_007} for the cosmological constant, in the optimized scheme, looks as follows:
\begin{align}
&\beta_{\Kk}^{\dyn}(\tg^{\dyn},\KkD;d)
= (\aDz-2)\,\KkD +   \frac{4d\,\rhoG\,\tg^{\dyn}}{(4\pi)^{\frac{d}{2}-1}\Gamma(d\slash2 +2)} 
 \label{eqn:appx_BetaD_023} \\
&\qquad +  \frac{2(d+1)\,\tg^{\dyn}}{(4\pi)^{\frac{d}{2}-1}d\Gamma(d\slash2)(1-2\KkD)^2} \,
 \Big\{ 2(d-4)\, \KkD \,\big(1-\tfrac{\aDz}{(d+2)}\big)
-\tfrac{(d-6) d}{(d+2)}\big(1-\tfrac{\aDz}{(d+4)}\big)
\Big\} \nonumber
\end{align}
The  reduction to $\rhoG=1=\rhoP$ is omitted here, since this expression contains no factor of $\rhoP$, and $\rhoG$ occurs only trivially in the second term on the RHS. So no further simplification can be achieved.

\paragraph{The anomalous dimension related to $\tg^{\background}$:}
Eq. \eqref{eqn:appx_BetaD_008} for the choice \eqref{eqn:appx_BetaD_020}  results in
\begin{align}
&\eta^{\background}(\tg^{\dyn},\KkD,\tg^{\background};d)=\label{eqn:appx_BetaD_024}\\ \nonumber
&\phantom{\eta^{\background}}=  
 \frac{(1-2\KkD)^{-1}}{(4\pi)^{\frac{d}{2}-1}\Gamma(d\slash2)}\, \Big\{
\tfrac{(d+1)}{3} (d-\aDz)
-\frac{2}{3}\tfrac{(d-4)(d+1)}{(d-2)}\, \KkD \, \frac{(d-\aDz)}{ (1-2\KkD)}
\,   \\
&\phantom{\eta^{\background}}\quad \phantom{\quad + (4\pi)^{-\frac{d}{2}}\epsilon\,\, }
+ \big(\tfrac{(d-6)(d+1)}{3} -16\tfrac{(d-1)}{(d-2)}\rhoP\big)\, \frac{(1-\aDz\slash(d+2))}{(1-2\KkD)}  \nonumber \\
%
&\phantom{\eta^{\background}}\quad \phantom{\quad + (4\pi)^{-\frac{d}{2}}\epsilon\,\, }+16\tfrac{(d-4)(d-1)}{(d-2)}\rhoP\, \KkD \, \frac{(1-\aDz\slash(d+2))}{(1-2\KkD)^2} 
 \nonumber \\
&\phantom{\eta^{\background}}\quad \phantom{\quad + (4\pi)^{-\frac{d}{2}}\epsilon\,\, } -8\tfrac{(d-6)(d-1) d}{(d^2-4)}\,  \rhoP\,
 \frac{(1-\aDz\slash(d+4))}{ (1-2\KkD)^{2}}  \Big\}\,\,\tg^{\background} \nonumber \\ &\quad \qquad
-\frac{4}{(4\pi)^{\frac{d}{2}-1}\Gamma(d\slash2)} \rhoG \,\Big\{ \frac{(d+2)}{3}     + \frac{4(d+4)}{d(d+2)}\rhoP  \Big\}\,\tg^{\background} \nonumber
\end{align}
This lengthy expression reduces slightly when we switch off the separation given by the $\rho$-parameters. For $\rhoG=1=\rhoP$ we obtain
\begin{align}
&\eta^{\background}(\tg^{\dyn},\KkD,\tg^{\background};d)=\label{eqn:appx_BetaD_024B}\\ \nonumber
&\phantom{\eta^{\background}}=  
 \frac{(1-2\KkD)^{-1}}{(4\pi)^{\frac{d}{2}-1}\Gamma(d\slash2)}\, \Big\{
\tfrac{(d+1)}{3} (d-\aDz)
-\frac{2}{3}\tfrac{(d-4)(d+1)}{(d-2)}\, \KkD \, \frac{(d-\aDz)}{ (1-2\KkD)}
\,   \\
&\phantom{ + (4\pi)^{-\frac{d}{2}}\epsilon\,\, }
+ \big(\tfrac{(d-6)(d+1)(d-2)-16(d-1)}{3(d-2)} \big)\, \frac{(1-\aDz\slash(d+2))}{(1-2\KkD)}  \nonumber \\
%
& \phantom{ + (4\pi)^{-\frac{d}{2}}\epsilon\,\, }+16\tfrac{(d-4)(d-1)}{(d-2)}\, \KkD \, \frac{(1-\aDz\slash(d+2))}{(1-2\KkD)^2} 
-8\tfrac{(d-6)(d-1) d}{(d^2-4)}\,  \,
 \frac{(1-\aDz\slash(d+4))}{ (1-2\KkD)^{2}}  \Big\}\,\,\tg^{\background} 
\nonumber \\ &\quad\phantom{\eta^{\background}}
-\frac{4}{(4\pi)^{\frac{d}{2}-1}\Gamma(d\slash2)} \left(\frac{d^3+4d^2+16d+48}{3d(d+2)}\right)\tg^{\background} \nonumber
\end{align}
The last term of \eqref{eqn:appx_BetaD_024B} contains the ghost contributions.

\paragraph{The beta-function of $\KkB$:}
For the cosmological constant in the $\background$-sector we only write down the result prior to setting $\rhoG=1$:
\begin{align}
 \beta_{\Kk}^{\background}(\tg^{\dyn},\KkD,\tg^{\background},\KkB;d) &= (\eta^{\background}-2)\KkB 
- \frac{8 (d+4) \rhoG }{(4\pi)^{\frac{d}{2}-1}(d+2)\Gamma(d\slash2)} \, \tg^{\background} 
\label{eqn:appx_BetaD_025} \\
&\quad
+ \frac{(d+1)(1-2\KkD)^{-2}}{(4\pi)^{\frac{d}{2}-1}\Gamma(d\slash2+1)}
\Big\{	\tfrac{2d}{(d+2)}\big((d-2) -\tfrac{(d-1)}{(d+4)}\aDz\big) \nonumber\\
&\quad \phantom{+ \frac{(d+1)(1-2\KkD)^{-2}}{(4\pi)^{\frac{d}{2}-1}\Gamma(d\slash2+1)} \quad}
-4 (d-2)\, (1-\aDz\slash(d+2))\KkD 
\Big\} \,\tg^{\background}\nonumber
\end{align}
This completes the list of beta-functions in the $\{\dyn,\background\}$-description.

\paragraph{The anomalous dimension related to $\tg^{(0)}$:}
For the above made choice of the shape function the anomalous dimension for $\tg^{(0)}$ assumes the form
\begin{align}
 \eta^{(0)}&=
+ \frac{2\, (1-2\KkD)^{-1}}{(4\pi)^{\frac{d}{2}-1}\Gamma(d\slash2)}\, \Big\{\frac{(d+1)}{6 } (d-\aDz)
- 2\frac{(d-1)}{(d+2)}\frac{((d+2)-\aDz)}{(1-2\KkD)} \rhoP\Big\}\tg^{(0)} \nonumber \\&\quad
-\frac{4 }{(4\pi)^{\frac{d}{2}-1}\Gamma(d\slash2)} \rhoG \Big\{  \frac{d}{3}      +\frac{4}{d}\rhoP \Big\}\tg^{(0)}
\label{eqn:appx_BetaD_026}
\end{align}
Setting $\rhoG=1=\rhoP$ this further simplifies to
\begin{align}
 \eta^{(0)}&=
 \frac{4\, }{(4\pi)^{\frac{d}{2}-1}\Gamma(d\slash2)}\, \Big\{\frac{(d+1)}{12 } \frac{(d-\aDz)}{(1-2\KkD)}
- \frac{(d-1)}{(d+2)}\frac{((d+2)-\aDz)}{(1-2\KkD)^2}
 -\Big(\frac{d^2+12}{3d} \Big) \Big\}\tg^{(0)}
\label{eqn:appx_BetaD_026B}
\end{align}
which can also be derived using the identity $\eta^{(0)}\slash \tg^{(0)} = \eta^{\background}\slash \tg^{\background}+ \aDz\slash \tg^{\dyn}$ directly.

\paragraph{The beta-function of $\Kk^{(0)}$:}
For the scale-dependence of the cosmological constant $\Kk^{(0)}$ we have to consult eq. \eqref{eqn:appx_BetaD_010} which yields
\begin{align}
\beta_{\Kk}^{(0)}
&= (\eta^{(0)}-2)\Kk^{(0)}+  \tg^{(0)}\,  \frac{\tfrac{d}{2}(d+1)}{(4\pi)^{\frac{d}{2}-1}\Gamma(d\slash2+2)}\frac{\big((d+2)-\aDz\big)}{(1-2\KkD)}   -\tg^{(0)}\frac{4d\,\rhoG}{(4\pi)^{\frac{d}{2}-1}\Gamma(d\slash2+1)}   
\label{eqn:appx_BetaD_027}
\end{align}
Omitting the separation given by the $\rho$-parameters we find
\begin{align}
\beta_{\Kk}^{(0)}
&= (\eta^{(0)}-2)\Kk^{(0)}+  \tg^{(0)}\,  \frac{\tfrac{d}{2}\big((d-3+8\KkD)(d+2)-(d+1)\aDz\big)}{(4\pi)^{\frac{d}{2}-1}\Gamma(d\slash2+2)(1-2\KkD)} 
\label{eqn:appx_BetaD_028}
\end{align}

The higher level beta-functions are described by the $\dyn$ couplings given in eqs. \eqref{eqn:appx_BetaD_021}, \eqref{eqn:appx_BetaD_022} and \eqref{eqn:appx_BetaD_023}.

											    


\subsection{The semiclassical approximation} \label{sec:AppSemiclassical}
In this appendix we present an explicit solution to the RG equations which is valid approximately provided $\tg_k^{\cix}\ll1$ and $\KkD_k\ll1$.
(The magnitude of $\KkB_k$ plays no role for the validity.) We refer to it as the semiclassical approximation since in an $(k\slash m_{\text{Pl}})$-expansion it describes the leading deviations from the strictly classical behavior with exactly constant dimensionful couplings $G_k^{\cix}\equiv G_0^{\cix}$ and $\Kkbar_k^{\cix}\equiv\Kkbar_0^{\cix}$, respectively.

\noindent {\bf Newton constants.} The $k$-dependence of the Newton constants is covered by the corresponding anomalous dimension: $k\partial_k G_k^{\cix}= \eta^{\cix}\, G_k^{\cix}$ for $\cix\in\{\dyn,\,\background,\,(0)\}$. 
If $\tg_k^{\cix}\ll1$ we may expand in $\tg_k^{\cix}$ and retain the linear term only. Returning to dimensionful variables all  anomalous dimensions have the following form then:
\begin{align}
\eta^{\cix}= B_1^{\cix}(\KkbartD_{k}\slash k^2)\,G_k^{\cix} 
\label{eqn:appx_BetaD_030}
\end{align}
The  solution to the ensuing RG equation with initial conditions posed at $k=0$ is given by
\begin{align}
\frac{1}{G_k^{\cix}}&= \frac{1}{G_{0}^{\cix}} - \int_{0}^k\md k^{\prime}\, B_1^{\cix}(\KkbartD_{k}\slash k^{\prime\, 2})\,k^{\prime\,(d-1)}
\label{eqn:appx_BetaD_031}
\end{align}
The next step in the approximation consists in expanding the integrand of \eqref{eqn:appx_BetaD_031} in small values of $\KkD_k\equiv \Kkbar_k^{\dyn}\slash k^2$. In the lowest order  we have $B_1^{\cix}(\KkbartD_{k}\slash k^{2})=B_1^{\cix}(0)+ \Order{\KkbartD_{k}\slash k^2}$ for which the integration can be performed easily. Thus to leading order for $\tg_k^{\cix}\ll 1$ and $\KkD_k\ll1$:
\begin{align}
 G_k^{\cix}&=  \frac{G_0^{\cix}}{1+\omega_d^{\cix}\, G_0^{\cix}\,k^{d-2}} 
 = G_0^{\cix}\big[ 1-\omega_d^{\cix}\,G_0^{\cix}\, k^{d-2}\big]
\label{eqn:appx_BetaD_032}
\end{align}
Here we have defined the coefficients $\omega_d^{\cix}\equiv-B_1^{\cix}(0)\slash(d-2)$ which are analogous to those in \cite{mr} and its later generalizations. 

For the $\dyn$ Newton coupling the coefficient $\omega_d^{\dyn}$ follows  from  \eqref{eqn:appx_BetaD_004}:
\begin{align}
\omega_d^{\dyn}&=
\frac{1}{(4\pi)^{\frac{d}{2}-1}}\,\Big\{\tfrac{(d^4-7d^3+4d^2+12d)-8d(d-2)\rhoG}{6(d-2)^2}  \, \ThrfA{2}{(d\slash 2)}{0}
 -\tfrac{2d\big((d-6)(d-1)d + 8\rhoG \big)}{(d-2)^2} \rhoP  \ThrfA{3}{(d+2)\slash2}{0} 
 \Big\}\,, \nonumber \\
\omega_d^{\dyn}&= \frac{1}{(4\pi)^{\frac{d}{2}-1}}\,\Big\{\tfrac{(d^3+7d^2-74d-48)}{6(d-2)}  \, \ThrfA{2}{(d\slash 2)}{0}
 -\tfrac{d(2d^2-10d-8)}{(d-2)}  \ThrfA{3}{(d+2)\slash2}{0} 
 \Big\}
\label{eqn:appx_BetaD_033}
\end{align}
The first equation contains the separation between contributions of different origin, the second relaxes this division and sets $\rhoG=1=\rhoP$.
For the $\background$ Newton constant, we find  from eq. \eqref{eqn:appx_BetaD_007}, with and without the $\rho$-factors, respectively:
\begin{align}
\omega_d^{\background}
&=    \frac{1}{(4\pi)^{\frac{d}{2}-1}}\, \Big\{-\tfrac{ ( d(d+1) - 4d \rhoG)}{3(d-2)} \,\ThrfA{1}{(d-2)\slash 2}{0}
+\tfrac{2 d ((d-6) (d-1) d+8 \rhoG) \rhoP}{(d-2)^2}  \ThrfA{3}{(d+2)\slash2}{0}
 \nonumber \\
&\quad \phantom{ \frac{1}{(4\pi)^{\frac{d}{2}-1}}\, \Big\{-}
-\tfrac{(d-2) d ((d-5) d-6-8\rhoG)-48 ((d-1)+(d-4) \rhoG) \rhoP}{6 (d-2)^2} \, \ThrfA{2}{(d\slash 2)}{0} 
 \Big\}  \label{eqn:appx_BetaD_034} \\
\omega_d^{\background}&=    \frac{1}{(4\pi)^{\frac{d}{2}-1}}\, \Big\{-\tfrac{d(d-3)}{3(d-2)} \,\ThrfA{1}{(d-2)\slash 2}{0}
-\tfrac{(d^3-7d^2+74d+48)}{6(d-2)}  \, \ThrfA{2}{(d\slash 2)}{0}
+\tfrac{d(2d^2-10d-8)}{(d-2)}  \ThrfA{3}{(d+2)\slash2}{0} 
 \Big\}\nonumber
\end{align}

In the level description the $\omega$-coefficient  is determined by eq. \eqref{eqn:appx_BetaD_009}:
\begin{align}
\omega_d^{(0)}&=
  \tfrac{2}{(4\pi)^{\frac{d}{2}-1}}\, \Big\{-\frac{d (d+1 - 4 \rhoG) }{3(d-2)} \,\ThrfA{1}{(d-2)\slash 2}{0}
+\tfrac{2 (d(d-1)+4 \rhoG)}{(d-2)}\rhoP\, \ThrfA{2}{(d\slash 2)}{0}\Big\}  \nonumber \\
\omega_d^{(0)}&=
  \tfrac{2}{(4\pi)^{\frac{d}{2}-1}}\, \Big\{-\frac{d(d-3)}{6(d-2)} \,\ThrfA{1}{(d-2)\slash 2}{0}
+\tfrac{(d^2-d+4)}{(d-2)}\, \ThrfA{2}{(d\slash 2)}{0}\Big\} \label{eqn:appx_BetaD_035}
\end{align}
In the last equation we set again  $\rhoG=1=\rhoP$ to simplify the result.

\noindent {\bf Cosmological constants.}
The RG equations for all cosmological constants $\Kk_k^{\cix}$ occuring in the present  truncation have the  structure
\begin{align}
 \partial_t \Kk^{\cix}_k=\big(\eta^{\cix}-2\big)\Kk^{\cix}_k + A^{\cix}(\Kk_k^{\dyn},\tg_k^{\dyn})\,\tg^{\cix}_k\label{eqn:appx_BetaD_036}
\end{align}
where $ A^{\cix}(\Kk_k^{\dyn},\tg_k^{\dyn})
\equiv A_1^{\cix}(\Kk_k^{\cix})-\tfrac{1}{2}\aDz\, A_2^{\cix}(\Kk_k^{\cix})$. 
Here we  define $A_1^{\cix}\equiv  A^{\cix}(\Kk_k^{\dyn})\big|_{\aDz=0}$ to be the  contribution that remains after omitting $\aDz$ on the RHS of the flow equations. The dimensionful analog of eq. \eqref{eqn:appx_BetaD_036} can be solved formally  to yield
\begin{align}
 \frac{\Kkbar^{\cix}_k}{G^{\cix}_k}&=\frac{\Kkbar^{\cix}_0}{G^{\cix}_0}+ \int_0^{k}\md k^{\prime} A^{\cix}(\Kk_{k^{\prime}}^{\dyn},\tg_{k^{\prime}}^{\dyn})\,k^{\prime\,d-1}  \label{eqn:appx_BetaD_037}
\end{align}
For $\KkD_k\ll1$ we expand  $A_1^{\cix}(\KkbartD_{k}\slash k^2)=A_1^{\cix}(0)+\Order{\KkbartD_{k}\slash k^2}$  and   solve the integral  to leading order. Furthermore, we also insert the approximate result for Newtons coupling \eqref{eqn:appx_BetaD_032} into eq. \eqref{eqn:appx_BetaD_037}. We finally obtain
\begin{align}
 \Kkbar_k^{\cix}&= \frac{G_k^{\cix}}{G_0^{\cix}} \Big( \Kkbar_0^{\cix} + \nu_d^{\cix}\,G_0^{\cix}\,k^d+ \Order{\KkbartD_{0}\slash k^2} \Big)
\nonumber \\
&= \Kkbar_0^{\cix} + \nu_d^{\cix}\,G_0^{\cix}\,k^d+ \Order{\KkbartD_{0}\slash k^2,\, G_0^{\cix\, 2} k^{d-2}} \label{eqn:appx_BetaD_038}
\end{align}
whereby we defined the coefficients $ \nu^{\cix}_d \equiv A_1^{\cix}(0)\slash d$, again following the conventions used in earlier publications \cite{cosmo1}.

For the $\dyn$ cosmological constant $\nu_d^{\cix}$ can be deduced from eq. \eqref{eqn:appx_BetaD_006}:
\begin{align}
\nu_d^{\dyn} &= -\frac{1}{(4\pi)^{\frac{d}{2}-1}} \,
 \Big\{ 
\tfrac{(d-6)}{2} (d+1) -4 \rhoG
\Big\} \ThrfA{2}{(d+2)\slash 2}{0}\,, \nonumber\\
\nu_d^{\dyn} &=
-\frac{(d-7)(d+2)}{2(4\pi)^{\frac{d}{2}-1}} \,
 \ThrfA{2}{(d+2)\slash 2}{0}
\label{eqn:appx_BetaD_039}
\end{align}
For the $\background$ cosmological coupling we find from eq. \eqref{eqn:appx_BetaD_008}
\begin{align}
\nu_d^{\background} &= \frac{1}{(4\pi)^{\frac{d}{2}-1}}
 \Big\{ 
\big((d+1)-4\rhoG\big)\ThrfA {1}{(d\slash 2)}{0} 
-\big(
\tfrac{(d-6)}{2} (d+1) -4 \rhoG\big)\ThrfA{2}{(d+2)\slash 2}{0}
\Big\}\,, \nonumber \\
\nu_d^{\background}&= \frac{1}{(4\pi)^{\frac{d}{2}-1}}
 \Big\{ 
(d-3)\ThrfA {1}{(d\slash 2)}{0} 
-
\tfrac{(d-7)(d+2)}{2}\ThrfA{2}{(d+2)\slash 2}{0}
\Big\} 
\label{eqn:appx_BetaD_040}
\end{align}
and  in the level-description we obtain from eq. \eqref{eqn:appx_BetaD_010}:
\begin{align}
\nu_d^{(0)} &= \frac{1}{(4\pi)^{\frac{d}{2}-1}}
\big\{
(d+1)-4\rhoG\big\} \ThrfA {1}{(d\slash 2)}{0} \,, \nonumber \\
\nu_d^{(0)} &= \frac{(d-3)}{(4\pi)^{\frac{d}{2}-1}}
\, \ThrfA {1}{(d\slash 2)}{0} 
\label{eqn:appx_BetaD_041}
\end{align}
As always, the first equation of the above pairs retains arbitrary $\rho$'s, while the second is the true final result with $\rhoG=1=\rhoP$.

\subsection[Expanded beta-functions in \texorpdfstring{$d=2+\epsilon$}{d = 2+epsilon}]{Expanded beta-functions in $\bm{d=2+\epsilon}$}
Besides the most relevant spacetime  dimension $d=4$, it is instructive to consider the case of $d=2+\epsilon$. 
Here, we can study the dynamical effects that add to the trivial topological 2-dimensional theory by `turning on'  $\epsilon$.
The results we obtain give rise to certain universal terms that can be ascribed to the universal value $\ThrfA{n+1}{n}{0}=1\slash \Gamma(n+1)$ which is obtained for any shape function $R^{(0)}$. 
In the sequel we display the beta-functions obtained by inserting $d=2+\epsilon$ into the general results and expanding in $\epsilon$, thereby discarding contributions of order $\epsilon$ and higher.
If the limit $\epsilon\rightarrow0$ is finite, the result is a single term of order $\epsilon^0$, if not, there will be additional pole terms in $1\slash\epsilon$.

\paragraph{The anomalous dimension related to $\tg^{\dyn}$:}
Instead of giving the full expression of the anomalous dimension $\aDz$, it is  more useful for further investigations to present the numerator and denominator functions $B_{1,2}^{\dyn}$ expanded in $\epsilon$ separately. We obtain for the numerator  in eq. \eqref{eqn:appx_BetaD_003}:
\begin{align}
\kAD(\KkD)&=
-\frac{8}{\epsilon}\Big\{ 
2 \ThrfA{3}{2}{-2 \KkD}
-2 \rhoP (\ThrfA{2}{1}{-2 \KkD} 
 +\KkD \big( \ThrfA{2}{0}{-2 \KkD }-2   \ThrfA{3}{1}{-2 \KkD } \big)
\Big\}           \nonumber \\
&\quad 
-\frac{4}{3}\Big\{
-2\rhoG(1+3\rhoP) - \KkD(-2+\log(64\pi^3))\ThrfA{2}{0}{-2\KkD} \label{eqn:appx_BetaD2_001}  \\
&\quad \qquad \quad
+3\big(-1 + \rhoP(-5 +\log(16\pi^2))\big)\ThrfA{2}{1}{-2\KkD} \nonumber \\
&\quad \qquad \quad
+ 3 \Big[4\KkD \rhoP (-2 +\log(4\pi))\ThrfA{3}{1}{-2\KkD} 
+\KkD \ThrfAn{2}{0}{-2\KkD}{1}
 \nonumber \\
&\quad \qquad \qquad \quad
 -2 \rhoP \big((-7+\log(16\pi^2))\ThrfA{3}{2}{-2\KkD}
-\rhoG \ThrfAn{2}{1}{0}{1} 
\nonumber \\
&\quad \qquad  \qquad \quad
\qquad \quad
+2\KkD \ThrfAn{3}{1}{-2\KkD}{1} +2 \rhoG \ThrfAn{3}{2}{0}{1} 
\nonumber \\
&\quad \qquad  \qquad \quad
\qquad \quad
+ \ThrfAn{2}{1}{-2\KkD}{1} - 2\ThrfAn{3}{2}{-2\KkD}{1}	\big)
\Big]
\Big\}   
+\Order{\epsilon} \nonumber
\end{align}
We see that $\kAD$ gives rise to a Laurent series with a singular part $\propto 1\slash \epsilon$.
The function that appears in the denominator of \eqref{eqn:appx_BetaD_003} assumes the form
\begin{align}
\kBD(\KkD)&=
\frac{4}{\epsilon}\Big\{ 
2 \ThrfA{3}{2}{-2 \KkD}
-2 \rhoP (\ThrfA{2}{1}{-2 \KkD} 
 +\KkD \big( \ThrfA{2}{0}{-2 \KkD }-2   \ThrfA{3}{1}{-2 \KkD } \big)
\Big\}           \nonumber \\
&\quad 
+\frac{2}{3}\Big\{
 - \KkD(-2+\log(64\pi^3))\ThrfA{2}{0}{-2\KkD} \label{eqn:appx_BetaD2_002}  \\
&\quad \qquad \quad
+3\big(-1 + \rhoP(-5 +\log(16\pi^2))\big)\ThrfA{2}{1}{-2\KkD} \nonumber
  \\
&\quad \qquad \quad
+ 3 \Big[4\KkD \rhoP (-2 +\log(4\pi))\ThrfA{3}{1}{-2\KkD} 
+\KkD \ThrfAn{2}{0}{-2\KkD}{1}
 \nonumber \\
&\quad \qquad \qquad \quad
 -2 \rhoP \big((-7+\log(16\pi^2))\ThrfA{3}{2}{-2\KkD}
+2\KkD \ThrfAn{3}{1}{-2\KkD}{1}
\nonumber \\
&\quad \qquad  \qquad \quad
\qquad \quad
+ \ThrfAn{2}{1}{-2\KkD}{1} - 2\ThrfAn{3}{2}{-2\KkD}{1}	\big)
\Big]
\Big\}   +\Order{\epsilon}
\nonumber
\end{align}
Obviously $\kBD$, too, is again divergent in $\epsilon$. However, due to the relation $\aDz=\frac{\kAD(\KkD;2+\epsilon)\,\tg^{\dyn}}{1-\kBD(\KkD;2+\epsilon) \,\tg^{\dyn}}$, the anomalous dimension is finite in the limit $\epsilon\rightarrow0$:
\begin{align}
 &\aDz(\tg^{\dyn},\KkD) \label{eqn:appx_BetaD2_003} \\
&\qquad= 2\frac{\rhoP\big( \ThrfA{2}{1}{-2\KkD} -2 \ThrfA{3}{2}{-2\KkD} +2\KkD\ThrfA{3}{1}{-2\KkD}\big)-2 \KkD \ThrfA{2}{0}{-2\KkD}}{
\rhoP\big( \ThrfB{2}{1}{-2\KkD} -2 \ThrfB{3}{2}{-2\KkD} +2\KkD\ThrfB{3}{1}{-2\KkD}\big)-2 \KkD \ThrfB{2}{0}{-2\KkD}
} + \Order{\epsilon} \nonumber
\end{align}

\paragraph{The beta-function of $\KkD$:}
The running of the $\dyn$ cosmological constant expanded in terms of $\epsilon$ is given by
\begin{align}
\beta_{\Kk}^{\dyn}(\tg^{\dyn},\KkD)
&= (\aDz-2)\,\KkD +   8\,\rhoG \,\ThrfA {2}{2}{0}\, \tg^{\dyn}
 \label{eqn:appx_BetaD2_004} \\
&\quad +  3
 \big( -4\, \KkD \,\qA{2}{1}{-2\KkD}
+4\qA{2}{ 2}{-2\KkD}  
\big)  \,\tg^{\dyn} + \Order{\epsilon}\nonumber
\end{align}
Notice that the anomalous dimension in the beta-function has to be understood as the expanded version of eq. \eqref{eqn:appx_BetaD2_003}.

\paragraph{The anomalous dimension related to $\tg^{\background}$:}
\begin{align}
 \eta^{\background}(\tg^{\dyn},\KkD,\tg^{\background}) &=
\frac{8}{\epsilon}\Big\{ 
4 \qA{3}{2}{-2 \KkD}
-2 \rhoP \ThrfA{2}{1}{-2 \KkD} 
 +\KkD \big( \qA{2}{0}{-2 \KkD }-4   \qA{3}{1}{-2 \KkD } \big)
\Big\}    \tg^{\background}       \nonumber \\
&\quad 
-\frac{1}{3}\Big\{
 4 \KkD(-2+\log(64\pi^3))\qA{2}{0}{-2\KkD}  - 6 \qA{1}{0}{-2\KkD}\nonumber \\
&\quad \qquad \quad
+12\big(1 -2 \rhoP(-3 +\log(4\pi))\big)\qA{2}{1}{-2\KkD} 
 \label{eqn:appx_BetaD2_005} \\
&\quad \qquad \quad
- 12 \Big[4\KkD \rhoP (-2 +\log(4\pi))\qA{3}{1}{-2\KkD} 
+\KkD \ThrfAn{2}{0}{-2\KkD}{1}
 \nonumber \\
&\quad \qquad \qquad \quad
 -2 \rhoP \big((-7+\log(16\pi^2))\qA{3}{2}{-2\KkD}
+2\KkD \qAn{3}{1}{-2\KkD}{1}
\nonumber \\
&\quad \qquad  \qquad \quad
\qquad \quad
+ \qAn{2}{1}{-2\KkD}{1} - 2\qAn{3}{2}{-2\KkD}{1}	\big)
\Big] \nonumber \\
&\quad \phantom{-\frac{2}{3}} \quad \,
+8 \Big(2+\rhoP\big( 6-3\qAn{2}{1}{0}{1}+6 \qAn{3}{2}{0}{1} \big)\Big)\rhoG 
\Big\}  + \Order{\epsilon} \nonumber
\end{align}
This series contains a divergent part $\propto 1\slash\epsilon$ for the anomalous dimension $\eta^{\background}$ itself. It happens to  vanish however for $\KkD=0$. 

\paragraph{The beta-function of $\KkB$:}
\begin{align}
 \beta_{\Kk}^{\background}(\tg^{\dyn},\KkD,\tg^{\background},\KkB) &= (\eta^{\background}-2)\KkB - 8 \rhoG \big(\ThrfA {1}{1}{0} +  \,\ThrfA {2}{ 2}{0}  \big)\, \tg^{\background}\label{eqn:appx_BetaD2_006}  \\
&\quad 
+3
\big(		
2\, \qA{1}{1}{-2\KkD}
+4\, \KkD \, \qA{2}{1}{-2\KkD}  
-4 \qA{2}{ 2}{-2\KkD}
\big) \,\tg^{\background}+ \Order{\epsilon}\nonumber 
\end{align}
Notice that while $\beta_{\Kk}^{\background}$ contains no explicit $1\slash\epsilon$-poles the anomalous dimension $\eta^{\background}$ to be used in \eqref{eqn:appx_BetaD2_006} does have such poles.

\paragraph{The anomalous dimension related to $\tg^{(0)}$:}
\begin{align}
 &\eta^{(0)}(\tg^{(p)},\Kk^{(p)})
\label{eqn:appx_BetaD2_007} \\
&\qquad=
2\, \big( \qA{1}{0}{-2\KkD}
- 2\rhoP\, \qA{2}{1}{-2\KkD}\big)\,\tg^{(0)} 
%
-4\rhoG \big(  \tfrac{2}{3}    +2\rhoP \big)\tg^{(0)}+ \Order{\epsilon}
\nonumber 
\end{align}

\paragraph{The beta-function of $\Kk^{(0)}$:}
\begin{align}
\beta_{\Kk}^{(0)}(\tg^{(p)},\Kk^{(p)})
&= (\eta^{(0)}-2)\Kk^{(0)}+ 6\, \qA{1}{1}{-2\KkD}  \tg^{(0)}  -8 \rhoG \ThrfA {1}{1}{0} \tg^{(0)} +\Order{\epsilon}
\end{align}
Note that in the level-description neither $\eta^{(0)}$ nor $\beta_{\Kk}^{(0)}$ has explicit $1\slash\epsilon$-poles.
\subsection[Dimension \texorpdfstring{$(d=3)$}{(d = 3)}]{Dimension $\bm{(d=3)}$} \label{app:BetaD3}
Next, we apply the results of Appendix \ref{sec:BetaD} to the special case of $d=3$. We proceed in the usual manner, starting with the $\dyn$ couplings, then consider the $\background$-sector, and finally take a look at the language description of the beta-functions.

\paragraph{The anomalous dimension related to $\tg^{\dyn}$:}
The structure of the anomalous dimension follows eq. \eqref{eqn:appx_BetaD_003}, with the numerator function
\begin{align}
\kAD(\KkD;3)&= \frac{3}{2 \sqrt{\pi}}\,\Big\{(2+12\rhoP) \, \ThrfA{2}{3\slash 2}{-2\KkD}+16\rhoP\, \KkD\, \ThrfA{3}{3\slash 2}{-2\KkD}
\, \nonumber \\
&\phantom{\quad + (4\pi)^{-\frac{d}{2}}\epsilon\,\, }-\tfrac{8}{3}\, \KkD \, \ThrfA{2}{1\slash 2}{-2\KkD}
-36   \rhoP\,
    \, \ThrfA{3}{5\slash2}{-2\KkD} 
 \, \Big\} \nonumber \\ &\quad
+\frac{2}{\sqrt{\pi}} \rhoG \,\Big\{ \big(1-4\rhoP\big) \ThrfA{2}{3\slash 2}{0} + 12 \rhoP\ThrfA{3}{5\slash 2}{0} \Big\}\,
\label{eqn:appx_BetaD3_004}
\end{align}
We see that $\kAD$ receives contributions of both gravitons and ghosts, of both dia- and para-magnetic nature.
The denominator function $\kBD$ is unaffected by the ghosts:
\begin{align}
\kBD(\KkD;3)&= \frac{3}{4\sqrt{\pi}}\,\Big\{(2+12\rhoP) \, \ThrfB{2}{3\slash 2}{-2\KkD}+16\rhoP\, \KkD\, \ThrfB{3}{3\slash 2}{-2\KkD}
\, \nonumber \\
& \phantom{\quad + (4\pi)^{-\frac{d}{2}}\epsilon\,\, }-\tfrac{8}{3}\, \KkD \, \ThrfB{2}{1\slash 2}{-2\KkD}
-36   \rhoP\,
    \ThrfB{3}{5\slash2}{-2\KkD} 
 \, \Big\} \label{eqn:appx_BetaD3_005}
\end{align}

\paragraph{The beta-function of $\KkD$:}
\begin{align}
\beta_{\Kk}^{\dyn}(\tg^{\dyn},\KkD;3)
&= (\aDz-2)\,\KkD +  \tg^{\dyn} \frac{6\,\rhoG}{\sqrt{\pi}} \,\ThrfA {2}{5\slash 2}{0} 
 \label{eqn:appx_BetaD3_006} \\
&\quad + \tg^{\dyn} \frac{1}{\sqrt{\pi}} \,
 \Big\{ -4\, \KkD \,\qA{2}{3\slash 2}{-2\KkD}
+9\qA{2}{5\slash 2}{-2\KkD}  
\Big\} \nonumber
\end{align}

\paragraph{The anomalous dimension related to $\tg^{\background}$:}
\begin{align}
\eta^{\background}(\tg^{\dyn},\KkD,\tg^{\background};3)&=
 \frac{3}{2\sqrt{\pi}}\, \Big\{
\frac{4}{3} \qA{1}{1\slash 2}{-2\KkD}
+\tfrac{8}{3}\, \KkD \, \qA{2}{1\slash 2}{-2\KkD}
- \big(2 +16\rhoP\big)\, \qA{2}{3\slash 2}{-2\KkD}
\, \nonumber  \\
&\quad \phantom{\frac{3}{2\sqrt{\pi}}\, \Big\{ }
-16\rhoP\, \KkD \, \qA{3}{3\slash 2}{-2\KkD}
+36\,  \rhoP\,
   \qA{3}{5\slash2}{-2\KkD} \, \Big\}\,\,\tg^{\background} \nonumber \\ &\quad
-\frac{2}{\sqrt{\pi}} \rhoG \,\Big\{  \ThrfA{1}{1\slash 2}{0}    + \big(1-2\rhoP\big) \ThrfA{2}{3\slash 2}{0} + 12 \rhoP\ThrfA{3}{5\slash 2}{0} \Big\}\,\tg^{\background} \label{eqn:appx_BetaD3_007}
\end{align}

\paragraph{The beta-function of $\KkB$:}
\begin{align}
 \beta_{\Kk}^{\background}(\tg^{\dyn},\KkD,\tg^{\background},\KkB;3) &= (\eta^{\background}-2)\KkB 
- \frac{6}{\sqrt{\pi}} \rhoG \Big\{\ThrfA {1}{3\slash 2}{0} +  \,\ThrfA {2}{5\slash 2}{0}  \Big\}\, \tg^{\background} \label{eqn:appx_BetaD3_008}
 \\
&\quad 
+ \frac{1}{\sqrt{\pi}}
\Big\{		
6\, \qA{1}{3\slash 2}{-2\KkD}
+4\, \KkD \, \qA{2}{3\slash 2}{-2\KkD}  
- 9  \qA{2}{5\slash 2}{-2\KkD}
\Big\} \,\tg^{\background}
\nonumber
\end{align}

\paragraph{The anomalous dimension related to $\tg^{(0)}$:}
\begin{align}
 \eta^{(0)}(\tg^{(p)},\Kk^{(p)};3)&=
+ \frac{3}{\sqrt{\pi}}\, \Big\{\tfrac{2}{3} \qA{1}{1\slash 2}{-2\KkD}
- 2\rhoP\, \qA{2}{3\slash 2}{-2\KkD}\Big\}\tg^{(0)} \nonumber \\&\quad
-\frac{2 }{\sqrt{\pi}} \rhoG \Big\{   \ThrfA{1}{1\slash 2}{0}    +2\rhoP \,\ThrfA{2}{3\slash 2}{0}\Big\}\tg^{(0)}
\end{align}

\paragraph{The beta-function of $\Kk^{(0)}$:}
\begin{align}
\beta_{\Kk}^{(0)}(\tg^{(p)},\Kk^{(p)};3)
&= (\eta^{(0)}-2)\Kk^{(0)}+  \tg^{(0)}\,  \frac{6}{\sqrt{\pi}} 
 \, \qA{1}{3\slash 2}{-2\KkD}  -\tg^{(0)}\frac{6}{\sqrt{\pi}}  \rhoG \ThrfA {1}{3\slash 2}{0} 
\end{align}
In the level-description, the RG equations for the higher couplings ($p\geq1$) involve $\eta^{(p)}=\eta^{\dyn}$ and $\beta^{(p)}_{\Kk}=\beta_{\Kk}^{\dyn}$.
\end{appendix}

\end{document}